\newif\iffigs\figstrue
\documentclass[a4paper,11pt]{article}
\usepackage{latexsym,amssymb,lscape,graphics}
\usepackage{graphicx}        
\usepackage{longtable}
\usepackage{multirow}
\usepackage{color}
\usepackage{slashed,epsfig}
\usepackage{amsfonts}

\newcommand{\e}{\textrm{e}}

\newtheorem{definizione}{Definition}[section]

\newcommand{\bd}{\begin{definizione}}
\newcommand{\ed}{\end{definizione}}

\def\IC{\relax\,\hbox{$\inbar\kern-.3em{\rm C}$}}
\def\IG{\relax\,\hbox{$\inbar\kern-.3em{\rm G}$}}
\def\IB{\relax{\rm I\kern-.18em B}}
\def\ID{\relax{\rm I\kern-.18em D}}
\def\IL{\relax{\rm I\kern-.18em L}}
\def\IF{\relax{\rm I\kern-.18em F}}
\def\IH{\relax{\rm I\kern-.18em H}}
\def\II{\relax{\rm I\kern-.17em I}}
\def\IN{\relax{\rm I\kern-.18em N}}
\def\IP{\relax{\rm I\kern-.18em P}}
\def\IQ{\relax\,\hbox{$\inbar\kern-.3em{\rm Q}$}}
\def\bfzero{\relax\,\hbox{$\inbar\kern-.3em{\rm 0}$}}
\def\IK{\relax{\rm I\kern-.18em K}}
\def\IG{\relax\,\hbox{$\inbar\kern-.3em{\rm G}$}}
 \font\cmss=cmss10 \font\cmsss=cmss10 at 7pt
\def\IR{\relax{\rm I\kern-.18em R}}
\def\ZZ{\relax\ifmmode\mathchoice
{\hbox{\cmss Z\kern-.4em Z}}{\hbox{\cmss Z\kern-.4em Z}}
{\lower.9pt\hbox{\cmsss Z\kern-.4em Z}} {\lower1.2pt\hbox{\cmsss
Z\kern-.4em Z}}\else{\cmss Z\kern-.4em Z}\fi}
\def\bfone{\relax{\rm 1\kern-.35em 1}}

\def\inbar{\vrule height1.5ex width.4pt depth0pt}
\def\bfzero{\relax{\rm I\kern-.18em 0}}
\def\bfone{\relax{\rm 1\kern-.35em 1}}
\def\twomat#1#2#3#4{\left(\begin{array}{cc}
 {#1}&{#2}\nonumber \\ {#3}&{#4}\nonumber \\
\end{array}
\right)}
\def\twovec#1#2{\left(\begin{array}{c}
{#1}\nonumber \\ {#2}\nonumber \\
\end{array}
\right)}
\def\o#1#2{{{#1}\over{#2}}}
\DeclareFontFamily{U}{rsf}{} \DeclareFontShape{U}{rsf}{m}{n}{
  <5> <6> rsfs5 <7> <8> <9> rsfs7 <10-> rsfs10}{}
\DeclareMathAlphabet\Scr{U}{rsf}{m}{n}

\def\e{\epsilon} \def\c{\gamma}
 
\def\L{\Lambda} 
\def\T{T}

\renewcommand{\arraystretch}{1.3}
\newcommand{\ft}[2]{{\textstyle\frac{#1}{#2}}}
\def\tilde{\widetilde}

\def\1bar{1\hskip -.275cm -}
\def\2bar{2\hskip -.275cm -}
\def\3bar{3\hskip -.275cm -}

\newsavebox{\uuunit}
\sbox{\uuunit}
                 {\setlength{\unitlength}{0.825em}
                      \begin{picture}(0.6,0.7)
                                      \thinlines
                                      \put(0,0){\line(1,0){0.5}}
                                      \put(0.15,0){\line(0,1){0.7}}
                                      \put(0.35,0){\line(0,1){0.8}}
                                     \multiput(0.3,0.8)(-0.04,-0.02){10}{\rule{0.5pt}{0.5pt}}
                      \end {picture}}

\makeatletter \@addtoreset{equation}{section} \makeatother

\def\bfone{\relax{\rm 1\kern-.35em 1}}

\def\bfone{\relax{\rm 1\kern-.35em 1}}
\font\cmss=cmss10 \font\cmsss=cmss10 at 7pt

\newcommand{\slal}{\mathfrak{sl}}

\begin{document}
\begin{titlepage}
\vskip 0.2cm
\begin{center}
{\Large {\bf Integrability of Supergravity Black Holes \\
and \\
new Tensor Classifiers of Regular and Nilpotent Orbits
}}\\[1cm]
{\large Pietro Fr\'e$^{a}$\footnote{Presently Prof. Fr\'e holds the office of Scientific Counselor  of the Italian Embassy in the Russian Federation.}, Alexander S. Sorin$^{b}$ and Mario Trigiante$^{c}$}
{}~\\
\quad \\
{{\em $^{a}$
 Dipartimento di Fisica Teorica, Universit\'a di Torino,}}
\\
{{\em $\&$ INFN - Sezione di Torino}}\\
{\em via P. Giuria 1, I-10125 Torino, Italy}~\quad\\
{\tt fre@to.infn.it}
{}~\\
\quad \\
{{\em $^{b}$ Bogoliubov Laboratory of Theoretical Physics,}}\\
{{\em Joint Institute for Nuclear Research,}}\\
{\em 141980 Dubna, Moscow Region, Russia}~\quad\\
{\tt sorin@theor.jinr.ru}
{}~\\
\quad \\
{{\em $^e$  Dipartimento di Fisica Politecnico di Torino,}}\\
{\em C.so Duca degli Abruzzi, 24, I-10129 Torino, Italy}~\quad\\
{\tt mario.trigiante@gmail.com}
\quad \\
\end{center}
~{}
\begin{abstract}
In this paper we apply in a systematic way a previously developed integration algorithm of the relevant Lax equation to the construction of spherical symmetric, asymptotically flat black hole solutions of  $\mathcal{N}=2$ supergravities   with symmetric Special Geometry. Our main goal is the classification of these black-holes according to the $\mathrm{H}^\star$ orbits in which the space of possible Lax operators decomposes. By  $\mathrm{H}^\star$ one denotes the isotropy group of the coset $\mathrm{U_{D=3}/H^\star}$ which appears in the time-like dimensional reduction of supergravity from $D=4$ to $D=3$ dimensions. The main result of our investigation is the construction of three universal tensors, extracted from quadratic and quartic powers of the Lax operator, that are capable of classifying both regular and nilpotent  $\mathrm{H}^\star$ orbits of Lax operators. Our tensor based classification is compared, in the case of the simple one-field model $S^3$, to the algebraic classification of nilpotent orbits and it is shown to provide a simple discriminating method. In particular we present a detailed  analysis of the $S^3$ model, constructing explicitly  its solutions and discussing the Liouville integrability of the corresponding dynamical system. By means of the Kostant-representation of a generic Lie algebra element, we were able to develop an algorithm which produces the necessary number of hamiltonians in involution required by Liouville integrability of  generic orbits. The degenerate orbits correspond to extremal black-holes and are nilpotent. We present an in depth discussion of their identification and of the construction of the corresponding supergravity solutions. We dwell on the relation between $\mathrm{H}^\star$ orbits and critical points of the geodesic potential showing that there is correspondence yet not   one-to-one. Finally we present the conjecture that our newly identified tensor classifiers are universal and able to label all  regular and nilpotent orbits in all homogeneous symmetric Special Geometries.
\end{abstract}
\end{titlepage}
\tableofcontents
\newpage
\section{Introduction}
\paragraph{Historical Background}
The topic of spherically symmetric, asymptotically flat extremal black hole solutions in
supergravity has a history of more than sixteen years. In the mid
nineties a broad interest was raised by two almost parallel
discoveries:
\begin{enumerate}
\item The attractor mechanism in BPS black-holes, where the scalar fields of the supergravity multiplets
flow to fixed values $\phi^i_{fix}$ at the event-horizon,
 independent from the boundary values at infinity
 $\phi^i_{\infty}$ and solely  determined by the quantized electromagnetic charges $\left\{p^\Lambda \, , \, q_\Sigma\right\}$ of
 all present gauge fields  \cite{ferrarakallosh,Gibbons:1996af}.
 The area of the horizon, interpreted as the black-hole entropy, is then universally given as
 $\mbox{Area}_H \propto \sqrt{\mathfrak{I}_4(p,q)}$, where $\mathfrak{I}_4(p,q)$ is the unique
 quartic symplectic invariant of the unified duality group $\mathrm{U_{D=4}}$.  \item
The first statistical interpretation of black-hole entropy.  The
horizon area of BPS supergravity black-holes can be interpreted as
$\mbox{Area}_H \, = \, \log{N_s}$ where $N_s$ denotes the number of
string theory microstates that correspond to the same classical
solution of the effective supergravity lagrangian
\cite{stromingerBH}.
\end{enumerate}
These two discoveries have a strong conceptual link pivoted around the interpretation of the entropy as the square root of the quartic symplectic invariant. Indeed the quantized charges provide the clue to construct $D$-brane configurations yielding the considered  black-hole solution and on its turn these $D$-brane constructions provide the means to single out the underlying string microstates. This is a particular instance of the general deep relation between the continuous $\mathrm{U}$-duality symmetries of supergravity and the exact discrete dualities mapping different string theories and different string vacua into each other. The group of string dualities was conjectured to be the restriction to integers $\mathrm{U(\mathbb{Z})}$ of the supergravity group \cite{dualitiessg}.
\par
In view of these perspectives the search and analysis of supergravity BPS black hole solutions was extensively pursued
in the  nineties in all versions of extended supergravity \cite{olderBHliterature}. The basic tool in these analyses was the use of the first order  \textit{Killing spinor equations} obtained by imposing that a certain fraction of the original supersymmetry should be preserved by the classical solution \cite{firstorderSUSY,Behrndt:1997ny,iomatteo,ortino}. Allied tool in this  was the use of the harmonic function construction of $p$-brane solutions of higher dimensional supergravities (see for instance \cite{harmfunpbrane} and references therein). In parallel to this study of classical supergravity solutions an extended investigation of the black-hole microstates within string theory \cite{microstateliterature} was pursued.
\par
The  bridge between the two aspects of the problem, namely the
macroscopic and the microscopic one, was constantly provided by the geometric
and algebraic structure of supergravity theories dictating the
properties of the $\mathrm{U}$-duality group and of the
supersymmetry field dependent central charges $Z^A$. In this
context the richest and most interesting case of study is that of
$\mathcal{N}=2$ supergravity where the geometric structure of the
scalar sector, i.e. \textit{Special K\"ahler Geometry} \cite{SKG,Andrianopoli:1996cm,specHomgeo},
on one side provides  a challenging mathematical framework to
formulate and investigate all the fundamental questions about
black-hole construction and properties, on the other side it
directly relates these latter to string-compactifications on
three-folds of vanishing first Chern class, i.e. Calabi-Yau
threefolds \cite{CYandBH} or their singular orbifold limits
\cite{orbifoldlimits}.
\paragraph{The second wave of interest and the \textit{fake-superpotential}.}
Renewed interest in the topics of spherically symmetric supergravity
black-holes and a new wave of extended research activities developed
in the last decade as soon as it was realized that the attractor
mechanism is not limited to the BPS black-holes but occurs also for
the non BPS ones \cite{criticalrefs}. In this context there emerged
the concept of \textit{fake-superpotential} \cite{fakeprepo,fakeprepoHJ,Lopes Cardoso:2007ky,fakeprepo2}. The first order
differential equations obtained by imposing the existence of Killing
spinors are just particular instance of a more general class of ``gradient-flow''
equations where the radial flow of the scalar fields (including the
warp-factor that defines the four-dimensional metric) is ruled by:
\begin{equation}\label{equaW}
    \frac{d \phi^i}{dr} \, = \,G^{ij}(\phi) \frac{\partial}{\partial \phi^j} \, \mathcal{W}\left( \phi\right)
\end{equation}
where $\mathcal{W}\left( \phi\right)$ is a suitable real function of the real scalar fields (\textit{fake-superpotential}). In the case of $\mathcal{N}=2$  extremal BPS black-hole this latter is given by:
\begin{equation}\label{truepotential}
    \mathcal{W}\left( \phi\right) \, \propto \,\sqrt{Z(\phi)\, {\bar Z}(\phi)}
\end{equation}
where $Z(\phi)$ denotes the complex field-dependent central charge well defined in terms of special geometry. For various instances of non BPS attractors other ad-hoc constructions of the fake-superpotential were presented in the literature \cite{fakeprepo,fakeprepoHJ,fakeprepo2}.
\paragraph{A bell of integrability}
The most relevant point in these new developments is that equation (\ref{equaW}) is reminiscent of the Hamilton-Jacobi formulation of classical mechanics (see standard textbooks like \cite{HJLi}, for a discusson of this issue also in relation to Liouville integrability). This fact was first observed and exploited in \cite{fakeprepoHJ} in the context of supergravity black holes to derive important general properties of $\mathcal{W}$ like its duality invariance. Considering the radial variable as an euclidian time, the fake prepotential plays the role of the principal Jacobi function while the set of all fields $\phi^i$ is assimilated to the coordinates of phase-space. This opens an entirely new perspective on the nature of the black-hole construction problem and rings a bell of integrability. Indeed the existence of the fake-superpotential, alias Jacobi function, is guaranteed for a system of $2n$ dynamical variables $\phi^i$ equipped with an underlying Poisson structure, namely with a Poisson bracket:
\begin{equation}\label{parentesipesce}
    \left\{ \phi^i \, , \, \phi^j\right \} \, = \, - \, \left\{ \phi^j \, , \, \phi^i\right \}
\end{equation}
if this latter is \textit{Liouville integrable}, namely if there exist $n$ hamiltonian  functions $\mathfrak{H}^\alpha(\phi)$ in involution
\begin{equation}
    \left\{ \mathfrak{H}^\alpha \, , \, \mathfrak{H}^\beta \right \} \, = \, 0 \quad \quad \forall \, \alpha,\beta\label{parentesipesce2}
\end{equation}
whose set includes the hamiltonian $\mathfrak{H}_0$ defining the field equations of the dynamical system:
\begin{equation}\label{evoluzia}
    \frac{d \phi^i}{dr} \, = \, \left\{ \mathfrak{H}_0 \, , \, \phi^i \right\}
\end{equation}
Clearly, in order for the above remarks to make sense, the crucial issue is the existence of a Poissonian structure and of a hamiltonian allowing to recast the supergravity field equations  into the form of a dynamical system.
\paragraph{Time-like reductions and the D=3 approach to supergravity black-holes}
A positive algorithmic answer to the issue raised above, namely whether  black-hole equations might be put into the form of a dynamical system came with the development of the $D=3$ approach to black-hole solutions \cite{Breitenlohner:1987dg,pioline,Gaiotto:2007ag,Bergshoeff:2008be,Bossard:2009at}.
\par
The fundamental algebraic root of this development is located in the so named $c$-map \cite{cmappa} from Special K\"ahler Manifolds of complex dimension $n$ to quaternion manifolds of real dimension $4n+4$:
\begin{equation}\label{cmappus}
    \mbox{c-map} \quad : \quad \mathcal{SK}_{n} \, \rightarrow \,  \mathcal{Q}_{(4n+4)}
\end{equation}
This latter follows from the systematic procedure of dimensional reduction from a $D=4,\mathcal{N}=2$ supergravity theory to a $D=3$ $\sigma$-model endowed with $\mathcal{N}=4$ three-dimensional supersymmetry. Naming $z^i$ the scalar fields that fill the special K\"ahler manifold $\mathcal{SK}_{n}$ and $g_{i{\bar \jmath}}$ its metric the $D=3$ $\sigma$-model which encodes all the supergravity field equations after dimensional reduction on a space-like direction admits, as target manifold, a quaternionic manifold whose $4n+4$ coordinates we name as follows:
\begin{equation}\label{finnico}
    \underbrace{\{U,a\}}_{2}\, \bigcup \,\underbrace{\{ z^i\}}_{2n} \, \bigcup\, \underbrace{\mathbf{Z} \, = \, \{ Z^\Lambda \, , \, Z_\Sigma \}}_{2n+2}
\end{equation}
and whose quaternionic metric has the following general form:
\begin{equation}\label{quatermetric}
    ds^2_{\mathcal{Q}} \, = \, \frac{1}{4} \, \left [ d{U}^2+2\,g_{i{\bar \jmath}}\,d{z}^i\,d{{\bar z}}^{\bar \jmath}
+ \e^{-2\,U}\,(d{a}+{\bf Z}^T\mathbb{C}d{{\bf
Z}})^2\,-\,2 \, e^{-U}\,d{{\bf
Z}}^T\,\mathcal{M}_4(z,{\bar z})\,d{{\bf Z}}\right ]
\end{equation}
In equation (\ref{quatermetric}), $\mathbb{C}$ denotes the $(2n+2)\times(2n+2)$  antisymmetric metric defined over the fibers of the symplectic bundle characterizing special geometry, while the \textit{negative definite}, $(2n+2)\times(2n+2)$ matrix $\mathcal{M}_4(z,{\bar z})$ is an object uniquely defined by the geometric set up of special geometry (see sect.\ref{solvapara} for details on $\mathcal{M}_4$).
\par
The brilliant discovery related with the $D=3$ approach to
supergravity black-holes consists in the following. The radial
dependence of all the relevant functions parameterizing the
supergravity solution can be viewed as the field equations of
another one-dimensional $\sigma$-model where the evolution
parameter $\tau$ is actually a monotonic function of the radial
variable $r$ and where the target manifold is a
\textit{pseudo-quaternionic manifold}
$\mathcal{Q}^\star_{(4n+4)}$ related to the quaternionic manifold
$\mathcal{Q}_{(4n+4)}$ in the following way. The coordinates of
$\mathcal{Q}^\star_{(4n+4)}$ are the same as those displayed in
eq.(\ref{finnico}). The metric of $\mathcal{Q}^\star_{(4n+4)}$
differs from that displayed in eq.(\ref{quatermetric}) only by a
crucial change of sign:
\begin{equation}\label{pseudoquatermetric}
    ds^2_{\mathcal{Q}^\star} \, = \, \frac{1}{4} \, \left [ d{U}^2+2\,g_{i{\bar \jmath}}\,d{z}^i\,d{{\bar z}}^{\bar \jmath}
+ \e^{-2\,U}\,(d{a}+{\bf Z}^T\mathbb{C}d{{\bf
Z}})^2\,+\,2 \, e^{-U}\,d{{\bf
Z}}^T\,\mathcal{M}_4(z,{\bar z})\,d{{\bf Z}}\right ]
\end{equation}
The new metric is non-euclidian and it has the following signature:
\begin{equation}\label{segnaturaqstarro}
    \mbox{sign}\,\left(ds^2_{\mathcal{Q}^\star}\right) \, = \, \left(\underbrace{+\, , \,  \dots \, , \, + \,}_{2n+2} , \underbrace{\, - \, , \, \dots \, , \, - }_{2n+2}\right )
\end{equation}
The general result quoted above is obtained by performing a dimensional reduction on a time-like direction.
\begin{description}
      \item[a] The first important consequence of the $D=3$ approach is that by means of it we have introduced a lagrangian description of our system and, consequently, through standard procedures, also a hamiltonian one with associated Poisson brackets.
      \item[b] The second important consequence is that the indefinite signature (\ref{segnaturaqstarro}) introduces a clear-cut distinction between non-extremal and extremal black-holes. As solutions of the $\sigma$-model defined by the metric
          (\ref{pseudoquatermetric}), all black-holes correspond to geodesics: The non-extremal ones   to time-like geodesics, while the extremal black-holes are associated with light-like ones. Space-like geodesics produce supergravity solutions with naked singularities \cite{Breitenlohner:1987dg}.
      \item[c] The third important consequence is the group theoretical
       interpretation of the sign change leading from the metric
      (\ref{quatermetric}) to the metric (\ref{pseudoquatermetric}) in those cases where the Special Manifold $\mathcal{SK}_n$ is a symmetric space $\mathrm{U_{D=4}/H_{D=4}}$. In those instances also the quaternionic manifold defined by the metric
      (\ref{quatermetric}) is a symmetric coset manifold:
      \begin{equation}\label{cosettusD3}
        \frac{\mathrm{U_{D=3}}}{\mathrm{H_{D=3}}}
      \end{equation}
      where $\mathrm{H_{D=3}}\subset \mathrm{U_{D=3}}$ is the \textit{maximal compact subgroup} of the $\mathrm{U}$-duality group, in three dimensions $\mathrm{U_{D=3}}$.
      The change of sign in the metric (\ref{segnaturaqstarro}) simply turns the coset (\ref{cosettusD3}) into a new one:
      \begin{equation}\label{cosettusD3bis}
        \frac{\mathrm{U_{D=3}}}{\mathrm{H_{D=3}^\star}}
      \end{equation}
      where $\mathrm{H_{D=3}}^\star \subset \mathrm{U_{D=3}}$ is another \textit{non-compact maximal subgroup} of the $\mathrm{U}$-duality group whose Lie algebra $\mathbb{H}^\star$ happens to be a different real form of the complexification of the Lie algebra $\mathbb{H}$ of $\mathrm{H_{D=3}}$. That such a different real form always exists within $\mathrm{U_{D=3}}$ is one of the group theoretical miracles of supergravity.
    \end{description}
\subsection{The Lax pair description}
Once the problem of black-holes is reformulated in terms of geodesics within the coset manifold (\ref{cosettusD3bis}) a rich spectrum of additional mathematical techniques becomes available for its study and solution.
\par
The most relevant of these techniques is the Lax pair representation of the supergravity field equations. According to a formalism that we review in the present paper, the fundamental evolution equation takes the following form:
\begin{equation}\label{primolasso}
    \frac{d}{d\tau} \, L(\tau) \, + \, \left[ W(\tau) \, , \, L(\tau)\right] \, = \, 0
\end{equation}
where the so named Lax operator $L(\tau)$ and the connection $W(\tau)$ are  Lie algebra elements of $\mathbb{U}$ respectively lying
in the orthogonal subspace $\mathbb{K}$ and in the subalgebra $\mathbb{H}$ in relation with the decomposition:
\begin{equation}\label{finofresco}
    \mathbb{U} \, = \, \mathbb{H} \oplus \mathbb{K}
\end{equation}
As it was proven by us in \cite{sahaedio, Fre':2007hd, noiultimo,
marioetal} and \cite{Chemissany:2010zp}, both for the case of the
coset (\ref{cosettusD3}) and the coset (\ref{cosettusD3}), the Lax
pair representation (\ref{primolasso}) allows for the construction
of an explicit integration algorithm which provides the finite
form of any supergravity solution in terms of two initial
conditions, the Lax $L_0 =L(0)$ at $\tau=0$ and the solvable coset
representative $\mathbb{L}_0 \, = \, \mathbb{L}(0)$ at the same
instant. Since the evolution of the Lax operator occurs via a
similarity transformation of $L_0$ by means of a time evolving
element of the subgroup $\mathrm{H}^\star$, it follows that the
space of all possible solutions splits into disjoint subspaces
classified by the $\mathbb{H}^\star$ orbits within $\mathbb{K}$
which, in every $\mathcal{N}=2$ supergravity based on homogeneous
symmetric special geometries, is a well defined irreducible
representation of $\mathbb{H}^\star$.
\subsection{The problems addressed in this paper}
The main problem addressed in this paper is that of the
classification of the $\mathbb{H}^\star$-orbits in $\mathbb{K}$,
scanning the physical properties of the corresponding supergravity
solutions. Furthermore we devote much attention to the relation
between the classification of $\mathbb{H}^\star$-orbits and the
classification of fixed points of the so called geodesic potential
that governs the attractor mechanism. Our accessory goal is that of
illustrating the physical content of the integration algorithm that
we presented in previous papers. We do this through the very much
detailed and in depth study of one model, the simplest non trivial
instance of $\mathcal{N}=2$ supergravity coupled to just one vector
multiplet with non-vanishing Yukawa couplings: the $S^3$-model. In this case the
duality group in three-dimensions is $\mathrm{U_{D=3}}=\mathrm{G_{(2,2)}}$ and the
relevant subgroup is $\mathrm{H}^\star_{D=3} \, = \,
\mathrm{SL(2,\mathbb{R}) \times SL(2,\mathbb{R})}$. The $D=3$ analysis of the $S^3$ mode was performed in \cite{Gaiotto:2007ag} and the corresponding nilpotent orbits were studied in \cite{bruxelles}. One of our results consists in rederiving  these results by using the novel method of tensor classifiers, see below.
\par
Since we aimed at writing a paper that might be readable by members of both the supergravity/superstring community and of the community of mathematical physicists active in the field of integrable dynamical systems, we tried to explain  all the main concepts, definitions and mathematical structures used in our constructions and arguments. We provided many explicit examples which we hope might be useful not only as illustrations but also per se.
\subsection{New results derived in this paper}
\paragraph{Tensor classifiers.} The main result presented in this paper is the discovery of a certain number of \textit{tensor classifiers} of $\mathbb{H}^\star$-orbits. These are $\mathbb{H}^\star$ covariant tensors constructed out of  powers of the Lax operator $L^n$ which can either  vanish or not, depending on the chosen $\mathbb{H}^\star$-orbit for $L$ and, being symmetric matrices, are also intrinsically characterized by their rank and by their signature. This approach is meant to be  alternative to the standard classification of the nilpotent orbits based on the Konstant-Sekiguchi theorem \cite{nilorbits}. We present here a complete set of such tensor classifiers able to discriminate all the regular and nilpotent orbits of the $\mathrm{G_{(2,2)}}$-model. Although our explicit construction is limited to this case study, we advocate that it follows a general pattern and can be easily generalized to all $\mathcal{N}=2$ supergravities based on symmetric spaces.
\par
By means of our new classifiers we were able to single out, not only
the nilpotent orbits leading to extremal black-holes, but also the
diagonalizable ones leading to non-extremal black-hole solutions. As
a byproduct of our classification we prove that the equation
\begin{equation}\label{cubicequa}
    L^3 \, = \, v^2 L
\end{equation}
 $v^2$ being  the extremality parameter,
 first given in \cite{Bossard:2009at} as a necessary condition for regularity,
is indeed not a sufficient one. In fact, for non-extremal solutions
with $v^2>0$, it does not define a single orbit, but rather its
locus splits in two or more orbits, separated by the tensor
classifiers and only one of them is the true Schwarzschild orbit of
regular solutions.
\par
Similarly we show that there is no one to one correspondence
between orbits and fixed points of the geodesic potential. Each
orbit can admit one or two type of fixed points.  Yet, what seems to
be true is that for each orbit only one type of fixed points is
reached by the corresponding solution, while the other type is associated with a solution which breaks down
before reaching its targeted fixed point:  An intrinsic singularity
of the metric occurs at a finite value of the parameter $\tau$
when the scalar field is still far from its destination.
\par
Tensor classifiers also allows for a characterization of
supersymmetric solutions (BPS solutions). In fact one of the
tensors ($\mathcal{T}^{xy}$) vanishes if and only if a fraction of
supersymmetry is preserved by the solution.
\paragraph{Breaking solutions}
The above explained mechanism is what occurs with the solutions
generated by Lax operators belonging to nilpotent orbits of higher
degree of nilpotency, namely $L^n=0$, $L^{n-1} \ne 0$ for $n>3$, if $L$ is in the fundamental representation
of ${\rm U}_{D=3}\neq {\rm E}_8$, or $n>5$ if $L$ is int eh adjoint representation of ${\rm U}_{D=3}$.
In this cases the corresponding geodesic potential admits fixed
points but they are never reached since the solution breaks done
at finite values of $\tau$. On the contrary the same fixed point
sits in other orbits whose corresponding solution attains the
targeted fixed point. The explanation of this at first sight
paradoxical fact resides in that the Lax operator contains more
information than the pure electromagnetic charges which determine
the geodesic potential. In particular it contains information
about the scalar charges and two Lax operator that have the same
electromagnetic charges may differ by the values of the scalar charges.
 The latter decide whether the scalar fields will or will
not attain their target.\par The fact that regular extremal
solutions are defined by Lax operators whose degree $n$ of nilpotency
is contained within the aforementioned bounds is consistent with the arguments
given in \cite{Gaiotto:2007ag,Bossard:2009at}, though we could not
find in the literature a detailed analysis of the solutions with
$n>3$ and of their singularities.

\paragraph{Kostant decomposition and Liouville integrability}
Another result presented in this paper concerns the explicit
construction of the required number of hamiltonians in involution
that guarantee Liouville integrability of the dynamical system
described by Lax equation eq.(\ref{primolasso}). We found an
algorithm to construct such hamiltonians  that is based on the so
called Kostant normal form of Lie algebra elements. Once a Lax
operator is put into Kostant form, all its matrix elements are
constants of motion and a simple procedure based on determinants
allows us to find the rational functions of these matrix elements
that provide the required number of functionally independent
commuting hamiltonians.
\paragraph{Scaling limits}
Having classified not only the nilpotent but also the regular orbits we present an analysis of the extremality limit in terms of Lax operators. We show how sending the extremality parameter to zero defines a double scaling limit in the parameter space that characterizes a regular Lax operator, the result of which is finite and constructs a nilpotent Lax operator from a regular diagonalizable one.
\subsection{Guide to reading}
Given the length of this paper, we think that a short guide to its content might help the reader considerably.
\begin{enumerate}
  \item In section 2 we review the general set up of $\mathcal{N}=2$ supergravity, in particular we summarize for non expert readers the mathematical definition and the algebraic structure of Special K\"ahler Geometry. We also recall the general form of the decomposition of the $\mathrm{U_{D=3}}$ Lie algebra with respect to its $\mathrm{U_{D=4}}$ subalgebra.
  \item In section 3 we specialize the general concepts illustrated in the previous section to the particular case of the master example whose study is the main task of our paper, namely the so named $S^3$ model based on a single vector multiplet and on a prepotential of cubic type.
  \item In section 4 we review the general formulae that encode the fields describing a supergravity solution into a coset representative lying in the solvable Borel subgroup of $\mathrm{U_{D=3}}$.
  \item In section 5 we recall the principles of the attractor mechanism and we review in some detail the fixed point structure associated with each nilpotent orbit of the $S^3$ model.
  \item In section 6 we review the explicit integration algorithm of Lax equation
  \item Section 7 introduces the definition of $\mathrm{H}^\star$, presents the Poissonian structure defined over the Borel subalgebra and discusses the construction of the required number of Liouville involutive hamiltonians by using the Kostant normal form of Lie algebra elements.
  \item Section 8, which is the true heart of the paper, introduces  the new tensor classifiers. Then it presents a simple and general method to construct standard representatives that are abstractly upper triangular. The catch of the method is the diagonalization of the adjoint action of a new Cartan subalgebra chosen inside $\mathbb{H}^\star$. The positive root step operators with respect to this new Cartan subalgebra lie part in $\mathbb{H}^\star$ and part in $\mathbb{K}$. Selecting the subset of those in $\mathbb{K}$ and taking linear combinations thereof, we are able to construct standard representatives of each nilpotent orbit and classify the latter.
  \item Section 9 scans the explicit form of all the Black Holes constructed in all the regular and nilpotent orbits.
  \item Section 10 presents two  examples of non extremal solutions constructed with the integration algorithm: in both cases the metric is the Reissner Nordstr\"om non extremal one. Then the extremality limit is performed and the two solutions degenerate into the extremal Reissner Nordstr\"om solutions of BPS and non BPS type respectively. At the same time a double scaling limit defined on the Lax operator retrieves the standard  representative of the BPS and non BPS nilpotent orbits.
      \item In Section 11 we apply a general  construction, developed in \cite{Bergshoeff:2008be}, of representatives of the nilpotent orbits corresponding to regular extremal black holes which are characterized by the least number of independent parameters (generating solutions). We recover the same results, in terms of tensor classifiers, found in the previous analysis, in which other representatives of the same orbits were considered. Through suitable limits we find from the Lax matrices of the regular generating solutions, those of the small black holes. As a byproduct of this analysis we prove the following useful property: If $E,E'$ are shift operators in $\mathbb{K}$ corresponding to orthogonal roots, $E+E'$ and $E-E'$ lie in the same $H^\mathbb{C}$-orbit, $H^\mathbb{C}$ being the complexification of $H^\star$ (or $H$). As a consequence of this  these matrices generate geodesics with the same supersymetry properties.
  \item We end with some concluding remarks which include a comment on the mathematical analogy between the general problem addressed in the present paper and that of studying the duality orbits in $D=4$ of two-centered solutions, addressed in \cite{Ferrara:2010ug} and \cite{Andrianopoli:2011gy}.
\end{enumerate}
\section{Recalling the general set up of $\mathcal{N}=2$ supergravity}
In this paper we are specifically interested in  spherically symmetric, asymptotically flat black-hole solutions of $D=4$ supergravity.
The most relevant case, which is also that of the specific example  we plan to treat in full-fledged completeness, corresponds
to $\mathcal{N}=2$ supersymmetry which leads to scalar manifolds  endowed with \textit{special K\"ahler geometry}. Yet one
very relevant point is the following. For  $D=4$ \textit{ungauged supergravities} the bosonic
lagrangian admits a general form which we presently discuss. To a large extent the integrability properties and the
actual construction of black-hole solutions via time-like dimensional reduction to $D=3$ depend only on such general form
of the bosonic lagrangian and on the algebraic structure of its group of duality symmetries.
\subsection{$D=4$ supergravity and its duality symmetries}
\label{d4sugrasym}
The aforementioned general form of the $D=4$ supergravity lagrangian is the following:
\begin{eqnarray}
\mathcal{L}^{(4)} &=& \sqrt{|\mbox{det}\, g|}\left[\frac{R[g]}{2} - \frac{1}{4}
\partial_{\hat{\mu}}\phi^a\partial^{\hat{\mu}}\phi^b h_{ab}(\phi) \,
+ \,
\mbox{Im}\mathcal{N}_{\Lambda\Sigma}\, F_{\hat{\mu}\hat{\nu}}^\Lambda
F^{\Sigma|\hat{\mu}\hat{\nu}}\right] \nonumber\\
&&+
\frac{1}{2}\mbox{Re}\mathcal{N}_{\Lambda\Sigma}\, F_{\hat{\mu}\hat{\nu}}^\Lambda
F^{\Sigma}_{\hat{\rho}\hat{\sigma}}\epsilon^{\hat{\mu}\hat{\nu}\hat{\rho}\hat{\sigma}}\,,
\label{d4generlag}
\end{eqnarray}
where $F_{\hat{\mu}\hat{\nu}}^\Lambda\equiv (\partial_{\hat{\mu}}A^\Lambda_{\hat{\nu}}-\partial_{\hat{\nu}}A^\Lambda_{\hat{\mu}})/2$.
In principle the effective theory described by the lagrangian (\ref{d4generlag}) can be obtained by compactification on suitable internal manifolds  from $D=10$ supergravity or $11$--dimensional M-theory, however,
how we stepped down from $D=10,11$ to $D=4$ is not necessary to specify at this level. It is implicitly encoded
in the number of residual supersymmetries that we consider. If $N_Q
=32$ is maximal it means that we used toroidal compactification.
Lower values of $N_Q$ correspond to compactifications on manifolds of restricted
holonomy, Calabi Yau three-folds, for instance, or orbifolds.
\par
In eq.(\ref{d4generlag}) $\phi^a$ denotes the whole set of $n_S$ scalar fields
parametrizing the scalar manifold $ \mathcal{M}_{scalar}^{D=4}$
which, for $N_Q \ge 8$, is necessarily a coset manifold:
\begin{equation}
  \mathcal{M}_{scalar}^{D=4} \, =
  \,\frac{\mathrm{U_{D=4}}}{\mathrm{H}_{D=4}}
\label{cosettoquando}
\end{equation}
For $N_Q \le 8$ eq.(\ref{cosettoquando}) is not obligatory but it is
possible. Particularly in the $\mathcal{N}=2$ case, i.e. for $N_Q =8$, a large
variety of homogeneous special K\"ahler or quaternionic manifolds
\cite{specHomgeo} fall into the set up of the present general discussion.
The fields $\phi^a$ have $\sigma$--model interactions
dictated by the metric $h_{ab}(\phi)$ of $\mathcal{M}_{scalar}^{D=4}$. The theory includes also $n$ vector
fields $A_{\hat{\mu}}^\Lambda$ for which
\begin{equation}
  \mathcal{F}^{\pm| \Lambda}_{\hat{\mu}\hat{\nu}} \equiv \ft 12
  \left[{F}^{\Lambda}_{\hat{\mu}\hat{\nu}} \mp \, {\rm i} \, \frac{\sqrt{|\mbox{det}\, g|}}{2}
  \epsilon_{\hat{\mu}\hat{\nu}\hat{\rho}\hat{\sigma}} \, F^{\Lambda|\hat{\rho}\hat{\sigma}} \right]
\label{Fpiumeno}
\end{equation}
denote the self-dual (respectively antiself-dual) parts of the field-strengths. As displayed in
eq.(\ref{d4generlag}) they are non minimally coupled to the scalars via the symmetric complex matrix
\begin{equation}
  \mathcal{N}_{\Lambda\Sigma}(\phi)
  ={\rm i}\, \mbox{Im}\mathcal{N}_{\Lambda\Sigma}+ \mbox{Re}\mathcal{N}_{\Lambda\Sigma}
\label{scriptaenna}
\end{equation}
which transforms projectively under $\mathrm{U_{D=4}}$. Indeed the field strengths ${F}^{
\Lambda}_{\mu\nu}$ plus their magnetic duals fill up a $2\,
n$--dimensional symplectic representation of $\mathrm{\mathbb{U}_{D=4}}$
which we call by the name of $\mathbf{W}$.
\par
Following the notations and the conventions of \cite{myparis}, we rephrase the above statements
by asserting that there is always a symplectic embedding of the duality group
$\mathrm{U}_{D=4}$,
\begin{equation}
  \mathrm{U}_{D=4} \mapsto \mathrm{Sp(2n, \mathbb{R})} \quad ; \quad
  n = n_V \, \equiv \, \mbox{$\#$ of vector fields}
\label{sympembed}
\end{equation}
so that for each element $\xi \in \mathrm{U}_{D=4}$ we have its
representation by means of a  suitable real symplectic matrix:
\begin{equation}
  \xi \mapsto \Lambda_\xi \equiv \left( \begin{array}{cc}
     A_\xi & B_\xi \\
     C_\xi & D_\xi \
  \end{array} \right)
\label{embeddusmatra}
\end{equation}
satisfying the defining relation:
\begin{equation}
  \Lambda_\xi ^T \, \left( \begin{array}{cc}
     \mathbf{0}_{n \times n}  & { \mathbf{1}}_{n \times n} \\
     -{ \mathbf{1}}_{n \times n}  & \mathbf{0}_{n \times n}  \
  \end{array} \right) \, \Lambda_\xi = \left( \begin{array}{cc}
     \mathbf{0}_{n \times n}  & { \mathbf{1}}_{n \times n} \\
     -{ \mathbf{1}}_{n \times n}  & \mathbf{0}_{n \times n}  \
  \end{array} \right)
\label{definingsympe}
\end{equation}
which implies the following relations on the $n \times n$ blocks:
\begin{eqnarray}
A^T_\xi \, C_\xi - C^T_\xi \, A_\xi & = & 0 \nonumber\\
A^T_\xi \, D_\xi - C^T_\xi \, B_\xi& = & \mathbf{1}\nonumber\\
B^T_\xi \, C_\xi - D^T_\xi \, A_\xi& = & - \mathbf{1}\nonumber\\
B^T_\xi \, D_\xi - D^T_\xi \, B_\xi & =  & 0 \label{symplerele}
\end{eqnarray}
Under an element of the duality groups the field strengths transform
as follows:
\begin{equation}
  \left(\begin{array}{c}
     \mathcal{F}^+ \\
     \mathcal{G}^+ \
  \end{array} \right)  ^\prime \, = \,\left( \begin{array}{cc}
     A_\xi & B_\xi \\
     C_\xi & D_\xi \
  \end{array} \right) \,  \left(\begin{array}{c}
     \mathcal{F}^+ \\
     \mathcal{G}^+ \
  \end{array} \right) \quad ; \quad \left(\begin{array}{c}
     \mathcal{F}^- \\
     \mathcal{G}^- \
  \end{array} \right)  ^\prime \, = \,\left( \begin{array}{cc}
     A_\xi & B_\xi \\
     C_\xi & D_\xi \
  \end{array} \right) \,  \left(\begin{array}{c}
     \mathcal{F}^- \\
     \mathcal{G}^- \
  \end{array} \right)
\label{lucoidale1}
\end{equation}
where, by their own definitions:
\begin{equation}
    \mathcal{G}^+ = \mathcal{N} \, \mathcal{F}^+ \quad ; \quad \mathcal{G}^- = \overline{\mathcal{N}} \,
    \mathcal{F}^-
\label{lucoidale2}
\end{equation}
and the complex symmetric matrix $\mathcal{N}$ transform as follows:
\begin{equation}
  \mathcal{N}~^\prime = \left(  C_\xi + D_\xi \, \mathcal{N}\right) \, \left( A_\xi + B_\xi \,\mathcal{N}\right)
  ^{-1}
\label{Ntransfa}
\end{equation}
The supergravity spherically symmetric black holes we want to
consider correspond to exact solutions of the field equations
derived from the lagrangian\footnote{ Since we are going to use
many of the results of the recent paper \cite{Chemissany:2010zp}
it is convenient to make contact with the notations of that paper
which are slightly different from the present ones, which are
consistent with those used for instance in
(\cite{Andrianopoli:1996cm}. In \cite{Chemissany:2010zp}, the
general form of the $d=4$ action is written as follows:
$$
S_4=\int\Bigl(\ft{1}{2}\star R_4 - \ft{1}{2}\,G_{rs}\star
d\phi^r\wedge d\phi^s-\ft{1}{2}\mu_{IJ}\star G^I\wedge
G^J+\ft{1}{2}\nu_{IJ}G^I\wedge G^J\Bigr)
$$
The metric of the $D=4$ scalar manifold is named $G_{ij}(\phi)$ rather than $h_{ij}$ and the two are related by a rescaling: $2\,G_{rs}  =  h_{rs}$. Similarly the field strengths of the $n_V$ vector fields are named $G^I_{\mu\nu}$ rather than $F^{\Lambda}_{\mu\nu}$ and have been endowed with a different normalization due to different conventions for $p$-form components. In the conventions adopted in the present paper a $p$-form is $A_p= A_{\mu_1\ldots\mu_p} \, dx^{\mu_1}\wedge\ldots\wedge \, dx^{\mu_p}$ while in paper \cite{Chemissany:2010zp}
the convention $A_p= \ft {1}{p!} \, A_{\mu_1\ldots\mu_p} \, dx^{\mu_1}\wedge\ldots\wedge \, d x^{\mu_p}$ was used. As a result the final correspondence is $G^I_{\mu\nu} \, = \, 2\, F^{\Lambda}_{\mu\nu}$, the indices $I$ and $\Lambda$ being identified and running on the same set of values namely $n_V$. Finally the symmetric matrices $\mu_{IJ}$ and $\nu_{IJ}$ have to be identified with the real and imaginary parts, respectively, of the complex matrix $\mathcal{N}$ according to the precise correspondence $\mu_{IJ} \, =  \,-
\mbox{Im} \mathcal{N}_{\Lambda\Sigma}$ (positive definite), $\nu_{IJ} \, = \,
\mbox{Re} \mathcal{N}_{\Lambda\Sigma}$.} in eq.(\ref{d4generlag}).
\par
The lagrangian (\ref{d4generlag}) can be dimensionally reduced to
$D=3$ over a space or the time direction and after that all vector
fields can be dualized to scalars. Such a reduction scheme is
named  the Ehlers reduction. The resulting $D=3$ theory is made
purely of scalar fields that span a new $\sigma$-model. In the
case the original $D=4$ scalar manifold is a coset manifold, as
specified in eq. (\ref{cosettoquando}), then the $D=3$  target
space is also a  coset manifold:
\begin{equation}
  \mathcal{M}_{target}^{D=3} \, =
  \,\frac{\mathrm{U_{D=3}}}{\mathrm{H}_{D=3}}
\label{cosettotre}
\end{equation}
where the (necessarily) non-compact group Lie $\mathrm{U_{D=3}}$ enlarges the original non-compact group $\mathrm{U_{D=4}}$ according to precise rules recalled in the next subsection. Under such conditions the complete integrability of the system can be established and the actual solutions can be constructed using the powerful mathematical techniques discussed in \cite{sahaedio,Fre':2007hd,Chemissany:2010zp,noiultimo,Fre:2009dg,CFS}. In this paper the mentioned techniques will be further extended in relation with the construction of conserved hamiltonians and illustrated within the chosen master example that we plan to treat in some detail.
\par
It is very important to recall that the difference between the
dimensional reduction over a space direction and that over the
time direction resides uniquely in the nature of the denominator
group ${\mathrm{H}_{D=3}}\subset \mathrm{U_{D=3}}$ mentioned in
eq.(\ref{cosettotre}). In the case of space-reductions
${\mathrm{H}_{D=3}}$ is the unique maximal compact subgroup
${\mathrm{H}_{mc}}\subset\mathrm{U_{D=3}}$ of the non-compact
numerator group. In the case of time-reductions the denominator
group is the  unique and always existing non-compact subgroup
${\mathrm{H}_{mc}^\star}\subset\mathrm{U_{D=3}}$, which lives in
$\mathrm{U_{D=3}}$ and corresponds to a different non-compact real
section of the complexification of its maximal compact subgroup
${\mathrm{H}_{mc}}$. In order to avoid misunderstandings from the
part of readers who are not supergravity specialists it is worth
recalling that the algebraic structures  we rely on are quite
specific and issue from the severe constraints of supersymmetry:
The naive conclusion that the main involved mathematical structure
is just the coset structure is too hasty and may lead to wrong
statements.  First of all the numerator group $\mathrm{U_{D=3}}$
is always a non-compact one. Furthermore it is not any non-compact
group, rather it is precisely that (or that family of groups)
predicted by the relevant supersymmetry. Secondly the denominator
group is the unique maximally compact subgroup determined  by the
specific real form of $\mathrm{U_{D=3}}$, which, on its turn is
determined by $\mathrm{U_{D=4}}$. Thirdly the fact that
$\mathrm{U_{D=3}}$ contains as subgroup a different real form of
its maximal compact subgroup is not a generic fact, rather a
peculiar property of those groups $\mathrm{U_{D=3}}$ which
supersymmetry predicts.
\par
It is also important, mostly for the benefit of those readers who are not supergravity specialists, to make a clear distinction between those aspects of the considered lagrangian model (\ref{d4generlag}) that are general and those that are specific to the case where the $D=4$ scalar manifold is a symmetric coset manifold.
\par
As we already mentioned, when supersymmetry is larger than
$\mathcal{N}=2$ the scalar manifold is always a symmetric coset
space. For $\mathcal{N}=2$, on the other hand, the prediction of
supersymmetry is that $\mathcal{M}_{scalar}^{D=4}$, spanned by the
scalar fields in the vector multiplets,  should be a special K\"
ahler manifold $\mathcal{SK}_n$, $n$ being the number of
considered vector multiplets\footnote{For simplicity we do not
envisage the inclusion of hypermultiplets which would span
additional quaternionic manifolds.}. Special K\" ahler manifolds
are a vast category of spaces that typically are not cosets and
may admit no continuous group of isometries, as it happens, for
instance, in the case  of moduli spaces of K\"ahler structure or
complex structure deformations of Calabi-Yau threefolds.
Nevertheless there exists a subclass of special K\" ahler
manifolds that are also symmetric spaces. For those manifolds the
special K\" ahler structure and the group structure coexist and
are tight together in a specific way that is mandatory to
consider. Our master example falls in that class.
\subsection{Short summary of Special K\"ahler Geometry}
Special K\"ahler geometry in special coordinates was introduced
in 1984--85 by B. de Wit et al.
and E. Cremmer et al. (see pioneering papers in \cite{SKG}), where the coupling of
$\mathcal{N}=2$ vector multiplets to $\mathcal{N}=2$ supergravity was fully determined. The
more intrinsic definition of special K\"ahler geometry in terms of
symplectic bundles is due to Strominger (1990), who
obtained it in connection with the moduli spaces of
Calabi--Yau compactifications, (see ref.s in \cite{SKG}). The coordinate-independent description
and derivation of special K\"ahler geometry in the context of $\mathcal{N}=2$
supergravity is due to Castellani, D'Auria, Ferrara
and to D'Auria, Ferrara, Fre' (1991)(see ref.s in \cite{SKG}).
\par
Let us summarize the relevant concepts and definitions
\subsubsection{Hodge--K\"ahler manifolds}
\def\mom{{M(k, \IC)}}
Consider a {\sl line bundle} ${\cal L}
{\stackrel{\pi}{\longrightarrow}} {\cal M}$ over a K\"ahler
manifold ${\cal M}$. By definition this is a holomorphic vector
bundle of rank $r=1$. For such bundles the only available Chern
class is the first:
\begin{equation}
c_1 ( {\cal L} ) \, =\, \o{i}{2\pi}
\, {\bar \partial} \,
\left ( \, h^{-1} \, \partial \, h \, \right )\, =
\, \o{i}{2\pi} \,
{\bar \partial} \,\partial \, \mbox{log} \,  h
\label{chernclass23}
\end{equation}
where the 1-component real function $h(z,{\bar z})$ is some
hermitian fibre metric on ${\cal L}$. Let $\xi (z)$ be a
holomorphic section of the line bundle ${\cal L}$: noting that
under the action of the operator ${\bar
\partial} \,\partial \, $ the term $\mbox{log} \left ({\bar \xi}({\bar z})
\, \xi (z) \right )$ yields a vanishing contribution, we conclude that
the formula in eq.(\ref{chernclass23})  for the first Chern class can be
re-expressed as follows:
\begin{equation}
c_1 ( {\cal L} ) ~=~\o{i}{2\pi} \,
{\bar \partial} \,\partial \, \mbox{log} \,\parallel \, \xi(z) \, \parallel^2
\label{chernclass24}
\end{equation}
where $\parallel \, \xi(z) \, \parallel^2 ~=~h(z,{\bar z}) \,
{\bar \xi}({\bar z}) \,
\xi (z) $ denotes
the norm of the holomorphic section $\xi (z) $.
\par
Eq.(\ref{chernclass24}) is the starting point for the definition
of Hodge--K\"ahler manifolds. A K\"ahler manifold ${\cal M}$ is a
Hodge manifold if and only if there exists a line bundle ${\cal L}
{\stackrel{\pi}{\longrightarrow}} {\cal M}$ such that its first
Chern class equals the cohomology class of the K\"ahler two-form
$\mathrm{K}$:
\begin{equation}
c_1({\cal L} )~=~\left [ \, \mathrm{K} \, \right ]
\label{chernclass25}
\end{equation}
\par
In local terms this means that there is a holomorphic section $\xi
(z)$ such that we can write
\begin{equation}
\mathrm{K}\, =\, \o{i}{2\pi} \, g_{ij^{\star}} \, dz^{i} \, \wedge
\, d{\bar z}^{j^{\star}} \, = \, \o{i}{2\pi} \, {\bar \partial}
\,\partial \, \mbox{log} \,\parallel \, \xi (z) \,
\parallel^2
\label{chernclass26}
\end{equation}
Recalling the local expression of the K\"ahler metric
in terms of the K\"ahler potential
$ g_{ij^{\star}}\, =\, {\partial}_i \, {\partial}_{j^{\star}}
{\mathcal{K}} (z,{\bar z})$,
it follows from eq.(\ref{chernclass26}) that if the
manifold ${\cal M}$ is a Hodge manifold,
then the exponential of the K\"ahler potential
can be interpreted as the metric
$h(z,{\bar z}) \, = \, \exp \left ( {\cal K} (z,{\bar z})\right )$
on an appropriate line bundle ${\cal L}$.
\par
\subsubsection{Connection on the line bundle}
On any complex line bundle ${\cal L}$ there is a canonical hermitian connection defined as :
\begin{equation}
\begin{array}{ccccccc}
{\theta}& \equiv & h^{-1} \, \partial  \, h = {\o{1}{h}}\, \partial_i h \,
dz^{i} &; &
{\bar \theta}& \equiv & h^{-1} \, {\bar \partial}  \, h = {\o{1}{h}} \,
\partial_{i^\star} h  \,
d{\bar z}^{i^\star} \cr
\end{array}
\label{canconline}
\end{equation}
For the line-bundle advocated by the Hodge-K\"ahler structure we have
\begin{equation}
\left  [ \, {\bar \partial}\,\theta \,  \right ] \, = \,
c_1({\cal L}) \, = \, [\mathrm{K}]
\label{curvc1}
\end{equation}
and since the fibre metric $h$ can be identified with the
exponential of the K\"ahler potential we obtain:
\begin{equation}
\begin{array}{ccccccc}
{\theta}& = &  \partial  \,{\cal K} =  \partial_i {\cal K}
dz^{i} & ; &
{\bar \theta}& = &   {\bar \partial}  \, {\cal K} =
\partial_{i^\star} {\cal K}
d{\bar z}^{i^\star}\cr
\end{array}
\label{curvconline}
\end{equation}
To define special K\"ahler geometry,  in addition to the afore-mentioned line--bundle
${\cal L}$ we need a flat holomorphic vector bundle ${\cal SV}
\, \longrightarrow \, {\cal M}$ whose sections play an important role in the construction of the supergravity Lagrangians. For reasons
intrinsic to such constructions the rank of the vector bundle ${\cal SV}$ must be $2\, n_V$ where $n_V$ is the total number of vector fields in the theory. If we have $n$-vector multiplets the total number of vectors is $n_V = n+1$ since, in addition to the vectors of the vector multiplets, we always have the graviphoton sitting in the graviton multiplet. On the other hand the total number of scalars is $2 n$. Suitably paired into $n$-complex fields $z^i$, these scalars span the $n$ complex dimensions of the base manifold ${\cal M}$ to the rank $2n+2$ bundle ${\cal SV}
\, \longrightarrow \, {\cal M}$.
\par
In the sequel we make extensive use of covariant derivatives with
respect to the canonical connection of the line--bundle ${\cal
L}$. Let us review its normalization. As it is well known there
exists a correspondence between line--bundles and
$\mathrm{U(1)}$--bundles. If $\mbox{exp}[f_{\alpha\beta}(z)]$ is
the transition function between two local trivializations of the
line--bundle ${\cal L} {\stackrel{\pi}{\longrightarrow}} {\cal
M}$, the transition function in the corresponding principal
$\mathrm{U(1)}$--bundle ${\cal U} \, \longrightarrow {\cal M}$ is
just $\mbox{exp}[{\rm i}{\rm Im}f_{\alpha\beta}(z)]$ and the
K\"ahler potentials in two different charts are related by: ${\cal
K}_\beta = {\cal K}_\alpha + f_{\alpha\beta}   + {\bar
{f}}_{\alpha\beta}$. At the level of connections this
correspondence is formulated by setting: $\mbox{
$\mathrm{U(1)}$--connection}   \equiv   {\cal Q} \,  = \,
\mbox{Im} \theta = -{\o{\rm i}{2}}   \left ( \theta - {\bar
\theta} \right)$. If we apply this formula to the case of the
$\mathrm{U(1)}$--bundle ${\cal U} \, \longrightarrow \, {\cal M}$
associated with the line--bundle ${\cal L}$ whose first Chern
class equals the K\"ahler class, we get:
\begin{equation}
{\cal Q}  =    -{\o{\rm i}{2}} \left ( \partial_i {\cal K}
dz^{i} -
\partial_{i^\star} {\cal K}
d{\bar z}^{i^\star} \right )
\label{u1conect}
\end{equation}
 Let now
 $\Phi (z, \bar z)$ be a section of ${\cal U}^p$.  By definition its
covariant derivative is $ \nabla \Phi = (d + i p {\cal Q}) \Phi $
or, in components,
\begin{equation}
\begin{array}{ccccccc}
\nabla_i \Phi &=&
 (\partial_i + {1\over 2} p \partial_i {\cal K}) \Phi &; &
\nabla_{i^*}\Phi &=&(\partial_{i^*}-{1\over 2} p \partial_{i^*} {\cal K})
\Phi \cr
\end{array}
\label{scrivo2}
\end{equation}
A covariantly holomorphic section of ${\cal U}$ is defined by the equation:
$ \nabla_{i^*} \Phi = 0  $.
We can easily map each  section $\Phi (z, \bar z)$
of ${\cal U}^p$
into a  section of the line--bundle ${\cal L}$ by setting:
\begin{equation}
\tilde{\Phi} = e^{-p {\cal K}/2} \Phi  \,   .
\label{mappuccia}
\end{equation}
  With this position we obtain:
\begin{equation}
\begin{array}{ccccccc}
\nabla_i    \tilde{\Phi}&    =&
(\partial_i   +   p   \partial_i  {\cal K})
\tilde{\Phi}& ; &
\nabla_{i^*}\tilde{\Phi}&=& \partial_{i^*} \tilde{\Phi}\cr
\end{array}
\end{equation}
Under the map of eq.(\ref{mappuccia}) covariantly holomorphic sections
of ${\cal U}$ flow into holomorphic sections of ${\cal L}$
and viceversa.
\subsubsection{Special K\"ahler Manifolds}
We are now ready to give the first of two equivalent definitions of special K\"ahler
manifolds:
\bd
A Hodge K\"ahler manifold is {\bf Special K\"ahler (of the local type)}
if there exists a completely symmetric holomorphic 3-index section $W_{i
j k}$ of $(T^\star{\cal M})^3 \otimes {\cal L}^2$ (and its
antiholomorphic conjugate $W_{i^* j^* k^*}$) such that the following
identity is satisfied by the Riemann tensor of the Levi--Civita
connection:
\begin{eqnarray}
\partial_{m^*}   W_{ijk}& =& 0   \quad   \partial_m  W_{i^*  j^*  k^*}
=0 \nonumber \\
\nabla_{[m}      W_{i]jk}& =&  0
\quad \nabla_{[m}W_{i^*]j^*k^*}= 0 \nonumber \\
{\cal R}_{i^*j\ell^*k}& =&  g_{\ell^*j}g_{ki^*}
+g_{\ell^*k}g_{j i^*} - e^{2 {\cal K}}
W_{i^* \ell^* s^*} W_{t k j} g^{s^*t}
\label{specialone}
\end{eqnarray}
\label{defspecial}
\ed
In the above equations $\nabla$ denotes the covariant derivative with
respect to both the Levi--Civita and the $\mathrm{U(1)}$ holomorphic connection
of eq.(\ref{u1conect}).
In the case of $W_{ijk}$, the $\mathrm{U(1)}$ weight is $p = 2$.
\par
Out of the $W_{ijk}$ we can construct covariantly holomorphic
sections of weight 2 and - 2 by setting:
\begin{equation}
C_{ijk}\,=\,W_{ijk}\,e^{  {\cal K}}  \quad ; \quad
C_{i^\star j^\star k^\star}\,=\,W_{i^\star j^\star k^\star}\,e^{  {\cal K}}
\label{specialissimo}
\end{equation}
The flat bundle mentioned in the previous subsection apparently does not appear in this definition of special geometry.
Yet it is there. It is indeed the essential ingredient in the second definition whose equivalence to the first we shall
shortly provide.
\par
Let ${\cal L} {\stackrel{\pi}{\longrightarrow}} {\cal M}$ denote
the complex line bundle whose first Chern class equals the
cohomology class of the K\"ahler form $\mathrm{K}$ of an
$n$-dimensional Hodge--K\"ahler manifold ${\cal M}$. Let ${\cal
SV} \, \longrightarrow \,{\cal M}$ denote a holomorphic flat
vector bundle of rank $2n+2$ with structural group
$\mathrm{Sp(2n+2,\mathbb{R})}$. Consider   tensor bundles of the
type ${\cal H}\,=\,{\cal SV} \otimes {\cal L}$. A typical
holomorphic section of such a bundle will be denoted by ${\Omega}$
and will have the following structure:
\begin{equation}
{\Omega} \, = \, {\twovec{{X}^\Lambda}{{F}_ \Sigma} } \quad
\Lambda,\Sigma =0,1,\dots,n
\label{ololo}
\end{equation}
By definition
the transition functions between two local trivializations
$U_i \subset {\cal M}$ and $U_j \subset {\cal M}$
of the bundle ${\cal H}$ have the following form:
\begin{equation}
{\twovec{X}{ F}}_i=e^{f_{ij}} M_{ij}{\twovec{X}{F}}_j
\end{equation}
where   $f_{ij}$ are holomorphic maps $U_i \cap U_j \, \rightarrow
\,\IC $
while $M_{ij}$ is a constant $\mathrm{Sp(2n+2,\mathbb{R})}$ matrix. For a consistent
definition of the bundle the transition functions are obviously
subject to the cocycle condition on a triple overlap:
$e^{f_{ij}+f_{jk}+f_{ki}} = 1 $ and $ M_{ij} M_{jk} M_{ki} = 1 $.
\par
Let ${\rm i}\langle\ \vert\ \rangle$ be the compatible
hermitian metric on $\cal H$
\begin{equation}
{\rm i}\langle \Omega \, \vert \, \bar \Omega \rangle \, \equiv \,-
{\rm i} \Omega^\T \twomat {0} {\bfone} {-\bfone}{0} {\bar \Omega}
\label{compati}
\end{equation}
\bd
We say that a Hodge--K\"ahler manifold ${\cal M}$
is {\bf special K\"ahler} if there exists
a bundle ${\cal H}$ of the type described above such that
for some section $\Omega \, \in \, \Gamma({\cal H},{\cal M})$
the K\"ahler two form is given by:
\begin{equation}
\mathrm{K}= \o{\rm i}{2\pi}
 \partial \bar \partial \, \mbox{\rm log} \, \left ({\rm i}\langle \Omega \,
 \vert \, \bar \Omega
\rangle \right )
\label{compati1} .
\end{equation}
\ed
From the point of view of local properties, eq.(\ref{compati1})
implies that we have an expression for the K\"ahler potential
in terms of the holomorphic section $\Omega$:
\begin{equation}
{\cal K}\,  = \,  -\mbox{log}\left ({\rm i}\langle \Omega \,
 \vert \, \bar \Omega
\rangle \right )\,
=\, -\mbox{log}\left [ {\rm i} \left ({\bar X}^\Lambda F_\Lambda -
{\bar F}_\Sigma X^\Sigma \right ) \right ]
\label{specpot}
\end{equation}
The relation between the two definitions of special manifolds is
obtained by introducing a non--holomorphic section of the bundle
${\cal H}$ according to:
\begin{equation}
V \, = \, \twovec{L^{\Lambda}}{M_\Sigma} \, \equiv \, e^{{\cal K}/2}\Omega
\,= \, e^{{\cal K}/2} \twovec{X^{\Lambda}}{F_\Sigma}
\label{covholsec}
\end{equation}
so that eq.(\ref{specpot}) becomes:
\begin{equation}
1 \, = \,  {\rm i}\langle V  \,
 \vert \, \bar V
\rangle  \,
= \,   {\rm i} \left ({\bar L}^\Lambda M_\Lambda -
{\bar M}_\Sigma L^\Sigma \right )
\label{specpotuno}
\end{equation}
Since $V$ is related to a holomorphic section by eq.(\ref{covholsec})
it immediately follows that:
\begin{equation}
\nabla_{i^\star} V \, = \, \left ( \partial_{i^\star} - {\o{1}{2}}
\partial_{i^\star}{\cal K} \right ) \, V \, = \, 0
\label{nonsabeo}
\end{equation}
On the other hand, from eq.(\ref{specpotuno}), defining:
\begin{eqnarray}
U_i  & = &  \nabla_i V  =   \left ( \partial_{i} + {\o{1}{2}}
\partial_{i}{\cal K} \right ) \, V   \equiv
\twovec{f^{\Lambda}_{i} }{h_{\Sigma\vert i}}\nonumber\\
{\bar U}_{i^\star}  & = &  \nabla_{i^\star}{\bar V}  =   \left ( \partial_{i^\star} + {\o{1}{2}}
\partial_{i^\star}{\cal K} \right ) \, {\bar V}   \equiv
\twovec{{\bar f}^{\Lambda}_{i^\star} }{{\bar h}_{\Sigma\vert i^\star}}
\label{uvector}
\end{eqnarray}
it follows that:
\begin{equation}
\nabla_i U_j  = {\rm i} C_{ijk} \, g^{k\ell^\star} \, {\bar U}_{\ell^\star}
\label{ctensor}
\end{equation}
where $\nabla_i$ denotes the covariant derivative containing both
the Levi--Civita connection on the bundle ${\cal TM}$ and the
canonical connection $\theta$ on the line bundle ${\cal L}$.
In eq.(\ref{ctensor}) the symbol $C_{ijk}$ denotes a covariantly
holomorphic (
$\nabla_{\ell^\star}C_{ijk}=0$) section of the bundle
${\cal TM}^3\otimes{\cal L}^2$ that is totally symmetric in its indices.
This tensor can be identified with the tensor of eq.(\ref{specialissimo})
appearing in eq.(\ref{specialone}).
Alternatively, the set of differential equations:
\begin {eqnarray}
&&\nabla _i V  = U_i\\
 && \nabla _i U_j = {\rm i} C_{ijk} g^{k \ell^\star} U_{\ell^\star}\\
 && \nabla _{i^\star} U_j = g_{{i^\star}j} V\\
 &&\nabla _{i^\star} V =0 \label{defaltern}
\end{eqnarray}
with V satisfying eq.s (\ref{covholsec}, \ref {specpotuno}) give yet
another definition of special geometry.
In particular it is easy to find eq.(\ref{specialone})
as integrability conditions of(\ref{defaltern})
\subsubsection{The vector kinetic matrix $\mathcal{N}_{\Lambda\Sigma}$ in special geometry}
In the bosonic supergravity action (\ref{d4generlag}) we do not see sections of any symplectic bundle over the scalar manifold
but we see the real and imaginary parts of the  matrix $\mathcal{N}_{\Lambda\Sigma}$ necessary in order to write the kinetic terms
of the vector fields. Special geometry enters precisely at this level, since it is utilized to define such a matrix.
Explicitly $\mathcal{N}_{\Lambda\Sigma}$ which, in relation with its interpretation in the case of Calabi-Yau threefolds, is named
the {\it period matrix}, is defined by means of the following relations:
\begin{equation}
{\bar M}_\Lambda = {\bar {\cal N}}_{\Lambda\Sigma}{\bar L}^\Sigma \quad ;
\quad
h_{\Sigma\vert i} = {\bar {\cal N}}_{\Lambda\Sigma} f^\Sigma_i
\label{etamedia}
\end{equation}
which can be solved introducing the two $(n+1)\times (n+1)$ vectors
\begin{equation}
f^\Lambda_I = \twovec{f^\Lambda_i}{{\bar L}^\Lambda} \quad ; \quad
h_{\Lambda \vert I} =  \twovec{h_{\Lambda \vert i}}{{\bar M}_\Lambda}
\label{nuovivec}
\end{equation}
and setting:
\begin{equation}
{\bar {\cal N}}_{\Lambda\Sigma}= h_{\Lambda \vert I} \circ \left (
f^{-1} \right )^I_{\phantom{I} \Sigma}
\label{intriscripen}
\end{equation}
As a consequence of its definition the matrix ${\cal N}$ transforms,
under diffeomorphisms of the base K\"ahler manifold, exactly as it
is requested by the rule in eq.(\ref{Ntransfa}).
Indeed this is the very reason
why the structure of special geometry has been introduced. The
existence of the symplectic bundle ${\cal H} \, \longrightarrow \,
{\cal M}$ is required in order to be able to pull--back the action
of the diffeomorphisms on the field
strengths and to construct the kinetic matrix ${\cal N}$.
\subsection{The Gaillard-Zumino formula for $\mathcal{N}_{\Lambda\Sigma}$ in the coset case}
In the case of theories based on scalar manifolds that are symmetric coset spaces, independently from the fact that they
have $\mathcal{N}=2$ or  higher supersymmetry, there is always the symplectic embedding mentioned in eq.(\ref{sympembed}).
In terms of such an embedding one can write a general formula for the period matrix $\mathcal{N}_{\Lambda\Sigma}$ which was first derived by Gaillard and Zumino in 1981 \cite{gaizum}.
\par
Let $\mathbb{L}(\phi)$ be the coset representative in the chosen parametrization of the symmetric manifold $\mathrm{U_{D=4}/H}$. By definition for each choice of the $\phi$ fields, $\mathbb{L}(\phi) \in \mathrm{U_{D=4}/H}$ is a group element and as such it maps to a
symplectic $(2n+2)\times (2n+2)$ matrix as follows:
\begin{equation}
\label{pregaia}
\mathbb{L}(\phi) \, \mapsto \,
\left(\begin{array}{c|c}
A(\phi) & B(\phi) \\
\hline
C(\phi) & D(\phi) \\
\end{array}
\right)
\end{equation}
Setting:
\begin{eqnarray}
  \mathbf{f} &=& \frac{1}{\sqrt{2}} \, \left( A(\phi) - {\rm i} \, B(f) \right )\\
  \mathbf{h} &=& \frac{1}{\sqrt{2}} \, \left( C(\phi) - {\rm i} \, D(f) \right )
\end{eqnarray}
the period matrix with the correct transformation property (\ref{Ntransfa}) is obtained by setting:
\begin{equation}\label{gaiazumaforma}
    \overline{\mathcal{N}}(\phi)\, = \, \mathbf{h} \, \mathbf{f}^{-1}
\end{equation}
As the reader can see eq.(\ref{gaiazumaforma}) has the same structure as eq.(\ref{intriscripen}) used to define $\mathcal{N}$ in the case of special geometry. This means that when we are dealing with a special K\"ahler coset manifold the two definitions should agree and we  need to construct the holomorphic section of the symplectic bundle which
provides the correspondence between the two definitions. This is precisely the task that we face in the master example we want to consider.
\subsection{General structure of the $\mathbb{U}_{D=4}$ Lie algebra}
Upon toroidal dimensional reduction from $D=4$ to $D=3$ and then
full--dualization of the vector fields, we obtain a $\sigma$-model
on a target manifold $\mathrm{U_{D=3}/H_{D=3}}$. The Lie algebra
$\mathbb{U}_{D=3}$ of the numerator group has a universal
structure  in the following sense. It  always contains, as
subalgebra, the duality algebra $\mathbb{U}_{D=4}$ of the parent
supergravity theory in $D=4$ and the $\mathrm{sl(2,\mathbb{R})_E}$
algebra which is produced by the dimensional reduction of pure
gravity. Furthermore, with respect to this subalgebra
$\mathbb{U}_{D=3}$ admits the following universal decomposition,
holding for all $\mathcal{N}$-extended supergravities:
\begin{equation}
\mbox{adj}(\mathbb{U}_{D=3}) =
\mbox{adj}(\mathbb{U}_{D=4})\oplus\mbox{adj}(\mathrm{SL(2,\mathbb{R})_E})\oplus
W_{(2,\mathbf{W})}
\label{gendecompo}
\end{equation}
where $\mathbf{W}$ is the {\bf symplectic} representation of
$\mathbb{U}_{D=4}$ to which the electric and magnetic field
strengths are assigned. Indeed the scalar fields associated with
the generators of $W_{(2,\mathbf{W})}$ are just those coming from
the vectors in $D=4$. Denoting the generators of
$\mathbb{U}_{D=4}$ by $T^a$, the generators of
$\mathrm{SL(2,\mathbb{R})_E}$ by $\mathrm{L^x}$ and denoting by
$W^{iM}$ the generators in $W_{(2,\mathbf{W})}$, the commutation
relations that correspond to the decomposition (\ref{gendecompo})
have the following general form:
\begin{eqnarray}
\nonumber && [T^a,T^b] = f^{ab}_{\phantom{ab}c} \, T^c  \\
\nonumber && [L^x,L^y] = f^{xy}_{\phantom{xy}z} \, L^z , \\
&&\nonumber [T^a,W^{iM}] = (\Lambda^a)^M_{\,\,\,N} \, W^{iN},
\\ \nonumber && [L^x, W^{iM}] = (\lambda^x)^i_{\,\, j}\, W^{jM}, \\
&&[W^{iM},W^{jN}] = \epsilon^{ij}\, (K_a)^{MN}\, T^a + \,
\mathbb{C}^{MN}\, k_x^{ij}\, L^x \label{genGD3pre}
\end{eqnarray}
where the $2 \times 2$ matrices $(\lambda^x)^i_j$, are the canonical generators of $\mathrm{SL(2,\mathbb{R})}$
in the fundamental, defining representation:
\begin{equation}
  \lambda^3 = \left(\begin{array}{cc}
     \ft 12 & 0 \\
     0 & -\ft 12 \
  \end{array} \right) \quad ; \quad \lambda^1 = \left(\begin{array}{cc}
     0 & \ft 12  \\
     \ft 12 & 0\
  \end{array} \right) \quad ; \quad \lambda^2 = \left(\begin{array}{cc}
     0 & \ft 12  \\
     -\ft 12 & 0\
  \end{array} \right)
\label{lambdax}
\end{equation}
while $\Lambda^a$ are the generators
of $\mathbb{U}_{D=4}$ in the symplectic representation $\mathbf{W}$. By
\begin{equation}
  \mathbb{C}^{MN} \equiv \left( \begin{array}{c|c}
     \mathbf{0}_{n\times n} & \mathbf{1}_{n\times n} \\
     \hline
     -\mathbf{1}_{n\times n} & \mathbf{0}_{n\times n} \
  \end{array}\right)
\label{omegamatra}
\end{equation}
we denote the antisymmetric symplectic metric in $2n$ dimensions, $n=n_V$
being the number of vector fields in $D=4$ as we have already stressed. The symplectic character
of the representation $\mathbf{W}$ is asserted by the identity:
\begin{equation}
  \Lambda^a\, \mathbb{C} + \mathbb{C}\, \left( \Lambda^a \right )^T = 0
\label{Lamsymp}
\end{equation}
The fundamental doublet representation of $\mathrm{SL(2,\mathbb{R})}$
is also symplectic and we have denoted by $\epsilon^{ij}= \left( \begin{array}{cc}
  0 & 1 \\
  -1 & 0
\end{array}\right) $ the
$2$-dimensional symplectic metric, so that:
\begin{equation}
    \lambda^x\, \epsilon + \epsilon\, \left( \lambda^x \right )^T = 0,
\label{lamsymp}
\end{equation}
The matrices
$\left(K_a\right)^{MN}=\left(K_a\right)^{NM}$ and
$\left(k_x\right)^{ij}=\left(k_y\right)^{ji}$ are just symmetric matrices
in one-to-one correspondence with the generators of $\mathbb{U}_{D=4}$ and
$\mathrm{SL(2,\mathbb{R})}$, respectively. Implementing Jacobi
identities, however we find the following relations:
\begin{eqnarray}
  && \nonumber K_a\Lambda^c +
\Lambda^c K_a = f^{bc}_{\phantom{bc}a}K_b, \quad k_x\lambda^y + \lambda^y k_x
= f^{yz}_{\phantom{yz}x}k_z,
\label{jacobrele}
\end{eqnarray}
which admit the unique solution:
\begin{equation}
    K_a = \alpha \, \mathbf{g}_{ab} \,\Lambda^b\mathbb{C}, \quad ; \quad k_x
= \beta \, \mathbf{g}_{xy} \, \lambda^y \epsilon
\label{uniquesolutK&k}
\end{equation}
where $\mathbf{g}_{ab}$, $\mathbf{g}_{xy}$ are the Cartan-Killing metrics
on the algebras $\mathbb{U}_{D=4}$ and $\mathrm{SL(2,\mathbb{R})}$, respectively
and  $\alpha$ and $\beta$ are two arbitrary constants. These latter
can always be reabsorbed into the normalization of the generators
$W^{iM}$ and correspondingly set to one. Hence the algebra
(\ref{genGD3pre}) can always be put into the following elegant form:
\begin{eqnarray}
\nonumber && [T^a,T^b] = f^{ab}_{\phantom{ab}c} \, T^c  \\
\nonumber && [L^x,L^y] = f^{xy}_{\phantom{xy}z} \, L^z , \\ &&\nonumber
[T^a,W^{iM}] = (\Lambda^a)^M_{\,\,\,N} \, W^{iN},
\\ && [L^x, W^{iM}] = (\lambda^x)^i_{\,\, j}\, W^{jM}, \\
\nonumber &&[W^{iM},W^{jN}] =
\epsilon^{ij}\, (\Lambda_a)^{MN}\, T^a + \, \mathbb{C}^{MN}\, \lambda_x^{ij}\, L^x
\label{genGD3}
\end{eqnarray}
where we have used the convention that symplectic indices are raised
and lowered with the symplectic metric, while adjoint representation
indices are raised and lowered with the Cartan-Killing metric.
\section{Introducing the example of the $S^3$ model}
The master example we consider in this paper is the simplest
possible case of vector multiplet coupling in $\mathcal{N}=2$
supergravity: we just introduce one vector multiplet. This means
that we have two vector fields in the theory and one complex
scalar field $z$. This scalar field parameterizes a
one-dimensional special K\"ahler manifold which, in our choice,
will be the  complex lower half-plane endowed with the standard
Poincar\'e metric. In other words\footnote{The special overall
normalization of the Poincar\'e metric is chosen in order to match
the general definitions of special geometry applied to the present
case.}:
\begin{equation}\label{poincaretto}
    g_{z{\bar z}} \partial^\mu z \, \partial_\mu {\bar z} \, = \,  \frac{3}{4} \, \frac{1}{(\mathrm{Im} z)^2} \, \partial^\mu z \, \partial_\mu {\bar z}
\end{equation}
is the $\sigma$-model part of the Lagrangian (\ref{d4generlag}). From the point of view of geometry the lower half-plane is the symmetric coset manifold $\mathrm{SL(2,R)/SO(2)} \, \sim \, \mathrm{SU(1,1)/U(1)}$ which admits a standard solvable parametrization as it follows. Let:
\begin{equation}\label{standSL2}
    L_0 \, = \, \ft 12 \left(
                         \begin{array}{cc}
                           1 & 0 \\
                           0 & -1 \\
                         \end{array}
                       \right) \quad ; \quad L_+ \, = \, \ft 12 \left(
                         \begin{array}{cc}
                           0 & 1 \\
                           0 & 0 \\
                         \end{array}
                       \right)\quad ; \quad L_- \, = \, \ft 12 \left(
                         \begin{array}{cc}
                           0 & 0 \\
                           1 & 0 \\
                         \end{array}
                       \right)
\end{equation}
be the standard three generators of the $\slal(2,\mathbb{R})$ Lie algebra satisfying the commutation relations $\left[L_0,L_\pm\right]=\pm L_\pm$ and $\left[ L_+,L_-\right] = 2 L_0$. The coset manifold $\mathrm{SL(2,R)/SO(2)}$ is metrically equivalent with the solvable group manifold generated by $L_0$ and $L_+$. Correspondingly we can introduce the coset representative:
\begin{equation}\label{cosetsl2}
    \mathbb{L}_4(\phi,y) \, = \, \exp[y \, L_1] \, \exp[\varphi \, L_0] \, = \, \left(
\begin{array}{ll}
 e^{\varphi /2} & e^{-\varphi /2}
   y \\
 0 & e^{-\varphi /2}
\end{array}
\right)
\end{equation}
Generic group  elements of $\mathrm{SL(2,\mathbb{R})}$ are just $2\times 2$ real matrices with determinant one:
\begin{equation}\label{sl2Rdefi}
    \mathrm{SL(2,\mathbb{R})} \, \ni \, \mathfrak{A} \, = \, \left(
                                                    \begin{array}{cc}
                                                      a & b  \\
                                                      c & d \\
                                                    \end{array}
                                                  \right) \quad ; \quad ad-bc=1
\end{equation}
and their action on the lower half-plane is defined by usual fractional linear transformations:
\begin{equation}\label{frac}
    \mathfrak{A} \quad : \quad z \, \rightarrow \, \frac{a \, z \, + \, b}{c \, z \, + d}
\end{equation}
The correspondence between the lower complex half-plane $\mathbb{C}_-$ and the solvable -parameterized coset (\ref{cosetsl2}) is easily established observing that the entire set of $\mathrm{Im} z <0$ complex numbers is just the orbit of the number ${\rm i}$ under the action of $\mathbb{L}(\phi,y)$:
\begin{equation}\label{orbitando1}
    \mathbb{L}_4(\phi,y) \quad : \quad  \,{\rm i} \, \rightarrow \, \frac{-e^{\varphi /2} \, {\rm i} \, + \, e^{-\varphi /2} \, y}{ e^{-\varphi /2}}  \, = \, y \, - \, {\rm i} e^{\varphi}
\end{equation}
This simple argument shows that we can rewrite the coset representative $\mathbb{L}(\phi,y)$ in terms of the complex scalar field $z$ as follows:
\begin{equation}\label{cosettuszetus}
    \mathbb{L}_4(z) \, = \, \left(
\begin{array}{ll}
 \sqrt{|\mbox{Im{\it z} }|} &
   \frac{\mbox{Re{\it z}} }{\sqrt{|\mbox{Im{\it z} }|}} \\
 0 & \frac{1}{\sqrt{|\mbox{Im{\it z} }|}}
\end{array}
\right)
\end{equation}
The issue of special K\"ahler geometry becomes clear at this stage. If we did not have vectors in the game, the choice of the coset metric would be sufficient and nothing more would have to be said. The point is that we still have to define the kinetic matrix of the vector and for that the symplectic bundle is necessary. On the same base manifold $\mathrm{SL(2,\mathbb{R})/SO(2)}$ we have different special structures which lead to different physical models and to different duality groups $\mathrm{U_{D=3}}$ upon reduction to $D=3$. The special structure is determined by the choice of the symplectic embedding $\mathrm{SL(2,\mathbb{R})}\, \rightarrow\, \mathrm{Sp(4,\mathbb{R})}$.
The symplectic embedding that defines our master model and which eventually leads to the duality group $\mathrm{U_{D=3}}\, = \,\mathrm{G_{2(2)}}$ is cubic and it is described in the following subsection.
\subsection{The cubic special K\"ahler structure on $\mathrm{SL(2,\mathbb{R})/SO(2)}$ }
The group $\mathrm{SL(2,\mathbb{R})}$ is also locally isomorphic to $\mathrm{SO(1,2)}$ and the fundamental representation of the first corresponds to the spin $J =\ft 12$ of the latter. The spin $J=\ft 32$ representation is obviously four-dimensional and, in the $\mathrm{SL(2,\mathbb{R})}$ language, it corresponds to a symmetric three-index tensor $t_{abc}$. Let us explicitly construct the $4\times 4$ matrices of such a representation. This is easily done by choosing an order for the four independent components of the symmetric tensor $t_{abc}$. For instance we can identify the four axes of the representation with $t_{111},t_{112},t_{122},t_{222}$. So doing, the image of the group element $\mathfrak{A}$ in the cubic symmetric tensor product representation is the following $4\times 4$ matrix:
\begin{equation}\label{cubembed}
    \mathcal{D}_3\left(\mathfrak{A}\right) \, = \, \left(
\begin{array}{llll}
 a^3 & 3 a^2 b & 3 a b^2 & b^3 \\
 a^2 c & d a^2+2 b c a & c b^2+2 a d b & b^2 d \\
 a c^2 & b c^2+2 a d c & a d^2+2 b c d & b d^2 \\
 c^3 & 3 c^2 d & 3 c d^2 & d^3
\end{array}
\right)
\end{equation}
By explicit evaluation we can easily check that:
\begin{equation}\label{presymple}
     \mathcal{D}_3^T\left(\mathfrak{A}\right) \, \widehat{\mathbb{C}}_4 \, \mathcal{D}_3\left(A\right) \, = \,\widehat{\mathbb{C}}_4 \quad \mbox{where} \quad \widehat{\mathbb{C}}_4 \, = \, \left(
\begin{array}{llll}
 0 & 0 & 0 & 1 \\
 0 & 0 & -3 & 0 \\
 0 & 3 & 0 & 0 \\
 -1 & 0 & 0 & 0
\end{array}
\right)
\end{equation}
Since $\widehat{\mathbb{C}}_4$ is antisymmetric, equation (\ref{presymple}) is already a clear indication that the triple symmetric representation defines a symplectic embedding. To make this manifest it suffices to change basis. Consider the matrix:
\begin{equation}\label{smatra}
    S \, = \, \left(
\begin{array}{llll}
 0 & 1 & 0 & 0 \\
 -\frac{1}{\sqrt{3}} & 0 & 0 & 0 \\
 0 & 0 & \frac{1}{\sqrt{3}} & 0 \\
 0 & 0 & 0 & 1
\end{array}
\right)
\end{equation}
and define:
\begin{equation}\label{pongo}
    \Lambda\left( \mathfrak{A}\right ) \, = \, S^{-1} \, D_3\left ( \mathfrak{A}\right ) S
\end{equation}
We can easily check that:
\begin{equation}\label{presymple2}
     \Lambda^T\left(\mathfrak{A}\right) \, {\mathbb{C}}_4 \, \Lambda\left(\mathfrak{A}\right) \, = \,{\mathbb{C}}_4 \quad \mbox{where} \quad {\mathbb{C}}_4 \, = \, \left(
\begin{array}{llll}
 0 & 0 & 1 & 0 \\
 0 & 0 & 0 & 1 \\
 -1 & 0 & 0 & 0 \\
 0 & -1 & 0 & 0
\end{array}
\right)
\end{equation}
So we have indeed constructed a standard symplectic embedding  $\mathrm{SL(2,\mathbb{R})}\mapsto\mathrm{Sp(4,\mathbb{R})} $ whose explicit form is the following:
\begin{eqnarray}\label{sympobeddo}
    \mathfrak{A} =\left(\begin{array}{cc}a & b  \\c & d \\ \end{array} \right)
                                                  & \mapsto &  \left(
\begin{array}{ll|ll}
 d a^2+2 b c a & -\sqrt{3} a^2 c & -c b^2-2 a d b & -\sqrt{3} b^2 d \\
 -\sqrt{3} a^2 b & a^3 & \sqrt{3} a b^2 & b^3 \\
 \hline
 -b c^2-2 a d c & \sqrt{3} a c^2 & a d^2+2 b c d & \sqrt{3} b d^2 \\
 -\sqrt{3} c^2 d & c^3 & \sqrt{3} c d^2 & d^3
\end{array}
\right) \equiv \Lambda\left( \mathfrak{A}\right)\nonumber\\
\end{eqnarray}
The $2\times 2$ blocks $A,B,C,D$ of the $4 \times 4$ symplectic matrix $\Lambda\left( \mathfrak{A}\right)$ are easily readable from eq.(\ref{sympobeddo}) so that, assuming now that the matrix $\mathfrak{A}(z)$ is the coset representative of the manifold $\mathrm{SU(1,1)/U(1)}$, we can apply the Gaillard-Zumino formula (\ref{gaiazumaforma}) and obtain the explicit form of the kinetic matrix $\mathcal{N}_{\Lambda\Sigma}$:
\begin{equation}\label{Nbarra}
    \overline{\mathcal{N}} \, = \, \left(
\begin{array}{ll}
 -\frac{2 a c-i b c+i a d+2 b d}{a^2+b^2} & -\frac{\sqrt{3} (c+i d) (a c+b
   d)}{(a-i b) (a+i b)^2} \\
 -\frac{\sqrt{3} (c+i d) (a c+b d)}{(a-i b) (a+i b)^2} & -\frac{(c+i d)^2 (2
   a c+i b c-i a d+2 b d)}{(a-i b) (a+i b)^3}
\end{array}
\right)
\end{equation}
Inserting the specific values of the entries $a,b,c,d$ corresponding to the coset representative (\ref{cosettuszetus}), we get the explicit dependence of the kinetic period matrix on the complex scalar field $z$:
\begin{equation}\label{periodoSL2}
    \overline{\mathcal{N}}_{\Lambda\Sigma}(z) \, = \, \left(
\begin{array}{ll}
 -\frac{3 z+{\bar z}}{2 z
   {\bar z}} & -\frac{\sqrt{3}
   (z+{\bar z})}{2 z {\bar z}^2}
   \\
 -\frac{\sqrt{3} (z+{\bar z})}{2
   z {\bar z}^2} & -\frac{z+3
   {\bar z}}{2 z {\bar z}^3}
\end{array}
\right)
\end{equation}
This might conclude the determination of the lagrangian of our master example, yet we have not yet seen the special K\"ahler structure induced by the cubic embedding. Let us present it.
\par
The key point is the construction of the required holomorphic symplectic section $\Omega(z)$. As usual the transformation properties of a geometrical object indicate the way to build it explicitly. For consistency we should have that:
\begin{equation}\label{omegatrasfo}
    \Omega\left( \frac{a\, z\, + \, b}{c \, z\, + \, d}\right) \, = \, f(z) \, \Lambda(\mathfrak{A}) \, \Omega(z)
\end{equation}
where $\Lambda(\mathfrak{A})$ is the symplectic  representation (\ref{sympobeddo}) of the considered $\mathrm{SL(2,\mathbb{R})}$ matrix
$\left(
         \begin{array}{cc}
           a & b \\
           c & d \\
         \end{array}
       \right)
 $ and $f(z)$ is the associated transition function for that line-bundle whose Chern-class is the K\"ahler class of the base-manifold. The identification of the symplectic fibres with the cubic symmetric representation provide the construction mechanism of $\Omega$. Consider a vector $\left(\begin{array}{c}
                                      v_1 \\
                                      v_2
                                    \end{array}
  \right)$ that transforms in the fundamental doublet representation of $\mathrm{SL(2,\mathbb{R})}$. On one hand we can identify the complex coordinate $z$ on the lower half-plane as $z=v_1/v_2$, on the other we can construct a symmetric three-index tensor taking the tensor products of three $v_i$, namely: $t_{ijk} \, = \, v_i \, v_j \, v_k$. Dividing  the resulting tensor by $v_2^3$ we obtain a four vector:
  \begin{equation}\label{presecta}
    \widehat{\Omega}(z) \, = \, \frac{1}{v_2^3}\, \left (\begin{array}{c}
                                         v_1^3 \\
                                         v_1^2 \, v_2 \\
                                         v_1 \, v_2^2 \\
                                          v_2^3
                                       \end{array}
    \right)\, = \, \left (\begin{array}{c}
                                         z^3 \\
                                         z^2 \\
                                         z \\
                                         1
                                       \end{array}
    \right)
  \end{equation}
Next, recalling the change of basis (\ref{smatra},\ref{pongo}) required to put the cubic representation into a standard symplectic form we set:
\begin{equation}\label{standasec}
    \Omega(z) \, = \, S \, \widehat{\Omega}(z) \, = \, \left(
\begin{array}{l}
 -\sqrt{3} z^2 \\
 z^3 \\
 \sqrt{3} z \\
 1
\end{array}
\right)
\end{equation}
and we can easily verify that this object transforms in the appropriate way. Indeed we obtain:
\begin{equation}\label{trasfo}
    \Omega\left( \frac{a\, z\, + \, b}{c \, z\, + \, d}\right) \, = \, (c\,z\,+\,d)^{-3} \, \Lambda(\mathfrak{A}) \, \Omega(z)
\end{equation}
The pre-factor $(c\,z\,+\,d)^{-3}$ is the correct one for the prescribed line-bundle. To see this let us first calculate the K\"ahler potential and the K\"ahler form. Inserting (\ref{standasec}) into eq.(\ref{specpot}) we get:
\begin{eqnarray}
{\cal K}&  = &  -\mbox{log}\left ({\rm i}\langle \Omega \,
 \vert \, \bar \Omega
\rangle \right )\,
=\, -\log \left(-{\rm i}
   (z-{\bar z})^3\right) \nonumber\\
   \mathrm{K} & = & \frac{\rm i}{2\pi} \, \partial \, \bar{\partial} \, \mathcal{K} \, = \, \frac{\rm i}{2\pi} \, \frac{3}{(\mbox{Im} z)^2} dz \wedge d{\bar z}
\label{specpot2}
\end{eqnarray}
This shows that the constructed symplectic bundle leads indeed to the standard Poincar\'e metric and the exponential of the K\"ahler potential transforms with the prefactor $(c\,z\,+\,d)^{3}$ whose inverse appears in eq.(\ref{trasfo}).
\par
To conclude let us show that the special geometry definition of the period matrix $\mathcal{N}$ agrees with the Gaillard-Zumino definition holding true for all symplectically embedded cosets. To this effect we calculate the necessary ingredients:
\begin{equation}\label{Ui}
    \nabla_z V(z) \, = \, \exp\left[ \frac{\mathcal{K}}{2}\right]\, \left(\partial_z \Omega(z) \, + \, \partial_z \mathcal{K} \, \Omega(z) \right) \, = \,\left(
\begin{array}{l}
 \frac{\sqrt{3} z (z+2
   {\bar z})}{(z-{\bar z})
   \sqrt{-i (z-{\bar z})^3}} \\
 -\frac{3 z^2
   {\bar z}}{(z-{\bar z})
   \sqrt{-i (z-{\bar z})^3}} \\
 -\frac{\sqrt{3} (2
   z+{\bar z})}{(z-{\bar z})
   \sqrt{-i (z-{\bar z})^3}} \\
 -\frac{3}{(z-{\bar z}) \sqrt{-i
   (z-{\bar z})^3}}
\end{array}
\right) \, \equiv \, \left(\begin{array}{c}
                             f^\Lambda_z \\
                             \hline
                             h_{\Sigma z}
                           \end{array}\right)
\end{equation}
Then according to equation (\ref{nuovivec}) we obtain:
\begin{eqnarray}
f^\Lambda_I & = & \left(
\begin{array}{ll}
 \frac{\sqrt{3} z (z+2
   {\bar z})}{(z-{\bar z})
   \sqrt{-i (z-{\bar z})^3}} &
   -\frac{2 \sqrt{6}
   {\bar z}^2}{(-i
   (z-{\bar z}))^{3/2}} \\
 -\frac{3 z^2
   {\bar z}}{(z-{\bar z})
   \sqrt{-i (z-{\bar z})^3}} &
   \frac{2 \sqrt{2}
   {\bar z}^3}{(-i
   (z-{\bar z}))^{3/2}}
\end{array}
\right) \nonumber\\
h_{\Lambda \vert I} & = & \left(
\begin{array}{ll}
 -\frac{\sqrt{3} (2
   z+{\bar z})}{(z-{\bar z})
   \sqrt{-i (z-{\bar z})^3}} &
   \frac{2 \sqrt{6} {\bar z}}{(-i
   (z-{\bar z}))^{3/2}} \\
 -\frac{3}{(z-{\bar z}) \sqrt{-i
   (z-{\bar z})^3}} & \frac{2
   \sqrt{2}}{(-i
   (z-{\bar z}))^{3/2}}
\end{array}
\right)
\label{nuovivec2}
\end{eqnarray}
and applying definition (\ref{intriscripen}) we exactly retrieve the same form of $\mathcal{N}_{\Lambda\Sigma}$ as given in eq.(\ref{periodoSL2}).
\par
For completeness and also for later use we calculate the remaining items pertaining to special geometry, in particular the symmetric $C$-tensor. From the general definition (\ref{ctensor}) applied to the present one-dimensional case we get:
\begin{equation}\label{cidefiqui}
    \nabla_z \, U_z \, = \, {\rm i} \, C_{zzz} \, h^{zz^\star} {\bar U}_{z^\star} \quad \Rightarrow \quad C_{zzz} \, = \, -\frac{6 {\rm i}}{(z-z^\star)^3}
\end{equation}
As for the standard Levi-Civita connection we have:
\begin{equation}\label{Levicivitaconno}
    \Gamma^z_{zz} \, = \, \frac{2}{z-z^\star} \quad ; \quad \Gamma^{z^\star}_{z^\star z^\star} \, = \, - \frac{2}{z-z^\star} \quad ; \quad \mbox{all other components vanish}
\end{equation}

This concludes our illustration of the cubic special K\"ahler structure on $\mathrm{SL(2,\mathbb{R})/SO(2)}$.
\subsection{The quartic invariant}\label{tqi}
In the cubic spin $j=\ft 32$ of  $\mathrm{SL(2,\mathbb{R})}$ there is a quartic invariant which plays an important role in the discussion of black-holes. As it happens for  all the other supergravity models, the quartic invariant of the symplectic vector of magnetic and electric charges:
\begin{equation}\label{chargevector}
    \mathcal{Q} \, = \, \left ( \begin{array}{c}
                                   p^\Lambda \\
                                   q_\Sigma
                                 \end{array}
     \right )
\end{equation}
is related to the entropy of the extremal black-holes, the latter being its square root. The origin of the quartic invariant is easily understood in terms of the symmetric tensor $t_{ijk}$. Using the $\mathrm{SL(2,\mathbb{R})}$-invariant
antisymmetric symbol $\epsilon^{ij}$ we can construct an invariant order four polynomial in the tensor $t_{ijk}$ by writing:
\begin{equation}\label{invariantus}
    \mathfrak{I}_4 \, \propto\, \epsilon^{ai}\, \epsilon^{bj} \,\epsilon^{pl}\, \epsilon^{qm} \,  \epsilon^{kr}\, \epsilon^{cn} \, t_{abc} \, t_{ijk}\,t_{pqr} \, t_{lmn}
\end{equation}
If we use the standard basis $t_{111},t_{112},t_{122},t_{222}$, we rotate it with the matrix (\ref{smatra}) and we identify the components of the resultant vector with those of the charge vector $\mathcal{Q}$ the explicit form of the invariant quartic polynomial is the following one:
\begin{equation}\label{Jinv}
 \mathfrak{I}_4  \, = \,  \frac{1}{3\sqrt{3}} q_2 p_1^3+\frac{1}{12} q_1^2 p_1^2-\frac{1}{2} p_2 q_1
   q_2 p_1-\frac{1}{3\sqrt{3}} p_2 q_1^3-\frac{1}{4} p_2^2 q_2^2
\end{equation}
where we have also chosen a specific overall normalization which turns out to be convenient in the sequel.\par
Let us now comment of the physical meaning of the above charges:  $p^\Lambda,\,q_\Lambda$ are related to the
$D6$, $D4$,$D0$, $D2$-brane charges, to be denoted by $P^0,P^1,Q_0,Q_1$, respectively (see \cite{Bellucci:2007eh} \cite{Bellucci:2006ib}), as follows:
\begin{eqnarray}
P^0&=&\frac{p_2}{\sqrt{2}}\,\,;\,\,\,P^1=\frac{p_1}{\sqrt{6}}\,\,;\,\,\,Q_0=\frac{q_2}{\sqrt{2}}\,\,;\,\,\,
Q_1=\sqrt{\frac{3}{2}}\,q_1\,.
\end{eqnarray}
In terms of $P^\Lambda,\,Q_\Lambda$ the quartic invariant reads (see also \cite{Ferrara:2010ug}) :
\begin{eqnarray}
 \mathfrak{I}_4 & = &-(Q_0P^0)^2-2\,Q_0P^0Q_1P^1+\frac{1}{3}\,(Q_1P^1)^2+4Q_0(P^1)^3-\frac{4}{27}\,P^0(Q_1)^3\,.
\end{eqnarray}
\subsection{Connection to the standard parametrization of the $S^3$ model}\label{ctsps3m}
In the previous subsection we have defined the correspondence between the charges used in the present work and those $(P^\Lambda,\,Q_\Lambda)$ which are directly connected with the brane interpretation. The latter correspond to a more standard choice of the holomorphic symplectic section in terms of a complex scalar $S$:
\begin{eqnarray}
\hat{\Omega}(S)&=&\left(\matrix{1\cr S\cr -S^3\cr 3\,S^2}\right)
\end{eqnarray}
In terms of $S$ the prepotential $F(S)$ has the simple form:
$F(S)=S^3$. The relation between $S$ and $z$ is given by the
following isometry:
\begin{eqnarray}
S&=&-\frac{1}{z}
\end{eqnarray}
up to a symplectic transformation in the fiber.

\subsection{The Lie algebra $\mathfrak{g}_{2(2)}$ as the $\mathbb{U}_{D=3}$ Lie algebra of our master example}
The complex Lie algebra $\mathfrak{g}_2(\mathbb{C})$ has rank two and it is defined by the $2\times 2$ Cartan matrix encoded in the following Dynkin diagram:
\begin{center}
\begin{picture}(110,30)
\put (-60,20){$\mathfrak{g}_2$}
\put (10,23){\circle {10}}
\put (15,25){\line (1,0){20}}
\put (15,23){\line (1,0){20}}
\put (20.5,19){{\LARGE$>$}}
\put (15,20.5){\line (1,0){20}}
\put (40,23){\circle {10}}
\put (65,21){$=\quad\quad\left (\begin{array}{cc}
  2 & -3\\
  -1 & 2
\end{array} \right)$}
\end{picture}
\end{center}
The $\mathfrak{g}_2$ root system $\Delta$ consists of the following six positive roots plus their negatives:
\begin{equation}
\label{g2rootsystem}
\begin{array}{rclcrcl}
\alpha_1&=&(1,0)&;&\alpha_2&=&\frac{\sqrt{3}}{2}\,(-\sqrt{3},1)\\
\alpha_3 \, =\, \alpha_1+\alpha_2&=&\frac{1}{2}\,(-1,\sqrt{3}) &;&
\alpha_4 \, = \, 2\,\alpha_1+\alpha_2 &=&
\frac{1}{2}\,(1,\sqrt{3}) \\
\alpha_5 \, = \, 3\,\alpha_1+\alpha_2&=&\frac{\sqrt{3}}{2}\,(\sqrt{3},1)&;&\alpha_6 \, = \,  3\,\alpha_1+2\,\alpha_2 &=&
(0,\sqrt{3})\
\end{array}
\end{equation}
The $g_{2(2)}$ Lie algebra is the non-compact maximally split section of $\mathfrak{g}_2(\mathbb{C})$.  As for all maximally split algebras the Cartan generators $H_i$ and the step operators $E^\alpha$ associated with each root $\alpha$ can be chosen completely real in all representations. Furthermore we can always construct bases where the Cartans $H_i$ are diagonal matrices, the step operators $E^\alpha$ associated with positive roots  $\alpha >0 $ are upper triangular matrices and the step operators $E^{-\alpha}$ are the lower triangular transposed of the former.
\par
In the fundamental $7$-dimensional representation the explicit form of the $\mathfrak{g}_{2(2)}$-generators with the above properties is presented hereby. Naming  $\{H_1,\,H_2\}$  the  Cartan generators along the two ortho-normal
directions and adopting the standard Cartan--Weyl normalizations:
\begin{eqnarray}
[E^\alpha,E^{-\alpha}]&=&\alpha^i\,H_i\,\,,\,\,\,[H_i,E^\alpha]=\alpha^i\,E^\alpha\,.
\end{eqnarray}
we have:
\begin{equation}
{\scriptsize
\begin{array}{lllclll}
H_1&=&\left(\matrix{ \frac{1}
   {2} & 0 & 0 & 0 & 0 & 0 & 0 \cr 0 & - \frac{1}{2}  & 0 & 0 & 0 & 0 & 0 \cr 0 & 0 & 1 &
   0 & 0 & 0 & 0 \cr 0 & 0 & 0 & 0 & 0 & 0 & 0 \cr 0 &
   0 & 0 & 0 & -1 & 0 & 0 \cr 0 & 0 & 0 & 0 & 0 &
    \frac{1}{2} & 0 \cr 0 & 0 & 0 & 0 & 0 & 0 & -
     \frac{1}{2} \cr  }\right) &;& H_2 &=& \left(\matrix{ \frac{{\sqrt{3}}}
   {2} & 0 & 0 & 0 & 0 & 0 & 0 \cr 0 & \frac{{
       \sqrt{3}}}{2} & 0 & 0 & 0 & 0 & 0 \cr 0 & 0 &
   0 & 0 & 0 & 0 & 0 \cr 0 & 0 & 0 & 0 & 0 & 0 & 0 \cr
   0 & 0 & 0 & 0 & 0 & 0 & 0 \cr 0 & 0 & 0 & 0 & 0 &
   -\frac{{\sqrt{3}}}
   {2} & 0 \cr 0 & 0 & 0 & 0 & 0 & 0 & -\frac{{
        \sqrt{3}}}{2} \cr  }\right)\\
\end{array}
}
\end{equation}
\begin{equation}
{
\scriptsize
\begin{array}{lllclll}
E^{\alpha_1}&=&\left(\matrix{ 0 & \frac{1}
   {{\sqrt{2}}} & 0 & 0 & 0 & 0 & 0 \cr 0 & 0 & 0 &
   0 & 0 & 0 & 0 \cr 0 & 0 & 0 & 1 & 0 & 0 & 0 \cr 0 &
   0 & 0 & 0 & 1 & 0 & 0 \cr 0 & 0 & 0 & 0 & 0 & 0 &
   0 \cr 0 & 0 & 0 & 0 & 0 & 0 & \frac{1}
   {{\sqrt{2}}} \cr 0 & 0 & 0 & 0 & 0 & 0 & 0 \cr
   }\right) &;& E^{\alpha_2}&=&\left(\matrix{ 0 & 0 & 0 & 0 & 0 & 0 & 0 \cr 0 & 0 & {
     \sqrt{\frac{3}{2}}} & 0 & 0 & 0 & 0 \cr 0 & 0 &
   0 & 0 & 0 & 0 & 0 \cr 0 & 0 & 0 & 0 & 0 & 0 & 0 \cr
   0 & 0 & 0 & 0 & 0 & -{\sqrt{\frac{3}
       {2}}} & 0 \cr 0 & 0 & 0 & 0 & 0 & 0 & 0 \cr 0 &
   0 & 0 & 0 & 0 & 0 & 0 \cr  }\right) \\
 \end{array}
}
\end{equation}
\begin{equation}
{
\scriptsize
\begin{array}{lllclll}
   E^{\alpha_1+\alpha_2}&=&\left(\matrix{ 0 & 0 & \frac{1}
   {{\sqrt{2}}} & 0 & 0 & 0 & 0 \cr 0 & 0 & 0 &
    -1 & 0 & 0 & 0 \cr 0 & 0 & 0 & 0 & 0 & 0 & 0 \cr
   0 & 0 & 0 & 0 & 0 &
    -1 & 0 \cr 0 & 0 & 0 & 0 & 0 & 0 & \frac{1}
   {{\sqrt{2}}} \cr 0 & 0 & 0 & 0 & 0 & 0 & 0 \cr 0 &
   0 & 0 & 0 & 0 & 0 & 0 \cr  }\right) &;& E^{2\alpha_1+\alpha_2}&=&\left(\matrix{ 0 & 0 & 0 &
    -1 & 0 & 0 & 0 \cr 0 & 0 & 0 & 0 & \frac{1}
   {{\sqrt{2}}} & 0 & 0 \cr 0 & 0 & 0 & 0 & 0 & -
     \frac{1}{{\sqrt{2}}}  & 0 \cr 0 & 0 & 0 &
   0 & 0 & 0 & 1 \cr 0 & 0 & 0 & 0 & 0 & 0 & 0 \cr 0 &
   0 & 0 & 0 & 0 & 0 & 0 \cr 0 & 0 & 0 & 0 & 0 & 0 &
   0 \cr  }\right)\\
   \end{array}
   }
\end{equation}
\begin{equation}
{
\scriptsize
\begin{array}{lllclll}
E^{3\alpha_1+\alpha_2}&=&\left(\matrix{ 0 & 0 & 0 & 0 &-
{\sqrt{\frac{3}
      {2}}} & 0 & 0 \cr 0 & 0 & 0 & 0 & 0 & 0 & 0 \cr
   0 & 0 & 0 & 0 & 0 & 0 & -{\sqrt{\frac{3}
      {2}}} \cr 0 & 0 & 0 & 0 & 0 & 0 & 0 \cr 0 & 0 &
   0 & 0 & 0 & 0 & 0 \cr 0 & 0 & 0 & 0 & 0 & 0 & 0 \cr
   0 & 0 & 0 & 0 & 0 & 0 & 0 \cr  }\right) &;&
   E^{3\alpha_1+2\alpha_2}&=&\left(\matrix{ 0 & 0 & 0 & 0 & 0 & -{\sqrt{\frac{3}
      {2}}} & 0 \cr 0 & 0 & 0 & 0 & 0 & 0 & -{\sqrt
    {\frac{3}{2}}} \cr 0 & 0 & 0 & 0 & 0 & 0 & 0 \cr
   0 & 0 & 0 & 0 & 0 & 0 & 0 \cr 0 & 0 & 0 & 0 & 0 &
   0 & 0 \cr 0 & 0 & 0 & 0 & 0 & 0 & 0 \cr 0 & 0 & 0 &
   0 & 0 & 0 & 0 \cr  }\right)\\
   \end{array}
   }
\end{equation}
The connection between the $\mathfrak{g}_{2(2)}$ Lie algebra and
our master supergravity model will be established if we can show
that it admits the decomposition (\ref{gendecompo}) and can be put
into the form (\ref{genGD3pre}) where $W_{2,\mathbf{}W}$ is the
doublet of $\mathrm{SL(2,\mathbb{R})_E}$ and $\mathbf{W}$ is the
triple symmetric representation of the original
$\mathrm{SL(2,\mathbb{R})}$ in $D=4$. From a group theoretical
viewpoint such conditions are indeed satisfied since we have:
\begin{equation}\label{decompongone}
    \mbox{adj} \left[\mathfrak{g}_{2(2)}\right] \, = \, \left(\mbox{adj} \left[\slal(2,\mathbb{R})_E \right], \,\mathbf{1}\right) \, \oplus \,
    \left( \mathbf{1}\, , \, \mbox{adj} \left[\slal(2,\mathbb{R})\right]\right) \, \oplus \, \left (\mathbf{2}\, , \, \mathbf{4}\right)
\end{equation}
Explicitly the $\mathfrak{g}_{2(2)}$ Lie algebra can be cast into
the form (\ref{genGD3pre}) in the following way.
\par
First we single out the two relevant $\slal(2,\mathbb{R})$
subalgebras. The Ehlers algebra is associated with the highest
root and we have:
\begin{equation}\label{ehlersalg}
    L_0^E \,= \, \frac{1}{\sqrt{3}} \, H_2 \quad ; \quad L_\pm^E \, = \, \sqrt{\frac 23} \, E^{\pm (3\alpha_1+2\alpha_2)}
\end{equation}
while the original $\mathbb{U}_{D=4} \, = \, \slal(2,\mathbb{R})$
is associated with the first simple root orthogonal to the highest
one and we have:
\begin{equation}\label{d4alg}
    L_0 \,= \, H_1 \quad ; \quad L_\pm \, = \, \sqrt{2} \, E^{\pm \alpha_1}
\end{equation}
Then we can arrange the remaining eight generators in the  tensor $W^{i\beta}$ as follows:
\begin{eqnarray}
  W^{1M} &=& \sqrt{\frac 23} \, \left( E^{\alpha_1+\alpha_2} \, , \, E^{\alpha_2} \, , \, E^{2\alpha_1 + \alpha_2}
  \, , \, E^{3\alpha_1 + \alpha_2} \right) \nonumber \\
  W^{2M} &=& \sqrt{\frac 23} \, \left( - E^{-2 \alpha_1-\alpha_2} \, , \,- E^{-3\alpha_1-\alpha_2} \, , \, E^{-\alpha_1 - \alpha_2}
  \, , \, E^{- \alpha_2} \right) \label{wgenni}
\end{eqnarray}
Calculating the commutators of $W^{iM}$ with the generators of the two $\slal(2)$ algebras we find:
\begin{eqnarray}\label{turnotto}
\left [ L_0^E \, , \, \left(\begin{array}{c}
                                  W^1 \\
                                  W^2
                                \end{array} \right)\right] & = &
\left(\begin{array}{c|c}
\ft 12 \, \mathbf{1} & 0 \\
\hline
0 & - \ft 12 \, \mathbf{1}
\end{array} \right) \,
\left(\begin{array}{c}
W^1 \\
W^2
\end{array} \right)\nonumber\\
\left [ L_+^E \, , \, \left(\begin{array}{c}
                                  W^1 \\
                                  W^2
                                \end{array} \right)\right]
                                & = & \left(\begin{array}{c|c}
                                0 & 0 \\
                                \hline
                                -\mathbf{1} & 0\
                                \end{array} \right) \,\left(\begin{array}{c}
                                  W^1 \\
                                  W^2
                                \end{array} \right)\nonumber\\
\left [ L_-^E \, , \, \left(\begin{array}{c}
                                  W^1 \\
                                  W^2
                                \end{array} \right)\right] & = &
\left(\begin{array}{c|c}
0 & -\mathbf{1} \\
\hline
0 & 0\
\end{array} \right) \,
\left(\begin{array}{c} W^1 \\
W^2
\end{array} \right)
\end{eqnarray}
and:
\begin{eqnarray}\label{turnotto2}
\left [ L_0  \, , \, \left(\begin{array}{c}
                                  W^1 \\
                                  W^2
                                \end{array} \right)\right] & = & - \, \left(\begin{array}{c|c}
U_0 & 0 \\
\hline
0 & U_0
\end{array} \right) \,
\left(\begin{array}{c}
W^1 \\
W^2
\end{array} \right)\nonumber\\
\left [ L_\pm  \, , \, \left(\begin{array}{c}
                                  W^1 \\
                                  W^2
                                \end{array} \right)\right] & = & - \,\left(\begin{array}{c|c}
U_\pm & 0 \\
\hline
0 & U_\pm
\end{array} \right) \,
\left(\begin{array}{c}
W^1 \\
W^2
\end{array} \right)
\end{eqnarray}
where:
\begin{eqnarray}
\label{32repre}
  U_0 &=& \left(
\begin{array}{llll}
 \frac{1}{2} & 0 & 0 & 0 \\
 0 & \frac{3}{2} & 0 & 0 \\
 0 & 0 & -\frac{1}{2} & 0 \\
 0 & 0 & 0 & -\frac{3}{2}
\end{array}
\right) \nonumber\\
  U_+ &=& \left(
\begin{array}{llll}
 0 & 0 & -2 & 0 \\
 -\sqrt{3} & 0 & 0 & 0 \\
 0 & 0 & 0 & \sqrt{3} \\
 0 & 0 & 0 & 0
\end{array}
\right) \nonumber\\
  U_- &=& \left(
\begin{array}{llll}
 0 & -\sqrt{3} & 0 & 0 \\
 0 & 0 & 0 & 0 \\
 -2 & 0 & 0 & 0 \\
 0 & 0 & \sqrt{3} & 0
\end{array}
\right)
\end{eqnarray}
which are the generators of $\mathrm{SL(2,\mathbb{R})}$ in the symplectic embedding (\ref{sympobeddo}) as it can be easily verified by considering the embedding of a group element infinitesimally closed to the identity:
\begin{equation}\label{generisl2}
 \left(\begin{array}{cc}
         a & b \\
         c & d
       \end{array}
  \right)  \, = \,  \left(\begin{array}{cc}
            1+\ft 12 \,\epsilon_0 & \epsilon_+ \\
            \epsilon_- & 1-\ft 12 \, \epsilon_0
          \end{array}
    \right)
\end{equation}
and collecting the matrix coefficients of the first order terms in $\epsilon_0$ and $\epsilon_\pm$.
\subsection{The $\mathfrak{g}_{(2,2)}$ Lie algebra in terms of Chevalley triples}
For later use it is convenient to rewrite the commutation relations of the $\mathfrak{g}_{(2,2)}$ in terms of triples of Chevalley generators as it was done in \cite{bruxelles}, whose results we want to compare with ours.
\par
Since the algebra has rank two there are two fundamental triples of Chevalley generators:
\begin{equation}\label{fundatriplets}
    \left(\mathcal{H}_1,e_1,f_1\right ) \quad ; \quad \left(\mathcal{H}_2,e_2,f_2\right )
\end{equation}
with the following commutation relations:
\begin{equation}\label{fundacommu}
    \begin{array}{ccccccc}
       \left[\mathcal{H}_2,e_2\right]=2e_2 & \null &\left[\mathcal{H}_1,e_2\right]=-3e_2& \null & \left[\mathcal{H}_2,f_2\right]=-2f_2 & \null & \left[\mathcal{H}_1,f_2\right]=3f_2  \\
       \left[\mathcal{H}_2,e_1\right]=-e_1 & \null &\left[\mathcal{H}_1,e_1\right]=
       2e_1& \null & \left[\mathcal{H}_2,f_1\right]=f_1 & \null & \left[\mathcal{H}_1,f_1\right]=-2f_1 \\
       \left[e_2,f_2\right]=\mathcal{H}_2 & \null &\left[e_2,f_1\right]=0 & \null &
       \left[e_1,f_1\right]=\mathcal{H}_1 & \null & \left[e_1,f_2\right]=0 \\
     \end{array}
\end{equation}
The remaining basis elements are defined as follows:
\begin{equation}\label{ancora}
    \begin{array}{ccccccc}
       e_3=\left[e_1,e_2\right]& \null & e_4=\frac{1}{2}\,\left[e_1,e_3\right] & \null &
       e_5=\frac{1}{3}\,\left[e_4,e_1\right] & \null & e_6=\left[e_2,e_5\right] \\
       f_3=\left[f_2,f_1\right]& \null & f_4=\frac{1}{2}\,\left[f_3,f_1\right] & \null &
       f_5=\frac{1}{3}\,\left[f_1,f_4\right] & \null & f_6=\left[f_5,f_2\right] \\
     \end{array}
\end{equation}
and satisfy the following Serre relations:
\begin{equation}\label{serrerele}
    \left [ e_2,e_3\right] \,=\,\left [ e_5,e_1\right] \,=\,\,\left [ f_2,f_3\right] \,=\,\left [ f_5,f_1\right] \,=\,0
\end{equation}
The Chevalley form of the commutation relation is obtained from the standard Cartan Weyl basis introducing the following identifications:
\begin{equation}\label{chevallabasa}
    \begin{array}{ccccccc}
       e_1 & = & \sqrt{2} E^{\alpha_1}& ; & e_2 & = &\sqrt{\ft 23} E^{\alpha_2} \\
       e_3 & = & \sqrt{2} E^{\alpha_3} & ; &  e_4 & =& \sqrt{2} E^{\alpha_4} \\
       e_5 & = & \sqrt{\ft 23} E^{\alpha_5} & ; & e_6 & = & \sqrt{\ft 23} E^{\alpha_6}\\
       f_1 & = & \sqrt{2} E^{-\alpha_1}& ; & f_2 & = &\sqrt{\ft 23} E^{-\alpha_2} \\
       f_3 & = & \sqrt{2} E^{-\alpha_3} & ; &  f_4 & = &\sqrt{2} E^{-\alpha_4} \\
       f_5 & = & \sqrt{\ft 23} E^{-\alpha_5} &  ; & f_6 & = & \sqrt{\ft 23} E^{-\alpha_6}
     \end{array}
\end{equation}
and\footnote{Note that we are using a slightly different notation with respect to \cite{bruxelles}: Denoting by bold symbols the Chevalley generators used in that reference we have the following correspondence:
\begin{eqnarray}
e_1&=& {\bf e}_2\,;\,\,e_2={\bf e}_1\,;\,\,e_3=-{\bf e}_3\,;\,\,e_4={\bf e}_4/2\,;\,\,e_5={\bf e}_5/6\,;\,\,e_6={\bf e}_6/6\,;\,\, \mathcal{H}_1={\bf h}_2\,;\,\, \mathcal{H}_2={\bf h}_1\,.
\end{eqnarray} }
\begin{equation}\label{cartanucci}
    \mathcal{H}_1 \, = \, 2 \alpha_1\cdot H \quad ; \quad \mathcal{H}_2 \, = \, \ft 23 \, \alpha_2\cdot H
\end{equation}
\section{Solvable parametrization of the coset and Supergravity fields in Black-Hole configurations}
\label{solvapara}
Let us now summarize the structure of the $D=3$ $\sigma$-model which encodes the fields of time-like dimensionally reduced $\mathcal{N}=2$ supergravity coupled to $n$ vector multiplets. In this discussion we assume that, in $D=4$, the necessary special K\"ahler manifold is a symmetric space:
\begin{equation}\label{specoset}
    \mathcal{SK}_n \, = \, \frac{\mathrm{U_{D=4}}}{\mathrm{H_{D=4}}}
\end{equation}
so that also the $D=3$ target manifold is a coset manifold, actually the Wick rotation of a quaternionic symmetric coset space:
\begin{equation}\label{quatespace}
    {\mathcal{Q}}^\star_{4n+4} \, = \, \frac{\mathrm{U_{D=3}}}{\mathrm{H^\star_{D=3}}}
\end{equation}
The real dimension of  ${\mathcal{Q}}^\star_{4n+4}$ is $4n+4$. This is the total number of supergravity degrees of freedom and, correspondingly, of radial functions  parameterizing a spherically symmetric Black-Hole configuration.
\par
In the following table we summarize the naming, numbering,  and interpretation of the $D=3$ scalar fields both in the general case and in the master example based on the Lie algebra ${\mathfrak{g}_{2(2)}}$.
\begin{center}
\begin{tabular}{|l|cc||cc|}
  \hline
  \null & Generic &\null & \null&${\mathfrak{g}_{2(2)}}$ \\
  \cline{1-2}\cline{4-5}
  \hline
  warp factor & $U(\tau)$ & $1$ & 1 & \null \\
  Taub Nut field & $a(\tau)$ & 1 & 1 & \null \\
   D=4 scalars& $\phi^i$ & $2n$ & 2 & $(\varphi,y) \mapsto z={\rm i} \, e^\varphi +y$\\
  Scalars from vectors & $Z^M(\tau) \, = \, \left(Z^\Lambda(\tau)\, , \, Z_\Sigma(\tau) \right)$ & 2n+2 & 4 & \null \\
  \hline
  \textbf{Total} & \null & 4n+4 & 8 & \null \\
  \hline
\end{tabular}
\end{center}
As explained at length in \cite{Chemissany:2010zp} and in the previous literature on this topic after reduction to $D=3$ we consider only those solutions of the $\sigma$-model that depend on a single coordinate which in the case of space-reductions is time,
while in the presently considered case of time-reduction is a radial coordinate $r$. It is actually convenient to use a parameter $\tau$ which is the inverse of the radial one: $\tau \propto 1/r$.
\par
The $D=4$ solution of supergravity is then parameterized in the following way in terms of the $\sigma$-model fields\footnote{Note that, in order to retrieve the notations used, for instance, in \cite{Chemissany:2010zp}, in all the formulas below one should replace $U\rightarrow 2U$. }. For the metric we have:
\begin{equation}\label{metricona}
    ds_{(4)}^2 \, = \, - \, e^{U(\tau)} \, \left(dt + A_{KK}\right)^2 \, + \, e^{-\, U(\tau)}\left[ e^{4\,A(\tau)} \, d\tau^2 \, + \,
    e^{2\,A(\tau)}\left( d\theta^2 + \sin^2\theta \, d\phi^2\right)\right]
\end{equation}
where $e^{2\,A(\tau)}$ is a shorthand notation for the following function:
\begin{equation}\label{funzia}
    e^{2\,A(\tau)} \, = \, \left\{ \begin{array}{cc}
                                    \frac{v^2}{\sinh^2(v\tau)} & \mbox{if $v^2 >0$} \\
                                    \frac{1}{\tau^2} & \mbox{if $v^2 = 0$}
                                  \end{array}
    \right.
\end{equation}
The parameter $v^2$ mentioned in the above formula is one of the conserved charges of the dynamical model and it is named the extremality parameter. Its geometrical interpretation within the framework of the $\sigma$-model is very simple and clear. In terms of the scalar fields mentioned in the above table the metric of the target manifold ${\mathcal{Q}}^\star_{4n+4}$ takes the following general form:
\begin{eqnarray} \label{geodaction}
ds^2_{\mathcal{Q}} &=& \frac{1}{4} \, \left [ d{U}^2+\,h_{rs}\,d{\phi}^r\,d{\phi}^s
+ \e^{-2\,U}\,(d{a}+{\bf Z}^T\mathbb{C}d{{\bf
Z}})^2\,+\,2 \, e^{-U}\,d{{\bf
Z}}^T\,\mathcal{M}_4\,d{{\bf Z}}\right ]\\
\mathcal{M}_4 & = &
\left(\begin{array}{c|c}
 { \mathrm{Im}}\mathcal{N}^{-1} & { \mathrm{Im}}\mathcal{N}^{-1}\,{\mathrm{Re}}\mathcal{N} \\
\hline
{\mathrm{Re}}\mathcal{N}\,{ \mathrm{Im}}\,\mathcal{N}^{-1} & {\mathrm{Im}}\mathcal{N}\,
+\, {\mathrm{Re}}\mathcal{N} \, { \mathrm{Im}}\mathcal{N}^{-1}\, {\mathrm{Re}}\mathcal{N} \
\end{array}\right) \label{quaternionic}\\
\mathcal{M}_4^{-1} & = &
\left(\begin{array}{c|c}
{\mathrm{Im}}\mathcal{N}\,
+\, {\mathrm{Re}}\mathcal{N} \, { \mathrm{Im}}\mathcal{N}^{-1}\, {\mathrm{Re}}\mathcal{N} & \, -{\mathrm{Re}}\mathcal{N}\,{ \mathrm{Im}}\,\mathcal{N}^{-1}\\
\hline
-\, { \mathrm{Im}}\mathcal{N}^{-1}\,{\mathrm{Re}}\mathcal{N}  & { \mathrm{Im}}\mathcal{N}^{-1} \
\end{array}\right) \label{inversem4}
\end{eqnarray}
where $\mathbb{C}^{MN}$ is the symplectic invariant metric on the
fibres of the special geometry symplectic bundle and
$\mathcal{N}_{\Lambda\Sigma}$ is the kinetic matrix of the
vectors. One-dimensional solutions of the $\sigma$-model are just
geodesics of the above metric which has the following indefinite
signature
\begin{equation}
\mbox{sign}\left[ds^2_{\mathcal{Q}}\right] \, = \, \left(\underbrace{+,\dots,+}_{2+2n},\underbrace{-,\dots ,-}_{2n+2}\right)
\end{equation}
since the matrix $\mathcal{M}_4 $ is negative definite. Hence the geodesics can be time-like, null-like or space-like depending on the
three possible cases:
\begin{equation}\label{beddamatre}
    \dot{U}^2+\,h_{rs}\,\dot{\phi}^r\,\dot{\phi}^s
+ \e^{-2\,U}\,(\dot{a}+{\bf Z}^T\mathbb{C}\dot{{\bf
Z}})^2\,+\,2 \, e^{-\,U}\,\dot{{\bf
Z}}^T\,\mathcal{M}_4\,\dot{{\bf Z}} \, = \, \left\{ \begin{array}{ccc}
                                                      v^2 & > & 0 \\
                                                      v^2 & = & 0 \\
                                                      - \, v^2 & < & 0
                                                    \end{array}
 \right.
\end{equation}
where the dot denotes derivative with respect to the affine parameter $\tau$. Space-like geodesics correspond to unphysical solutions with naked singularities and are excluded. Time-like geodesics correspond to non-extremal black-holes while null-like geodesics yield extremal black-holes.
\subsection{General properties of the $d=4$ metric} Before proceeding to the specific structure of our considered master model, it is convenient to summarize some general properties of the $d=4$ metric in eq.(\ref{metricona}). First we consider the case of non extremal black-holes  $v^2 > 0$ and in particular the Schwarzschild solution which, as we are going to demonstrate is the unique representative of the whole $\mathrm{G_{(2,2)}}$ orbit of regular black-hole solutions.
\paragraph{The Schwarzschild case}
Consider the case where the function $\exp[-U(\tau)]$ and the extremality parameter are  the following ones:
\begin{equation}\label{schwarzacasa}
    \exp[-U(\tau)] \, = \, \exp[-\alpha \, \tau] \quad ; \quad v^2 \, = \, \frac{\alpha^2}{4}
\end{equation}
Introducing the following position:
\begin{equation}\label{posiziona}
    \tau \, = \, \frac{\log\left[1-\frac{2m}{r}\right]}{2m} \quad ; \quad \alpha \, = \, 2 \, m
\end{equation}
the reader can immediately verify that the metric (\ref{metricona}) at $A_{KK}\, = \, 0$ is turned into the standard Schwarzschild metric:
\begin{equation}\label{schwarazata}
    ds^2_{Schw} \, = -\,\left(1-\frac{2m}{r}\right) \, dt^2 \, + \, \left(1-\frac{2m}{r}\right)^{-1} \, dr^2 \, + \, r^2 \,\left( d\theta^2 + \sin^2\theta \, d\phi^2\right)
\end{equation}
\paragraph{The extremal Reissner Nordstr\"om case}
Consider now the following choices:
\begin{equation}\label{reissacasa}
    \exp[-U(\tau)] \, = \, \left(1 + q \, \tau\right) \quad ; \quad v^2 \, = \, 0
\end{equation}
Introducing the following position:
\begin{equation}\label{posiziona2}
    \tau \, = \, \frac{1}{r-q}
\end{equation}
by means of elementary algebra the reader can verify that the metric (\ref{metricona}) at $A_{KK}\, = \, 0$ is turned into the extremal Reissner Nordstr\"om metric:
\begin{equation}\label{extremata}
    ds^2_{RNext} \, = -\,\left(1-\frac{q}{r}\right)^2 \, dt^2 \, + \, \left(1-\frac{q}{r}\right)^{-2} \, dr^2 \, + \, r^2 \,\left( d\theta^2 + \sin^2\theta \, d\phi^2\right)
\end{equation}
which follows from the non extremal one:
\begin{equation}\label{nonextremata}
    ds^2_{RN} \, = -\,\left(1\, - \, \frac{2m}{r}+\frac{q^2}{r^2}\right) \, dt^2 \, + \, \left(1\, - \, \frac{2m}{r}+\frac{q^2}{r^2}\right)^{-1} \, dr^2 \, + \, r^2 \,\left( d\theta^2 + \sin^2\theta \, d\phi^2\right)
\end{equation}
when the mass is equal to the charge: $m=q$.
\par
It follows from the discussion of this simple example that the extremal black-hole metrics (\ref{metricona}) are all suitable deformations of the extremal Reissner Nordstr\"om metric, just as the regular black-hole metrics are suitable deformations of the Schwarzschild one.
\paragraph{Curvature of the extremal spaces}
In order to facilitate the discussion of the various solutions found by means of the integration method discussed in further sections, it is useful to consider the general form of the Riemann tensor associated with the metrics
(\ref{metricona}) in the extremal case.
To this effect we introduce the vielbein 1-forms:
\begin{eqnarray}
  E^0 &=& \exp \left [ \frac{U}{2}\right] \, dt \nonumber\\
  E^1 &=& \exp \left [ - \frac{U}{2}\right] \, \frac{d\tau}{\tau^2}\nonumber\\
  E^2 &=& \exp \left [ - \frac{U}{2}\right] \, \frac{1}{\tau} \, d\theta\nonumber\\
  E^3 &=& \exp \left [ - \frac{U}{2}\right] \, \frac{1}{\tau} \, \sin[\theta] \, d\phi \label{vielbeine}
\end{eqnarray}
and the corresponding spin connection:
\begin{equation}\label{spinaconna}
    dE^a \, + \, \omega^{ab} \, \wedge \, E^c \, \eta_{bc} \, = \, 0
\end{equation}
Defining the curvature 2-form in the standard way:
\begin{equation}\label{curva2forma}
  \mathfrak{R}^{ab} \,= \,  d\omega^{ab} \, + \omega^{ac} \, \wedge \, \omega^{db} \, \eta_{cd}
\end{equation}
we find that it is diagonal :
\begin{eqnarray}\label{riemanno}
    \mathfrak{R}^{01} & = & \mathcal{C}_1 \, E^0 \, \wedge \, E^1\nonumber \\
    \mathfrak{R}^{02} & = & \mathcal{C}_2 \, E^0 \, \wedge \, E^2\nonumber \\
    \mathfrak{R}^{03} & = & \mathcal{C}_2\, E^0 \, \wedge \, E^3\nonumber \\
    \mathfrak{R}^{12} & = & \mathcal{C}_3 \, E^1 \, \wedge \, E^2\nonumber \\
    \mathfrak{R}^{13} & = & \mathcal{C}_3\, E^1 \, \wedge \, E^3\nonumber \\
    \mathfrak{R}^{23} & = & \mathcal{C}_4 \, E^3 \, \wedge \, E^4
\end{eqnarray}
and involves  four independent differential expressions in the function $U(\tau)$, namely
\begin{eqnarray}
  \mathcal{C}_1(\tau )&=& -\frac{1}{4} e^{U(\tau )} \tau
   ^3 \left(\tau  U'(\tau )^2+2
   U'(\tau )+\tau  U''(\tau
   )\right) \nonumber\\
  \mathcal{C}_2(\tau ) &=& \frac{1}{8} e^{U(\tau )} \tau
   ^3 U'(\tau ) \left(\tau
   U'(\tau )+2\right) \nonumber\\
  \mathcal{C}_3(\tau ) &=& \frac{1}{4} e^{U(\tau )} \tau
   ^3 \left(U'(\tau )+\tau
   U''(\tau )\right) \nonumber\\
  \mathcal{C}_4(\tau )&=& -\frac{1}{8} e^{U(\tau )} \tau
   ^3 U'(\tau ) \left(\tau
   U'(\tau )+4\right) \label{curvinediU}
\end{eqnarray}
We will consider the behavior of these four independent component of the Riemann tensor in the various solutions
\subsection{Specific form of the Special Geometry data in the $S^3$ model}
In the case of our master model the relevant formulae discussed above specialize as follows. The scalar metric is the standard Poincar\'e metric on the lower half-plane, namely:
\par
\begin{equation}\label{poincametrica}
    h_{rs}\,\dot{\phi}^r\,\dot{\phi}^s \, = \, 4 g_{zz^\star}\,dz\,dz^\star \, = \, 3 \, \frac{dz \, d\bar{z}}{\left(\mathrm{Im}z\right)^2}
\end{equation}
while the explicit form of the matrix $\mathcal{M}_4$ is the following one:
\begin{equation}\label{m4matrona}
\mathcal{M}_4 \, = \,    -\left(
\begin{array}{llll}
 -\frac{\left(\mathrm{Re}z^2+\mathrm{Im}z^2\right) \left(3 \mathrm{Re}z^2+\mathrm{Im}z^2\right)}{\mathrm{Im}z^3} &
   \frac{\sqrt{3} \mathrm{Re}z \left(\mathrm{Re}z^2+\mathrm{Im}z^2\right)^2}{\mathrm{Im}z^3} & \frac{3
   \mathrm{Re}z^3+2 \mathrm{Im}z^2 \mathrm{Re}z}{\mathrm{Im}z^3} & \frac{\sqrt{3} \mathrm{Re}z^2}{\mathrm{Im}z^3} \\
 \frac{\sqrt{3} \mathrm{Re}z \left(\mathrm{Re}z^2+\mathrm{Im}z^2\right)^2}{\mathrm{Im}z^3} &
   -\frac{\left(\mathrm{Re}z^2+\mathrm{Im}z^2\right)^3}{\mathrm{Im}z^3} & -\frac{\sqrt{3} \mathrm{Re}z^2
   \left(\mathrm{Re}z^2+\mathrm{Im}z^2\right)}{\mathrm{Im}z^3} & -\frac{\mathrm{Re}z^3}{\mathrm{Im}z^3} \\
 \frac{3 \mathrm{Re}z^3+2 \mathrm{Im}z^2 \mathrm{Re}z}{\mathrm{Im}z^3} & -\frac{\sqrt{3} \mathrm{Re}z^2
   \left(\mathrm{Re}z^2+\mathrm{Im}z^2\right)}{\mathrm{Im}z^3} & -\frac{3 \mathrm{Re}z^2+\mathrm{Im}z^2}{\mathrm{Im}z^3} &
   -\frac{\sqrt{3} \mathrm{Re}z}{\mathrm{Im}z^3} \\
 \frac{\sqrt{3} \mathrm{Re}z^2}{\mathrm{Im}z^3} & -\frac{\mathrm{Re}z^3}{\mathrm{Im}z^3} & -\frac{\sqrt{3}
   \mathrm{Re}z}{\mathrm{Im}z^3} & -\frac{1}{\mathrm{Im}z^3}
\end{array}
\right)
\end{equation}
\par
To complete the illustration of the metric (\ref{metricona}) we still have to explain the meaning of the one-form $A_{KK}$. This latter is the Kaluza-Klein vector, whose field strength $F_{KK} \, = \, dA_{KK}$ is related to the $D=3$ $\sigma$-model scalars by dualization, as follows:
\begin{equation}\label{dualKK}
    F_{KK}^{ij} \, = \, - \, \frac{e^{-4\,U}}{\sqrt{|\mbox{det}g_3|}} \, \epsilon^{ijk} \, \left(\partial_k a \, + \, Z^\Lambda \, \partial_k \, Z_\Lambda \, - \, Z_\Sigma \, \partial_k \, Z^\Sigma \right)
\end{equation}
In eq.(\ref{dualKK}) $g_3$ denotes the three-dimensional metric. Using as $D=3$ coordinates the parameter $\tau$ and the two Euler angles $\theta,\varphi$, when the scalars depend only on $\tau$ we find that the only non vanishing component of $F_{KK}^{ij}$ is the following one:
\begin{equation}\label{feroceTaubNut}
    F_{KK|\theta\varphi}\, = \,  \, g_{\theta\theta} \, g_{\varphi\varphi} \, F_{KK}^{\theta\varphi} \,= \, - \, \sin\theta\,\underbrace{\left[
    e^{-2\,U} \,\left(\dot a \, + \, Z^\Lambda \, \dot{Z}_\Lambda \, - \, Z_\Sigma \, \dot{Z}^\Sigma\right)\right]}_{ \mathbf{n} \,=\,
    \mbox{Taub NUT charge}}
\end{equation}
As we are going to see later, the combination of derivatives under-braced in equation (\ref{feroceTaubNut}) is a constant of motion of the Lax flows and is named $\mathbf{n}$, the Taub-NUT charge. The fact that $\mathbf{n}$ is a constant is very important and obligatory in order for the dualization formulae to make sense. Indeed the Kaluza-Klein field strength $F_{KK}$ satisfies the Bianchi identity only in force of the constancy of $\mathbf{n}$. In view of this the Kaluza-Klein vector is easily determined and reads:
\begin{equation}\label{KKvector}
    A_{KK} \, = \, 2\,\mathbf{n}\, \cos\theta \, d\varphi
\end{equation}
The field-strength two-form is instead:
\begin{equation}\label{fildaforza}
    F_{KK} \, =\, - 2\, \mathbf{n} \, \sin\theta \, d\theta \, \wedge \, d\varphi
\end{equation}
This concludes the illustration of the metric.
\par
We still have to describe the parametrization of the gauge fields by means of the $\sigma$-model scalar fields. This is done in complete analogy to the case of the Kaluza-Klein vector. From the dimensional reduction procedure it follows that the $D=4$ field-strength two-forms are the following ones:
\begin{equation}\label{FLAM}
    F^\Lambda \, = \, \hat{F}^\Lambda \, + \, dZ^\Lambda \, \wedge \, \left(dt \, + \, A_{KK}\right)
\end{equation}
where $\hat{F}^\Lambda$ lives in three-dimensions and has the following components:
\begin{equation}\label{FLAij}
    \left(\hat{F}^\Lambda\right)^{ij} \, = \, \frac{e^{-2U}}{\sqrt{|\mbox{det} g_3|}} \, \left(\mbox{Im}\mathcal{N}^{-1}\right)^{\Lambda\Sigma} \, \epsilon^{ijk}\left(\partial_k Z_\Sigma \, + \, \mbox{Re}\mathcal{N}_{\Sigma\Gamma} \, \partial_k Z^\Gamma\right)
\end{equation}
This means that in the case of fields depending only on $\tau$ we find:
\begin{equation}\label{fildini}
    \left(\hat{F}^\Lambda\right)^{\theta\varphi} \, = \, \sin\theta \, \underbrace{\left[e^{-2U} \left(\mbox{Im}\mathcal{N}^{-1}\right)^{\Lambda\Sigma} \, \left(\dot{Z}_\Sigma \, + \, \mbox{Re}\mathcal{N}_{\Sigma\Gamma} \, \dot{Z}^\Gamma\right) \right]}_{p^\Lambda \, = \, \mbox{magnetic charges}}
\end{equation}
Similarly to the case of the Kaluza-Klein vector, the combinations of derivatives and fields under-braced in the above formula are constants of motion of the Lax flows and have the interpretation of magnetic charges. Indeed the magnetic charges are just the upper $n+1$ components of the full $2n+2$ vector of magnetic and electric charges.  This latter is defined as  follows:
\begin{equation}\label{cariche}
    \mathcal{Q}^M  \, = \, \sqrt{2} \,\left[ e^{-\,U} \, \mathcal{M}_4 \, \dot{Z} \, - \, \mathbf{n} \, \mathbb{C} \, Z\right]^M \, = \, \left(\begin{array}{c}
                                                                p^\Lambda \\
                                                                e_\Sigma
                                                              \end{array}
    \right)
\end{equation}
and all of its components are constants of motion.
\par
In view of this the final form of the $D=4$ field-strengths is the following one:
\begin{equation}\label{finalfildone}
    F^\Lambda \, = \, 2 \, p^\Lambda \, \sin\theta \, d\theta \wedge d\varphi \, + \, \dot{Z}^\Lambda d\tau \wedge
    \left(dt + 2\mathbf{n} \, \cos\theta \, d\varphi\right)
\end{equation}
This concludes the review of the oxidation formulae that allow to
write all the fields of $D=4$ supergravity corresponding to a
black-hole solution in terms of the fields parameterizing the
$D=3$ $\sigma$-model. It remains to be seen how such fields appear
in the coset representative $\mathbb{L}(\tau)$ for which we are
able to write Lax equations and solve them. The relation between
$U,a,\phi^i,Z^M$ is fully general and is encoded in the solvable
parametrization of the coset representative. Explicitly we set:
\begin{equation}\label{genaforma}
    \mathbb{L}(\Phi) \, = \, \exp\left[ - a \, L_+^E \right]\, \exp\left[ \sqrt{2}\, Z^M \,  \mathcal{W}_M \right] \,
    \mathbb{L}_4(\phi) \, \exp\left[ U \, L_0^E \right]
\end{equation}
where $L_0^E,L_\pm^E$ are the generators of the Ehlers group and $\mathcal{W}^M \equiv W^{1M}$; furthermore $\mathbb{L}_4(\phi)$ is the coset representative of the $D=4$ scalar coset manifold immersed in the $\mathrm{U_{D=3}}$ group.
\subsection{Extraction of the scalar fields from the coset representative $\mathbb{L}$ in the $S^3 \Leftrightarrow\mathfrak{g}_{2(2)}$ case}
Let us now consider the three-dimensional $\mathfrak{g}_{2(2)}$ description of the $SL(2,R)$ model with cubic embedding and assume that the $7\times 7$ upper triangular matrix
$\mathbb{L}(\tau)$ is the coset representative, solution of the dynamical problem. How do we extract from it the relevant scalar fields?
The answer is given by the following iterative procedure.
\par
First of all we can determine the warp factor $U$ by means of the following simple formula:
\begin{equation}\label{warpfactor}
    U (\tau)\, = \, \log \left[\ft 12 \, \mbox{Tr} \left (\mathbb{L}(\tau) \, L_+^E \, \mathbb{L}^{-1}(\tau)\, L_-^E \right)\right]
\end{equation}
Secondly we obtain the fields $\varphi$ and $y$ as follows:
\begin{eqnarray}
  \varphi(\tau) &=& -\log\left[\ft 16 \mbox{Tr} \left (\mathbb{L}^{-1}(\tau)\, L_+ \,\mathbb{L}(\tau)\, L_- \right)\right]\nonumber \\
  y(\tau) &=& \frac{\mbox{Tr} \left (\mathbb{L}^{-1}(\tau)\, L_0 \,\mathbb{L}(\tau)\, L_- \right)}{\mbox{Tr} \left (\mathbb{L}^{-1}(\tau)\, L_+ \,\mathbb{L}(\tau)\, L_- \right)}\label{fiytau}
\end{eqnarray}
and from this result we can reconstruct the behavior of the $D=4$ complex  scalar:
\begin{equation}\label{ztau}
    z(\tau) \, = \, -{\rm i} \exp \left[\varphi(\tau)\right ] \, + \, y(\tau)
\end{equation}
The knowledge of $U,\varphi,y$  allows to define:
\begin{equation}\label{omtau}
    \Omega (\tau)\, = \, \mathbb{L}(\tau)\, \exp \left[ - \, U \, L_0^E \right] \, \exp \left [- \varphi L_0\right]\, \, \exp \left [- y L_+\right]
\end{equation}
from which we extract the $Z^M$ fields by means of the following formula:
\begin{equation}\label{ZMtau}
    Z^M(\tau) \, = \, \frac{1}{2\sqrt{2}} \, \mbox{Tr} \left[ \Omega(\tau)\, \mathcal{W}_M^T\right]
\end{equation}
where $T$ means transposed. Finally the knowledge of $Z^M(\tau)$ allows to extract the $a$ field by means of the following trace:
\begin{equation}\label{atau}
    a(\tau) \, = \,-\, \ft 12 \mbox{Tr} \left[\Omega (\tau) \, \exp\left[-\sqrt{2} \, Z^M(\tau) \, \mathcal{W}_M\right]  \, L_+^E\right]
\end{equation}

\section{Attractor mechanism, the entropy and other special geometry invariants}
One of the most important features of supergravity black-holes  is
the attractor mechanism discovered in the nineties by Ferrara and
Kallosh for the case of BPS solutions \cite{ferrarakallosh} and in
recent time extended to non-BPS cases \cite{criticalrefs}.
According to this mechanism the evolving scalar fields $z^i(\tau)$
flow to fixed values at the horizon  of the black-hole ($\tau = -
\infty$), which do not depend from their initial values at
infinity radius ($\tau=0$) but only on the electromagnetic charges
$p,q$.
\par
In order to establish the connection of the quartic invariant $\mathfrak{I}_4$ defined in eq.(\ref{Jinv}) with the black-hole entropy and review the attractor mechanism, we must briefly recall the essential items of   black hole field equations in the \textit{geodesic potential approach} \cite{Gibbons:1996af}. In this framework we do not consider all the fields listed in the table after eq. (\ref{quatespace}). We  introduce only the warp factor $U(\tau)$ and the original scalar fields of $D=4$ supergravity. The information about  vector gauge fields is encoded solely in the set of electric and magnetic charges $\mathcal{Q}$ defined by eq.(\ref{chargevector}) and retrieved in eq.(\ref{cariche}).
Under these conditions the correct field equations for an $\mathcal{N}=2$ black-hole are derived from the geodesic one dimensional field-theory described by the following lagrangian:
\begin{eqnarray}
S_{eff} & \equiv & \int \, {\cal L}_{eff}(\tau) \, d\tau \quad ;
\quad \tau = -\frac{1}{r} \nonumber\\
 {\cal L}_{eff}(\tau ) & = & \ft 14 \,\left( \frac{dU}{d\tau} \right)^2 +
 g_{ij^\star} \, \frac{dz^i}{d\tau} \,  \frac{dz^{j^\star}}{d\tau} + e^{U}
 \, V_{BH}(z, {\bar z} , \mathcal{Q})
 \label{effact}
\end{eqnarray}
where the geodesic potential $V(z, {\bar z} , \mathcal{Q})$ is defined by the following formula in terms of the matrix
$\mathcal{M}_4$ introduced in eq.(\ref{geodaction}):
\begin{equation}\label{geopotentissimo}
    V_{BH}(z, {\bar z}, \mathcal{Q})\, = \, \ft 14 \, \mathcal{Q}^t \, {\cal M}_4^{-1}\left( {\cal N}\right) \, \mathcal{Q}
\end{equation}
The effective lagrangian (\ref{effact}) is derived from the
$\sigma$-model lagrangian (\ref{quaternionic}) upon substitution
of the first integrals of motion corresponding to the
electromagnetic charges (\ref{cariche}) under the condition that
the Taub-NUT charge, defined in (\ref{feroceTaubNut}),
vanishes\footnote{As we are going to see later, each orbit of Lax
operators always contains representatives such that the Taub-NUT
charge is zero. Alternatively from a dynamical system point of
view the Taub-NUT charge can be annihilated by setting a
constraint which is consistent with the hamiltonian and which
reduces the dimension of the system by one unit. The problem of
black hole physics is therefore equivalent to the sigma model
based on an appropriate codimension one hypersurface in the coset
manifold $\mathrm{G}/\mathrm{H}^\star$.} ($\mathbf{n}=0$). Indeed,
when the Taub--NUT charge $\mathbf{n}$ vanishes, which will be our
systematic choice, we can invert the above mentioned  relations,
expressing the derivatives of the $Z^M$ fields in terms of the
charge vector $\mathcal{Q}^M$ and the inverse of the matrix
$\mathcal{M}_4$. Upon substitution in the $D=3$ sigma model
lagrangian (\ref{geodaction}) we obtain the effective lagrangian
for the $D=4$ scalar fields $z^i$ and the warping factor $U$ given
by eq.s(\ref{effact}-\ref{potenzialusgeodesicus}).
\par
The important thing is that, thanks to various identities of special geometry, the effective geodesic potential admits the following alternative representation:
\begin{eqnarray}\label{potenzialusgeodesicus}
V_{BH}(z, {\bar z}, \mathcal{Q})&= &   -\,\ft 12 \,\left( \vert Z \vert ^2 +  \vert Z_i \vert ^2 \right)\equiv -\,\ft 12 \, \left (
   Z \, {\bar Z} +   Z_i g^{i\bar \jmath} \, \bar Z_{\bar \jmath} \right)
\end{eqnarray}
where the symbol $Z$ denotes the complex scalar field valued central charge of the supersymmetry algebra:
\begin{equation}\label{centralcharge}
    Z \, \equiv \, V^T \, \mathbb{C} \, \mathcal{Q} \, = \,  \, M_\Sigma \, p^\Sigma \, -\, L^\Lambda \, q_\Lambda
\end{equation}
and $Z_i$ denote its covariant derivatives:
\begin{eqnarray}
  Z_i &=& \nabla_i \, Z \, = \, U_i \, \mathbb{C} \, \mathcal{Q} \quad ; \quad Z^{\bar \jmath} =  g^{{\bar \jmath}  i} Z_i \nonumber \\
  \bar{Z}_{\bar \jmath} &=& \nabla_{\bar \jmath} \, Z \, = \, \bar{U}_{\bar \jmath} \, \mathbb{C} \, \mathcal{Q} \quad ; \quad
  {\bar Z}^i = g^{  i{\bar \jmath}} \,{\bar Z}_{\bar \jmath}
\end{eqnarray}
Eq.(\ref{potenzialusgeodesicus}) is a result in special geometry
whose proof can be found in several articles and reviews of the
late nineties\footnote{See for instance the lecture notes
\cite{richardandme}.}.
\subsection{Critical points of the geodesic potential and attractors}
The structure of the geodesic potential illustrated above allows for a detailed discussion of its critical points, which are relevant for the asymptotic behavior of the scalar fields.
\par
By definition, critical points correspond to those values of  $z^i$ for which the first derivative of the potential vanishes: $\partial_i V_{BH} \, = \, 0$. Utilizing the fundamental identities of special geometry  and eq.(\ref{potenzialusgeodesicus}), the vanishing derivative condition of the potential can be reformulated as follows:
\begin{equation}\label{criticus}
    0 \, = \, 2 \, Z_i \, {\bar Z} \, + \, {\rm i} C_{ijk} \, {\bar Z}^j\, {\bar Z}^k
\end{equation}
From this equation it follows that there are three possible types of critical points:
\begin{equation}\label{trecretini}
    \begin{array}{cccccccccccl}
      Z_i & = & 0 & ; & Z  & \ne & 0 & ; & \null &\null & \null &\mbox{BPS attractor} \\
      Z_i & \ne & 0 & ; & Z & = & 0 & ; & {\rm i} \, C_{ijk} \, \bar{Z}^j \, \bar{Z}^k & = & 0 & \mbox{non BPS attractor I} \\
      Z_i & \ne & 0 & ; & Z & \ne & 0 & ; & {\rm i} \, C_{ijk} \, \bar{Z}^j \, \bar{Z}^k & = & - \, 2 \, Z_i \, {\bar Z} & \mbox{non BPS attractor II} \\
    \end{array}
\end{equation}
It should be noted that in the case of one-dimensional special geometries, like the $S^3$-model, only BPS attractors and non BPS attractors of type II are possible. Indeed non BPS attractors of type I are forbidden  unless $C_{zzz}$ vanishes identically.
\par
In order to characterize the various type of attractors, the authors
of \cite{itanteA} and \cite{itanteB} introduced a certain number of
special geometry invariants that obey different and characterizing
relations at attractor points of different type. They are defined as
follows. Let us introduce the symbols:
\begin{equation}\label{fischietto}
    N_3 \, \equiv \, C_{ijk} \, \bar{Z}^i \, \bar{Z}^j \, \bar{Z}^k \quad ; \quad \bar{N}_3 \, \equiv \, C_{i^\star j^\star k^\star} \,
    Z^{i^\star} \, Z^{j^\star} \, Z^{k^\star}
\end{equation}
and let us set:
\begin{equation}
\begin{array}{ccccccc}
  i_1 & = & Z\, {\bar Z} & ; & i_2 & = & Z_i \, {\bar Z}_{\bar \jmath} \, g^{i{\bar \jmath}} \\
  i_3 & = & \ft 16 \, \left (Z \, N_3 \, + \, {\bar Z} {\bar N}_3 \right)& ; & i_4 & = & {\rm i} \, \ft 16 \, \left (Z \, N_3 \, - \, {\bar Z} {\bar N}_3 \right)  \\
  i_5  & = & C_{ijk} \, C_{{\bar \ell}{\bar m} {\bar n}} \,  {\bar Z}^{j} \, \, {\bar Z}^{k} \,{ Z}^{\bar m} \,{ Z}^{\bar n}
   \, g^{i {\bar \ell}}& ; & \null & \null & \null
\end{array}
\end{equation}
An important identity  satisfied by the above invariants, that depend both on the scalar fields $z^i$ and the charges $(p,q)$, is the following one:
\begin{equation}\label{i4inva}
    \mathfrak{I}_4 (p,q) \, = \, \ft 14 (i_1 - i_2)^2 + i_4 - \ft 14 \, i_5
\end{equation}
where $\mathfrak{I}_4 (p,q)$ is the quartic symplectic invariant that depends only on the charges (see eq.(\ref{Jinv})). This means that in the above combination the dependence on the fields $z^i$ cancels identically.
\par
In the case of the one-dimensional $S^3$ model there are two additional identities \cite{itanteB}  that read as follows:
\begin{equation}\label{addident}
    i_2^2 \, = \, \ft 34 \, i_5\,;\,\,i_3^2+i_4^2=4 i_1\left(\frac{i_2}{3}\right)^3 \quad ; \quad \mbox{for the $S^3$ model}
\end{equation}
In \cite{itanteA} it was proposed that the three types of critical
points can be characterized by the following relations among the
above invariants holding at the attractor point:
\paragraph{At BPS attractor points} we have:
\begin{equation}\label{forfettoBPS}
    i_1 \, \ne \, 0 \quad ; \quad i_2 = i_3 = i_4 = i_5 = 0 \quad ; \quad
\end{equation}
\paragraph{At non BPS attractor points of type I} we have:
\begin{equation}\label{forfettoI}
    i_2 \, \ne \, 0 \quad ; \quad i_1 = i_3 = i_4 = i_5 = 0
\end{equation}
\paragraph{At non BPS attractor points of type II} we have:
\begin{equation}\label{forfettoII}
    i_2 \, = \, 3 i_1 \quad ; \quad i_3 = 0 \quad ; \quad i_4 = - 2 \, i_1^2  \quad ; \quad i_5 \, = \, 12 \, i_1^2
\end{equation}
These relations follow from the definition of the critical point with the use of standard special geometry manipulations.
Their values resides in that they inform us in a simple way about the nature of  the black-hole solution we are considering.
Indeed they provide a partial classification of solution orbits since, given a configuration of charges $(p,q)$, whose structure depends, as we are going to see, from the choice of an $\mathrm{H}^\star$ orbit for the Lax operator, we can calculate the possible critical points of the corresponding geodesic potential and find out to which type they belong. We might expect several different critical points for each $(p,q)$-choice, yet it turns out that there is only one and it always belongs to the same type for all elements of the same $\mathrm{H}^\star$ orbit. This fact, whose \textit{a priori proof} has still to be given, implies that a classification of attractor points is also a partial classification of Lax operator orbits.
We shall come back  on this crucial issue later on. Yet it is appropriate to emphasize the word \textit{partial classification}.
Although the type of fixed point is the same for each element of the same orbit we should by no means assume that fixed point types select orbits. Indeed there are Lax operators belonging to different $\mathrm{H}^\star$ orbits that have the same electromagnetic charges and therefore define the same fixed point. Furthermore the fact that a Lax operator defines certain charges and hence an associated fixed point does not imply that the solution generated by such Lax will necessarily reach that fixed point. As we explicitly show later on, the solution can break up at a finite value of $\tau$, stopping before the fixed point is attained. Hence the classification of fixed points is not a classification of $\mathrm{H}^\star$ orbits although the two classifications have partial relations to each other.
\subsection{Fixed scalars at BPS attractor points}\label{bpsharmonicgeneral}
In the case of BPS attractors we can find the explicit expression in terms of the (p,q)-charges for the scalar field fixed values   at the critical point.
\par
By means of standard special geometry manipulations the BPS critical point equation
\begin{equation}\label{fixedpoints}
    \nabla_j Z \, = \, 0 \quad ; \quad \nabla_{\bar \jmath}\bar Z \, = \, 0
\end{equation}
can be rewritten in the following celebrated form which, in the late nineties, appeared in numerous research and review papers (see for instance \cite{richardandme}):
\begin{eqnarray}
p^\Lambda & = & \mbox{i}\left( Z_{fix}\,{\bar L}^\Lambda_{fix}
- {\bar Z}_{fix}\,L^\Lambda_{fix} \right)\\
q_\Sigma & = & \mbox{i}\left( Z_{fix}\,{\bar M}_\Sigma^{fix}
- {\bar Z}_{fix}\,M_\Sigma^{fix} \right)\\
\label{minima}
\end{eqnarray}
Using the explicit form of the symplectic section $\Omega(z)$ given in eq.(\ref{standasec}), we can easily solve eq.s (\ref{minima}) for the $S^3$ model and obtain the following fixed scalars:
\begin{equation}\label{fixedzfiel}
    z_{fixed} \, = \, -\frac{p_1 q_1+3 p_2 q_2+{\rm i } \,6 \,\sqrt{\mathfrak{I}_4(p,q)}}{2
   \left(q_1^2+\sqrt{3} p_1 q_2\right)}
\end{equation}
where $\mathfrak{I}_4(p,q)$ is the quartic invariant defined in eq.(\ref{Jinv}). In fact, following \cite{Behrndt:1997ny,iomatteo}, one can give the BPS solution in a closed form by replacing in the expression  (\ref{fixedzfiel}) $ z_{fixed}$ the quantized charges with harmonic functions
\begin{eqnarray}
q_\Lambda &\rightarrow & H_\Lambda\equiv h_\Lambda - \sqrt{2}\,q_\Lambda \,\tau\,\,;\,\,\,p^\Lambda \rightarrow  H^\Lambda\equiv h^\Lambda - \sqrt{2}\,p^\Lambda \,\tau
\end{eqnarray}
The same substitution allows to describe the radial evolution of the warp factor:
\begin{eqnarray}
e^{-U}&=&\frac{1}{2}\,\sqrt{\mathfrak{I}_4(H^\Lambda,H_\Lambda)}
\end{eqnarray}
The constants $h^\Lambda,\,h_\Lambda$ in the harmonic functions are subject to two conditions: one originates from the requirement of asymptotic flatness ($\lim_{\tau\rightarrow 0^-} e^U=1$), while the other reads $h^\Lambda q_\Lambda-h_\Lambda p^\Lambda=0$. The remaining two free parameters are fixed by the choice of the value of $z$ at radial infinity.
The fixed value $S_{fixed}$ of the scalar $S=-1/z$, defining the conventional parametrization of the $D=4$ manifold discussed in subsection \ref{ctsps3m}, reads:
\begin{eqnarray}
S_{fixed}&=&-\frac{1}{z_{fixed}}=\frac{p_1 q_1+3 p_2 q_2-{\rm i } \,6 \,\sqrt{\mathfrak{I}_4(p,q)}}{2
   \left(p_1^2-\sqrt{3} p_1 q_2\right)}=\frac{P^1 Q_1+3 P^0 Q_0-3\,{\rm i }  \,\sqrt{\mathfrak{I}_4(P,Q)}}{2
   \left(3\,(P^1)^2-Q_0\,P^1\right)}
\end{eqnarray}
where the charges $P^\Lambda,\,Q_\Lambda$ were defined in section \ref{tqi}.\par
By replacing the fixed values (\ref{fixedzfiel}) into the expression (\ref{potenzialusgeodesicus}) for the potential we find:
\begin{equation}\label{fantastico}
    V_{BH}\left(z_{fixed}, {\bar z}_{fixed} \, , \,\mathcal{Q}\right)\, = \, - \, \sqrt{\mathfrak{I}_4(p,q)}
\end{equation}
The above result implies that the horizon area in the case of an extremal  BPS black-hole is proportional to the square root of $\mathfrak{I}_4(p,q)$ and, as such, depends only on the charges. The argument goes as follows.
\par
Consider the behavior of the warp factor $\exp[-U]$ in the vicinity of the horizon, when $\tau \, \rightarrow \, - \, \infty$.
For regular black-holes the near horizon metric must factorize as follows:
\begin{equation}\label{nearhorizon}
    ds^2_{\mbox{near hor.}} \, \approx \, \underbrace{- \, \frac{1}{r_H^2 \, \tau^2} \, dt^2 \, + \, {r_H^2} \, \left(\frac{d\tau}{\tau} \right)^2}_{\mbox{$\mathrm{AdS}_2$ metric}}
    \, + \, \underbrace{r_H^2 \, \left(d\theta^2 \, \sin^2\theta \, d\phi^2\right)}_{\mbox{$\mathrm{S}^2$ metric}}
\end{equation}
where $r_H$ is the Schwarzschild radius defining the horizon. This implies that the asymptotic behavior of the warp factor, for $\tau \, \rightarrow \, - \, \infty$ is the following one:
\begin{equation}\label{asympo}
    \exp[-U] \, \sim \, r_H^2\,\tau^2
\end{equation}
In the same limit the scalar fields go to their fixed values and their derivatives become essentially zero. Hence near the horizon we have:
\begin{equation}\label{fortesoffia}
    \left(\dot{U}\right)^2 \, \approx \, \frac{4}{\tau^2} \quad ; \quad g_{ij^\star} \, \frac{dz^i}{d\tau} \,  \frac{dz^{j^\star}}{d\tau} \,\approx \, 0 \quad ; \quad e^{U}
 \, V_{BH}(z, {\bar z} , \mathcal{Q}) \,\approx \, \frac{1}{r^2_H \, \tau^2} \, V\left(z_{fixed}, {\bar z}_{fixed} \, , \,\mathcal{Q}\right)
\end{equation}
Since for extremal black-holes the sum of the above three terms vanishes (see eq.(\ref{beddamatre})), we conclude that:
\begin{equation}\label{frescobal1}
    r_H^2 \, = \, - \,V_{BH}\left(z_{fixed}, {\bar z}_{fixed} \, , \,\mathcal{Q}\right)
\end{equation}
which yields
\begin{equation}\label{area}
    \mathrm{Area}_H \, = \, 4\, \pi \, r_H^2 \, = \, 4\, \pi \,\sqrt{\mathfrak{I}_4(p,q)}
\end{equation}
\subsection{Fixed values for non BPS attractors}
In the case of non-BPS attractor points it is much more difficult to
provide general rules and an analysis case by case is mandatory. For
this reason we focus on the case of the $S^3$ model and we
anticipate the results of the orbit classification provided in
sections \ref{classificazia} and \ref{scanning} in order to study
the associated attractor structures. We begin with extremal
black-holes that correspond to nilpotent orbits of Lax operators. We exclude from this preliminary analysis  the orbits to be denoted by $\mathrm{NO'_3}$ and $\mathrm{NO'_4}$, see later discussion.
\subsubsection{The nilpotent orbit $\mathrm{NO_1}$}
Referring to eq.(\ref{qinp}) we see that at vanishing Taub-NUT charge the first nilpotent orbit is characterized by the following structure of the electro-magnetic charge vector:
\begin{equation}\label{Qorbit1}
    \mathcal{Q} \, = \, \left\{p_1,p_2,\frac{p_1^2}{\sqrt{3}
   p_2},-\frac{p_1^3}{3 \sqrt{3} p_2^2}\right\}
\end{equation}
Inserting this values in the expression for the quartic invariant (\ref{Jinv}) we find that in this case it vanishes $\mathfrak{I}_4=0$. On the other hand inserting (\ref{Qorbit1}) into the expression for $\nabla_zZ$ we see that there is no value of the field $z$ for which $Z_z$ vanishes. Hence no BPS attractor point exists in this case. On the other
hand inserting the $Q$-vector (\ref{Qorbit1}) into eq.(\ref{criticus}) we find a non-BPS type II critical value at
\begin{equation}\label{zetafisson01}
    z_{fix} \, = \, -\frac{\sqrt{3} p_2}{p_1}
\end{equation}
This point, however, is on the real axis and therefore it lies on the boundary of the Lobachevskiy-Poincar\'e lower half-plane. In the Poincar\'e metric it lies at an infinite distance from all interior points. Evaluating the $i$-invariants at this limiting point we find:
\begin{equation}\label{invariO1}
    i_1 \, = \, i_2 \, = \, i_3 \, = \, i_4 \, = \, i_5 \, = \, 0
\end{equation}
We conclude that from the attractor view point, the $\mathrm{NO}_1$
orbit  is characterized by the vanishing of all special geometry
invariants and has a non-BPS attractor point on the real axis at
infinite distance. The corresponding solutions will describe
\textit{small black holes}.
\subsubsection{The nilpotent orbit $\mathrm{NO_2}$}
\label{ninpotinaNO2} From the point of view of electromagnetic
charges $(p,q)$ the second nilpotent orbit is characterized only
by the constraint that the quartic invariant should be zero. In
this sense it is a deformation of the first nilpotent orbit that
removes the second identity satisfied by the charges. This
observation provides a convenient way of parameterizing the charge
vector, suitable for the analysis of critical points and
attractors. We can set:
\begin{equation}\label{Qorbit2}
    \mathcal{Q} \, = \, \left\{p_1,p_2,\frac{\left(1-\gamma ^2\right) p_1^2}{\sqrt{3} p_2},\frac{\left(-2 \gamma ^3+3 \gamma ^2-1\right) p_1^3}{3 \sqrt{3} p_2^2}\right\}
\end{equation}
For arbitrary values of the two magnetic charges $p_1,p_2$ and for arbitrary values of the deformation parameter $\gamma$ the quartic invariant is zero. For $\gamma=0$ the charge vector of the second nilpotent orbit degenerates into that of the first.
By straightforward algebraic manipulations, we can verify that the critical point equation (\ref{criticus}) takes, upon substitution of eq.(\ref{Qorbit2}) the following form:
\begin{equation}\label{equizia}
    0 \, = \, \frac{1}{216 y^4 p_2^4} \, A(x,y,\gamma)
\end{equation}
where the numerator is given by
\begin{eqnarray}
  A(x,y,\gamma) &=& 6 \left(i (x-i y)^4 (x+i y)^2 (\gamma -1)^4 (2 \gamma +1)^2 p_1^6+\right.\nonumber\\
  &&\left. 2 \sqrt{3} (x-i y)^3 \left(-3 i x^2+4 y x+i y^2\right) (\gamma -1)^3 \left(2 \gamma ^2+3
   \gamma +1\right) p_2 p_1^5 \right. \nonumber\\
   &&\left.+3 i (x-i y)^2 (\gamma -1)^2 \left(3 \left(3 \gamma ^2+10 \gamma +5\right) x^2 +2 i y \left(3 \gamma ^2+10 \gamma +5\right)
   x\right.\right.\nonumber\\
   &&\left.\left.+y^2 (\gamma +1)^2\right) p_2^2 p_1^4+12 \sqrt{3} \left(i \left(\gamma ^3-6 \gamma ^2+5\right) x^3+y \left(\gamma ^3-6 \gamma ^2+5\right) x^2\right.\right.\nonumber\\
   &&\left.\left.-i y^2
   \left(\gamma ^2-1\right) x-y^3 \left(\gamma ^2-1\right)\right) p_2^3 p_1^3\right.\nonumber\\
   &&\left.+9 \left(-3 i \left(2 \gamma ^2-5\right) x^2+2 y \left(5-2 \gamma ^2\right)
   x+i y^2\right) p_2^4 p_1^2\right.\nonumber\\
   &&\left.+18 \sqrt{3} (3 i x+y) p_2^5 p_1+27 i p_2^6\right)\label{axygamma}
  \end{eqnarray}
the symbols $x$ and $y$  respectively denoting, the real and imaginary parts of the complex field $z$.
With simple manipulations one can derive the result that for real $x$ and $y$ the only zero of the function $A(x,y,\gamma)$ is given by:
\begin{equation}\label{findone}
    y\, = \, 0 \quad; \quad x \, = \,-\frac{\sqrt{3} p_2}{2 \gamma  p_1+p_1}
\end{equation}
Once again this point is on the boundary of the scalar field domain and moreover it is also a zero of the denominator in eq.(\ref{equizia}). In order to make sense it is necessary that approaching this point from the lower half-plane the limit should be a true zero of the critical point equation. Hence we consider
\begin{eqnarray}\label{caroide}
    &\frac{1}{216 \epsilon^4 p_2^4} \, A(-\frac{\sqrt{3} p_2}{2 \gamma  p_1+p_1},\epsilon,\gamma)
    \,=\,&\nonumber\\
    &-\frac{\left(-2 \gamma ^2+\gamma +1\right)^4 \epsilon ^4 p_1^4+4 i \sqrt{3} \gamma  \left(-2 \gamma ^2+\gamma +1\right)^3 \epsilon ^3 p_2 p_1^3+12 \gamma
   ^2 \left(-2 \gamma ^2+\gamma +1\right)^2 \epsilon ^2 p_2^2 p_1^2-72 i \sqrt{3} \gamma ^3 \left(2 \gamma ^2-\gamma -1\right) \epsilon  p_2^3 p_1-81 \gamma ^4
   p_2^4}{36 (2 \gamma  \epsilon +\epsilon )^2 \frac{p_2^4}{p_1^2}}&\nonumber\\
\end{eqnarray}
and we can verify that:
\begin{equation}\label{duelimiti}
    \lim_{\epsilon \rightarrow 0}\frac{1}{216 \epsilon^4 p_2^4} \, A(-\frac{\sqrt{3} p_2}{2 \gamma  p_1+p_1},\epsilon,\gamma)
    \, = \, \cases{0 \quad \,\,\,\mbox{If $\gamma=0$}\cr
    \infty \quad \mbox{If $\gamma\ne0$}\cr}
\end{equation}
We conclude that there is a critical point on the boundary only for the first nilpotent orbit. This finds confirmation in the calculation of the invariants $i_1 \, \dots, i_5$. While at $\gamma = 0$ they are all zero, at $\gamma\ne 0$ they are all divergent
while $z$ approaches the would be critical value of eq. ($\ref{findone}$).
\par
We conclude that for the second nilpotent orbit there is no
attractor point. Not even on the boundary. The metric is in any case
that of a \textit{small black hole}.
\subsubsection{The nilpotent orbits $\mathrm{NO_3}$ and $\mathrm{NO_4}$}
As we shall demonstrate later on, in the case of the $S^3$ model,
the classification of nilpotent $\mathrm{H}^\star$ orbits for the
Lax operator includes two large orbits one of which contains BPS
attractors while the second contains non-BPS attractors of type
II. At vanishing Taub-NUT charge convenient representatives for
both cases are characterized by the charge vector of the following
form:
\begin{equation}\label{pqcasi}
    \left \{\begin{array}{c}
              p_1 \\
              p_2 \\
              q_1 \\
              q_2
            \end{array}
     \right \} \, = \, \left \{\begin{array}{c}
              0 \\
              p \\
              \sqrt{3} q \\
              0
            \end{array}
     \right \}  \,\Leftarrow\,\cases{ p > 0 \, , \, q< 0 \quad \mbox{or} \quad  p < 0 \, , \, q> 0 \quad \mbox{BPS}\cr
     p > 0 \, , \, q> 0 \quad \mbox{or} \quad  p < 0 \, , \, q< 0 \quad \mbox{non BPS} \cr}
\end{equation}
Postponing details of the supergravity solution to sections \ref{antiBPSsino} and \ref{BPSsino} let us consider the solution of the attractor equations (\ref{criticus}) with the above charge vector.
\paragraph{Non BPS case} For $p$ and $q$ having the same sign it is easily verified that there is no solution of the equation $Z_z =0$ and hence no BPS attractor point. On the other hand there is a solution of  the critical point equation (\ref{criticus}) with both $Z_z \ne 0$ and $Z\ne0$. It corresponds to the following simple fixed value:
\begin{equation}\label{zetafissuNBPS}
    z_{fixed} \, = \, -{\rm i} \, \sqrt{\frac{p}{q}}
\end{equation}
With such fixed value the $i$-invariant take the following values:
\begin{equation}\label{invinonbibis}
\left\{i_1,i_2,i_3,i_4,i_5\right\}\, = \,    \left\{\frac{1}{2}
   \sqrt{\frac{p}{q}}
   q^2\, ,\,\frac{3}{2}
   \sqrt{\frac{p}{q}}
   q^2\, ,\,0\, ,\,-\frac{p q^3}{2},3 p
   q^3\right\}
\end{equation}
which satisfy the relations (\ref{forfettoII}) characterizing a non-BPS attractor point of type II. Furthermore the quartic invariant $\mathfrak{I}_4(p,q) \, = \, - p\, q^3 < 0$ is negative in this case and we expect that the horizon area will be proportional to $\sqrt{-\mathfrak{I}_4}$. This will indeed be the case.
\paragraph{BPS case} If $p$ and $q$ have opposite signs there is just one solution of the equation $Z_z \, = \, 0$ with $Z\ne 0$. Hence we a have a BPS attractor. The fixed point is:
\begin{equation}\label{cogni}
    z_{fixed} \, = \,- {\rm i} \, \sqrt{-\frac{p}{q}}
\end{equation}
which perfectly fits the general formula (\ref{fixedzfiel}). Moreover calculating the $i$-invariants at the fixed  point we obtain:
\begin{equation}\label{invinonbibis2}
\left\{i_1,i_2,i_3,i_4,i_5\right\}\, = \,    \left\{ 2 \sqrt{-p \, q^3}\, ,\,0\, ,\,0\, ,\,0\, , \, 0\right\}
\end{equation}
which  fulfills the relations (\ref{forfettoI}) proper of the BPS attractors.
\subsubsection{The largest nilpotent orbit $\mathrm{NO}_5$}
\label{larghezza}
As pointed out in section \ref{frintola} a representative of the fifth nilpotent orbit at vanishing Taub-NUT charge can be chosen such that the coresponding charges are the following ones:
\begin{equation}\label{frunta}
   \left(
\begin{array}{l}
 p_1\\
 p_2 \\
 q_1 \\
 q_2
\end{array}
\right)
 \, = \, \left(
\begin{array}{l}
 - \sqrt{3}\\
 \sqrt{3} \\
 1 \\
 -1
\end{array}
\right)
\end{equation}
With such a datum we look for the possible critical points. There are none of BPS type but there exists one of non BPS type II. It corresponds to the following point of the lower half-plane:
\begin{equation}\label{fissuspunctus}
    z_{fixed} \, = \, \frac{1}{17} \sqrt{3} \left(1-2
   \sqrt[3]{2}+4
   2^{2/3}\right)-\frac{1}{17} i
   \sqrt{6 \left(24-31
   \sqrt[3]{2}+28 2^{2/3}\right)}
\end{equation}
Calculating the corresponding $i$-invariants at the fixed point we find:
\begin{equation}\label{comptibo}
 \left\{i_1,i_2,i_3,i_4,i_5\right\}\, = \,     \left\{\frac{1}{\sqrt{3}},\sqrt{3
   },0,-\frac{2}{3},4\right\}
\end{equation}
which do indeed satisfy the relations (\ref{forfettoII}) proper of
the non BPS attractors of type II. Hence we might naively assume
that the solution generated by this Lax operator flows at such a
critical point. However it is not so. As we demonstrate in later
sections the solution generated by Lax operators of the fifth
nilpotent orbit are broken solutions describing small black holes
and they do not flow to fixed points. On the contrary we can
construct Lax operators of the third nilpotent orbit that have the
same charges and correctly flow to that non BPS fixed point.
\section{The integration algorithm and the Lax equation}
\label{integralgoritmo}
In section \ref{solvapara} we have seen how the explicit form of
the supergravity fields can be extracted from a solvable coset
representative $\mathbb{L}(\tau)$ which satisfies the field
equations of the associated $\sigma$-model. This means that at all
times $\mathbb{L}(\tau)$  is an element of the solvable upper
triangular group, in the present case of the Borel subgroup of
$\mathrm{G_{2(2)}}$. At the same time the corresponding left
invariant one-form:
\begin{equation}\label{fregata}
    \Sigma (\tau) \, \equiv \, \mathbb{L}^{-1}(\tau)\, \frac{d}{d\tau} \mathbb{L}(\tau)
\end{equation}
decomposes as follows:
\begin{equation}\label{finocchio}
    \Sigma (\tau) \, = \, L(\tau) \, \oplus \, W(\tau)
\end{equation}
where:
\begin{eqnarray}
  W(\tau) \, \in \, \mathbb{H}^\star&\Rightarrow & \eta \, W^T(\tau) + W(\tau) \eta \, = \,0 \nonumber\\
  L(\tau) \, \in \, \mathbb{K} &\Rightarrow & \eta \, L^T(\tau) - L(\tau) \eta \, = \,0 \nonumber\\
\end{eqnarray}
moreover we have that
\begin{equation}\label{Rmatrice}
    W(\tau) \, = \, L_>(\tau) \, - \, L_<(\tau)
\end{equation}
where $L_>$ and $L_<$ respectively denote the upper and lower triangular parts of the Lax operator $L(\tau)$ which is requested to satisfy the Lax equation:
\begin{equation}\label{laxequation}
    \frac{d}{d\tau} L(\tau) \, = \, \left [ W(\tau) \, , \, L(\tau) \right ]
\end{equation}
The initial conditions for the solution of such a problem are given by specifying both the Lax operator $L_0$ at a reference time $\tau = 0$ and the coset representative $\mathbb{L}_0$ at the same reference time. In \cite{Fre:2009dg}
and \cite{noiultimo}  a close analytical algorithm was presented that
provides the solution of the above problem and hence of supergravity in terms of the given initial conditions. Let us recall the conclusive formulae of that algorithm.
\par
Given $L_0$ let us define the following matrix-function:
\begin{equation}\label{Cfunzia}
    \mathcal{C}(\tau)\, : = \, \exp\left[ - 2 \, \tau \, L_0\right]
\end{equation}
and the following determinants:
\begin{equation}\label{DktN}
    \mathfrak{D}_{i}(\mathcal{C}) \, := \, \mbox{Det} \, \left ( \begin{array}{ccc}
    \mathcal{C}_{1,1}(\tau) & \dots & \mathcal{C}_{1,i}(\tau)\\
    \vdots & \vdots & \vdots \\
    \mathcal{C}_{i,1}(\tau) & \dots & \mathcal{C}_{i,i}(\tau)
    \end{array}\right)  \, , \quad
    \mathfrak{D}_{0}(\tau):=1 \,.
    \end{equation}
The matrix elements of the inverse of the coset representative satisfying the required second order equation are given by the following compact and very elegant formula:
\begin{eqnarray}\label{solutioncosetrepr}
\left(\mathbb{L}(\tau)^{-1}\right)_{ij}
 &\equiv&\frac{1}{\sqrt{\mathfrak{D}_i(\mathcal{C})\mathfrak{D}_{i-1}(\mathcal{C})}}\mathrm{Det}\left(\begin{array}{cccc} \mathcal{C}_{1,1}(\tau)&\dots
&\mathcal{C}_{1,i-1}(\tau)&
(\mathcal{C}(\tau)\mathbb{L}(0)^{-1})_{1,j}\\
\vdots&\vdots&\vdots&\vdots\\
\mathcal{C}_{i,1}(\tau)&\dots &
\mathcal{C}_{i,i-1}(\tau)& (\mathcal{C}(\tau)\mathbb{L}(0)^{-1})_{i,j}\\
\end{array}\right)\nonumber
\end{eqnarray}
where $\mathbb{L}(0) \, \equiv \, \mathbb{L}_0$   encodes the value of the coset representative at the reference time and hence the second set of boundary conditions.
\subsection{Consideration on Liouville integrability and $\mathrm{H}^\star$ orbits}
The very fact that we can write a closed form integral shows that the considered dynamical system is integrable. Hence its Liouville integrability should be guaranteed by the existence of the appropriate number of conserved hamiltonians in involution. The derivation and study of such hamiltonians is not only a matter of principle but also a very important tool in the classification of the possible supergravity solutions. Given the very structure of the integration algorithm the primary goal of such a classification consists of the classification of orbits of initial
Lax operators $L_0$ under the action of the subgroup $\mathrm{H}^\star$. Indeed an important property of the integration algorithm is that the value of the Lax operator $L(\tau)$ at  time $\tau$ is given by the following conjugation of the initial Lax operator $L_0$:
\begin{equation}\label{algoronegeneral1}
    L(\tau) \, = \,
   \mathcal{Q}(\mathcal{C})\, L_0\,
   \left(\mathcal{Q}(\mathcal{C})\right)^{-1}\,
\end{equation}
where
\begin{equation}\label{staidentro}
    \mathcal{Q}(\mathcal{C}) \in \mathrm{H}^\star
\end{equation}
and its matrix elements have the following explicit form:
\begin{eqnarray} \label{Gensolut2}
\mathcal{Q}_{ij}(\mathcal{C}) & \equiv
&\frac{1}{\sqrt{\mathfrak{D}_i(\mathcal{C})\mathfrak{D}_{i-1}(\mathcal{C})}}
\, \mathrm{Det} \, \left(\begin{array}{cccc}
\mathcal{C}_{1,1}(\tau)&\dots &\mathcal{C}_{1,i-1}(\tau)& (\mathcal{C}^{\frac{1}{2}}(\tau))_{1,j}\\
\vdots&\vdots&\vdots&\vdots\\
\mathcal{C}_{i,1}(\tau)&\dots &
\mathcal{C}_{i,i-1}(\tau)& (\mathcal{C}^{\frac{1}{2}}(\tau))_{i,j}\\
\end{array}\right)
\end{eqnarray}
Hence if the Lax operator belongs to a given $\mathrm{H}^\star$-orbit at some time it belongs to the same orbit at all times and classification of orbits is a classification of solutions. Therefore if we arrive at a classification of the $\mathrm{H}^\star$ orbits by means of Liouville hamiltonians and associated structures such a classification will also be a valuable classification of supergravity solutions.
\par
Alternatively the evolving Lax operator can be written as a Borel transform of the Lax operator at the initial time.
Instead of eq.(\ref{algoronegeneral1}) one can also write:
\begin{equation}\label{sborellando}
    L(\tau) \, = \, \chi_{>}(\tau) \, L_0 \, \chi_{>}^{-1}(\tau)
\end{equation}
where:
\begin{equation}\label{chipiu}
    \chi_{>}(\tau) \, \in \, \mbox{Borel}\left (\mathrm{U_{D=3}}\right)
\end{equation}
is an upper triangular matrix belonging to the Borel subgroup of the $D=3$ $\mathrm{U}$-duality group.
The explicit form of the matrix $\chi_{>}(\tau) $ is the following one:
\begin{equation}\label{fammiricordo}
    (\chi_{>}(\tau))_{i,j} \, = \,
\frac{1}{\sqrt{\mathfrak{D}_i(\mathcal{C})\mathfrak{D}_{i-1}(\mathcal{C})}}
\, \mathrm{Det} \, \left(\begin{array}{cccc}
\mathcal{C}_{1,1}(\tau)&\dots &\mathcal{C}_{1,i-1}(\tau)&
\mathcal{C}_{1,j}(\tau)\\
\vdots&\vdots&\vdots&\vdots\\
\mathcal{C}_{i,1}(\tau)&\dots &
\mathcal{C}_{i,i-1}(\tau)& \mathcal{C}_{i,j}(\tau)\\
\end{array}\right)
\end{equation}
The above information provides a tool to construct the requested number of hamiltonians in involution as we show later on.
\section{The $\mathbb{H}^\star$ subalgebra and the underlying integrable dynamical system}
In order to study the formal structure of the dynamical system encoded in the Lax pair representation discussed in the previous section, we need to analyze the algebraic structure of the $\mathbb{H}^\star$-decomposition of the $\mathfrak{g}_{2(2)}$ Lie algebra which defines the coset manifold of the relevant $\sigma$-model. This will enable us to interpret the group theoretical structure of the conserved hamiltonians which characterize the various orbits of possible Lax operators and provide Liouville integrability of the dynamical system defined on each of them.
\subsection{Definition of $\mathbb{H}^\star$}
The starting point of our group theoretical analysis is
the $7$-dimensional metric which is invariant with respect to the $\mathbb{H}^\star$ subalgebra:
\begin{equation}\label{Hstarra}
   \mathfrak{g}_{2(2)}\, \supset \, \mathbb{H}^\star \, \equiv \, \slal(2,\mathbb{R}) \, \oplus \, \slal(2,\mathbb{R})
\end{equation}
in the fundamental representation of $\mathfrak{g}_{2(2)}$. Such a metric is the  following one:
\begin{equation}\label{etag22}
    \eta \, = \, \left(
\begin{array}{lllllll}
 -1 & 0 & 0 & 0 & 0 & 0 & 0 \\
 0 & -1 & 0 & 0 & 0 & 0 & 0 \\
 0 & 0 & 1 & 0 & 0 & 0 & 0 \\
 0 & 0 & 0 & 1 & 0 & 0 & 0 \\
 0 & 0 & 0 & 0 & 1 & 0 & 0 \\
 0 & 0 & 0 & 0 & 0 & -1 & 0 \\
 0 & 0 & 0 & 0 & 0 & 0 & -1
\end{array}
\right)
\end{equation}
Consequently the generators of the stability subalgebra $\mathbb{H}^\star$, whose  abstract structure is mentioned in eq.(\ref{Hstarra}), are those elements  $\mathfrak{h} \in \mathfrak{g}_{2(2)}$  that satisfy the following equation:
\begin{equation}\label{helementi}
    \eta \, \mathfrak{h} \, + \, \mathfrak{h}^T \, \eta \, = \, 0
\end{equation}
The  complementary subspace $\mathbb{K}$ defined by the orthogonal decomposition:
\begin{equation}\label{ortodecompo}
    \mathfrak{g}_{2(2)} \, = \, \mathbb{H}^\star \, \oplus \, \mathbb{K}
\end{equation}
contains all those elements $\mathfrak{k} \in \mathfrak{g}_{2(2)}$ that fulfill the opposite condition:
\begin{equation}\label{kelementi}
    \eta \, \mathfrak{k} \, - \, \mathfrak{k}^T \, \eta \, = \, 0
\end{equation}
It is convenient to introduce both for $\mathbb{H}^\star$ and for
$\mathbb{K}$ a basis of generators which makes references to the
standard Cartan-Weyl basis of the ambient algebra, or even better
to the Chevalley triples.
\par
As basis of the $\mathbb{H}^\star$ subalgebra we use the following
six linear combinations of step operators that satisfy the
required condition (\ref{helementi}):
\begin{equation}\label{hgenitori}
    \begin{array}{lll}
 h_1 & = & {e_2}+{f_2} \\
 h_2 & = & {e_1}-{f_1} \\

 h_3 & = & {e_3}+{f_3} \\
 h_4 & = & {e_4}+{f_4} \\
 h_5 & = & {e_5}+{f_5} \\
 h_6 & = & {e_6}-{f_6}
\end{array}
\end{equation}
while as basis of the $\mathbb{K}$ complementary subspace we use the  following eight combinations that satisfy eq.(\ref{kelementi}):
\begin{equation}\label{Kgeneri}
    \begin{array}{lclll}
 \mathbf{K}_1 &=&  \mathcal{H}_1&\null&\null\\
  \mathbf{K}_2 &=&   \mathcal{H}_2&\null&\null\\
 \mathbf{K}_3 &=& k_1 & = & {e_2}-{f_2} \\
\mathbf{K}_4 &=&  k_2 & = & {e_1}+{f_1} \\
\mathbf{K}_5 &=&  k_3 & = & {e_3}-{f_3} \\
\mathbf{K}_6 &=&  k_4 & = & {e_4}-{f_4} \\
\mathbf{K}_7 &=&  k_5 & = & {e_5}-{f_5} \\
\mathbf{K}_8 &=&  k_6 & = & {e_6}+{f_6}
\end{array}
\end{equation}
Using this basis, the Lax operator which, by definition, is an element of the $\mathbb{K}$-subspace can be written as follows:
\begin{equation}\label{loffo}
    L\, = \, \sum_{A=1}^8 \, t^A \, \mathbf{K}_A \, = \, \left(
\begin{array}{lllllll}
 t_2 & t_4 & t_5 & -\sqrt{2} t_6 & -t_7 & -t_8 & 0 \\
 t_4 & t_1-t_2 & t_3 & -\sqrt{2} t_5 & t_6 & 0 & -t_8 \\
 -t_5 & -t_3 & 2 t_2-t_1 & \sqrt{2} t_4 & 0 & -t_6 & -t_7 \\
 \sqrt{2} t_6 & \sqrt{2} t_5 & \sqrt{2} t_4 & 0 & \sqrt{2} t_4 & -\sqrt{2}
   t_5 & \sqrt{2} t_6 \\
 t_7 & -t_6 & 0 & \sqrt{2} t_4 & t_1-2 t_2 & -t_3 & t_5 \\
 -t_8 & 0 & t_6 & \sqrt{2} t_5 & t_3 & t_2-t_1 & t_4 \\
 0 & -t_8 & t_7 & -\sqrt{2} t_6 & -t_5 & t_4 & -t_2
\end{array}
\right)
\end{equation}
This introduces the first set of coordinates $t^A$ on $\mathbb{K}$.
\subsection{The Borel subalgebra and its Poissonian structure}
According to the viewpoint introduced in \cite{noiultimo} we  consider the canonical Poissonian structure defined on the solvable Lie algebra of the coset $Solv\left(\mathrm{G}/\mathrm{H}^\star\right) \, = \, \mathrm{Borel}\left(\mathfrak{g}_{2(2)}\right)$.
\par
A standard set of generators of this solvable Lie algebra is the following one:
\begin{equation}\label{furioso}
    \begin{array}{ccccccc}
         T_1 & = & H_1 & ; & T_2 & = & H_2 \\
         T_3 & = & E^{\alpha_1} & ; & T_4 & = & E^{\alpha_2}\\
         T_5 & = & E^{\alpha_3} & ; & T_6& = & E^{\alpha_4}\\
         T_7 & = & E^{\alpha_5} & ; & T_8& = & E^{\alpha_6}\\
       \end{array}
    \end{equation}
and the Lax operator (\ref{loffo}) can be alternatively defined as follows:
\begin{eqnarray}
 && \mathfrak{B} \,=\, \sum_{A=1}^8 \Phi^A \, T_A \, \in \, \mathrm{Borel}\left(\mathfrak{g}_{2(2)}\right)
  \quad ; \quad  L \, = \,  \mathfrak{B} \, + \, \eta \, \mathfrak{B}^T \, \eta \, = \nonumber\\
  &&  \left(
\begin{array}{lllllll}
 \Phi _1+\sqrt{3} \Phi _2 & \frac{\Phi _3}{\sqrt{2}} & \frac{\Phi
   _5}{\sqrt{2}} & -\Phi _6 & -\sqrt{\frac{3}{2}} \Phi _7 &
   -\sqrt{\frac{3}{2}} \Phi _8 & 0 \\
 \frac{\Phi _3}{\sqrt{2}} & \sqrt{3} \Phi _2-\Phi _1 & \sqrt{\frac{3}{2}}
   \Phi _4 & -\Phi _5 & \frac{\Phi _6}{\sqrt{2}} & 0 & -\sqrt{\frac{3}{2}}
   \Phi _8 \\
 -\frac{\Phi _5}{\sqrt{2}} & -\sqrt{\frac{3}{2}} \Phi _4 & 2 \Phi _1 & \Phi
   _3 & 0 & -\frac{\Phi _6}{\sqrt{2}} & -\sqrt{\frac{3}{2}} \Phi _7 \\
 \Phi _6 & \Phi _5 & \Phi _3 & 0 & \Phi _3 & -\Phi _5 & \Phi _6 \\
 \sqrt{\frac{3}{2}} \Phi _7 & -\frac{\Phi _6}{\sqrt{2}} & 0 & \Phi _3 & -2
   \Phi _1 & -\sqrt{\frac{3}{2}} \Phi _4 & \frac{\Phi _5}{\sqrt{2}} \\
 -\sqrt{\frac{3}{2}} \Phi _8 & 0 & \frac{\Phi _6}{\sqrt{2}} & \Phi _5 &
   \sqrt{\frac{3}{2}} \Phi _4 & \Phi _1-\sqrt{3} \Phi _2 & \frac{\Phi
   _3}{\sqrt{2}} \\
 0 & -\sqrt{\frac{3}{2}} \Phi _8 & \sqrt{\frac{3}{2}} \Phi _7 & -\Phi _6 &
   -\frac{\Phi _5}{\sqrt{2}} & \frac{\Phi _3}{\sqrt{2}} & -\Phi _1-\sqrt{3}
   \Phi _2
\end{array}
\right)\nonumber\\
\label{ridiculus}
\end{eqnarray}
introducing, in this way, a second set of coordinates $\left \{\Phi^A \right \}$ on $\mathbb{K}$.
\par
Given the structure constants of the solvable Borel algebra:
\begin{equation}\label{Borelstrucche}
    \left[ T_A \, , \, T_B \right ] \, = \, f_{AB}^{\phantom{AB} C} \, T_C
\end{equation}
and a suitable constant metric $\left( \,\, , \, \,\right )_{\mathbf{g} }$ defined over it:
\begin{equation}\label{fertilizzante}
    \mathbf{g}_{AB} \, = \, \left( T_A\, , \, T_B \right )_{\mathbf{g} }
\end{equation}
we arrive at the Poisson structure over the co-adjoint orbits of $\mathrm{Borel}(\mathfrak{g})$ by considering the dual structure constants:
\begin{equation}\label{fuud}
    f^{AB}_{\phantom{AB}C} \, \equiv \, \mathbf{g}^{AA^\prime} \, \mathbf{g}^{BB^\prime} \, \mathbf{g}_{CC^\prime} \, f_{A^\prime B^\prime}^{\phantom{A^\prime B^\prime} C^\prime}
\end{equation}
(where $\mathbf{g}^{AB}$ is the inverse of the metric $\mathbf{g}^{AB}$) and defining, for any two functions $F(\Phi)$ and $G(\Phi)$ of the generalized canonical coordinates $\Phi^A$ the following \textit{Poisson bracket}:
\begin{equation}\label{pescione}
    \left \{ F\, , \, G\right \} \, \equiv \, \frac {\partial F}{\partial \Phi^A} \, \frac {\partial G}{\partial \Phi^B}  \, f^{AB}_{\phantom{AB}C} \, \Phi^C
\end{equation}
The correct metric on the solvable Lie algebra is chosen on the basis of the following principle. By means of $\mathbf{g}_{AB}$ we can construct a quadratic hamiltonian:
\begin{equation}\label{quadraham}
    \mathfrak{H}_{quad} \, = \, \mbox{cost} \, \mathbf{g}_{AB} \, \Phi^A \, \Phi^B
\end{equation}
and then write the following evolution equations:
\begin{equation}\label{evolveque}
    \frac{d}{d\tau} \, \Phi^A \, = \, \left \{ \mathfrak{H}_{quad} \, , \, \Phi^A\right \}
\end{equation}
The requirement that eq.s(\ref{evolveque}) should reproduce the Lax equation (\ref{laxequation}) fixes the choice of the metric. In our case the appropriate choice is the following:
\begin{equation}\label{matricagab}
    \mathbf{g}_{AB} \, = \, \left(
\begin{array}{llllllll}
 3 & 0 & 0 & 0 & 0 & 0 & 0 & 0 \\
 0 & 3 & 0 & 0 & 0 & 0 & 0 & 0 \\
 0 & 0 & \frac{3}{2} & 0 & 0 & 0 & 0 & 0 \\
 0 & 0 & 0 & -\frac{3}{2} & 0 & 0 & 0 & 0 \\
 0 & 0 & 0 & 0 & -\frac{3}{2} & 0 & 0 & 0 \\
 0 & 0 & 0 & 0 & 0 & -\frac{3}{2} & 0 & 0 \\
 0 & 0 & 0 & 0 & 0 & 0 & -\frac{3}{2} & 0 \\
 0 & 0 & 0 & 0 & 0 & 0 & 0 & \frac{3}{2}
\end{array}
\right)
\end{equation}
Correspondingly the non vanishing components of the structure constants $f^{AB}_{\phantom{AB}C}$ are the following ones:
\begin{equation}\label{strutturoide }
    \begin{array}{ccccccc}
      f^{13}_{\phantom{12}3} & = & \frac 13 & ; & f^{14}_{\phantom{12}4}  & = & - \, \frac 12  \\
      f^{15}_{\phantom{12}5} & = & - \, \frac 16 & ; & f^{16}_{\phantom{12}6}  & = & \frac 16 \\
     f^{17}_{\phantom{12}7} & = & \frac 12  & ; & f^{24}_{\phantom{12}4}  & = & \frac{1}{2\sqrt{3}}  \\
      f^{25}_{\phantom{12}5} & = & \frac{1}{2\sqrt{3}}   & ; & f^{26}_{\phantom{12}6}  & = & \frac{1}{2\sqrt{3}}    \\
      f^{27}_{\phantom{12}7} & = & \frac{1}{2\sqrt{3}}   & ; & f^{34}_{\phantom{12}5}  & = & \sqrt{\frac 23}  \\
      f^{35}_{\phantom{12}6} & = & \frac{2\sqrt{2}}{3} & ; & f^{36}_{\phantom{12}7}  & = & -\sqrt{\frac 23}  \\
    \end{array}
\end{equation}
\subsection{Hamiltonians in involution and the Kostant decomposition}
\label{Kostdecompo} Having established the existence of an
underlying dynamical system we are interested in an algorithm that
constructs the appropriate number of conserved hamiltonians in
involution providing its Liouville integrability. This was already
stressed above. To this effect a useful mathematical tool happens
to be the Kostant decomposition of a generic Lie algebra element
of a maximally split simple Lie algebra $\mathfrak{g}$, whose
corresponding Lie Group we denote by $\mathrm{G}$. According to a
theorem demonstrated by Kostant (see \cite{GekShap}), for any
generic element ${g}\in \mathfrak{g}$ there exists an appropriate
element $\mathcal{B} \in \exp \left [
\mathrm{Borel}(\mathfrak{g})\right]$ of the Borel subgroup of
$\mathrm{G}$ such that\footnote{Note that the Kostant
decomposition of a generic element is of the presented form for
all classical and exceptional Lie algebras with the exception of
the $A_\ell$-series. Indeed it is only in the case of
$\slal(\ell+1)$ that there exist Casimirs, since the rank of the
Lie Poisson tensor is less than maximum. These non trivial
Casimirs show up in an extra term in the Kostant decomposition.
Since we deal with $\mathfrak{g}_{(2,2)}$ we do not further dwell
on these extra terms.}:
\begin{equation}\label{furbatina}
    \mathcal{B} \, g \, \mathcal{B}^{-1} \, = \, \sum_{\alpha >0} K_\alpha\left(g\right) \, E^{\alpha} \, + \,
    \sum_{i=1}^r \, E^{-\alpha_{orth}^i}
\end{equation}
The above equation requires some explanations. Let us recall that, given a Lie algebra of rank $r$, we can always find an ordered set of $r$ positive roots
\begin{equation}\label{ortorutte}
    \alpha_{orth}^r \, > \, \alpha_{orth}^{r - 1} \, > \,  \dots \, > \, \alpha_{orth}^{1}
\end{equation}
which are mutually orthogonal
\begin{equation}\label{ortogonalusi}
  \left (  \alpha_{orth}^i \, , \, \alpha_{orth}^j \right) \, = \, 0 \quad \mbox{if} \,\, i\ne j
\end{equation}
and $\alpha_{orth}^r$ is the highest root of the Lie algebra. In equation (\ref{furbatina}) the second sum is extended over the negative of such orthogonal roots. The first sum is instead extended over the set of all positive roots and $K_\alpha\left(\mathfrak{g}\right)$ are named the Kostant coefficients of the Lie algebra element $\mathfrak{g}$. The right hand side of eq.(\ref{furbatina}) is named the \textit{Kostant normal form} of $\mathfrak{g}$.
\par
The Borel transformation which puts an arbitrary element of the Lie algebra into its Kostant form can be explicitly constructed by means of a very nice iterative algorithm. We can describe this latter in general terms.
\par
First of all we observe that, given the highest root $\alpha_h$ of a rank $r$ Lie algebra $\mathfrak{g}^{(r)}$ , this latter singles out a unique rank $r-1$ orthogonal subalgebra $\mathfrak{g}_\bot^{(r-1)} \subset \mathfrak{g}^{(r)}$ defined as follows. The Cartan subalgebra $\mathcal{C}^{(r-1)}_ \bot \,\subset \,\mathfrak{g}_\bot^{(r-1)}$ is spanned by all combinations of Cartan generators that are orthogonal to the highest root:
\begin{equation}\label{cartanka}
    \mathcal{H} \, \in \, \mathcal{C}^{(r-1)} _ \bot \quad \Leftrightarrow \quad  \mathcal{H} \, \in \, \mathcal{C}^{(r)} \quad \mbox{and} \quad \alpha_h \left(\mathcal{H} \right )\, = \, 0
\end{equation}
Moreover  $\mathfrak{g}_\bot^{(r-1)}$ includes also all linear combinations of  step operators $E^{\pm \beta}$ associated with roots $\beta$  that are orthogonal to the highest root: $\beta \cdot \alpha_h \, = \, 0$. The sum of $\alpha_h$-orthogonal roots is also $\alpha_h$-orthogonal. Hence the described procedure defines a bona fide subalgebra.
\par
Iterating this argument we conclude that every simple Lie algebra $\mathfrak{g}^{(r)}$ admits a nested chain of $r-1$ orthogonal subalgebras:
\begin{equation}\label{nested}
    \mathfrak{g}^{(r)} \, \supset \, \mathfrak{g}_\bot^{(r-1)} \, \mathfrak{g}_\bot^{(r-2)} \, \supset \, \dots \, \supset \, \mathfrak{g}_\bot^{(1)}
\end{equation}
the last of which, $\mathfrak{g}_\bot^{(1)}$, is necessarily an $\slal(2)$ Lie algebra.
\par
Let us next consider the general decomposition of a generic Lie algebra element $g \, \in \, \mathfrak{g}^{(r)}$  with respect to the orthogonal subalgebra. In full generality we can write:
\begin{equation}\label{scumpungu}
    g \, = \, x_+ E^{\alpha_h} \, + \, x_- E^{-\alpha_h} \, +\, x_0 \, H_h \, + \, g_\bot \, + v_+ \, + \, v_-
\end{equation}
where $x_\pm$ and $x_0$ are numbers,
\begin{equation}\label{Hacca}
    H_h \, \equiv \, \alpha_h \cdot \mathcal{H}
\end{equation}
is the Cartan generator associated with the highest root,
\begin{equation}\label{funtor}
    g_\bot  \, \in \, \mathfrak{g}_\bot^{(r-1)}
\end{equation}
is an element of the orthogonal subalgebra and $v_\pm$ are defined as follows:
\begin{eqnarray}
  v_+ &=& \sum_{\beta} \nu_\beta \, E^\beta \nonumber\\
  v_- &=& \sum_{\beta} \nu_{-\beta} \, E^{\beta}\label{ziopippo}
\end{eqnarray}
having denoted by $\beta$ all those roots different from the
highest one $\alpha_h$ which are not orthogonal to it. In terms of
these objects we can define the following element of the Borel
subgroup of $G$ \footnote{The form of this Borel transformation
was discussed in \cite{GekShap}.}:
\begin{equation}\label{fungobat}
    \mbox{Borel}(\mathrm{G}) \, \ni \, \mathcal{B} \, = \, \exp \left[-\lg\left( x_- \right)\, H_h\right] \, \cdot \,
   \exp \left[-\lg \left(\frac{x_0}{x_-} \right) \, E^{\alpha_h} \right]
\, \cdot \, \exp \left[- {\frac{1}{x_-} }   \, \left[ E^{\alpha_h} \, , \, v_-\right]\right]
\end{equation}
The reason why $\mathcal{B}$ is in the Borel subgroup resides in that the commutator $\left[ E^{\alpha_h} \, , \, v_-\right]$ can produce only positive root step operators since, by hypothesis, $\alpha_h$ is the highest root.
Considering now the transformed object:
\begin{equation}\label{fatuccio}
    \tilde{g} \, = \,\mathcal{B} \, g \, \mathcal{B}^{-1}
\end{equation}
we can verify that its decomposition has the following structure:
\begin{equation}\label{scumpungu2}
    \tilde{g} \, = \, \tilde{x}_+ E^{\alpha_h} \, + \,  E^{-\alpha_h} \,  + \, \tilde{g}_\bot \, + \tilde{v}_+
\end{equation}
In other words we have succeeded in eliminating the components along the $E^{-\beta}$ step operators and in the direction of the Cartan generator $H_h$. Moreover we have put to one the coefficient of  $E^{-\alpha_h}$.
\par
At this point we can consider the element $\tilde{g}_\bot $ and decompose it with respect to the highest root of the orthogonal subalgebra $\mathfrak{g}_\bot^{(r-1)}$. A new Borel transformation will do the same job on  $\tilde{g}_\bot$ that we just did on $g$. The important thing is that this new Borel transformation, being inside the group generated by the orthogonal subalgebra, will not affect the already determined structure (\ref{scumpungu}). Hence by iteratively applying the above described procedure to all the orthogonal subalgebras we can  put any Lie algebra element in its canonical Kostant form defined by eq.(\ref{furbatina}).

\subsubsection{Relevance of the Kostant decomposition for Liouville integrability}
A very significant property of the Kostant  coefficients $K_\alpha
(\mathfrak{g})$ is the following one. If two Lie algebra elements
$\mathfrak{g}_1$ and $\mathfrak{g}_2$ differ by a similarity
transformation performed with an element of the Borel subgroup,
then the corresponding sets of Kostant coefficients are identical.
In formula we have:
\begin{equation}\label{identitone}
    \mbox{If} \, \mathcal{B} \, \in \, \mbox{Borel}(G) \, \quad \, \mbox{then} \, \quad \,  \forall \, {g} \, \in \, \mathfrak{g}\, \quad : \quad \, K_\alpha \left (\mathcal{B} \, g \, \mathcal{B}^{-1} \right) \, = \, K_\alpha \left ( g  \right)
\end{equation}
Let us now recall an important aspect of the integration algorithm
derived in \cite{noiultimo,Fre:2009dg},\cite{Chemissany:2010zp} and reviewed
in section \ref{integralgoritmo}.  According to equation
(\ref{sborellando}), at any istant of time $\tau$ the Lax operator
$L(\tau)$ is a Borel transform of the Lax operator at the initial
time. From this it follows that the Kostant coefficients of
$L(\tau)$ and those of $L_0$ are the same, in other words the
Kostant coefficients are a set of conserved hamiltonians equal in
number to the number of positive roots of the considered Lie algebra. Since
the Noether charge matrix $Q^{Noether}$ and the Lax operator
${L}(\tau)$ are related by the conjugation by the upper-triangular
coset-representative $\mathbb{L}(\tau)$:
\begin{equation}\label{noethercharge1}
    Q^{Noether} \, = \, \mathbb{L}(\tau) \, L(\tau) \, \mathbb{L}^{-1}(\tau)
\end{equation}
their Kostant normal forms are the same as well.
\par
Without any further effort we have proved that:
\begin{equation}\label{frullato}
\forall \alpha \, = \, \mbox{positive root} \quad : \quad \quad \left \{ \mathfrak{H}_{quad} \, , \, K_\alpha (L) \, \right\} \, = \, 0
\end{equation}
Naming:
\begin{equation}\label{numerorutta}
    n_r \, = \,\# \, \mbox{positive roots}
\end{equation}
the dimension of our dynamical system, which is the dimension of the corresponding
split coset, is given by:
\begin{equation}\label{dimcoseto}
    N\, = \, r+n_r
\end{equation}
Since $n_r > r$ we always have a number of conserved hamiltonians that is strictly larger than one half of the dimension of the dynamical space. Liouville integrability is established if among the rational functions of $K_\alpha (L)$ we can select $\frac{n_r+r}{2}$ functionally independent ones that are in involution.
\par
We have found an algorithm able to generate such involutive hamiltonians. Let us  name $\mathcal{KN}(L)$ the Kostant normal form of the Lax operator and let us consider the following polynomial function of $r$ variables:
\begin{equation}\label{cruccata}
    \mathcal{P}\left (\lambda, \mu_1,\dots \, \mu_{r-1}\right ) \, \equiv \, \mbox{Det} \left ( \mathcal{KN}(L) - \sum_{i=1}^{r-1} \mu_i \, E^{\alpha_{orth}^{r+1-i} }\, - \, \lambda \, \mathbf{1} \right)
\end{equation}
If we expand $\mathcal{P}\left (\lambda, \mu_1,\dots \,
\mu_{r-1}\right )$ in powers of its variables, among the
coefficients of such development we \textbf{conjecture} that we
can always retrieve the required number of involutive independent
hamiltonians.  Although we do not possess a formal proof of our
statement we have scanned several cases with several different Lie
algebras and we were always able to single out the required number
of Liouville hamiltonians. In the case of  the Lie algebra
$\mathfrak{g}_{2(2)}$ the required number is four and we have
found the following explicit result\footnote{This type of
construction was presented for the case of the complex
$\mathfrak{g}_2$ Lie algebra in \cite{GekShap} applied to the full
Kostant-Toda model.}:
\begin{eqnarray}
  \mathcal{P}\left (\lambda, \mu\right ) &=& \lambda^7 \, + \, \mathfrak{K}_1 \, \lambda^5 \, + \, \mathfrak{K}_1^2 \lambda^3
  \, + \,\mathfrak{K}_2 \lambda \, - \, 3 \, \lambda^5 \,\mu \, + \, \mathfrak{K}_1 \, \lambda^3 \,\mu\nonumber\\
  \null & & \, + \, \mathfrak{K}_3 \, \lambda \, \mu \, - \, \ft 94 \lambda^3 \, \mu^2 \, + \, \mathfrak{K}_4 \, \lambda \, \mu ^2
\end{eqnarray}
The four functions $\mathfrak{K}_i$, of the Lax entries $\Phi$, whose explicit expression is presented in appendix
\ref{papponaenorme}, are functionally independent and in involution, as we have explicitly checked:
\begin{equation}\label{involuziona}
    \left \{ \mathfrak{K}_i \, ,\, \mathfrak{K}_j \right \} \, = \, 0
\end{equation}
This proves Liouville integrability.

\subsection{Reduction of the dynamical system}
A very important concept in the theory of dynamical systems is that of semiinvariants. A function $\mathcal{S}(\Phi)$ of the dynamical variables is named a \textit{semiinvariant} if it happens that:
\begin{equation}\label{cacius}
    \left \{ \mathfrak{H}_{quad} \, , \, \mathcal{S} \right\} \, \sim \, \mathcal{S}
\end{equation}
The relevance of this concept is quite clear. Whenever we have a semiinvariant we can reduce the dynamical system by introducing the constrained surface:
\begin{equation}\label{constrasurface}
    \mathcal{S}(\Phi) \, = \, 0
\end{equation}
which, thanks to eq.(\ref{cacius}) is stable with respect to
dynamical evolution. In other words all dynamical trajectories
that cross the surface (\ref{constrasurface}) lie entirely on it.
In the case of the $\mathfrak{g}_{2(2)}$ model we are considering,
an important semiinvariant is provided by the field $\Phi_8$
associated with the highest root. By explicit evaluation we find:
\begin{equation}\label{firpo}
    \left \{ \mathfrak{H}_{quad} \, , \, \Phi_8 \right\} \, = \,4 \sqrt{3} \,
    \Phi_2 \, \Phi_8
\end{equation}
Hence we can define a consistent reduction of our dynamical system by imposing the constraint:
\begin{equation}\label{finalizzatoTaub}
    \Phi_8 \, = \, 0
\end{equation}
A Lax operator $L_0$ lying on this constrained surface automatically produces
a vanishing Taub-NUT charge since the latter is defined as the trace of the
Noether charge matrix:
\begin{equation}\label{noethercharge}
    Q^{Noether} \, = \, \mathbb{L}(\tau) \, L(\tau) \, \mathbb{L}^{-1}(\tau) \, = \, \mathbb{L}_0 \, L_0 \, \mathbb{L}_0^{-1}
\end{equation}
with the transposed of the highest root step operator:
\begin{equation}\label{tubolatta}
    \mathbf{n} \, = \, \mbox{Tr} \left (Q^{Noether} \, E^{-\alpha_6}\right )
\end{equation}
In equation (\ref{noethercharge}) $\mathbb{L}_0$ denotes the initial Lax operator which modifies only the asymptotic values of the scalar fields and therefore lies in the $D=4$ subgroup: $\mathbb{L}_0 \in \mathrm{U_{D=4}} \subset \mathrm{U_{D=3}}$. Such a group commutes with the Ehlers group and therefore with the highest root. Therefore if $\Phi_8 \, = \,0$ in $L_0$, no $\Phi_8$ will be produced in $Q^{Noether}$ and the Taub-NUT charge will stay zero, irrespectively of the scalar boundary conditions at infinity.
\par
The vanishing Taub-NUT shell corresponds therefore to a consistent
reduction of our dynamical system. On this shell the two rational
hamiltonians $\mathfrak{K}_3 \, \mathfrak{K}_4$ are ill-defined
sinve in both case $\Phi_8$ appears in the denominator (see eq.s
(\ref{hamil3},~\ref{hamil4}). Using their numerators, however, we
can construct a new non vanishing hamiltonian that commutes with
the two polynomial ones. Referring to the notation of the appendix
we can set:
\begin{equation}\label{newham}
    \overline{\mathfrak{K}}_3 \, = \, - \,
    \ft 12 \, \frac{\mathfrak{P}_5^2(\Phi)}{\mathfrak{P}_4(\Phi)}
\end{equation}
and verify that:
\begin{equation}\label{verifica}
    \left \{ \mathfrak{K}_1 \, , \, \overline{\mathfrak{K}}_3 \right \} \, = \, \left \{ \mathfrak{K}_2 \, , \, \overline{\mathfrak{K}}_3 \right \} \, = \, 0 \quad \mbox{at} \quad \Phi_8 \, = \, 0
\end{equation}

\subsection{Standard form of the $\mathbb{H}^\star$ decomposition}
Next task is that of reorganizing  both the generators of $\mathbb{H}^\star$ and those of the $\mathbb{K}$ subspace in such a way as to make the $\slal(2,\mathbb{R}) \oplus \slal(2,\mathbb{R})$ structure of the former manifest and put the latter in a standard basis for the representation $\left(\ft 12 \, , \, \ft 32\right)$ which it belongs to.
Both goals are obtained introducing the following linear combinations for $\mathbb{H}^\star$
\begin{equation}\label{phenix}
    \begin{array}{lllll}
  L_0^{(I)} &  = & \frac{1}{4} \left(h_3+h_5\right) &
    = & \frac{1}{4}
   ( {e_3}+ {e_5}+ {f_3}+ {f_5}) \\
  L_+^{(I)} &  = & \frac{1}{4}
   \left(-h_1-h_2+h_4+h_6\right) &  = & \frac{1}{4}
   (- {e_1}- {e_2}+ {e_4}+ {e_6}+ {f_1}- {
   f_2}+ {f_4}- {f_6}) \\
  L_-^{(I)} &  = & \frac{1}{4}
   \left(-h_1+h_2+h_4-h_6\right) &  = & \frac{1}{4}
   ( {e_1}- {e_2}+ {e_4}- {e_6}- {f_1}- {f_
   2}+ {f_4}+ {f_6}) \\
  L_0^{(II)} &  = & \frac{1}{4} \left(3 h_5-h_3\right)
   &  = & \frac{1}{4} (- {e_3}+3
    {e_5}- {f_3}+3  {f_5}) \\
  L_+^{(II)} &  = & \frac{1}{4} \left(3 h_1-h_2+h_4-3
   h_6\right) &  = & \frac{1}{4} (- {e_1}+3
    {e_2}+ {e_4}-3  {e_6}+ {f_1}+3
    {f_2}+ {f_4}+3  {f_6}) \\
  L_-^{(II)} &  = & \frac{1}{4} \left(3 h_1+h_2+h_4+3
   h_6\right) &  = & \frac{1}{4} ( {e_1}+3
    {e_2}+ {e_4}+3  {e_6}- {f_1}+3
    {f_2}+ {f_4}-3  {f_6})
\end{array}
\end{equation}
and
\begin{equation}\label{nuovigeni}
    \begin{array}{lll}
 \Lambda _{1,1} & = & -12 \mathbf{K}_1-4 \mathbf{K}_2-4 \mathbf{K}_5 \\
 \Lambda _{1,2} & = & -2 \sqrt{3} \mathbf{K}_3-2 \sqrt{3}
   \mathbf{K}_4-2 \sqrt{3} \mathbf{K}_6+2 \sqrt{3} \mathbf{K}_8 \\
 \Lambda _{1,3} & = & -6 \mathbf{K}_3-2 \mathbf{K}_4+2 \mathbf{K}_6-6 \mathbf{K}_8 \\
 \Lambda _{1,4} & = & -4 \sqrt{3} \mathbf{K}_1-4 \sqrt{3}
   \mathbf{K}_2-4 \sqrt{3} \mathbf{K}_7 \\
 \Lambda _{2,1} & = & -6 \mathbf{K}_3+2 \mathbf{K}_4+2 \mathbf{K}_6+6 \mathbf{K}_8 \\
 \Lambda _{2,2} & = & 4 \sqrt{3} \mathbf{K}_1+4 \sqrt{3}
   \mathbf{K}_2-4 \sqrt{3} \mathbf{K}_7 \\
 \Lambda _{2,3} & = & -12 \mathbf{K}_1-4 \mathbf{K}_2+4 \mathbf{K}_5 \\
 \Lambda _{2,4} & = & 2 \sqrt{3} \mathbf{K}_3-2 \sqrt{3}
   \mathbf{K}_4+2 \sqrt{3} \mathbf{K}_6+2 \sqrt{3} \mathbf{K}_8
\end{array}
\end{equation}
for $\mathbb{K}$.
\par
With these definitions we can check the standard commutation rules of the $\slal(2,\mathbb{R}) \oplus \slal(2,\mathbb{R})$ algebra:
\begin{eqnarray}
  \left [ L_0^{(I,II)} \, , \, L_\pm^{(I,II)} \right ]  &=& \pm L_\pm^{(I,II)} \nonumber\\
  \left [ L_+^{(I,II)} \, , \, L_-^{(I,II)} \right ]  &=& 2 L_0^{(I,II)} \nonumber\\
  \left [ L_i^{(I)} \, , \, L_j^{(II)} \right ]  &=& 0 \label{standarhstar}
\end{eqnarray}
On the other hand calculating the commutators of these $\mathbb{H}^\star$ generators with the $\Lambda_{\alpha|A}$ generators of $\mathbb{K}$ introduced in (\ref{nuovigeni}) we find
\begin{eqnarray}\label{cornetto}
\left [ L_0^I \, , \, \left(\begin{array}{c}
                                  \Lambda_{1} \\
                                  \Lambda_{2}
                                \end{array} \right)\right] & = &
\left(\begin{array}{c|c}
\ft 12 \, \mathbf{1} & 0 \\
\hline
0 & - \ft 12 \, \mathbf{1}
\end{array} \right) \,
\left(\begin{array}{c}
\Lambda_{1} \\
\Lambda_{2}
\end{array} \right)\nonumber\\
\left [ L_+^{I} \, , \, \left(\begin{array}{c}
                                  \Lambda_{1} \\
                                  \Lambda_{2}
                                \end{array} \right)\right]
                                & = & \left(\begin{array}{c|c}
                                0 & 0 \\
                                \hline
                                -\mathbf{1} & 0\
                                \end{array} \right) \,\left(\begin{array}{c}
                                  \Lambda_{1} \\
                                  \Lambda_{2}
                                \end{array} \right)\nonumber\\
\left [ L_-^I \, , \, \left(\begin{array}{c}
                                  \Lambda_{1} \\
                                  \Lambda_{2}
                                \end{array} \right)\right] & = &
\left(\begin{array}{c|c}
0 & -\mathbf{1} \\
\hline
0 & 0\
\end{array} \right) \,
\left(\begin{array}{c} \Lambda_{1} \\
                                  \Lambda_{2}
\end{array} \right)
\end{eqnarray}
and:
\begin{eqnarray}\label{cornetto2}
\left [ L_0^{II}  \, , \, \left(\begin{array}{c}
                                  \Lambda_{1} \\
                                  \Lambda_{2}
                                \end{array} \right)\right] & = & - \, \left(\begin{array}{c|c}
U_0 & 0 \\
\hline
0 & U_0
\end{array} \right) \,
\left(\begin{array}{c}
\Lambda_{1} \\
                                  \Lambda_{2}
\end{array} \right)\nonumber\\
\left [ L_\pm^{II}  \, , \, \left(\begin{array}{c}
                                  W^1 \\
                                  W^2
                                \end{array} \right)\right] & = & - \,\left(\begin{array}{c|c}
U_\pm & 0 \\
\hline
0 & U_\pm
\end{array} \right) \,
\left(\begin{array}{c}
\Lambda_{1} \\
                                  \Lambda_{2}
\end{array} \right)
\end{eqnarray}
in full analogy with eq.s (\ref{turnotto}) and (\ref{turnotto2}). This is correct since, from the abstract point of view, the $\mathfrak{g}_{2,(2)}$ algebra contains, up to conjugations, only one $\slal(2,\mathbb{R}) \oplus \slal(2,\mathbb{R})$ subalgebra and  the complementary subspace falls always into the $(\ft 12 \, , \, \ft 32)$ representation. Yet it is important to stress that, from the point of view of our constructions, the original $\slal(2)$ algebra of the $S^3$ model, times the Ehlers $\slal(2)$ Lie algebra are quite different from the $\slal(2,\mathbb{R}) \oplus \slal(2,\mathbb{R})$ Lie algebras defined above. The former are partly in the denominator algebra $\mathbb{H}^\star$, partly in the complementary subspace $\mathbb{K}$, while those displayed above are just the structure of $\mathbb{H}^\star$ itself.
\par
Following these redefinitions we can construct the Lax operator as:
\begin{equation}\label{Deltadefi}
    L \, = \, \Delta^{\alpha|A} \, \Lambda_{\alpha|A}
\end{equation}
which introduces a third set of coordinates $\Delta^{\alpha|A}$ as parameters of the eight-dimensional dynamical system.
\par
Final comparison of the three equivalent definitions (\ref{loffo}), (\ref{ridiculus}) and (\ref{Deltadefi}) provides the conversion table from one to the other of the coordinate sets $t$, $\Phi$ and $\Delta$. Such a table is displayed in the following equation.
\begin{equation}\label{fildiconvo}
    \begin{array}{|l|l|l|}
    \hline
    \hline
    \mbox{$\Phi$ - field } & \mbox{$t$ - field  } & \mbox{$\Delta^{\alpha|A}$ -field}\\
    \hline\hline
 \Phi ^1 & \frac{1}{2} \left(2 t^2-t^1\right) & 2
   \left(\Delta ^{1,1}-\sqrt{3} \Delta ^{1,4}+\sqrt{3} \Delta
   ^{2,2}+\Delta ^{2,3}\right) \\
   \hline
 \Phi ^2 & \frac{t^1}{2 \sqrt{3}} & -2 \left(\sqrt{3} \Delta
   ^{1,1}+\Delta ^{1,4}-\Delta ^{2,2}+\sqrt{3} \Delta
   ^{2,3}\right) \\
   \hline
 \Phi ^3 & \sqrt{2} t^4 & -2 \sqrt{2} \left(\sqrt{3} \Delta
   ^{1,2}+\Delta ^{1,3}-\Delta ^{2,1}+\sqrt{3} \Delta
   ^{2,4}\right) \\
   \hline
 \Phi ^4 & \sqrt{\frac{2}{3}} t^3 & -2 \sqrt{\frac{2}{3}}
   \left(\sqrt{3} \Delta ^{1,2}+3 \Delta ^{1,3}+3 \Delta
   ^{2,1}-\sqrt{3} \Delta ^{2,4}\right) \\
   \hline
 \Phi ^5 & \sqrt{2} t^5 & -4 \sqrt{2} (\Delta ^{1,1}-\Delta
   ^{2,3}) \\
   \hline
 \Phi ^6 & \sqrt{2} t^6 & 2 \sqrt{2} \left(-\sqrt{3} \Delta
   ^{1,2}+\Delta ^{1,3}+\Delta ^{2,1}+\sqrt{3} \Delta
   ^{2,4}\right) \\
   \hline
 \Phi ^7 & \sqrt{\frac{2}{3}} t^7 & -4 \sqrt{2} (\Delta
   ^{1,4}+\Delta ^{2,2}) \\
   \hline
 \Phi ^8 & \sqrt{\frac{2}{3}} t^8 & 2 \sqrt{\frac{2}{3}}
   \left(\sqrt{3} \Delta ^{1,2}-3 \Delta ^{1,3}+3 \Delta
   ^{2,1}+\sqrt{3} \Delta ^{2,4}\right)\\
   \hline
\end{array}
\end{equation}
\section{$\mathbb{H}^\star$ Invariants, Tensor Classifiers and Orbits}
Having singled out the appropriate basis in which the Lax operator $L$ is described by a two-index tensor
$\Delta^{\alpha|A}$ of the two $\mathrm{SL(2,\mathbb{R})}$ groups, further progress in the analysis of the invariants and irreducible tensors one can construct with the powers of $L$ is obtained by means of some standard group theory.
\par
In particular, recalling that $\mathrm{SL(2,\mathbb{R})} \sim \mathrm{SO(2,1)}$, irreducible representations can be characterized by the value of the angular momentum $j$ and tensor product of representations decompose according to standard rules.
\par
Let us consider the tensor product of two $j=\ft 32$ representations. We find:
\begin{eqnarray}\label{pazzolone}
    \left[\left(\mathbf{j=\frac{3}{2} }\right)\otimes \left(\mathbf{j=\frac{3}{2}}\right)\right]_{symm} & = & \underbrace{\mathbf{(j=3)}}_{7} \oplus \underbrace{\mathbf{(j=1)}}_{3}\nonumber\\
    \left[\left(\mathbf{j=\frac{3}{2} }\right)\otimes \left(\mathbf{j=\frac{3}{2}}\right)\right]_{antisym} & = & \underbrace{\mathbf{(j=2)}}_{5} \oplus \underbrace{\mathbf{(j=0)}}_{1}
\end{eqnarray}
Hence when we consider an antisymmetric tensor of the form:
\begin{equation}\label{coso1}
   {T}^{AB}\, = \, \epsilon_{\alpha\beta} \,\Delta^{\alpha|A} \, \Delta^{\beta|B}
\end{equation}
we know that it decomposes into a $j=2$ and a $j=0$ irreducible representation. This means that $T^{AB}$ can be converted into a symmetric tensor $\mathcal{T}^{xy}$ where $x,y=-,0,+$ are the vector indices of the $j=1$ representation, namely the adjoint of $\slal(2)$. In this basis the invariant metric is:
\begin{equation}\label{fingus}
    \eta_{xy} \, = \, \left(
\begin{array}{lll}
 0 & 0 & 1 \\
 0 & -2 & 0 \\
 1 & 0 & 0
\end{array}
\right)
\end{equation}
and the singlet representation is the $\eta$-trace: $\eta_{xy} \,
\mathcal{T}^{xy}$, while the $\eta$-traceless part corresponds to
the $j=2$ representation. Therefore what we just need are the
group theoretical conversion coefficients encoded in a tensor
$t^{xy}_{\phantom{xy}AB}$ such that\footnote{The normalization
with the prefactor $\frac{128}{\sqrt{3}}$ is chosen for future
convenience.}:
\begin{equation}\label{tensotto}
    \mathcal{T}^{xy} \, \equiv \,\frac{128}{\sqrt{3}}\, t^{xy}_{\phantom{xy}AB} \, \Delta^{\alpha|A}\,\Delta^{\beta|B}\,\epsilon_{\alpha\beta}
\end{equation}
transforms as a symmetric tensor in the vector indices. Having fixed our conventions for the $j=\ft 32$ and $j=1$ representations, the tensor $t^{xy}_{\phantom{xy}AB}$ is completely determined. It is symmetric in the indices $xy$ and antisymmetric in the indices $AB$, therefore it can be explicitly displayed as a set of six antisymmetric matrices:
\begin{equation}\label{mignottini}
    \begin{array}{ccccccc}
      t^{--} & = & \left(
\begin{array}{llll}
 0 & 0 & 0 & 0 \\
 0 & 0 & 0 & 0 \\
 0 & 0 & 0 & 1 \\
 0 & 0 & -1 & 0
\end{array}
\right) & ; & t^{-0} & = & \left(
\begin{array}{llll}
 0 & 0 & 0 & -\frac{1}{2} \\
 0 & 0 & 0 & 0 \\
 0 & 0 & 0 & 0 \\
 \frac{1}{2} & 0 & 0 & 0
\end{array}
\right) \\
      t^{-+} & = & \left(
\begin{array}{llll}
 0 & 0 & -\frac{1}{2 \sqrt{3}} & 0 \\
 0 & 0 & 0 & \frac{1}{2 \sqrt{3}} \\
 \frac{1}{2 \sqrt{3}} & 0 & 0 & 0 \\
 0 & -\frac{1}{2 \sqrt{3}} & 0 & 0
\end{array}
\right) & ; & t^{00} & = & \left(
\begin{array}{llll}
 0 & 0 & -\frac{1}{2 \sqrt{3}} & 0 \\
 0 & 0 & 0 & \frac{1}{2 \sqrt{3}} \\
 \frac{1}{2 \sqrt{3}} & 0 & 0 & 0 \\
 0 & -\frac{1}{2 \sqrt{3}} & 0 & 0
\end{array}
\right) \\
      t^{0+} & = & \left(
\begin{array}{llll}
 0 & 0 & 0 & 0 \\
 0 & 0 & \frac{1}{2} & 0 \\
 0 & -\frac{1}{2} & 0 & 0 \\
 0 & 0 & 0 & 0
\end{array}
\right) & ; & t^{++} & = & \left(
\begin{array}{llll}
 0 & 1 & 0 & 0 \\
 -1 & 0 & 0 & 0 \\
 0 & 0 & 0 & 0 \\
 0 & 0 & 0 & 0
\end{array}
\right)
    \end{array}
\end{equation}
The tensor $t^{xy}_{\phantom{xY}AB}$ has the property that $\eta_{xy}\,t^{xy}_{\phantom{xY}AB} \,=\,0$, hence it is the projector of the antisymmetric product on the pure $j=2$ irreducible representation. Using the $t$-coordinates in which it takes its simplest form, the $\mathcal{T}^{xy}$ tensor has the following explicit form:
\begin{eqnarray}
  \mathcal{T}^{--} &=& \frac{1}{3} \left(-t_1^2+2 \left(2 t_2+t_5+t_7\right) t_1-3
   t_2^2+t_3^2-3 t_4^2-3 t_6^2+t_8^2-2 t_3 t_4 \right.\nonumber\\
  \null &\null&\left. +2 t_3 t_6+6
   t_4 t_6-4 t_5 t_7-2 t_2 \left(3 t_5+t_7\right)+2 t_3 t_8-2
   t_4 t_8+2 t_6 t_8\right)\nonumber\\
  \mathcal{T}^{-0} &=& \frac{1}{3} \left(-t_3 t_5+3 t_4 t_5-3 t_6 t_5-t_8 t_5-t_3
   t_7+t_4 t_7+t_6 t_7\right.\nonumber\\
  \null &\null& \left.+t_1 \left(t_4-2 t_6-t_8\right)+t_7
   t_8+t_2 \left(-t_3+3 t_6+2 t_8\right)\right) \nonumber\\
  \mathcal{T}^{-+} &=& \frac{1}{9} \left(t_1^2-3 t_2^2-t_3^2-3 t_4^2-6 t_5^2+3
   t_6^2+2 t_7^2+t_8^2+6 t_3 t_6+6 t_4 t_8\right) \nonumber\\
  \mathcal{T}^{00} &=& \frac{1}{9} \left(t_1^2-3 t_2^2-t_3^2-3 t_4^2-6 t_5^2+3
   t_6^2+2 t_7^2+t_8^2+6 t_3 t_6+6 t_4 t_8\right)\nonumber\\
  \mathcal{T}^{0+} &=& \frac{1}{3} \left(t_3 t_5+3 t_4 t_5+3 t_6 t_5-t_8 t_5+t_3
   t_7+t_4 t_7-t_6 t_7+t_7 t_8 \right.\nonumber\\
  \null &\null&\left. +t_1 \left(-t_4-2
   t_6+t_8\right)-t_2 \left(t_3-3 t_6+2 t_8\right)\right) \nonumber\\
  \mathcal{T}^{++} &=& \frac{1}{3} \left(-t_1^2-2 \left(-2 t_2+t_5+t_7\right) t_1-3
   t_2^2+t_3^2-3 t_4^2-3 t_6^2+t_8^2+2 t_3 t_4\right.\nonumber\\
  \null &\null&\left. +2 t_3 t_6-6
   t_4 t_6-4 t_5 t_7+2 t_2 \left(3 t_5+t_7\right)-2 t_3 t_8-2
   t_4 t_8-2 t_6 t_8\right)
\end{eqnarray}
Next we consider the projector that from the symmetric product of two $j=\ft 12$ representations extracts a vector representation $j=1$. This projector essentially consists of appropriate linear combinations
of the Pauli matrices with one index raised by means of the $\epsilon$-symbol. Explicitly we have:
\begin{equation}\label{paffuto}
    \Pi^a_{\alpha\beta} \, = \, \left\{\left(
\begin{array}{ll}
 \frac{1}{2} & 0 \\
 0 & 0
\end{array}
\right)\, , \, \left(
\begin{array}{ll}
 0 & \frac{1}{4} \\
 \frac{1}{4} & 0
\end{array}
\right) \, , \, \left(
\begin{array}{ll}
 0 & 0 \\
 0 & \frac{1}{2}
\end{array}
\right)\right\}
\end{equation}
where the vector index $a$ enumerates the three matrices and
$\alpha\beta$ are instead the matrix indices. According to
equation (\ref{pazzolone}), the spin $j=1$ representation can be
extracted also from the symmetric product of two spin $j=\ft 32$
representations. This implies that we have also a projector
$\Sigma^x_{AB}$. In our conventions its explicit form is the
following one:
\begin{equation}\label{obesone}
    \Sigma^x_{AB} \, = \, \left\{\left(
\begin{array}{llll}
 0 & 0 & 0 & -\frac{3 \sqrt{3}}{10} \\
 0 & 0 & 0 & 0 \\
 0 & 0 & -\frac{3}{5} & 0 \\
 -\frac{3 \sqrt{3}}{10} & 0 & 0 & 0
\end{array}
\right)\, , \, \left(
\begin{array}{llll}
 0 & 0 & \frac{3}{20} & 0 \\
 0 & 0 & 0 & \frac{9}{20} \\
 \frac{3}{20} & 0 & 0 & 0 \\
 0 & \frac{9}{20} & 0 & 0
\end{array}
\right) \, , \, \left(
\begin{array}{llll}
 -\frac{3}{5} & 0 & 0 & 0 \\
 0 & 0 & \frac{3 \sqrt{3}}{10} & 0 \\
 0 & \frac{3 \sqrt{3}}{10} & 0 & 0 \\
 0 & 0 & 0 & 0
\end{array}
\right)\right\}
\end{equation}
Using these tensors we can construct a quadratic expression in the components of the Lax operator which transforms as a vector with respect to the first $\mathrm{SL(2)}$ group and as a vector also with respect to the second.
Precisely we set:
\begin{equation}\label{Wtensor}
    \mathcal{W}^{a|x} \, \equiv \, 1280 \, \Pi^a_{\alpha\beta} \, \Sigma^x_{AB} \, \Delta^{\alpha|A} \, \Delta^{\beta|B}
\end{equation}
The explicit calculation of this tensor is just a matter of substitution as it was the case for the $\mathcal{T}^{xy}$-tensor we presented above. We omit displaying the result which is lengthy and not particularly inspiring.
\par
Using $\mathcal{W}^{a|x}$ we can now construct two symmetric tensors, one carrying vector indices of the second group, the other carrying vector indices of the first. We set:
\begin{eqnarray}
  \mathfrak{T}^{xy} &=& \mathcal{W}^{a|x} \, \mathcal{W}^{b|y} \, \eta_{ab}\label{gluppo}\\
  \mathbb{T}^{ab} &=& \mathcal{W}^{a|x} \, \mathcal{W}^{b|y} \, \eta_{xy} \label{farnetico}
\end{eqnarray}
Having constructed the above objects we can now construct an irreducible sixth order invariant and an irreducible quadratic invariant by setting:
\begin{eqnarray}\label{fallo6}
   \mathfrak{ I}_6 & = & \mathfrak{T}^{xy} \, \mathfrak{T}^{zw} \, \eta_{xz} \, \eta_{yw} \nonumber\\
   \mathfrak{ I}_2 & = & - 96 \, \Delta^{\alpha|A}\,\Delta^{\beta|B} \, \epsilon_{\alpha\beta} \, \mathbb{C}_{AB}
\end{eqnarray}
where $\mathbb{C}_{AB}$ denotes the matrix elements of the symplectic metric $\mathbb{C}_4$ introduced in eq.(\ref{presymple}).
An immediate calculation shows that:
\begin{eqnarray}\label{invariantoni}
    \mathfrak{ I}_2 & = & \mathbf{g_{AB}} \, \Phi^{A} \, \Phi^{B} \, \equiv \, \mathfrak{h}_{2} \, = \, \frac{1}{4} \, \mbox{Tr} \,L^2\,\nonumber\\
    \mathfrak{ I}_6 & = & - 6\, \mbox{Tr} L^6 \, + \, \frac 12 \, \left(\mbox{Tr}L^2\right)^3
\end{eqnarray}
where $L$ denotes the Lax operator. Instead of the  irreducible invariant $I_6$ we can consider another combination of traces that emerges in the Kostant construction of  commuting hamiltonians. Recalling the result obtained in section \ref{Kostdecompo} we have:
\begin{equation}\label{chev6ham}
    \mathfrak{h}_6 \, = \, \frac{1}{6} \mbox{Tr} L^6 \, + \, \frac {1}{96} \, \left(\mbox{Tr}L^2\right)^3
\end{equation}
which implies:
\begin{eqnarray}\label{acconi}
    \mathfrak{h}_6 & = & -\,\frac{1}{36}  \mathfrak{I}_6 \, - \,\frac{56}{9}\mathfrak{ I}_2^3\nonumber\\
    \mathfrak{h}_2 & = &\mathfrak{I}_2
\end{eqnarray}
The two invariants of eq.(\ref{invariantoni}) or their combinations (\ref{acconi}) and the above defined tensor classifiers $\mathcal{T},\mathfrak{T},\mathbb{T}$ provide the means to separate one form the other the orbits of Lax operators, both nilpotent and diagonalizable.
\paragraph{Supersymmetry}
So far we have neglected the fermion sector of the three dimensional model, which is related to the bosonic sigma model by supersymmetry. The $D=3$ theory under consideration is characterized by four fermionic fields $\lambda^A$ whose supersymemtry variation on the geodesic background is expressed in terms of the Lax components
\cite{pioline}:
\begin{eqnarray}
\delta_\epsilon \lambda^A&\propto & \Delta^{\alpha,A}\,\epsilon_\alpha\,,
\end{eqnarray}
where $\epsilon_\alpha$ is a doublet of supersymmetry parameters.
BPS solutions are characterized by the property of preserving a fraction of supersymmetry, that is there exists
a spinor $\epsilon_\alpha$ satisfying the Killing spinor equation: $\delta_\epsilon \lambda^A=0$. Supersymmetry is thus preserved if and only if $\Delta^{\alpha|A}$ has a null-eigenvector: $\Delta^{\alpha|A}\,\epsilon_\alpha=0$. This in turn was shown in \cite{pioline} to be equivalent to the condition:
\begin{eqnarray}
\Delta^{\alpha|A}\Delta^{\beta|B}\epsilon_{\alpha\beta}&=&0\,\,\,\Leftrightarrow \,\,\, \mathcal{T}^{xy}\equiv 0\,.
\end{eqnarray}
We conclude that the solution is BPS if and only if $\mathcal{T}^{xy}\equiv 0$.

\subsection{The rotated Cartan-Weyl basis and the classification of regular and nilpotent orbits}
\label{classificazia}
The classification of nilpotent orbits of Lax operators for the non-compact coset:
\begin{equation}\label{cosetnostro}
\frac{\mathrm{U_{D=3}}}{\mathrm{H^\star_{D=3}}}    \, = \, \frac{\mathrm{G_{2,2}}}{\mathrm{SL(2,\mathbb{R})} \times \mathrm{SL(2,\mathbb{R})}}
\end{equation}
was pursued in \cite{bruxelles} using previous mathematical results on nilpotent orbits of the $\mathfrak{g}_{2}$ complex Lie algebra. The basic idea consists of choosing a new Cartan subalgebra for the the $\mathfrak{g}_{2,2}$ Lie algebra which entirely lies in the denominator subalgebra of the coset:
\begin{equation}\label{nuovacarta}
   \mathcal{C}_{new} \, \subset \, \mathbb{H}^\star
\end{equation}
Naming $H_{1,2}$ an orthogonal basis of two generators for such $\mathcal{C}_{new}$ their adjoint action can be diagonalized on the algebra and their common eigen-operators $E^{\pm i}$ ($i,1,\dots,6$) will necessarily correspond to the six positive and six  negative roots (\ref{g2rootsystem}) of the $\mathfrak{g}_2$ root
 system:
 \begin{equation}\label{corrispondo}
    \begin{array}{ccccccc}
       E^{\pm 1} & \Rightarrow & \pm \alpha_1 & ; &E^{\pm 2} & \Rightarrow & \pm\alpha_2\\
       E^{\pm 3} & \Rightarrow & \pm (\alpha_1 + \alpha_2 )& ; &E^{\pm 4} & \Rightarrow & \pm ( 2\alpha_1 +\alpha_2)\\
        E^{\pm 5} & \Rightarrow & \pm (3\alpha_1 + \alpha_2 )& ; &E^{\pm 6} & \Rightarrow & \pm ( 3\alpha_1 +2\alpha_2)\\
     \end{array}
 \end{equation}
 In this way, once a choice of $H_{1,2}$ has been made inside $\mathbb{H}^\star$, a  new  Cartan-Weyl basis can be constructed whose relation with the old one is unique and intrinsically defined by the Lie algebra structure. The property of the new basis is that the step operators $E^i$, which are necessarily nilpotent, either belong to $\mathbb{H}^\star$ or to $\mathbb{K}$. By this token one can choose as representatives of $\mathbb{H}^\star$ orbits in $\mathbb{K}$ step operators $E^i$ that lie in that subspace or combination thereof.
 \par
Our result is displayed  in the following table:
\begin{equation}\label{nuovabasa}
    \begin{array}{|c|c|}
    \hline
    \hline
    \mbox{New Cartan Weyl generators} & \mbox{their form in the $HK$-basis}\\
    \hline
    \hline
 H_1 & h_3 \\
 \hline
 H_2 & h_5 \\
 \hline
 E_1 & \frac{1}{2 \sqrt{2}}\,\left(k_3-3 \mathcal{H}_2-\mathcal{H}_1\right)
   \\
   \hline
 E_2 & -\frac{1}{4} \sqrt{\frac{3}{2}}
   \left(k_1+k_2+k_4-k_6\right) \\
   \hline
 E_3 & \frac{1}{4 \sqrt{2}}\,\left(-3 h_1+h_2-h_4+3 h_6\right) \\
 \hline
 E_4 & \frac{1}{4 \sqrt{2}}\, \left(-3 k_1+k_2+k_4+3 k_6\right) \\
 \hline
 E_5 & \frac{1}{4} \sqrt{\frac{3}{2}}
   \left(-h_1-h_2+h_4+h_6\right) \\
   \hline
 E_6 & \frac{1}{2} \sqrt{\frac{3}{2}}
   \left(k_5-\mathcal{H}_1-\mathcal{H}_2\right) \\
   \hline
 E_{-1} & -\frac{1}{2
   \sqrt{2}} \left(k_3+3 \mathcal{H}_2+\mathcal{H}_1\right)\\
   \hline
 E_{-2} & \frac{1}{4} \sqrt{\frac{3}{2}}
   \left(k_1-k_2+k_4+k_6\right) \\
   \hline
 E_{-3} & -\frac{1}{4 \sqrt{2}}\,\left(3 h_1+h_2+h_4+3 h_6\right) \\
 \hline
 E_{-4} & \frac{1}{4 \sqrt{2}}\,\left(3 k_1+k_2-k_4+3 k_6\right) \\
 \hline
 E_{-5} & \frac{1}{4} \sqrt{\frac{3}{2}}
   \left(-h_1+h_2+h_4-h_6\right) \\
   \hline
 E_{-6} & -\frac{1}{2} \sqrt{\frac{3}{2}}
   \left(k_5+\mathcal{H}_1+\mathcal{H}_2\right)\\
   \hline
\end{array}
\end{equation}
Note the difference between the above generators in the Cartan-Weyl basis and the Chevalley generators ${\bf H}_1,\,{\bf H}_2,{\bf E}_i$ of \cite{bruxelles}. The relation between the two bases is:
\begin{eqnarray}
{H}_1&=& 3{\bf H}_1+2{\bf H}_2\,;\,\,\,H_2={\bf H}_1\,;\,\,\,E_1=\frac{1}{2\sqrt{2}}\,{\bf E}_4\,;\,\,\,E_2=-\frac{1}{2\sqrt{6}}\,{\bf F}_5\,;\,\,\,E_3=-\frac{1}{\sqrt{2}}\,{\bf F}_2\,,\nonumber\\
E_4&=&\frac{1}{\sqrt{2}}\,{\bf E}_3\,;\,\,\,E_5=-\frac{1}{2\sqrt{6}}\,{\bf E}_6\,;\,\,\,E_6=-\sqrt{\frac{3}{2}}\,{\bf E}_1\,.
\end{eqnarray}
Although our expressions are different, yet in agreement with \cite{bruxelles}, we see that four step operators (for us $E_1,E_2,E_4, E_6$) lie in $\mathbb{K}$ while two lie in $\mathbb{H}^\star$, (for us $E_3,E_5$).
\par
Having established which step operators lie inside $\mathbb{K}$ we can now make use of a general theorem about solvable and nilpotent algebras which states that for every linear representation of such algebras one can find a basis where all its elements are upper triangular matrices \cite{helgason}. Transferred to our context this means that for each $\mathrm{H}^\star$-orbit of nilpotent $\mathbb{K}$-operators we can find at least one representative which is a linear combination of the step up operators lying in $\mathbb{K}$. Hence let us consider a generic linear combination of the four step operators $E^1,E^2,E^4,E^6$.
\begin{equation}\label{omu}
    \mathcal{O}_{\vec{\mu}} \, = \, \mu_1 \, E_1 \, + \, \mu_2 \, E_2 \, + \, \mu_4 E_4 \, + \, \mu_6 E_6
\end{equation}
and let us evaluate the subsequent powers of $\mathcal{O}_{\vec{\mu}}$. For generic coefficients $\vec{\mu}$ we find that:
\begin{equation}\label{settenillo}
    \mathcal{O}_{\vec{\mu}}^7 \, = \, \mathbf{0} \quad ; \quad \mathcal{O}_{\vec{\mu}}^n \, \ne \, \mathbf{0} \quad \mbox{for} \,\,\, n\, < \, 7
\end{equation}
The corresponding tensor classifiers of $\mathcal{O}_{\vec{\mu}}$ are the following ones:
\begin{eqnarray}
  \mathcal{T}^{xy} &=& \left(
\begin{array}{lll}
 0 & 0 & 0 \\
 0 & 0 & -\frac{\mu _1 \mu _2}{\sqrt{3}}
   \\
 0 & -\frac{\mu _1 \mu _2}{\sqrt{3}} &
   \frac{2 \mu _2 \mu _4}{\sqrt{3}}
\end{array}
\right) \label{tizio}\\
  \mbox{eigenval} \, \mathcal{T}^{xy} &=& \frac{\mu _2}{\sqrt{3}}\,\left\{0, \mu _4-\sqrt{\mu _1^2+\mu
   _4^2}, \mu _4+\sqrt{\mu _1^2+\mu
   _4^2}\right\} \label{eigentizio}\\
 \mathfrak{T}^{xy}  &=& \left(
\begin{array}{lll}
 0 & 0 & 0 \\
 0 & 0 & 0 \\
 0 & 0 & -\frac{27}{2} \mu _1^2 \mu _2^2
\end{array}
\right) \label{caio}\\
   \mbox{eigenval} \, \mathfrak{T}^{xy} &=& \left\{0,0,-\frac{27}{2} \mu _1^2 \mu
   _2^2\right\}\label{eigencaio} \\
{\mathbb{T}}^{ab} &=& \left(
\begin{array}{lll}
 0 & 0 & 0 \\
 0 & 0 & -\frac{9}{2} \sqrt{3} \mu _1^3
   \mu _2 \\
 0 & -\frac{9}{2} \sqrt{3} \mu _1^3 \mu
   _2 & \frac{9}{2} \mu _1^2 \left(3 \mu
   _4^2+4 \sqrt{3} \mu _1 \mu _6\right)
\end{array}
\right)\label{sempronio}\\
   \mbox{eigenval}\, {\mathbb{T}}^{ab} &=& \frac{9}{4}\,\mu_1^2 \left\{0,A+ \sqrt{A^2+12\,\mu_1^2\mu_2^2},A- \sqrt{A^2+12\,\mu_1^2\mu_2^2}\right\}\nonumber\\ \label{eigensempronio}
\end{eqnarray}
\begin{equation}
    W^{x|b} \, = \, \left(
\begin{array}{lll}
 0 & 0 & 0 \\
 0 & 0 & \frac{3}{2} \sqrt{3} \mu _1 \mu _2 \\
 -3 \mu _1^2 & \frac{3 \mu _1 \mu _4}{2} & -3 \left(\mu
   _4^2+\sqrt{3} \mu _1 \mu _6\right)
\end{array}
\right)\label{bivettornillo}
\end{equation}
where we have defined:
\begin{eqnarray}
A&\equiv& 3\mu_4^2+4\sqrt{3}\,\mu_1\mu_6
\end{eqnarray}
We see that, for generic $\mu_i$ the invariant signature of the three matrices $\mathcal{T},\,\mathfrak{T},\mathbb{T}$ is:
\begin{eqnarray}
 \mbox{eigenval} \, \mathcal{T}^{xy} &=& \left\{0, +, -\right\}\,\nonumber\\
  \mbox{eigenval} \, \mathfrak{T}^{xy} &=& \left\{0, 0, -\right\}\nonumber\\
   \mbox{eigenval} \, \mathbb{T}^{ab} &=& \left\{0, +, -\right\}\nonumber\\
\end{eqnarray}
The corresponding $H^*$-orbit consists of step-7 nilpotent generators and will be denoted by $\mathrm{NO_5}$.\par
Next we consider lower powers of  $\mathcal{O}_{\vec{\mu}}$ and we inquire under which conditions on the coefficients $\vec{\mu}$  they might vanish.
\par
We begin with the third power and we find:
\begin{equation}\label{powerthree}
    \mathcal{O}_{\vec{\mu}}^3 \, = \, \sqrt{3} \, \mu_1^2 \, \mu_2 \, \left(
\begin{array}{lllllll}
 0 & -\frac{1}{2 \sqrt{2}} & 0 &
   -\frac{1}{4} & 0 & 0 & 0 \\
 -\frac{1}{2 \sqrt{2}} & 0 & 0 & 0 &
   -\frac{1}{2 \sqrt{2}} & 0 & 0 \\
 0 & 0 & 0 & \frac{1}{4} & 0 &
   \frac{1}{2 \sqrt{2}} & 0 \\
 \frac{1}{4} & 0 & \frac{1}{4} & 0 &
   \frac{1}{4} & 0 & \frac{1}{4} \\
 0 & \frac{1}{2 \sqrt{2}} & 0 &
   \frac{1}{4} & 0 & 0 & 0 \\
 0 & 0 & -\frac{1}{2 \sqrt{2}} & 0 & 0 &
   0 & -\frac{1}{2 \sqrt{2}} \\
 0 & 0 & 0 & -\frac{1}{4} & 0 &
   -\frac{1}{2 \sqrt{2}} & 0
\end{array}
\right)
\end{equation}
Hence it follows that the operators of nilpotency three are characterized by the simple equation:
\begin{equation}\label{simplequation}
    \mu_1^2 \, \mu_2 \, = \, 0
\end{equation}
which obviously has two independent solutions, namely, either $\mu_1 \, = \, 0$ or $\mu_2\, = \, 0$.
\par
Calculating  the other powers $\mathcal{O}_{\vec{\mu}}^n$ for $n=4,5,6$ we see that they vanish only under the same condition (\ref{simplequation}), no other independent zeros showing up. Finally we consider the case of nilpotency $d_n \, =\,2$ and we see that the unique solution of the equation $\mathcal{O}_{\vec{\mu}}^2\, = \, 0$ is given by:
\begin{equation}\label{duenillo}
    \mu_1 \, = \, \mu_4 \, = \, 0
\end{equation}
Let us analyze the various cases separately.\par
{\underline{$\mu_1=0$}.} We find
\begin{eqnarray}
 \mbox{eigenval} \, \mathcal{T}^{xy} &=& \left\{0, 0, \frac{2}{\sqrt{3}}\mu_2\mu_4\right\}\,\,\Rightarrow \,\,\mbox{non-BPS}\,\nonumber\\
  \mbox{eigenval} \, \mathfrak{T}^{xy} &=& \left\{0, 0, 0\right\}\nonumber\\
   \mbox{eigenval} \, \mathbb{T}^{ab} &=& \left\{0, 0, 0\right\}\nonumber
\end{eqnarray}
while
\begin{equation}
    W^{x|b} \, = \, \left(
\begin{array}{lll}
 0 & 0 & 0 \\
 0 & 0 &0 \\
 0 & 0& -3 \mu
   _4^2
\end{array}\right)\neq {\bf 0}
\end{equation}
The non-zero eigenvalue of $\mathcal{T}$  can be either positive or negative, thus defining two $H^*$-orbits, to be denoted by $\mathrm{NO}_3,\,\mathrm{NO}'_3$, respectively:
\begin{eqnarray}
\mathrm{NO}_3&:&\mu_2\mu_4>0\nonumber\\
\mathrm{NO}'_3&:&\mu_2\mu_4<0
\end{eqnarray}
Representatives
 of these two orbits are:
 \begin{eqnarray}
 \mathrm{NO}_3&:&|\mu_2|\,E_2+|\mu_4|\,E_4\nonumber\\
\mathrm{NO}'_3&:&|\mu_2|\,E_2-|\mu_4|\,E_4
 \end{eqnarray}
 and the corresponding orbits were denoted by $O_{4K}',\,O_{4K}$ in \cite{bruxelles}.
In the sequel we shall work out representatives of $\mathrm{NO}_3$
which correspond to regular non-BPS solutions. In section
\ref{generat} we explicitly show that $\mathrm{NO}'_3$ contains
singular solutions.\par We can further set $\mu_4=0$, in which
case we obtain the orbit defined by the invariant properties:
\begin{eqnarray}
 \mbox{eigenval} \, \mathcal{T}^{xy} &=& \left\{0, 0, 0\right\}\,\,\Rightarrow \,\,\mbox{BPS}\,\nonumber\\
  \mbox{eigenval} \, \mathfrak{T}^{xy} &=& \left\{0, 0, 0\right\}\nonumber\\
   \mbox{eigenval} \, \mathbb{T}^{ab} &=& \left\{0, 0, 0\right\}\nonumber\\
   W^{x|a}={\bf 0}\nonumber
\end{eqnarray}
This orbit will be denoted by $\mathrm{NO_1}$ and its elements are step-2 nilpotent. It contains small black holes, as we shall show in the sequel.\par
{\underline{$\mu_2=0$}.} We find
\begin{eqnarray}
 \mbox{eigenval} \, \mathcal{T}^{xy} &=& \left\{0, 0,0\right\}\,\,\Rightarrow \,\,\mbox{BPS}\,\nonumber\\
  \mbox{eigenval} \, \mathfrak{T}^{xy} &=& \left\{0, 0, 0\right\}\nonumber\\
   \mbox{eigenval} \, \mathbb{T}^{ab} &=& \left\{0, 0,\frac{9}{2}\mu_1^2(4\sqrt{3}\mu_1\mu_6+3\mu_4^2)\right\}\nonumber\\
      W^{x|b} \, &=& \, \left(
\begin{array}{lll}
 0 & 0 & 0 \\
 0 & 0 &0 \\
 -3\,\mu^2_1 & 0& -3\sqrt{3} \mu_1\mu_6
\end{array}\right)\neq {\bf 0}
\end{eqnarray}
In this case we also have two possible signs for the non-vanishing
eigenvalue of $\mathbb{T}^{ab}$. Setting for the sake of
simplicity $\mu_4=0$, we can have two $H^*$-orbits
$\mathrm{NO}_4,\,\mathrm{NO}'_4$, corresponding to the cases
$\mu_1\mu_6>0$ and  $\mu_1\mu_6<0$, respectively:
 \begin{eqnarray}
\mathrm{NO}_4&:&\mu_1\mu_6>0\nonumber\\
\mathrm{NO}'_4&:&\mu_1\mu_6<0
\end{eqnarray}
 Just as for the previous  non-BPS orbits, while $\mathrm{NO}_4$ contains the known regular BPS solution, $\mathrm{NO}'_4$ contains singular solutions, as will be motivated in section \ref{generat}. Representatives
 of these two orbits are:
 \begin{eqnarray}
 \mathrm{NO}_4&:&|\mu_1|\,E_1+|\mu_6|\,E_6\nonumber\\
\mathrm{NO}'_4&:&|\mu_1|\,E_1-|\mu_6|\,E_6
 \end{eqnarray}
Since $E_1$ and $E_6$ correspond to orthogonal roots, these two representatives, and thus the corresponding two orbits, are mapped into one another by means of an $H^\mathbb{C}$-transformation, according to the general property proven in section \ref{generat}. These two representatives correspond to those given in \cite{bruxelles}
for the orbits $O_{3K},\,O_{3K}'$.\par
We can further take the limit $\mu_6\rightarrow 0$ obtaining the invariant properties:
\begin{eqnarray}
 \mbox{eigenval} \, \mathcal{T}^{xy} &=& \left\{0, 0,0\right\}\,\,\Rightarrow \,\,\mbox{BPS}\,\nonumber\\
  \mbox{eigenval} \, \mathfrak{T}^{xy} &=& \left\{0, 0, 0\right\}\nonumber\\
   \mbox{eigenval} \, \mathbb{T}^{ab} &=& \left\{0, 0,0\right\}\nonumber\\
      W^{x|b} \, &=& \, \left(
\begin{array}{lll}
 0 & 0 & 0 \\
 0 & 0 &0 \\
 -3\,\mu^2_1 & 0& 0
\end{array}\right)\neq {\bf 0}
\end{eqnarray}
which define a further orbit $\mathrm{NO}_2$ consisting of step-3 nilpotent matrices. This orbit will contain small black holes, as shown in the following.

In the following table we list five of the seven orbits discussed above, which we shall be mainly interested in, giving representatives thereof:
\begin{equation}\label{nillaorbita}
    \begin{array}{|c|l|r c l|}
    \hline
       \mbox{Orbit } & \mbox{Abstract } & \null &\mbox{Repr. at}& \null\\
       \mbox{name}& \mbox{Repres.} & \null &\mbox{ Taub-NUT}  &  = 0\\
       \null & \null & \null & \null &  \null\\
       \hline
       \mathrm{NO_1} &\mu_2 E_2 + \mu_6 E_6 & \mathcal{L}^{NO_1} & = & 2\, \sqrt{\ft 23}\, \mathcal{R}_1\, E_2 \, \mathcal{R}_1^{-1}\\
       \mathrm{NO_2} & \mu_4 E_4 + \mu_6 E_6 & \mathcal{L}^{NO_2} & = & 4 \, \sqrt{2}\,\mathcal{R}_2\, E_4 \, \mathcal{R}_2^{-1}\\
       \mathrm{NO_3} & |\mu_2| E_2 + |\mu_4| E_4 + \mu_6 E_6  & \mathcal{L}^{NO_3} & = & \mathfrak{L}^{(p|q)}_0\\
       \mathrm{NO_4} & \mu_1 E_1 + \mu_4 E_4 + \mu_6 E_6 &\mathcal{L}^{NO_4} & = & \widehat{\mathfrak{L}}^{(1|-1)}_0\\
       \mathrm{NO_5}& \mu_1 E_1 +\mu_2 E_2 + \mu_4 E_4 + \mu_6 E_6 & \mathcal{L}^{NO_5}& = & 2 \sqrt{\frac{2}{3}} \, \mathcal{R}_3\,\left ( E_1 \, + \, E_2\right ) \, \mathcal{R}_3^{-1}\\
       \hline
     \end{array}
\end{equation}
where for the $\mathrm{NO}_4$ orbit we take
$\mu_4^2>-\frac{4}{\sqrt{3}}\,\mu_1\mu_6$.
$\mathfrak{L}^{(p|q)}_0$, $ \widehat{\mathfrak{L}}^{(1|-1)}_0$ are
operators defined in later sections  and the elements of the
$\mathrm{H}^\star$ subgroup mentioned above have the following
explicit form:
\begin{eqnarray}
\mathrm{H}^\star \, \ni \, \mathcal{R}_1 &=& e^{\ft \pi 2 (L_1^I - L^I_{-1})}\nonumber \\
 \mathrm{H}^\star \, \ni \,  \mathcal{R}_2&=& e^{\ft \pi 2 (L_1^{II} - L^{II}_{-1})}\nonumber\\
\mathrm{H}^\star \, \ni \,   \mathcal{R}_3 &=& e^{  - \,\ft{\pi}{6}\, (L_+^{(I)} - L_-^{(I)} ) }\label{explicithstarro}
\end{eqnarray}
For later use convenience, in table (\ref{nillaorbita}) we have
already anticipated  a standard representative of each orbit at
vanishing Taub-NUT charge. \par
Summarizing we have found seven nilpotent $H^*$-orbits in $\mathbb{K}$: $\mathrm{NO_1},\,\mathrm{NO_2},\,\mathrm{NO_3},\,\mathrm{NO'_3},\,\mathrm{NO_4}$, $\mathrm{NO'_4},\,\mathrm{NO_5}$, each characterized by different $H^*$-invariant properties of the tensor classifiers.

\par
For each of these orbits we can study the structure of the stability subgroup and in this way determine its actual dimension. In the following section we shall focus on the ``unprimed'' orbits,  within which we choose
the simple representative of each orbit presented in the second column of table (\ref{nillaorbita}) as the Lax operator and we construct the corresponding supergravity solution, the value of the Taub-NUT charge in general is non-zero. Hence it is more convenient to choose different representatives corresponding to the vanishing Taub-NUT shell: $\mathbf{n}=0$. The third column of table (\ref{nillaorbita}) precisely provides such representatives. In the case of the first, second and fifth nilpotent orbits, the chosen representative is explicitly given as an $\mathrm{H}^\star$-rotation of one of the   representatives mentioned in the second column. As the reader can see
such a rotation is always a compact one belonging either to the first or the second $\mathrm{SL(2,\mathbb{R})}$ group. As we already mentioned, for the third and fourth orbit we preferred to choose another representative which  is directly constructed below (see sect.\ref{BPSsino}). Restricting the representatives to generate a vanishing Taub-NUT charge lowers each orbit dimension by one unit. In section \ref{generat} we shall consider representatives also of the primed orbits and comment on their four-dimensional interpretation.
\par
So let us scan the stability subgroups of the five nilpotent orbits $\mathrm{NO}_i$, $i=1,\dots, 5$.
\subsubsection{The very large nilpotent orbit $\mathrm{NO_5}$}
The tensor identifiers of this orbit can be summarized as follows:
\begin{eqnarray}
  \mathcal{T}^{xy}&\Rightarrow& \{0\, , \, + \, , \, - \}\nonumber\\
  \mathfrak{T}^{xy} &\Rightarrow &\{0\, , \, 0 \, , \, - \}\nonumber \\
  \mathbb{T}^{ab} &\Rightarrow& \{0\, , \, + \, , \, - \}\nonumber\\
  W^{x|q} &\ne & 0 \label{labelliNO5}
\end{eqnarray}
Calculating the elements of the $\mathbb{H}^\star$ subalgebra that commute with $\mu_1 E_1 + \mu_2 E_2 + \mu_4 E_4+ \mu_6 E_6$
we find that there are none as long as all the $\mu$-coefficients are different from zero. Hence the dimension of this orbit is as large as that of the $\mathrm{H}^\star$-group, namely six. The vanishing Taub-NUT shell inside $\mathrm{NO_5}$ is five-dimensional:
\begin{equation}\label{dimmashella5}
    \mbox{dim} \,\left[ \left(\mbox{$\mathbf{n}=0$ shell}\right) \, \subset \, \mathrm{NO_5}\right ] \, = \, 5
\end{equation}
\subsubsection{The  large nilpotent orbit $\mathrm{NO_4}$}
The tensor identifiers of this orbit can be summarized as
follows:
\begin{eqnarray}
  \mathcal{T}^{xy} \, = \, \mathbf{0} &\Rightarrow& \{0\, , \, 0\, , \,0\}\nonumber\\
  \mathfrak{T}^{xy} \, = \, \mathbf{0} &\Rightarrow &\{0\, , \, 0 \, , \, 0 \}\nonumber \\
  \mathbb{T}^{ab} &\Rightarrow& \{0\, , \, 0 \, , \,+ \}\nonumber\\
  W^{x|q} &\ne & {\bf 0} \label{labelliNO4}
\end{eqnarray}
Calculating the elements of the $\mathbb{H}^\star$ subalgebra that commute with $\mu_1 E_1 + \mu_4 E_4+ \mu_6 E_6$
we find that there is just a one-dimensional stability subalgebra generated by:
\begin{equation}\label{crippa1}
   S \, = \,   L_+^{I}
\end{equation}
whose degree of nilpotency is $n=2$. It follows that the stability subgroup of this orbit is the one-dimensional  translation group $\mathbb{R}$.
Hence the dimension of $\mathrm{NO_4}$ is five. The vanishing Taub-NUT shell inside $\mathrm{NO_4}$ is four-dimensional:
\begin{equation}\label{dimmashella4}
    \mbox{dim} \,\left[ \left(\mbox{$\mathbf{n}=0$ shell}\right) \, \subset \, \mathrm{NO_4}\right ] \, = \, 4
\end{equation}
\subsubsection{The  large nilpotent orbit $\mathrm{NO_3}$}
The tensor identifiers of this orbit can be summarized as follows:
\begin{eqnarray}
  \mathcal{T}^{xy}  &\Rightarrow& \{0\, , \, 0\, , \,+ \}\nonumber\\
  \mathfrak{T}^{xy} \, = \, \mathbf{0} &\Rightarrow &\{0\, , \, 0 \, , \, 0 \}\nonumber \\
  \mathbb{T}^{ab} &\Rightarrow& \{0\, , \, 0 \, , \, 0 \}\nonumber\\
  W^{x|q} &\ne & 0 \label{labelliNO3}
\end{eqnarray}
Calculating the elements of the $\mathbb{H}^\star$ subalgebra that commute with $\mu_2 E_2 + \mu_4 E_4+ \mu_6 E_6$
we find that also in this case there is a one-dimensional stability subalgebra that  is generated by another nilpotent element of $\mathbb{H}^\star$, namely:
\begin{equation}\label{crippa21}
  S \, = \,  \sqrt{3} \, \frac{\mu_4}{\mu_2} \,  L_+^{I} \, + \, L_+^{II}
\end{equation}
 It follows that also in this case the stability subgroup is a one-dimensional translation group $\mathbb{R}$, since the operator $S$ is nilpotent.
Hence also the dimension of $\mathrm{NO_3}$ is five. The vanishing Taub-NUT shell inside $\mathrm{NO_3}$ is four-dimensional:
\begin{equation}\label{dimmashella3}
    \mbox{dim} \,\left[ \left(\mbox{$\mathbf{n}=0$ shell}\right) \, \subset \, \mathrm{NO_3}\right ] \, = \, 4
\end{equation}
\begin{figure}[!hbt]
\begin{center}
\iffigs
 \includegraphics[height=80mm]{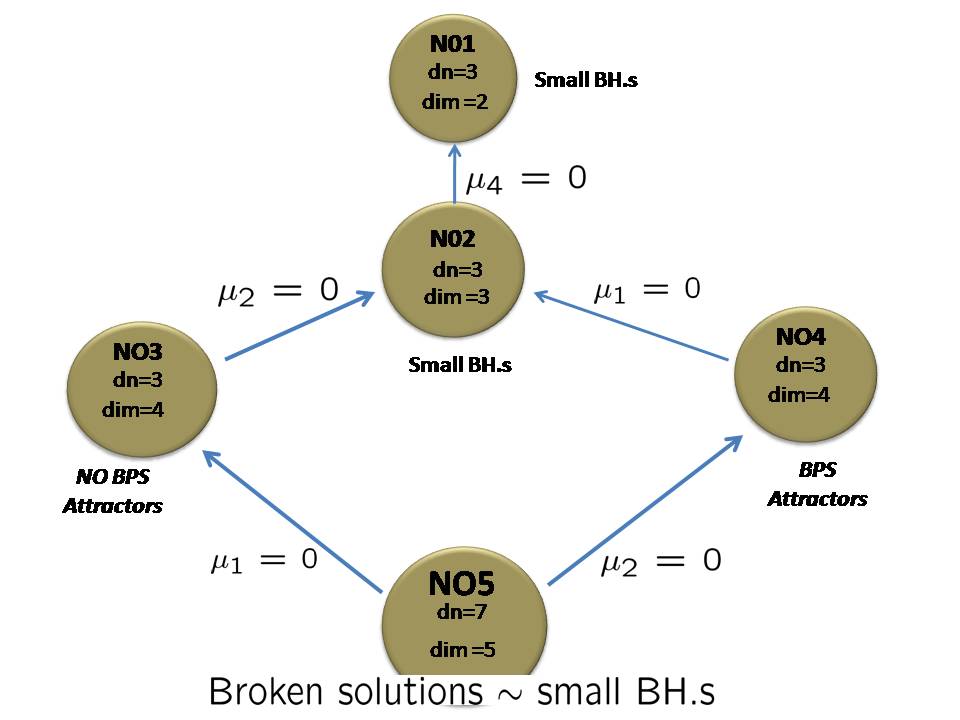}
\else
\end{center}
 \fi
\caption{\it
Organization of the $\mathrm{H}^\star$ orbits of nilpotent operators in the case of the $\mathfrak{g}_{2(2)}$ model.
The physical content of the orbits, anticipated in this picture is examined in the next section.}
\label{orbitando}
 \iffigs
 \hskip 1.5cm \unitlength=1.1mm
 \end{center}
  \fi
\end{figure}
\subsubsection{The  small nilpotent orbit $\mathrm{NO_2}$}
The tensor identifiers of this orbit can be summarized as follows:
\begin{eqnarray}
  \mathcal{T}^{xy} \, = \, \mathbf{0} &\Rightarrow& \{0\, , \, 0\, , \,0 \}\nonumber\\
  \mathfrak{T}^{xy} \, = \, \mathbf{0} &\Rightarrow &\{0\, , \, 0 \, , \, 0 \}\nonumber \\
  \mathbb{T}^{ab}\, = \, \mathbf{0} &\Rightarrow& \{0\, , \, 0 \, , \, 0 \}\nonumber\\
  W^{x|q} &\ne & 0 \label{labelliNO2}
\end{eqnarray}
Calculating the elements of the $\mathbb{H}^\star$ subalgebra that commute with $ \mu_4 E_4+ \mu_6 E_6$
we find that in this case there is a two dimensional stability subalgebra generated by the following combinations of $\mathbb{H}^\star$, standard generators:
\begin{eqnarray}
  S_0 &=& -L_0^I \, + \, L_0^{II} \, - \, \frac{\mu_6}{\sqrt{3} \, \mu_4} \, L_+^{II} \\
  S_{-1} &=& L_+^{I}\label{stabilino2}
\end{eqnarray}
which satisfy the commutation relation:
\begin{equation}\label{s1s0}
    \left [ S_0 \, , \, S_1 \right ] \, = \, - \, S_{-1}
\end{equation}
The operator $S_0$ is diagonalizable, while the operator $S_{-1}$ is nilpotent.
 It follows that  in this case the stability subgroup is the semidirect product $\mathrm{O(1,1)} \ltimes \mathbb{R}$.
Hence the dimension of $\mathrm{NO_2}$ is four. The vanishing Taub-NUT shell inside $\mathrm{NO_2}$ is three-dimensional:
\begin{equation}\label{dimmashella2}
    \mbox{dim} \,\left[ \left(\mbox{$\mathbf{n}=0$ shell}\right) \, \subset \, \mathrm{NO_2}\right ] \, = \, 3
\end{equation}
\subsubsection{The  very small nilpotent orbit $\mathrm{NO_1}$}
The tensor identifiers of this orbit can be summarized as follows:
\begin{eqnarray}
  \mathcal{T}^{xy} \, = \, \mathbf{0} &\Rightarrow& \{0\, , \, 0\, , \,0 \}\nonumber\\
  \mathfrak{T}^{xy} \, = \, \mathbf{0} &\Rightarrow &\{0\, , \, 0 \, , \, 0 \}\nonumber \\
  \mathbb{T}^{ab}\, = \, \mathbf{0} &\Rightarrow& \{0\, , \, 0 \, , \, 0 \}\nonumber\\
  W^{x|q} &= & 0 \label{labelliNO1}
\end{eqnarray}
Calculating the elements of the $\mathbb{H}^\star$ subalgebra that commute with $ \mu_2 E_2+ \mu_6 E_6$
we find that in this case there is a three dimensional stability subalgebra generated by the following combinations of $\mathbb{H}^\star$, standard generators:
\begin{eqnarray}
  S_{-3} &=& L_0^I \, +  \,\frac{\mu_6}{2 \, \mu_2} \, \left( L_+^{I} -L_-^{I}\right)\nonumber\\
  S_{0} &=& L_0^{II}\, +  \,\frac{3\,\mu_6}{2 \, \mu_2} \, \left( L_+^{I} +L_-^{I}\right)\nonumber\\
  S_{1} &=& L_+^{II}\label{stabilino1}
\end{eqnarray}
which satisfy the commutation relation:
\begin{eqnarray}\label{s1s0bis}
    \left [ S_0 \, , \, S_{-3} \right ] & = & - \, S_{-3}\nonumber\\
    \left [ S_0 \, , \, S_{1} \right ] & = &  \, S_{1}\nonumber\\
    \left [ S_{-3} \, , \, S_{1} \right ] & = & 0\nonumber\\
\end{eqnarray}
The operator $S_0$ is diagonalizable, while the operators $S_{-3}$,$S_{1}$ are nilpotent and commute among themselves.
 It follows that  in this case the stability subgroup is the semidirect product $\mathrm{O(1,1)} \ltimes \mathbb{R}^2$.
Hence the dimension of $\mathrm{NO_2}$ is three. The vanishing Taub-NUT shell inside $\mathrm{NO_2}$ is two-dimensional:
\begin{equation}\label{dimmashella1}
    \mbox{dim} \,\left[ \left(\mbox{$\mathbf{n}=0$ shell}\right) \, \subset \, \mathrm{NO_2}\right ] \, = \, 2
\end{equation}
The organization of nilpotent orbits is pictorially summarized in fig.\ref{orbitando}.
\section{Scanning of Supergravity solutions in the various orbits}
\label{scanning} Having characterized in a precise algebraic way the
space of nilpotent orbits of possible Lax operators it remains to be
seen what is their physical content, namely which type of
supergravity solutions is generated by the integration algorithm
starting from the initial Laxes of each orbit. In this section we
perform this task by examining  one by one the supergravity solutions
associated with each orbit and discussing their fundamental
properties. We do this starting from the regular orbits of
diagonalizable Laxes that are associated with non extremal
Black-Holes. In particular in connection with these latter we
analyse in careful detail the question of regularity of non-extremal
Black-Holes revising and making some of the statements
that appear in the literature  more precise. It is important to stress that, with the exception of the orbit $\mathrm{NO_5}$, the representatives of the other orbits that will be derived are known solutions.
In particular the regular  BPS  and non-BPS solutions (orbits $\mathrm{NO_4}$ and $\mathrm{NO_3}$ respectively)
were first derived in the context of cubic geometries in \cite{Behrndt:1997ny} and \cite{Lopes Cardoso:2007ky}, respectively.
\subsection{The regular orbits: non extremal Schwarzschild Black-Holes}
We begin our scanning with the regular orbit of diagonalizable Lax operators $L_0^{diag} \in \mathbb{K}$. Using a general Lie algebra theorem we know that any such element of $\mathbb{K}$ can be $\mathrm{H}^\star$-rotated into a Cartan element. Hence as general representative of non-extremal Black-Hole orbits we can take the following two-parameter diagonal matrix:
\begin{equation}\label{fraudolento}
    L_0^{diag}(\alpha,\beta) \, = \, \alpha \,\mathcal{H}_1 \, + \, \beta \, \mathcal{H}_2
\end{equation}
Evaluating the invariants and tensor structures of this operator we obtain:
\begin{eqnarray}
  \mathfrak{h}_2 &= & 2 \left(\alpha ^2-3 \beta  \alpha +3
   \beta ^2\right)\quad ; \quad \mathfrak{h}_6 = \beta ^2 \left(\alpha ^2-3 \beta
   \alpha +2 \beta ^2\right)^2 \nonumber\\
  \mathcal{T}^{xy} &=& \left(
\begin{array}{lll}
 \frac{1}{3} \left(-\alpha ^2+4
   \beta  \alpha -3 \beta
   ^2\right) & 0 & \frac{1}{9}
   \left(\alpha ^2-3 \beta
   ^2\right) \nonumber\\
 0 & \frac{1}{9} \left(\alpha
   ^2-3 \beta ^2\right) & 0 \\
 \frac{1}{9} \left(\alpha ^2-3
   \beta ^2\right) & 0 &
   \frac{1}{3} \left(-\alpha
   ^2+4 \beta  \alpha -3 \beta
   ^2\right)
\end{array}
\right)\nonumber\\
  \mathfrak{T}^{xy} &=& \left(
\begin{array}{lll}
 -\frac{9}{2} (\alpha -3 \beta
   ) (\alpha -\beta )^3 & 0 &
   \frac{9}{2} (\alpha -\beta
   )^2 \left(\alpha ^2-4 \beta
   \alpha +5 \beta ^2\right) \\
 0 & -\frac{9}{2} (\alpha -2
   \beta )^2 \beta ^2 & 0 \\
 \frac{9}{2} (\alpha -\beta )^2
   \left(\alpha ^2-4 \beta
   \alpha +5 \beta ^2\right) &
   0 & -\frac{9}{2} (\alpha -3
   \beta ) (\alpha -\beta )^3
\end{array}
\right)\nonumber\\
  \mathbb{T}^{ab} &=& \left(
\begin{array}{lll}
 -\frac{9}{2} (\alpha -3 \beta
   ) (\alpha -\beta )^3 & 0 &
   \frac{9}{2} (\alpha -\beta
   )^2 \left(\alpha ^2-4 \beta
   \alpha +5 \beta ^2\right) \\
 0 & -\frac{9}{2} (\alpha -2
   \beta )^2 \beta ^2 & 0 \\
 \frac{9}{2} (\alpha -\beta )^2
   \left(\alpha ^2-4 \beta
   \alpha +5 \beta ^2\right) &
   0 & -\frac{9}{2} (\alpha -3
   \beta ) (\alpha -\beta )^3
\end{array}
\right)\nonumber \\
  \mathcal{W}^{x|a} & = &\left(
\begin{array}{lll}
 \frac{3}{2} (\alpha -3 \beta )
   (\alpha -\beta ) & 0 &
   -\frac{3}{2} (\alpha -\beta
   )^2 \\
 0 & \frac{3}{2} (\alpha -2
   \beta ) \beta  & 0 \\
 -\frac{3}{2} (\alpha -\beta
   )^2 & 0 & \frac{3}{2}
   (\alpha -3 \beta ) (\alpha
   -\beta )
\end{array}
\right) \nonumber\\
\end{eqnarray}
Hence as long as the two parameters $\alpha$ and $\beta$ are generic the three tensor classifiers $\mathcal{T}^{xy}$,$\mathfrak{T}^{xy}$ and $\mathbb{T}^{ab}$ are all non-degenerate and possess three non vanishing eigenvalues.\par
In \cite{Bossard:2009at} it is stated that the Lax matrix (or, equivalently, the Noether charge matrix)
generating  regular solutions, in the fundamental representation of ${\rm U}_{D=3}$ (and for ${\rm U}_{D=3}\neq {\rm E}_8$), satisfies the following equation:
 \begin{equation}\label{cubicondo}
         L_0^3 = v^2 L_0
      \end{equation}
which was thus given as a necessary condition for regularity, which  includes also extremal solutions by taking the limit $v\rightarrow 0$. A similar quintic equation was written for $L_0$ in the adjoint representation of ${\rm U}_{D=3}$.
 In \cite{Chemissany:2010zp} eq. (\ref{cubicondo}), or its quintic version, was reformulated as the condition that all the  so called \textit{Chevalley hamiltonians}, which  are the  polynomial ($\mathrm{H}^\star$-invariant)
hamiltonians, except the quadratic one, should vanish.

In this section we demonstrate that  equation (\ref{cubicondo}), or the equivalent condition on the Chevalley hamiltonians, is not sufficient to guarantee the regularity of
the black-hole solution and its Schwarzschild character. There is
more than one orbit of diagonalizable Lax operators that are
selected by eq.(\ref{cubicondo}) and they are clearly distinguished
by the tensor classifiers. Only one of them is regular of finite
area and its definition can be stated in terms of $\mathrm{H}^\star$
invariant conditions by means of the tensor classifiers.
\par
First of all we observe that imposing the condition $\mathfrak{h}_6 \, = \,0$ we obtain three possible solutions:
\begin{equation}\label{ruperto}
    \mathfrak{h}_6 \, = \,0 \, \Rightarrow \, \left \{ \begin{array}{cccc}
                                                         \mbox{Schw.} & \beta & = & \frac{\alpha}{2}\\
                                                         \mbox{Dil1} & \beta & = & 0 \\
                                                         \mbox{Dil2} & \beta& = & \alpha
                                                       \end{array}
    \right.
\end{equation}
These three solutions belong to two different orbits of the group $\mathrm{H}^\star$ as it can be clearly established by analyzing the corresponding tensor classifiers. In the case of the solution which we named \textit{Schw}, since it will lead to Schwarzschild black-holes, the tensor $\mathcal{T}^{xy}$ results to be positive definite admitting three positive eigenvalues (signature $\{+,+,+\}$), while both tensors $\mathfrak{T}^{xy}$ and $\mathbb{T}^{ab}$ have rank one and possess one positive and two vanishing eigenvalues (signature $\{+,0,0\}$). On the contrary the two solutions named \textit{Dil1} and \textit{Dil2}, since the corresponding supergravity background involves an evolving dilaton field, are $\mathrm{H}^\star$ conjugate to each other but sit in a separate orbit with respect to the \textit{Schw} orbit. This can be clearly seen from the evaluation of the tensor classifiers. Also in this case the two tensors $\mathfrak{T}^{xy}$ and $\mathbb{T}^{ab}$ have rank one, yet they have signature  $\{-,0,0\}$ rather than $\{+,0,0\}$.
Even more significantly the tensor $\mathcal{T}^{xy}$ is no longer positive definite and has rather Lorentzian signature
$\{+,-,-\}$. Hence the signature of $\mathcal{T}^{xy}$ clearly separates the two orbits that are singled out by the condition
$\mathfrak{h}_6 \, = \,0$.
\par
Another intrinsic distinction between the two orbits \textit{Schw} and \textit{Dil} is provided by the structure of their stability subgroup. For the Schwarzschild orbit the stability subgroup is a compact $\mathrm{SO(2)} \subset \mathrm{SL(2,\mathbb{R})}\times \mathrm{SL(2,\mathbb{R})}$, while for the dilaton orbit it is a non-compact
$\mathrm{SO(1,1)} \subset \mathrm{SL(2,\mathbb{R})}\times \mathrm{SL(2,\mathbb{R})}$. Naming  the two standard representatives of the considered Lax orbits as it follows:
\begin{equation}
\begin{array}{rclcl}
  \mathfrak{S} &=& \alpha\left( \mathcal{H}_1\,+ \, \ft 12 \, \mathcal{H}_2\right) & = &\left(
\begin{array}{lllllll}
 \frac{\alpha }{2} & 0 & 0 & 0 & 0 &
   0 & 0 \\
 0 & \frac{\alpha }{2} & 0 & 0 & 0 &
   0 & 0 \\
 0 & 0 & 0 & 0 & 0 & 0 & 0 \\
 0 & 0 & 0 & 0 & 0 & 0 & 0 \\
 0 & 0 & 0 & 0 & 0 & 0 & 0 \\
 0 & 0 & 0 & 0 & 0 & -\frac{\alpha
   }{2} & 0 \\
 0 & 0 & 0 & 0 & 0 & 0 &
   -\frac{\alpha }{2}
\end{array}
\right) \\
\end{array}
\end{equation}
\begin{equation}
\begin{array}{rclcl}
  \mathfrak{D} &=& \alpha \, \mathcal{H}_1 & = &\left(
\begin{array}{lllllll}
 0 & 0 & 0 & 0 & 0 & 0 & 0 \\
 0 & \alpha  & 0 & 0 & 0 & 0 & 0 \\
 0 & 0 & -\alpha  & 0 & 0 & 0 & 0 \\
 0 & 0 & 0 & 0 & 0 & 0 & 0 \\
 0 & 0 & 0 & 0 & \alpha  & 0 & 0 \\
 0 & 0 & 0 & 0 & 0 & -\alpha  & 0 \\
 0 & 0 & 0 & 0 & 0 & 0 & 0
\end{array}
\right)
\end{array}\label{schwarzwaldo}
\end{equation}
we can verify that the stability subalgebra of both is one-dimensional and it is respectively generated by:
\begin{eqnarray}
    \mathfrak{t}_{Schw} & = & 3\left( L_+^I - L_-^I \right) + \left( L_+^{II} - L_-^{II} \right) \label{schwstab}\\
    \mathfrak{t}_{Dil} & = & 3\left( L_+^I + L_-^I \right) + \left( L_+^{II} + L_-^{II} \right) \label{dilstab}
\end{eqnarray}
the operator in eq.(\ref{schwstab})commuting with $\mathfrak{S}$, that in eq.(\ref{dilstab}) commuting instead with $\mathfrak{D}$.
\par
It remains to be seen how the condition of regularity of the black hole solution selects orbits.
\par
To this purpose we implement the integration algorithm with the general Lax operator $L_0^{diag}(\alpha,\beta)$ and the identity matrix as initial coset representative ($\mathbb{L}_0 \, = \, 1$). The result is provided by the following very simple data:
\begin{eqnarray}
  \exp[-U(\tau)] &=& - \, \alpha \, \tau \\
  z(\tau) &=& {\rm i} \, \exp[-(\alpha -2\beta)\, \tau] \\
  Z^A(\tau) &=& 0 \\
  a(\tau) &=& 0 \\
  v^2 & = & \alpha^2 - 3 \, \alpha \, \beta + 3\, \beta^2
\end{eqnarray}
The corresponding metric has the following form:
\begin{equation}\label{farwest}
    ds^2 \, = \, - \mathfrak{A}(\tau) \, dt^2 \, + \, \mathfrak{B}(\tau) \, d\tau^2 \, + \,\mathfrak{C}(\tau) \, \left(d\theta^2 \, + \, \sin^2\theta \, d\phi^2\right)
\end{equation}
where:
\begin{eqnarray}
  \mathfrak{A}(\tau)  &=& -e^{\alpha  \tau } \\
  \mathfrak{B}(\tau) &=& \frac{16 e^{4 \sqrt{\alpha
   ^2-3 \beta  \alpha +3 \beta
   ^2} \tau -\alpha  \tau }
   \left(\alpha ^2-3 \beta
   \alpha +3 \beta
   ^2\right)^2}{\left(-1+e^{2
   \sqrt{\alpha ^2-3 \beta
   \alpha +3 \beta ^2} \tau
   }\right)^4} \\
  \mathfrak{C}(\tau) &=& \frac{4 e^{2 \sqrt{\alpha ^2-3
   \beta  \alpha +3 \beta ^2}
   \tau -\alpha  \tau }
   \left(\alpha ^2-3 \beta
   \alpha +3 \beta
   ^2\right)}{\left(-1+e^{2
   \sqrt{\alpha ^2-3 \beta
   \alpha +3 \beta ^2} \tau
   }\right)^2}
\end{eqnarray}
Independently from the values of $\alpha$ and $\beta$ this metric is regular at $\tau \, = \, 0$. In order to be regular also at the horizon point, namely in the limit $\tau \to -\,\infty$ three very simple conditions have to be satisfied:
\begin{eqnarray}
  \alpha & \ge & 0 \\
  \alpha \, - \, 2 \, \beta  &\ge & 0\\
  2\sqrt{\alpha^2 -3\beta \,\alpha \, + \, 3\beta^2} &\ge & \alpha
\end{eqnarray}
By squaring we see that the last inequality is always satisfied when the first two are fulfilled. Hence all black holes arising from
generic diagonalizable Lax operators are regular yet the horizon area is always zero except for the case when $2\sqrt{\alpha^2 -3\beta \,\alpha \, + \, 3\beta^2} =\alpha$. Indeed we have:
\begin{equation}\label{limitini}
    \lim_{\tau \to -\infty} \, \mathfrak{C}(\tau) \, = \, \left\{ \begin{array}{ccccc}
                                                                   0 & \mbox{if} &2\sqrt{\alpha^2 -3\beta \,\alpha \, + \, 3\beta^2} > \alpha  & \null & \null\\
                                                                   \alpha^2 & \mbox{if} & 2\sqrt{\alpha^2 -3\beta \,\alpha \, + \, 3\beta^2} = \alpha& \Rightarrow & \beta=\frac{\alpha}{2}\\
                                                                 \end{array}
    \right.
\end{equation}
In this way we arrive at the conclusion that the only finite area regular black-holes are those of the \textit{Schw} orbit characterized not only by the vanishing of the Chevalley hamiltonian $\mathfrak{h}_6$ but also by the condition that the tensor classifier $\mathcal{T}^{xy}$ should be positive definite. When we set $\beta=\frac{\alpha}{2}$ the scalar field freezes to a constant, the extremality parameter becomes $v^2 \, = \, \frac{\alpha^2}{4}$ and, as we already shew in eq.s (\ref{schwarzacasa},\ref{posiziona},\ref{schwarazata}), the metric becomes the standard Schwarzschild metric.
\par
We guess that the same result should be true for all symmetric spaces in the classification of Special Geometries yet the above simple proof has to be redone in all instances in order to be completely sure.
\par
Having discussed also the regular non-extremal orbits we are in a position to summarize the entire spectrum of $\mathrm{H}^\star$ orbits. This is done in table \ref{tablizanilla}.
\renewcommand{\arraystretch}{1}
\begin{table}
\begin{center}
{\small
\begin{tabular}{||c||c|c|c|c|c|c|c|c|c||}
\hline\hline
Orbit & Order & Stand. & Stab. & Sign.& Sign.&Sign.& Bivect.& $\mathfrak{I}_4$& Dim.\\
\null & Nilp. & Repr. & subg. &$\mathcal{T}^{xy}$ & ${\mathfrak{T}^{xy}}$& ${\mathbb{T}^{ab}}$&$W^{a|x}$& at  & $\mathbf{n}=0$ \\
\null & \null & \null & \null &\null & \null& \null &\null &$\mathbf{n}=0$ & shell\\
\hline
\hline
Schw.&$\infty$& $\mathfrak{S }$&$\mathrm{O(2)}$&$\{+,+,+\}$&$\{+,0,0\}$&$\{+,0,0\}$&$\ne 0$&$\ne 0$&$4$
\\
\hline
Dil.  &$\infty$& $\mathfrak{D }$&$\mathrm{O(1,1)}$&$\{-,-,+\}$&$\{-,0,0\}$&$\{-,0,0\}$&$\ne 0$&$\ne 0$&$4$
\\
\hline
\hline
$\mathrm{NO}_1$ &$2$& $\mathcal{L}^{NO_1}$&$\mathrm{O(1,1)}\ltimes \mathbb{R}^2 $&$\{0,0,0\}$&$\{0,0,0\}$&$\{0,0,0\}$&$0$&$0$&$2$\\
\hline
$\mathrm{NO}_2$ &$3$& $\mathcal{L}^{NO_2}$&$\mathrm{O(1,1)}\ltimes \mathbb{R} $&$\{0,0,0\}$&$\{0,0,0\}$&$\{0,0,0\}$&$\ne 0$&$0$&$3$\\
\hline
$\mathrm{NO}_3$&$3$& $\mathcal{L}^{NO_3}$&$\mathbb{R} $&$\{0,0,+\}$&$\{0,0,0\}$&$\{0,0,0\}$&$\ne 0$&$< 0$&$4$\\
\hline
$\mathrm{NO}'_3$ &$3$&&$\mathbb{R} $&$\{0,0,-\}$&$\{0,0,0\}$&$\{0,0,0\}$&$\ne 0$&$> 0$&$4$\\
\hline
$\mathrm{NO}_4$ &$3$& $\mathcal{L}^{NO_4}$&$\mathbb{R}$&$\{0,0,0\}$&$\{0,0,0\}$&$\{0,0,+\}$&$\ne 0$&$> 0$&$4$\\
\hline
$\mathrm{NO}'_4$ &$3$& &$\mathbb{R}$&$\{0,0,0\}$&$\{0,0,0\}$&$\{0,0,-\}$&$\ne 0$&$< 0$&$4$\\
\hline
$\mathrm{NO}_5$ &$7$& $\mathcal{L}^{NO_5}$&$0$&$\{0,+,-\}$&$\{0,0,-\}$&$\{0,+,-\}$&$\ne 0$&$<0$&$5$\\
\hline
\end{tabular}
}
\end{center}
\caption{Classification of Regular and Nilpotent orbits of Lax operators in $\mathfrak{g}_{2,2}/\slal(2)\times \slal(2)$. In the above table $\mathrm{O(1,1)}\ltimes \mathbb{R}^n$ denotes the $n+1$ dimensional group that is the semidirect product of a dilatation group $\mathrm{O(1,1)}$ with an $n$-dimensional translation group $\mathbb{R}^n$. The ``primed'' orbits and their representatives will be discussed in section \ref{generat}.}
\label{tablizanilla}
\end{table}
\subsection{The smallest nilpotent orbit, $\mathrm{NO_1}$}
According to our previous results, the standard representative of the smallest nilpotent orbit is  the operator $E_2$ as given in eq.(\ref{nuovabasa}), yet by means of an $H^\star$ compact rotation we can bring this latter to a much simpler form which is better suited to illustrate the physical interpretation of the supergravity solutions encompassed by the $NO_1$ orbit. Explicitly we set:
\begin{eqnarray}\label{representn1}
    \mathfrak{L}^{NO_1} & = & 2\, \sqrt{\frac{2}{3}} \, \exp\left [ \frac{\pi}{2} \left(L_1^I -L_{-1}^I\right)\right] \, E_2
    \,\exp\left [ -\,\frac{\pi}{2} \left(L_1^I -L_{-1}^I\right)\right]\nonumber\\
    &=&\left(
\begin{array}{lllllll}
 1 & 0 & 0 & 0 & 1 & 0 & 0 \\
 0 & 0 & 0 & 0 & 0 & 0 & 0 \\
 0 & 0 & 1 & 0 & 0 & 0 & 1 \\
 0 & 0 & 0 & 0 & 0 & 0 & 0 \\
 -1 & 0 & 0 & 0 & -1 & 0 & 0 \\
 0 & 0 & 0 & 0 & 0 & 0 & 0 \\
 0 & 0 & -1 & 0 & 0 & 0 & -1
\end{array}
\right)
\end{eqnarray}
If we implement the integration algorithm with the boundary conditions $L_0\, = \, \mathfrak{L}^{NO_1}$, $\mathbb{L}_0\, = \, \mathbf{1}$ we obtain a very simple supergravity solution encoded in the following formulae:
\begin{eqnarray}
  U(\tau) &=& -\, \ft 12 \lg \left(1\,-\,2\,\tau \right) \\
  z(\tau) &=& -{\rm i} \, \frac{1}{\sqrt{1\,-\,2\,\tau}} \\
  \dot{Z}_1 \, = \, \dot{Z}_2 \, = \, \, \dot{Z}_3 \,&=& 0 \\
  \dot{Z}_4  &=& -\frac{\sqrt{2}}{(1-2 \tau
   )^2}
\end{eqnarray}
The corresponding physical charges are the following ones:
\begin{eqnarray}
  \mathbf{n} &=& 0 \nonumber\\
  \mbox{Mass} &=& 1 \nonumber\\
  \left (s_+ \, , \, s_0 \right) &=& \left (0 \, , \, 3 \right) \nonumber\\
  \left (p_1 \, , \, p_2 \, , \, q_1\, , \, q_2 \right) &=& \left (0 \, , \, 0 \, , \, 0\, , \, 2 \right) \label{caricheOn1}
\end{eqnarray}
With such charges the quartic invariant is equal to zero. On the other hand evaluating the limit which defines the area of the horizon:
\begin{equation}\label{vanisharea}
 \frac{1}{2\pi}\, \mbox{Area}_H \, \equiv \,  \lim_{\tau \, \rightarrow \, -\,\infty} \,\exp\left[ -\,U(\tau) \right]\,\frac{1}{\tau^2} \, = \, 0
\end{equation}
we discover that it vanishes.
\par
These properties pertain to the entire orbit. We prove this statement as follows.
\par
First of all we verify that the stability subgroup $\mathrm{O(1,1)}\ltimes \mathbb{R}^2$ of the operator $\mathfrak{L}^{NO_1}$ is generated by the following simple choice of generators:
\begin{eqnarray}
  \Omega_0 &=& 3 L_0^{I} \, - \,L_0^{II}\nonumber\\
  \Omega_1 &=& L_1^I \nonumber\\
  \Omega_2 &=& L_1^{II} \nonumber\\
  &&\left[\Omega_0\, , \, \left(\begin{array}{c}
                            \Omega_1 \\
                            \Omega_2
                          \end{array} \right)\, \right ] \, = \, \left(\begin{array}{c}
                            3\,\Omega_1 \\
                            -\,\Omega_2
                          \end{array} \right ) \quad; \quad \left[\Omega_1\, , \,\Omega_2\right ] \, = \, 0 \label{stabgrupON1}
\end{eqnarray}
which span an algebra isomorphic to the stability subalgebra of the standard representative (see eq.(\ref{stabilino1})).
This enormously facilitates the construction of the entire orbit, since as active part of the $\mathrm{H}^\star$ group on the operator $\mathfrak{L}^{NO_1}$ we can take the complementary $\mathrm{O(1,1)}\ltimes \mathbb{R}^2$ subgroup generated by the following operators:
\begin{eqnarray}
  \Xi_0 &=& 3 L_0^{I} \, + \,L_0^{II}\nonumber\\
  \Xi_1 &=& L_{-1}^I \nonumber\\
  \Xi_2 &=& L_{-1}^{II} \nonumber\\
  &&\left[\Xi_0\, , \, \left(\begin{array}{c}
                            \Xi_1 \\
                            \Xi_2
                          \end{array} \right)\, \right ] \, = \, \left(\begin{array}{c}
                            -3\,\Xi_1 \\
                            -\,\Xi_2
                          \end{array} \right ) \quad; \quad \left[\Xi_1\, , \,\Xi_2\right ] \, = \, 0 \label{stabgrupON1bis}
\end{eqnarray}
Hence we consider the three-parameter family of operators defined:
\begin{equation}\label{3parami}
    \mathfrak{L}^{NO_1}(\omega,x,y) \, = \, e^{ y\, \Xi_2  } \,e^{ x\, \Xi_1  }\,e^{ \ft 13 \, \lg \omega
    \, \Xi_0  }\, \mathfrak{L}^{NO_1} \, e^{ -\,\ft 13 \, \lg \omega \,\, \Xi_0  }\, e^{ -\,x\, \Xi_1  }
    \,e^{ -\, y\, \Xi_2  } \,
    \end{equation}
Calculating the Taub-NUT charge we verify that the surface where $n$ vanishes is singled out by the simple constraint:
\begin{equation}\label{nvanisha}
    x\,=\,\frac{y^3-3 y}{3 y^2-1}
\end{equation}
Furthermore the expression of the operator is considerably simplified if we slightly change parametrization by setting:
\begin{equation}\label{newparamu}
    \omega \,=\, (-1\, + \, 3\, y^2)\, \sigma
\end{equation}
Hence we consider the two parameter family of operators:
\begin{equation}\label{cosovo}
    \mathfrak{L}^{NO_1}(\sigma,y) \, \equiv \,\mathfrak{L}^{NO_1}\left((-1\, + \, 3\, y^2)\, \sigma,\frac{y^3-3 y}{3 y^2-1},y\right)
\end{equation}
The explicit form of $\mathfrak{L}^{NO_1}(\sigma,y)$ is in the
Appendix in eq.(\ref{orbacortissima}).
\par
Next we run the integration algorithm with initial conditions $L_0\,=\,\mathfrak{L}^{NO_1}(\sigma,y)$ and $\mathbb{L}_0 \, = \, \mathbf{\mathbf{L}}_0^{(\xi|\kappa)}$ defined in eq.(\ref{bfL0bps}). The result is a solution of the supergravity field equations which depends
on four parameters $(\sigma,y,\xi,\kappa)$. Explicitly we obtain the following result.
\paragraph{The metric}
\begin{equation}\label{Upiccolo}
    U(\tau) \, = \, -\frac{1}{2} \log \left(2 \sigma
   \tau
   \left(y^2+1\right)^3+1\right)
\end{equation}
\paragraph{The complex scalar field}
\begin{equation}\label{zpiccolo}
    z(\tau) \, =
\frac{\xi  \left(2
   \left(y^2+1\right) \sigma  \tau
\left(y^2-1\right)^2+1\right)-4
   y \left(y^4-1\right) \kappa
   \sigma  \tau }{2
   \left(y^2+1\right) \sigma  \tau
    \left(y^2-1\right)^2+1}\, - \, {\rm i} \,\frac{\kappa  \sqrt{2 \sigma  \tau
    \left(y^2+1\right)^3+1}}{2
   \left(y^2+1\right) \sigma  \tau
    \left(y^2-1\right)^2+1}
\end{equation}
\paragraph{The electromagnetic fields}
\begin{eqnarray}
  \dot{Z}^{1}(\tau) &=& \frac{4 \sqrt{6} y^2
   \left(\left(y^2-1\right) \kappa
   +2 y \xi \right) \sigma
   }{\kappa ^{3/2} \left(2 \sigma
   \tau
   \left(y^2+1\right)^3+1\right)^2
   }\nonumber\\
  \dot{Z}^{2}(\tau) &=& \frac{8 \sqrt{2} y^3 \sigma
   }{\kappa ^{3/2} \left(2 \sigma
   \tau
   \left(y^2+1\right)^3+1\right)^2
   }
\end{eqnarray}
\paragraph{The physical charges}
\begin{eqnarray}
  \mathbf{n} &=& 0 \\
  \mathrm{Mass} &=& -\left(y^2+1\right)^3 \sigma
\end{eqnarray}
\begin{eqnarray}
  s_+ &=& -\frac{12 y \left(y^4-1\right)
   \sigma }{\kappa } \\
  s_0 &=& -\frac{3 \left(y^2+1\right)
   \left(\left(y^4-6 y^2+1\right)
   \kappa +4 y \left(y^2-1\right)
   \xi \right) \sigma }{\kappa }
\end{eqnarray}
\begin{equation}\label{pqcaricata}
    \left(
\begin{array}{l}
 p_1 \\
 p_2 \\
 q_1 \\
 q_2
\end{array}
\right) \, = \,\left(
\begin{array}{l}
 -\frac{2 \sqrt{3}
   \left(y^2-1\right) \left(-\xi
   y^2+2 \kappa  y+\xi \right)^2
   \sigma }{\kappa ^{3/2}} \\
 -\frac{2 \left(-\xi  y^2+2 \kappa
    y+\xi \right)^3 \sigma
   }{\kappa ^{3/2}} \\
 \frac{2 \sqrt{3}
   \left(y^2-1\right)^2 \left(\xi
   y^2-2 \kappa  y-\xi \right)
   \sigma }{\kappa ^{3/2}} \\
 \frac{2 \left(y^2-1\right)^3
   \sigma }{\kappa ^{3/2}}
\end{array}
\right)
\end{equation}
\paragraph{Structure of the charges and attractor mechanism}
Observing the right hand side of eq.(\ref{pqcaricata}), we realize that in this orbit the electromagnetic charges satisfies the following two algebraic constraints:
\begin{eqnarray}
  q_1^2+\sqrt{3} p_1 q_2 &=& 0 \\
  p_1^3+3 \sqrt{3} p_2^2 q_2 &=& 0 \label{fortepiano}
\end{eqnarray}
which can be solved for $q_\Lambda$ in terms of $p^\Sigma$. Explicitly we have:
\begin{equation}\label{qinp}
 \left\{q_1,q_2\right\} \,= \,   \left\{\mp\frac{p_1^2}{\sqrt{3}
   p_2},-\frac{p_1^3}{3 \sqrt{3}
   p_2^2}\right\}
\end{equation}
Only the second branch of the above solution is consistent with eq.(\ref{pqcaricata}) from which
the constraints (\ref{fortepiano}) were derived. Restricting our attention to such a branch,  the two magnetic charges $p^\Sigma$ are identified by eq.(\ref{pqcaricata}) as it follows:
\begin{equation}\label{farango}
\left\{p_1,p_2\right\}\, = \,    \left\{-\frac{2 \sqrt{3}
   \left(y^2-1\right) \left(-\xi
   y^2+2 \kappa  y+\xi \right)^2
   \sigma }{\kappa
   ^{3/2}},-\frac{2 \left(-\xi
   y^2+2 \kappa  y+\xi \right)^3
   \sigma }{\kappa ^{3/2}}\right\}
\end{equation}
Eq.(\ref{farango}) can now be inverted expressing the parameters $y$ and $\sigma$ in terms of the charges $p^\Lambda$ and of the value of the scalar field at infinity $\kappa,\xi$. The explicit inversion of the above formulae is displayed in the Appendix in eq.(\ref{razzolin}).
\par
If we calculate the limiting value taken by complex scalar field when $\tau \rightarrow -\infty$ we find that it is always real and equal to:
\begin{equation}\label{cagnesco}
    \lim_{\tau \rightarrow -\,\infty} \,z(\tau) \, = \, z_{fix} \,= \, \left\{\frac{-\xi  y^2+2 \kappa
   y+\xi }{1-y^2}\right\}
\end{equation}
Substituting the values of $\sigma$ and $y$ as given by eq.(\ref{razzolin}) a miracle takes place. The dependence on $\kappa$ and $\xi$ drops out yielding:
\begin{equation}\label{zfissatus}
    z_{fix} \, = \, -\frac{\sqrt{3}
   p_2}{p_1}
\end{equation}
This is just the attractor mechanism. Independently from their values at infinity the scalar fields go to a fixed value at the horizon which depends only on the charges. The novelty, however, is that this horizon has a vanishing area. Indeed from the explicit form of the $U(\tau)$ function we obtain:
\begin{equation}\label{areanulla}
  \frac{1}{4\pi} \,\mbox{Area}_H \, = \,   \lim_{\tau \rightarrow \, - \,\infty} \, \frac{1}{\tau^2} \, \exp\left[-U(\tau)\right] \, = \,0
\end{equation}
This is consistent with the fact that the quartic invariant with such charges as those pertaining to this orbit, namely
$\left\{p_1,p_2,\frac{p_1^2}{\sqrt{
   3} p_2},-\frac{p_1^3}{3
   \sqrt{3} p_2^2}\right\}$, vanishes identically: $\mathfrak{I}_4 \, = \, 0$
   \subsection{Properties of the second small nilpotent orbit $\mathrm{NO_2}$}
   Also the second nilpotent orbits contains small black holes of vanishing horizon area, although the behavior of the metric coefficients is slightly different. For the case of this orbit we just examine a unique solution generated by the standard representative of the orbit:
 \begin{equation}\label{laxono2}
 L_0 \, = \,   \mathcal{L}^{NO_2} \, = \, \left(
\begin{array}{lllllll}
 2 & 0 & -2 & 0 & 0 & 0 & 0 \\
 0 & 4 & 0 & 2 \sqrt{2} & 0 & 0 & 0 \\
 2 & 0 & -2 & 0 & 0 & 0 & 0 \\
 0 & -2 \sqrt{2} & 0 & 0 & 0 & 2 \sqrt{2} & 0 \\
 0 & 0 & 0 & 0 & 2 & 0 & -2 \\
 0 & 0 & 0 & -2 \sqrt{2} & 0 & -4 & 0 \\
 0 & 0 & 0 & 0 & 2 & 0 & -2
\end{array}
\right)
 \end{equation}
 Using the identity matrix as initial condition at infinity $\mathbb{L}_0 \, = \, 1_{7\times 7}$, the algorithm produces the following solution:
 \paragraph{Metric}
 \begin{equation}\label{metraNO2}
    ds^2 \, = \, -\frac{1}{(1-4 \tau )^{3/2}} \, dt^2 \, + \,\frac{(1-4 \tau )^{3/2}}{\tau ^4} \, d\tau^2 \, + \, \frac{(1-4 \tau )^{3/2}}{\tau ^2} \, \left(d\theta^2 \, + \, \sin^2\theta \, d\phi^2 \right)
 \end{equation}
 \paragraph{Scalar field}
 \begin{equation}\label{scalaNO2}
    z(\tau) \, = \,- {\rm i} \sqrt{1 - 4\tau}
 \end{equation}
 \paragraph{Field Strengths}
 \begin{equation}\label{finotto}
   F^1 \, = \,8\sqrt{3} \sin\theta \, d\theta \, \wedge \, d\varphi -\frac{2 \sqrt{6}}{(1-4 \tau )^2} \, d\tau \,\wedge \, dt \quad \quad ; \quad \quad F^2 \, = \, 0
 \end{equation}
\paragraph{Charges}
 \begin{eqnarray}
  \mathbf{n} &=& 0 \\
  \mathrm{Mass} &=& 6
\end{eqnarray}
\begin{eqnarray}
  s_+ &=& 0\nonumber\\
  s_0 &=& -6
\end{eqnarray}
\begin{equation}\label{pqcaricataNO2}
    \left(
\begin{array}{l}
 p_1 \\
 p_2 \\
 q_1 \\
 q_2
\end{array}
\right) \, = \,\left(
\begin{array}{l}
 4\sqrt{3}  \\
 0 \\
 0\\
 0
\end{array}
\right)
\end{equation}
 According to the predictions of section \ref{ninpotinaNO2} we see that in the limit $\tau\to -\infty$ the scalar field diverges while the horizon area goes to zero:
 \begin{equation}\label{smallaarea}
    \frac{1}{4\pi} \, \mbox{Area}_H \, = \, \lim_{\tau\to -\infty}\, \frac{(1-4 \tau )^{3/2}}{\tau ^2} \, = \, 0
 \end{equation}
 Hence, as anticipated, also in this case we are in presence of small black-holes.
\subsection{The large  non-BPS nilpotent orbit $\mathrm{NO}_3$ and the attractor mechanism}
\label{antiBPSsino}
As  representative of the 3rd nilpotent orbit, instead of the original standard representative mentioned in table (\ref{nillaorbita}),   we take the following very simple  $\eta$-symmetric matrix:
\begin{equation}\label{lassandobps}
  \mathbb{K}  \, \ni \,\mathfrak{L}_0^{(p|q)} \,=\,\left(
\begin{array}{lllllll}
 q & 0 & 0 & \frac{q}{\sqrt{2}} & 0 & 0 & 0 \\
 0 & \frac{p+q}{2} & -\frac{p}{2} & 0 & -\frac{q}{2}
   & 0 & 0 \\
 0 & \frac{p}{2} & \frac{q-p}{2} & 0 & 0 &
   \frac{q}{2} & 0 \\
 -\frac{q}{\sqrt{2}} & 0 & 0 & 0 & 0 & 0 &
   -\frac{q}{\sqrt{2}} \\
 0 & \frac{q}{2} & 0 & 0 & \frac{p-q}{2} &
   \frac{p}{2} & 0 \\
 0 & 0 & -\frac{q}{2} & 0 & -\frac{p}{2} &
   \frac{1}{2} (-p-q) & 0 \\
 0 & 0 & 0 & \frac{q}{\sqrt{2}} & 0 & 0 & -q
\end{array}
\right)
\end{equation}
Calculating its corresponding invariants and tensor structures we find:
\begin{eqnarray}
  \mathfrak{h}_2 \, = \, \mathfrak{h}_6 &=& 0 \\
  \mathcal{T}^{xy} &=& \frac{1}{3} \,p\,q \,\left(\begin{array}{ccc}
                                              1 & 1 & 1 \\
                                              1 & 1  & 1 \\
                                              1 & 1 & 1
                                            \end{array}
  \right) \\
  \mathfrak{T}^{xy} &=& \mathbf{0} \\
  \mathbb{T}^{ab} &=& \mathbf{0} \\
  \mathcal{W}^{x|a} & = & -  \frac{3}{2} \,q^2 \,\left(\begin{array}{ccc}
                                              1 & 1 & 1 \\
                                              1 & 1  & 1 \\
                                              1 & 1 & 1
                                            \end{array}
  \right)\, \ne \, \mathbf{0}
\end{eqnarray}
Therefore the tensor $\mathcal{T}$ has two null eigenvalues and one non vanishing eigenvalue equal to $pq$. All such features correspond to those of the third  nilpotent orbit that, as we are going to see encompasses regular extremal non-BPS black-holes. \par
If we implement the integration algorithm utilizing $\mathfrak{L}_0^{(p|q)}$ as initial Lax operator $L_0$ and the identity matrix $\mathbf{1}_{7\times 7}$ as initial coset representative $\mathbb{L}_0$, we obtain a supergravity solution where the matrix of Noether charges $Q_{Noether}$ is $\mathfrak{L}_0^{(p|q)}$ itself. The corresponding physical charges calculated by means of their own definition are:
\begin{equation}
\begin{array}{rclcl}
  \mathbf{n} &\equiv& - \, \mbox{Tr} \left (Q_{Noether} \, L_1^E\right) &=& 0 \\
  \mbox{Mass}  &\equiv & \mbox{Tr} \left (Q_{Noether} \, L_0^E\right)&=& \frac{p+3 q }{2}\\
  \left(\begin{array}{c}
          s_+ \\
          s_0
        \end{array}
  \right) &\equiv& \mbox{Tr} \left (Q_{Noether} \, \begin{array}{c}L_1\\L_0\\ \end{array}\right)&=& \left(
\begin{array}{l}
 0 \\
 \frac{3 \left(q -p\right)}{2 }
\end{array}
\right)
\end{array} \label{fantastico1}
\end{equation}
and
\begin{equation}\label{carichetopo}
  \begin{array}{rclcl}
    \left( \begin{array}{c}
             p_1 \\
             p_2\\
             q_1 \\
             q_2
           \end{array}
    \right)  & \equiv &\mbox{Tr} \left(Q_{Noether} \, \begin{array}{c}
                                                      W^{1,1}\\
                                                     W^{1,2} \\
                                                      W^{1,3}\\
                                                      W^{1,4}
                                                    \end{array}
    \right) & = &\left(
\begin{array}{l}
 0 \\
 p \\
 \sqrt{3} q \\
 0
\end{array}
\right)
  \end{array}
\end{equation}
The generators of the solvable Lie algebra used in the above equations  were defined in  eq.s (\ref{ehlersalg}-\ref{d4alg}-\ref{wgenni}).
\par
The catch of the attractor mechanism consists of scanning the
space of boundary conditions of the scalar fields at $\tau=0$,
while keeping the topological electromagnetic charges $pq$ fixed.
From this point of view the formulation of boundary condition used
by the integration algorithm is not the best suited one. There,
for each choice of the Lax operator at $\tau = 0$, named $L_0$, we
consider all possible initial values of the scalar fields that are
encoded in the choice of the initial coset representative
$\mathbb{L}_0$. In the case of the $S^3$-model we parameterize the
boundary values of the complex $z$-field by means of the following
matrix
\begin{equation}\label{initolo}
    {\mathbf{L}}_0 ^{(\xi|\kappa)}\, = \, \exp\left [ \xi \, L_1 \right] \cdot \exp\left [ \log[\kappa] \, L_0 \right]
\end{equation}
which corresponds to:
\begin{equation}\label{zfiletto}
    z(0) \, = \, \xi \, - \, {\rm i} \, \kappa
\end{equation}
If we keep $L_0 \, = \, \mathfrak{L}_0^{(p|q)}$ and we just scan all
the possible asymptotic values of $z$ by setting $\mathbb{L}_0 =
\mathbf{L}_0 ^{(\xi|\kappa)}$, we produce supergravity solutions
that have different charges and no attractor mechanism can be seen.
Indeed the corresponding matrix of Noether charges is:
\begin{equation}\label{cavacchio}
    Q_{Noether}\,=\,{\mathbf{L}}_0 ^{(\xi|\kappa)}\, \mathfrak{L}_0^{(p|q)} \, \left( \mathbf{L}_0^{(\xi|\kappa)} \right)^{-1}
\end{equation}
and its traces with the $W^{1|M}$ generators will produce $pq$-charges different from those displayed in eq.(\ref{carichetopo}).
In order to vary the scalar boundary conditions at infinity keeping the same topological charges $pq$,  we have to consider a family of Lax operators, all belonging to the same orbit:
\begin{equation}\label{familylax}
    \mathfrak{L}_0^{(p|q)}(\xi,\kappa) \, \equiv \, \mathcal{O}(\xi ,\kappa)\,
    \mathfrak{L}_0^{(p|q)} \mathcal{O}^{-1}(\xi , \kappa)
\end{equation}
where $\mathcal{O}(\xi ,\kappa) \in \mathrm{H}^\star$ is a two-parameter continuous family of $\mathrm{H}^\star$-group elements such that setting $L_0 \, = \, \mathfrak{L}_0^{(p|q)}(\xi,\kappa)$ and $\mathbb{L}_0 \, = \,{\mathbf{L}}_0 ^{(\xi|\kappa)}$ we obtain supergravity solutions with fixed $pq$-charges. Mathematically such a condition on the group elements $\mathcal{O}(\xi ,\kappa)$ is formulated by setting:
\begin{equation}\label{fantistichino}
   \mbox{Tr} \left[{\mathbf{L}}_0 ^{(\xi|\kappa)} \, \mathcal{O}(\xi ,\kappa) \, \mathfrak{L}_0^{(p|q)} \, \mathcal{O}^{-1}(\xi , \kappa) \, \left( \mathbf{L}_0^{(\xi|\kappa)} \right)^{-1} \, W^{1,M} \right] \, = \, \left(
\begin{array}{l}
 0 \\
 p \\
 \sqrt{3} q \\
 0
\end{array}
\right)
\end{equation}
A priori it is not obvious that equation (\ref{fantistichino}) should admit general solutions, yet a little thought shows that this is not guaranteed yet might be possible. The basic consideration is that in all $\mathcal{N}=2$ supergravities compactified to $D=3$ over a time-direction   the following group theoretical miracle takes place:
\begin{equation}\label{miraculus}
    \mathrm{H}^{\star} \, \sim \, \mathrm{SL(2,\mathbb{R})}^E \, \times \, \mathrm{U}_{D=4}
\end{equation}
where the symbol $\sim$ denotes isomorphism.
Furthermore, as we already know, the Lax operator sits in a representation of $\mathrm{H}^{\star}$ isomorphic to the representation of $\mathrm{SL(2,\mathbb{R})}^E \, \times \, \mathrm{U}_{D=4}$ which the  generators $W^{\alpha|M}$ are assigned to. The space of asymptotic values of the scalar fields spans the original $D=4$ coset, metrically equivalent to the Borel subgroup of $\mathrm{U}_{\mathrm{D=4}}$:
\begin{equation}\label{metricaquiva}
    \frac{\mathrm{U_{D=4}}}{\mathrm{H_{D=4}}} \, \simeq \, \mathrm{Borel}\left (\mathrm{U_{D=4}}\right)
\end{equation}
Hence the pairing between the coset representatives $\mathbf{L}_0 ^{(\xi|\kappa)}$ and the $\mathrm{H}^\star$ elements $\mathcal{O}(\xi ,\kappa)$ defined by equation (\ref{fantistichino}) can be seen as a suitable immersion:
\begin{equation}\label{iotato}
  \iota \, : \,   \mathrm{Borel}\left (\mathrm{U_{D=4}}\right) \, \hookrightarrow \, \mathrm{H}^\star
\end{equation}
which can exist.
\par
The explicit construction of $\iota$ can be performed at the Lie algebra level considering infinitesimal deformations of the identity element:
\begin{eqnarray}\label{rumachku}
    \mathbf{L}_0 ^{(\xi|\kappa)} & \simeq & 1 \, + \, \delta\xi \, L_1 \, + \, \delta\kappa \,L_0\nonumber\\
    \mathcal{O}(\xi ,\kappa)& \simeq & 1 \, + \, \delta\xi \, \mathfrak{M}_1 \, + \, \delta\kappa \,\mathfrak{M}_0
\end{eqnarray}
where $\mathrm{H}^\star\, \ni \, \mathfrak{M}_{0,1}$ are elements of the stability subalgebra to be determined in such a way that:
\begin{eqnarray}
  \mbox{Tr}\left( \left[L_1\, , \, \mathfrak{L}_0^{(p|q)} \right] \, W^{1,M} \right)\, + \, \mbox{Tr}\left( \left[\mathfrak{M}_1\, , \, \mathfrak{L}_0^{(p|q)} \right] \, W^{1,M} \right) &=& 0 \nonumber\\
  \mbox{Tr}\left( \left[L_0\, , \, \mathfrak{L}_0^{(p|q)} \right] \, W^{1,M} \right)\, + \, \mbox{Tr}\left( \left[\mathfrak{M}_0\, , \, \mathfrak{L}_0^{(p|q)} \right] \, W^{1,M} \right) &=& 0 \label{grammafanta}
\end{eqnarray}
Indeed for $\delta\xi << 1$ and $\delta\kappa <<1$ eq.(\ref{fantistichino}) reduces to eq.(\ref{grammafanta}), which is algebraically uniquely solved by:
\begin{eqnarray}
  \mathfrak{M}_1 &=& L_0^{I} \, + \, \ft 12 \left(L_1^I \, - \, L_{-1}^I\right)\, - \, L_0^{II} \, + \, \ft 12 \left(L_1^{II} \, - \, L_{-1}^{II}\right)  \\
  \mathfrak{M}_0 &=& \ft 32 \left(L_1^I \, + \, L_{-1}^I\right)\, - \, \ft 14 \left(L_1^{II} \, + \, L_{-1}^{II}\right)
\end{eqnarray}
where the standard generators $L_x^{I,II}$ were defined in eq.(\ref{phenix}).
\par
Stepping up from the infinitesimal to the finite level we can verify that
\begin{equation}\label{Ogruppus}
    \mathcal{O}(\xi,\kappa) \, = \, \exp\left[\frac{\xi}{\kappa} \, \mathfrak{M}_1\right]\, \exp\left[\lg[\kappa] \, \mathfrak{M}_0\right]
\end{equation}
satisfies the required condition (\ref{fantistichino}). Explicitly we find:
\begin{eqnarray}
   &\mathfrak{ L}_0^{(p|q)}(\xi,\kappa) = &\nonumber\\
   &\left(
\begin{array}{lllllll}
 q \sqrt{\kappa } & \frac{q \xi }{\sqrt{\kappa }} &
   -\frac{q \xi }{\sqrt{\kappa }} & \frac{q \sqrt{\kappa
   }}{\sqrt{2}} & 0 & 0 & 0 \\
 \frac{q \xi }{\sqrt{\kappa }} & \frac{p+q \left(\kappa
   ^2+3 \xi ^2\right)}{2 \kappa ^{3/2}} & \frac{-3 q \xi
   ^2-p}{2 \kappa ^{3/2}} & \frac{\sqrt{2} q \xi
   }{\sqrt{\kappa }} & -\frac{q \sqrt{\kappa }}{2} & 0 &
   0 \\
 \frac{q \xi }{\sqrt{\kappa }} & \frac{3 q \xi ^2+p}{2
   \kappa ^{3/2}} & \frac{q \left(\kappa ^2-3 \xi
   ^2\right)-p}{2 \kappa ^{3/2}} & \frac{\sqrt{2} q \xi
   }{\sqrt{\kappa }} & 0 & \frac{q \sqrt{\kappa }}{2} & 0
   \\
 -\frac{q \sqrt{\kappa }}{\sqrt{2}} & -\frac{\sqrt{2} q
   \xi }{\sqrt{\kappa }} & \frac{\sqrt{2} q \xi
   }{\sqrt{\kappa }} & 0 & \frac{\sqrt{2} q \xi
   }{\sqrt{\kappa }} & \frac{\sqrt{2} q \xi
   }{\sqrt{\kappa }} & -\frac{q \sqrt{\kappa }}{\sqrt{2}}
   \\
 0 & \frac{q \sqrt{\kappa }}{2} & 0 & \frac{\sqrt{2} q
   \xi }{\sqrt{\kappa }} & \frac{-q \kappa ^2+3 q \xi
   ^2+p}{2 \kappa ^{3/2}} & \frac{3 q \xi ^2+p}{2 \kappa
   ^{3/2}} & -\frac{q \xi }{\sqrt{\kappa }} \\
 0 & 0 & -\frac{q \sqrt{\kappa }}{2} & -\frac{\sqrt{2} q
   \xi }{\sqrt{\kappa }} & \frac{-3 q \xi ^2-p}{2 \kappa
   ^{3/2}} & \frac{-q \kappa ^2-3 q \xi ^2-p}{2 \kappa
   ^{3/2}} & \frac{q \xi }{\sqrt{\kappa }} \\
 0 & 0 & 0 & \frac{q \sqrt{\kappa }}{\sqrt{2}} & \frac{q
   \xi }{\sqrt{\kappa }} & \frac{q \xi }{\sqrt{\kappa }}
   & -q \sqrt{\kappa }\\
\end{array}
\right)&\nonumber\\
\label{L0bps}
\end{eqnarray}
and
\begin{eqnarray}
  \mathbf{L}_0^{(\xi|\kappa)}&=& \left(
\begin{array}{lllllll}
 \sqrt{\kappa } & \frac{\xi }{\sqrt{\kappa }} & 0 & 0 & 0
   & 0 & 0 \\
 0 & \frac{1}{\sqrt{\kappa }} & 0 & 0 & 0 & 0 & 0 \\
 0 & 0 & \kappa  & \sqrt{2} \xi  & \frac{\xi ^2}{\kappa }
   & 0 & 0 \\
 0 & 0 & 0 & 1 & \frac{\sqrt{2} \xi }{\kappa } & 0 & 0 \\
 0 & 0 & 0 & 0 & \frac{1}{\kappa } & 0 & 0 \\
 0 & 0 & 0 & 0 & 0 & \sqrt{\kappa } & \frac{\xi
   }{\sqrt{\kappa }} \\
 0 & 0 & 0 & 0 & 0 & 0 & \frac{1}{\sqrt{\kappa }}
\end{array}
\right) \nonumber\\
  \label{bfL0bps}
\end{eqnarray}
The corresponding matrix of Noether charges
\begin{equation}\label{qnoetherBPS}
    Q_{Noether}^{(p|q)}(\xi \, , \, \kappa) \, = \, \mathbf{L}_0^{(\xi,\kappa)} \, {\mathfrak{L}}_0^{(p|q)}\, \left(\mathbf{L}_0^{(\xi,\kappa)}\right)^{-1}
\end{equation}
produces the following physical charges of the black-hole solutions pertaining to this pair of orbits:
\begin{equation}
\begin{array}{rcl}
  \mathbf{n} &=& 0 \\
  \mbox{Mass} &=& \frac{p+3 q \left(\kappa ^2+\xi ^2\right)}{2 \kappa
   ^{3/2}} \\
  \left(\begin{array}{c}
          s_+ \\
          s_0
        \end{array}
  \right) &=& \left(
\begin{array}{l}
 \frac{6 q \xi }{\kappa ^{3/2}} \\
 \frac{3 \left(q \left(\kappa ^2+\xi
   ^2\right)-p\right)}{2 \kappa ^{3/2}}
\end{array}
\right)
\end{array}\label{massotte}
\end{equation}
and
\begin{equation}\label{carichetopo2}
  \begin{array}{rcl}
    \left( \begin{array}{c}
             p_1 \\
             p_2\\
             q_1 \\
             q_2
           \end{array}
    \right) & = &\left(
\begin{array}{l}
 0 \\
 p \\
 \sqrt{3} q \\
 0
\end{array}
\right)
  \end{array}
\end{equation}
This procedure, which is successful for the construction of non-BPS solutions with only two electromagnetic-charges and arbitrary scalar field conditions at infinity needs to be generalized  in the case of the BPS orbit, allowing for non vanishing values of some of  the electric-magnetic potentials $Z^\Lambda,Z_\Lambda$ at infinity. The general solution one gets is the well known one illustrated in subsection \ref{bpsharmonicgeneral}, expressed in terms of Harmonic functions.
\subsubsection{The explicit supergravity solution as a function of its moduli}
Running the integration algorithm with the above adapted set of initial conditions we obtain the explicit form of the corresponding supergravity solution. It is as follows.
\paragraph{The metric}
The metric is defined by the function $U$ for which we obtain the following expression:
\begin{eqnarray}
  &\exp\left[U(\tau)\right] \, = \,&\nonumber\\
  &  \frac{\kappa ^{3/4}}{\sqrt{-q^3 \kappa ^3 \tau ^3-q^3
   \kappa  \xi ^2 \tau ^3+3 q^2 \kappa ^{5/2} \tau
   ^2+3 q^2 \sqrt{\kappa } \xi ^2 \tau ^2+p \left(q
   \sqrt{\kappa } \tau -1\right)^3 \tau -3 q \kappa
   ^2 \tau -3 q \xi ^2 \tau +\kappa ^{3/2}}}& \nonumber\\
   \label{eUbps}
\end{eqnarray}
\paragraph{The scalar field} The complex scalar field $z(\tau)$ has the following form:
\begin{eqnarray}
  &\mbox{Im} \, z(\tau) \, = \, &  \nonumber\\
  &
 - \frac{\sqrt[4]{\kappa } \sqrt{-q^3 \kappa ^3 \tau
   ^3-q^2 \kappa  \left(q \xi ^2+3 p\right) \tau ^3+3
   q^2 \kappa ^{5/2} \tau ^2+3 q \sqrt{\kappa }
   \left(q \xi ^2+p\right) \tau ^2-3 q \kappa ^2 \tau
   -\left(3 q \xi ^2+p\right) \tau +\kappa ^{3/2}
   \left(p q^3 \tau ^4+1\right)}}{\left(q
   \sqrt{\kappa } \tau -1\right)^2}&\nonumber\\
   \label{Immaparta}
\end{eqnarray}
\begin{eqnarray}
  \mbox{Re} \, z(\tau) &=& \frac{\xi }{\left(q \sqrt{\kappa } \tau -1\right)^2} \label{reallaparta}
\end{eqnarray}
\paragraph{The electromagnetic fields.} The explicit form of the two field strengths appearing in the $S^3$ model is completely determined by equation (\ref{finalfildone}). It suffices to know the magnetic charges $(p_1,p_2)\,=\,(0,p)$, the Taub-NUT charge $\mathbf{n}=0$ and the derivatives of the $Z^\Lambda(\tau)$ functions. We obtain
\begin{eqnarray}
&\dot{Z}^1(\tau)\, = \, &\nonumber\\
  &\frac{\sqrt{\frac{3}{2}} \xi  \left(q \left(2
   q^3 \tau ^3 \kappa ^{7/2}-6 q^2 \tau ^2
   \kappa ^3+6 q \tau  \kappa ^{5/2}-2 \kappa
   ^2+2 q^3 \xi ^2 \tau ^3 \kappa ^{3/2}-6 q^2
   \xi ^2 \tau ^2 \kappa +6 q \xi ^2 \tau
   \sqrt{\kappa }-3 \xi ^2\right)-p \left(q
   \sqrt{\kappa } \tau -1\right)^3 \left(3 q
   \sqrt{\kappa } \tau
   -1\right)\right)}{\left(q^3 \kappa ^3 \tau
   ^3+q^2 \kappa  \left(q \xi ^2+3 p\right)
   \tau ^3-3 q^2 \kappa ^{5/2} \tau ^2-3 q
   \sqrt{\kappa } \left(q \xi ^2+p\right) \tau
   ^2+3 q \kappa ^2 \tau +\left(3 q \xi
   ^2+p\right) \tau -\kappa ^{3/2} \left(p q^3
   \tau ^4+1\right)\right)^2}& \nonumber\\
   \label{zetino1}
\end{eqnarray}
\begin{eqnarray}
  &\dot{Z}^2(\tau)\, =\, &\nonumber\\
  &-\frac{\left(q \sqrt{\kappa } \tau -1\right)^2
   \left(p \left(q \sqrt{\kappa } \tau
   -1\right)^4+3 q \xi ^2\right)}{\sqrt{2}
   \left(q^3 \kappa ^3 \tau ^3+q^2 \kappa
   \left(q \xi ^2+3 p\right) \tau ^3-3 q^2
   \kappa ^{5/2} \tau ^2-3 q \sqrt{\kappa }
   \left(q \xi ^2+p\right) \tau ^2+3 q \kappa
   ^2 \tau +\left(3 q \xi ^2+p\right) \tau
   -\kappa ^{3/2} \left(p q^3 \tau
   ^4+1\right)\right)^2} &\nonumber\\
  \label{zetino2}
\end{eqnarray}
\paragraph{The fixed scalars at horizon and the entropy} Calculating the area of the horizon we find:
\begin{equation}\label{rollo}
\frac{1}{4\pi} \, \mbox{Area}_H \, \equiv\,   r^2_H \, = \lim_{\tau \rightarrow \, - \,\infty} \, \frac{1}{\tau^2} \, \exp\left[-U(\tau)\right] \, = \,\sqrt{p\,q^3}
\end{equation}
which makes sense only as long $p\,q^3 >0$ namely as long the $p,q$-charges are both positive or both negative. When this condition, which defines the physical branch of the solution, is fulfilled, eq.(\ref{rollo}) provides
the correct expected result for anti-BPS black-holes. Indeed, comparing with the definition of the quartic symplectic invariant in eq.(\ref{Jinv}) and with the form (\ref{carichetopo2}) of the electromagnetic charges of the present solution we see that:
\begin{equation}\label{grandioso}
    p\,q^3 \, = \,
                              - \mathfrak{I}_4 \quad \mbox{if $p,q$ have the same sign}
\end{equation}
This implies that the 3rd orbit  contains non-supersymmetric
extremal black-holes of finite area.
\par
Calculating now the limit of the scalar field at the horizon we find:
\begin{equation}\label{scalarefisso}
    \lim_{\tau \rightarrow \, - \,\infty} \, z(\tau) \, = \, -{\rm i} \, \frac{pq}{\sqrt{p\,q^3}} \, = \, \, -{\rm i}\, \sqrt{\frac{p}{q}}
\end{equation}
This is also the correct expected result. Comparing with eq.(\ref{zetafissuNBPS}) we see that the fixed scalar values exactly match
those predicted by general arguments at a non BPS fixed point of the geodesic potential.
\par
\begin{figure}[!hbt]
\begin{center}
\iffigs
 \includegraphics[height=50mm]{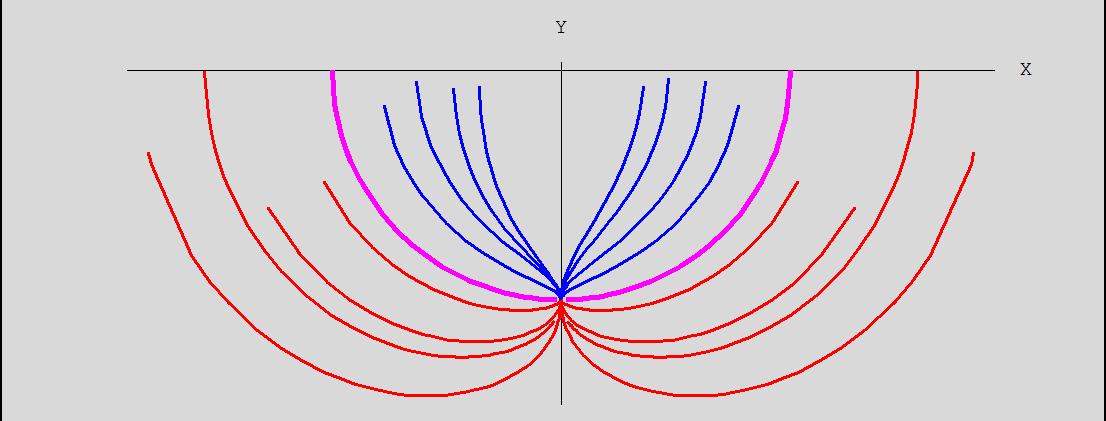}
\else
\end{center}
 \fi
\caption{\it Trajectories of scalar fields from infinity to the
horizon in the case of the black-hole with charges $(p=2,q=4)$
which belongs to the 3rd nilpotent non BPS orbit.}
\label{rollatina}
 \iffigs
 \hskip 1.5cm \unitlength=1.1mm
 \end{center}
  \fi
\end{figure}

In figure \ref{rollatina} we display the explicit behavior of the attractor mechanism by showing the trajectories of the scalar fields from their boundary value at infinity ($\tau = 0$) to their fixed value at the horizon. As we see, there are two type of trajectories, those where the boundary value at infinity lies out of the semicircle of radius $\rho = |z_{fix}|$ and those where the boundary value lies inside such circle. In the first case the trajectory escapes to some distant minimum and then bends to the attractor. In the second case the trajectory reaches the attractor passing through a flex point.
\subsection{The large BPS nilpotent orbit $\mathrm{NO}_4$ }
\label{BPSsino}

The discussion of the fourth nilpotent orbit which corresponds to BPS black-holes is a little bit harder than the discussion of the previous cases since, as we just observed above, in order to see the full fledged attractor mechanism we should introduce at least three non-vanishing electromagnetic charges. The general solution  becomes in that case algebraically too complicated to be managed and displayed. Hence we confine ourselves to solutions with only two non-vanishing electromagnetic charges. In this way we loose the axion field, namely the complex scalar $z$ remains purely imaginary along the whole trajectory. Yet the analysis we shall hereby present suffices to clarify the properties of the entire orbit.
\par
As representative of this orbit we take the following Lax operator depending on two parameters $(p,q)$:
\begin{equation}\label{bpslaxrep}
 \mathbb{K}  \, \ni \,   \widehat{\mathfrak{L}}_0^{(p|-q)} \, = \,\left(
\begin{array}{lllllll}
 q & 0 & 0 &
   -\frac{q}{\sqrt{2}} & 0 & 0
   & 0 \\
 0 & \frac{p+q}{2} &
   -\frac{p}{2} & 0 &
   \frac{q}{2} & 0 & 0 \\
 0 & \frac{p}{2} &
   \frac{q-p}{2} & 0 & 0 &
   -\frac{q}{2} & 0 \\
 \frac{q}{\sqrt{2}} & 0 & 0 &
   0 & 0 & 0 &
   \frac{q}{\sqrt{2}} \\
 0 & -\frac{q}{2} & 0 & 0 &
   \frac{p-q}{2} & \frac{p}{2}
   & 0 \\
 0 & 0 & \frac{q}{2} & 0 &
   -\frac{p}{2} & \frac{1}{2}
   (-p-q) & 0 \\
 0 & 0 & 0 &
   -\frac{q}{\sqrt{2}} & 0 & 0
   & -q
\end{array}
\right)
\end{equation}
It differs from the non-BPS operator (\ref{lassandobps}) only for some signs.
\par
Calculating its corresponding invariants and tensor structures we find:
\begin{eqnarray}
  \mathfrak{h}_2 \, = \, \mathfrak{h}_6 &=& 0 \\
  \mathcal{T}^{xy} &=& \mathbf{0}\\
  \mathfrak{T}^{xy} &=& \mathbf{0} \\
  \mathbb{T}^{ab} &=& \left(
\begin{array}{lll}
 18 p q^3 & -18 p q^3 & 18 p
   q^3 \\
 -18 p q^3 & 18 p q^3 & -18 p
   q^3 \\
 18 p q^3 & -18 p q^3 & 18 p
   q^3
\end{array}
\right)\\
  \mathcal{W}^{x|a} & = & \left(
\begin{array}{lll}
 -\frac{3}{2} q (p+q) &
   \frac{3}{2} q (q-p) &
   -\frac{3}{2} q (p+q) \\
 \frac{3}{2} q (p+q) &
   \frac{3}{2} (p-q) q &
   \frac{3}{2} q (p+q) \\
 -\frac{3}{2} q (p+q) &
   \frac{3}{2} q (q-p) &
   -\frac{3}{2} q (p+q)
\end{array}
\right)\, \ne \, \mathbf{0}
\end{eqnarray}
Therefore the tensor $\mathcal{T}$ identically vanishes, while the $\mathbb{T}$ tensor has two null eigenvalues and one non vanishing eigenvalue equal to $54 \,p q^3$. All these features, which are clearly inverted with respect to those pertaining to the non BPS orbit $\mathrm{NO}_3$, correspond instead to those of the fourth nilpotent orbit $\mathrm{NO}_4$ that, as we are going to see, encompasses regular extremal BPS black-holes.
\par
If we implement the integration algorithm utilizing $\widehat{\mathfrak{L}}_0^{(p|-q)}$ as initial Lax operator $L_0$ and the identity matrix $\mathbf{1}_{7\times 7}$ as initial coset representative $\mathbb{L}_0$, we obtain a supergravity solution where the matrix of Noether charges $Q_{Noether}$ is $\widehat{\mathfrak{L}}_0^{(p|-q)}$ itself. The corresponding physical charges calculated by means of their own definition are:
\begin{equation}
\begin{array}{rclcl}
  \mathbf{n} &\equiv& - \, \mbox{Tr} \left (Q_{Noether} \, L_1^E\right) &=& 0 \\
  \mbox{Mass}  &\equiv & \mbox{Tr} \left (Q_{Noether} \, L_0^E\right)&=& \frac{p+3 q }{2}\\
  \left(\begin{array}{c}
          s_+ \\
          s_0
        \end{array}
  \right) &\equiv& \mbox{Tr} \left (Q_{Noether} \, \begin{array}{c}L_1\\L_0\\ \end{array}\right)&=& \left(
\begin{array}{l}
 0 \\
 \frac{3 \left(q -p\right)}{2 }
\end{array}
\right)
\end{array} \label{fantasticobis}
\end{equation}
and
\begin{equation}\label{carichetopobis}
  \begin{array}{rclcl}
    \left( \begin{array}{c}
             p_1 \\
             p_2\\
             q_1 \\
             q_2
           \end{array}
    \right)  & \equiv &\mbox{Tr} \left(Q_{Noether} \, \begin{array}{c}
                                                      W^{1,1}\\
                                                     W^{1,2} \\
                                                      W^{1,3}\\
                                                      W^{1,4}
                                                    \end{array}
    \right) & = &\left(
\begin{array}{l}
 0 \\
 p \\
 -\sqrt{3} q \\
 0
\end{array}
\right)
  \end{array}
\end{equation}
Comparing the above equations with the corresponding ones for the non-BPS case $\mathrm{NO_3}$ we see that the only difference is the reversed  sign of the non vanishing electric charge. Let us however stress that the other charges remain the same. Hence the BPS case is not obtained from the non BPS one by changing $q\rightarrow -q$. The two solutions pertain to clearly separated orbits, as it is made evident by the vanishing of separate tensors in the two cases.
\par
Another important difference between the two cases is related with the behavior of $\widehat{\mathfrak{L}}_0^{(p|-q)}$ with respect to boundary conditions at infinity.
\par
Let us consider the analogue of eq.s(\ref{grammafanta}) for the BPS Lax operator, namely:
\begin{eqnarray}
  \mbox{Tr}\left( \left[L_1\, , \, \widehat{\mathfrak{L}}_0^{(p|-q)} \right] \, W^{1,M} \right)\, + \, \mbox{Tr}\left( \left[\mathfrak{M}_1\, , \, \widehat{\mathfrak{L}}_0^{(p|-q)} \right] \, W^{1,M} \right) &=& 0 \label{cetriolo1}\\
  \mbox{Tr}\left( \left[L_0\, , \, \widehat{\mathfrak{L}}_0^{(p|-q)} \right] \, W^{1,M} \right)\, + \, \mbox{Tr}\left( \left[\mathfrak{M}_0\, , \, \widehat{\mathfrak{L}}_0^{(p|-q)} \right] \, W^{1,M} \right) &=& 0 \label{sgramma}
\end{eqnarray}
While eq.(\ref{sgramma}) admits a unique non-trivial solution for $\mathfrak{M}_0$, no solution exists of the first equation (\ref{cetriolo1}). This means that we can exhibit two-charge BPS solutions with an arbitrary imaginary boundary value of the scalar field $z$ at infinity but its real part remains rigorously zero from infinity to the horizon.
\par
In view of these observations we introduce the analogue of eq.s (\ref{L0bps}) and (\ref{bfL0bps}) as follows
\begin{eqnarray}
   &\widehat{\mathfrak{ L}}_0^{(p|-q)}(\kappa) = &\nonumber\\
   & \left(
\begin{array}{lllllll}
 q \sqrt{\kappa } & 0 & 0 &
   -\frac{q \sqrt{\kappa
   }}{\sqrt{2}} & 0 & 0 & 0 \\
 0 & \frac{q \kappa ^2+p}{2
   \kappa ^{3/2}} &
   -\frac{p}{2 \kappa ^{3/2}}
   & 0 & \frac{q \sqrt{\kappa
   }}{2} & 0 & 0 \\
 0 & \frac{p}{2 \kappa ^{3/2}}
   & \frac{q \kappa ^2-p}{2
   \kappa ^{3/2}} & 0 & 0 &
   -\frac{q \sqrt{\kappa }}{2}
   & 0 \\
 \frac{q \sqrt{\kappa
   }}{\sqrt{2}} & 0 & 0 & 0 &
   0 & 0 & \frac{q
   \sqrt{\kappa }}{\sqrt{2}}
   \\
 0 & -\frac{q \sqrt{\kappa
   }}{2} & 0 & 0 & \frac{p-q
   \kappa ^2}{2 \kappa ^{3/2}}
   & \frac{p}{2 \kappa ^{3/2}}
   & 0 \\
 0 & 0 & \frac{q \sqrt{\kappa
   }}{2} & 0 & -\frac{p}{2
   \kappa ^{3/2}} & -\frac{q
   \kappa ^2+p}{2 \kappa
   ^{3/2}} & 0 \\
 0 & 0 & 0 & -\frac{q
   \sqrt{\kappa }}{\sqrt{2}} &
   0 & 0 & -q \sqrt{\kappa }
\end{array}
\right) &\nonumber\\
\label{NL0bps}
\end{eqnarray}
and
\begin{eqnarray}
  \mathbf{L}_0^{(\kappa)}&=& \left(
\begin{array}{lllllll}
 \sqrt{\kappa } & 0 & 0 & 0 &
   0 & 0 & 0 \\
 0 & \frac{1}{\sqrt{\kappa }}
   & 0 & 0 & 0 & 0 & 0 \\
 0 & 0 & \kappa  & 0 & 0 & 0 &
   0 \\
 0 & 0 & 0 & 1 & 0 & 0 & 0 \\
 0 & 0 & 0 & 0 &
   \frac{1}{\kappa } & 0 & 0
   \\
 0 & 0 & 0 & 0 & 0 &
   \sqrt{\kappa } & 0 \\
 0 & 0 & 0 & 0 & 0 & 0 &
   \frac{1}{\sqrt{\kappa }}
\end{array}
\right)\nonumber\\
  \label{NbfL0bps}
\end{eqnarray}
From the above data, by means of the integration algorithm we can extract the explicit form of the BPS solution with two charges.
\paragraph{The metric}
The metric is defined by the same  function $U$  obtained for the non-BPS orbit (see eq.(\ref{eUbps})) evaluated at $\xi=0$:
\begin{eqnarray}
  \exp\left[U(\tau)\right] & = &\frac{\kappa
   ^{3/4}}{\sqrt{-\left(\kappa
   ^{3/2}-p \tau \right)
   \left(q \sqrt{\kappa } \tau
   -1\right)^3}}
   \label{eUbpsbis}
\end{eqnarray}
To understand the difference between the two solutions the reader should keep in mind that the meaning of the parameter $q$ is now $q=-\frac{q_1}{\sqrt{3}}$ and not
$q=\frac{q_1}{\sqrt{3}}$ as in the non-BPS case.
\paragraph{The scalar field} The result for the complex scalar field $z(\tau)$ is similar to that for the metric, namely it is the same as that of the non-BPS case evaluated at $\xi=0$.
\begin{eqnarray}
\mbox{Im} \, z(\tau) & =  &  -\frac{\sqrt[4]{\kappa }
   \sqrt{\left(p \tau -\kappa
   ^{3/2}\right) \left(q
   \sqrt{\kappa } \tau
   -1\right)^3}}{\left(q
   \sqrt{\kappa } \tau
   -1\right)^2}
   \label{Immapartabis}\\
  \mbox{Re} \, z(\tau) &=& 0 \label{reallapartabis}
\end{eqnarray}
\paragraph{The electromagnetic fields.} The complete form of the $Z$-fields determining the electromagnetic field strengths is now given by:
\begin{eqnarray}
  Z^1(\tau) &=& 0 \\
  Z^2(\tau) &=& -\frac{p \tau }{\sqrt{2} \kappa ^{3/2}
   \left(\kappa ^{3/2}-p \tau \right)} \\
  Z_1(\tau) &=& -\frac{\sqrt{\frac{3}{2}} q \kappa  \tau
   }{q \sqrt{\kappa } \tau -1} \\
  Z_2(\tau) &=& 0
\end{eqnarray}
Note that also in this case the utilized functions are nothing else but those of the BPS case evaluated at $\xi=0$. Yet in this case there is also a crucial sign difference. The function $Z_1$ is changed of sign with respect to its analogue in the non-BPS case.
\paragraph{The fixed scalars at horizon and the entropy} Calculating the area of the horizon we find:
\begin{equation}\label{rollobis}
\frac{1}{4\pi} \, \mbox{Area}_H \, \equiv\,   r^2_H \, = \lim_{\tau \rightarrow \, - \,\infty} \, \frac{1}{\tau^2} \, \exp\left[-U(\tau)\right] \, = \,\sqrt{p\,q^3}
\end{equation}
which again makes sense only as long as $pq^3 > 0$. Observing the structure of the $p,q$-charges we see
that this time the magnetic and electric ones have to have opposite sign. Moreover $pq^3 =\mathfrak{I}_4$.
Hence we conclude that the this solution is indeed BPS as expected. The horizon area is:
\begin{equation}\label{rollobissone}
\frac{1}{4\pi} \, \mbox{Area}_H \, \equiv\,   r^2_H \, =  \, \sqrt{\mathfrak{I}_4}
\end{equation}
\subsection{Breaking solutions giving small black-holes}
\label{brekka} 
\begin{figure}[!hbt]
\begin{center}
\iffigs
 \includegraphics[height=45mm]{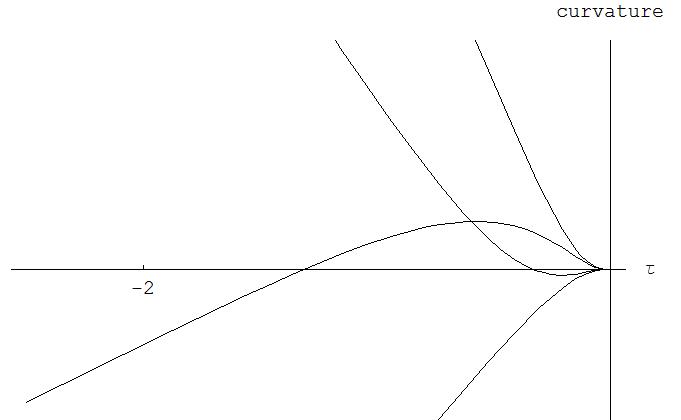}\includegraphics[height=45mm]{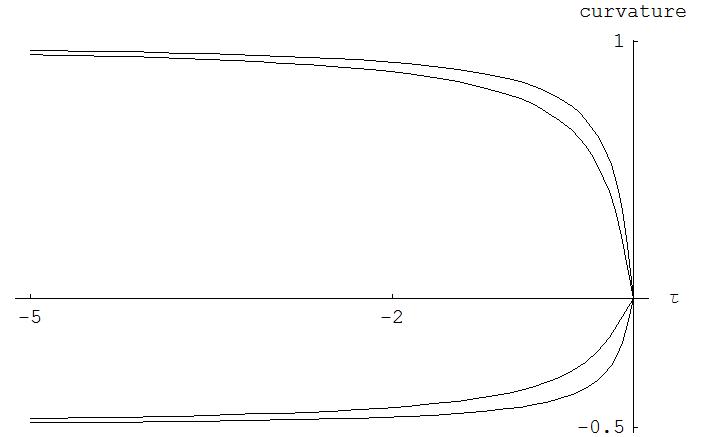}
\else
\end{center}
 \fi
\caption{\it Evolution of the four independent component of the
curvature for the non-BPS and BPS solutions with $\xi=0,\kappa=1$,
$q=2,p=\ft 1 8$. In the picture on the left we see the behavior of
the curvature near $\tau=0$ namely at asymptotic infinity where
they go all to zero. In the picture on the right we see the
asymptotic behavior for large negative $\tau$, namely near the
horizon where the curvatures go to their constant values and the
space degenerates into the direct product $\mathrm{AdS_2} \times
\mathrm{S^2}$}. \label{curvatine}
 \iffigs
 \hskip 1.5cm \unitlength=1.1mm
 \end{center}
  \fi
\end{figure}

\begin{figure}[!hbt]
\begin{center}
\iffigs
 \includegraphics[height=45mm]{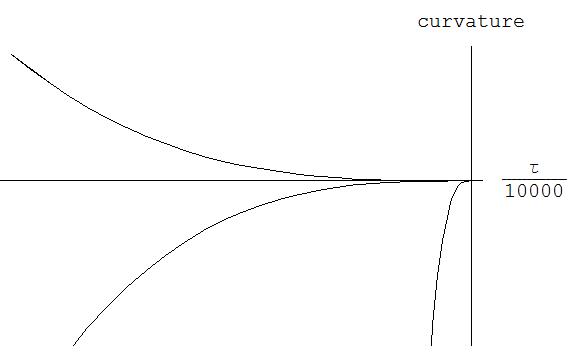}\includegraphics[height=45mm]{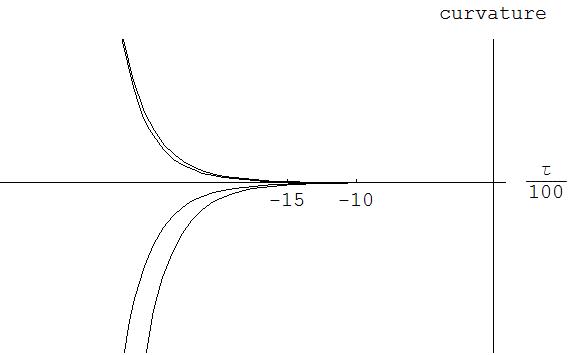}
\else
\end{center}
 \fi
\caption{\it
Evolution of the four independent component of the curvature for the non-BPS and BPS solutions with $\xi=0,\kappa=1$, $q=-2,p=\ft 1 8$. In the picture on the left we see the behavior of the curvature near $\tau=0$ namely at asymptotic infinity where they go all to zero. In the picture on the right we see the asymptotic behavior for  $\tau \rightarrow \ft{1}{q}$,  where they all go to infinity. The solution develops a singularity and the horizon area is zero.}
\label{curvatinebis}
 \iffigs
 \hskip 1.5cm \unitlength=1.1mm
 \end{center}
  \fi
\end{figure}

In order to understand the fate of solutions based on
Lax operators of higher degree of nilpotency and better grasp the
distinction between \textit{orbits of Lax operators} and
\textit{fixed points} of the potential it is convenient to study
more in depth the solution based on the metric (\ref{eUbpsbis}) and
the scalar field (\ref{Immapartabis},\ref{reallapartabis}). As long
as we do not mention the accompanying vector functions
$Z^\Lambda(\tau)$ we do not know whether (\ref{eUbpsbis}),
(\ref{Immapartabis},\ref{reallapartabis}) describe the non-BPS or
BPS solution. Yet in both cases $p,q$ are restricted to have the
same sign which means equal sign for $p_2,q_1$ in the non-BPS case
and opposite sign for the same charges in the BPS one. If we insert
the explicit form of the warp factor (\ref{eUbpsbis}) in the
expression (\ref{curvinediU}) for the independent component of the
Riemann tensor we can verify the following asymptotic behavior:
\begin{eqnarray}
  \lim_{\tau\rightarrow 0} \left\{ \mathcal{C}_1 (\tau) \, ,\, \mathcal{C}_2 (\tau)\, ,\, \mathcal{C}_3 (\tau) \, ,\, \mathcal{C}_4 (\tau) \right \} &=& \left\{ 0 \, ,\, 0\, ,\, 0\, ,\, 0 \right \} \label{bundaconda}\\
  \lim_{\tau\rightarrow - \infty} \left\{ \mathcal{C}_1 (\tau) \, ,\, \mathcal{C}_2 (\tau)\, ,\, \mathcal{C}_3 (\tau) \, ,\, \mathcal{C}_4 (\tau) \right \} &=& \frac{1}{\sqrt{p\,q^3}}\,\left\{- \ft 12 \, ,\, - \ft 12\, ,\, 1\, ,\, 1 \right \}  \label{gorizolimit}
\end{eqnarray}
In figure \ref{curvatine} we also present the behavior of the four functions in a numerical case-study where the approach to the asymptotic constant values at the horizon can be clearly seen.
Yet in this discussion there is a caveat. Suppose we wanted to
consider the same solution for values of $p,q$ that are of
opposite sign, for instance by giving a negative value to $q$.
This corresponds to the result of the integration algorithm for
the Lax operator $\widehat{\mathfrak{L}}^{(p|-q)}_0$ or
${\mathfrak{L}}^{(p,q)}_0$ at $q<0$, which is perfectly legitimate
since both of them are bona-fide nilpotent elements of the
$\mathbb{K}$-subspace for any values of $p$ and $q$. Furthermore,
calculating the electromagnetic charges that correspond to these
Laxes and studying the corresponding extrema of the potential we
conclude that there is a fixed point at $z_{fix} \, = \,- {\rm i}
\sqrt{\frac{p}{-q}}$ which in the case of the non-BPS orbit
$\mathrm{NO_3}$ is a BPS point and in the case of the BPS orbit
$\mathrm{NO_4}$ is instead a non BPS point. This appears somehow
paradoxical yet the question is: \textit{does the solution flow to
such fixed points?} The answer in this case is no! In figure
\ref{curvatinebis} we present the behavior of the four Riemann
curvature components in a numerical case-study where the  reason
why the scalar fields do not reach the fixed point becomes
evident. At a finite value of $\tau$, which corresponds to the
closest to zero real root of the polynomial under square root
appearing in $\exp[U(\tau)]$, all components of the Riemann tensor
diverge and a true singularity is developed at that point. It
follows that the solution breaks down at that point and the fields
cannot proceed further. The resulting solution has all the
features of a small black-hole. Indeed just as in small black-hole
solutions the entropy is zero and the scalar fields go to the
boundary of their moduli space. Indeed naming $\tau_0$ the finite
value of $\tau$ where the solution breaks we have:
\begin{eqnarray}
  \lim_{\tau \to \tau_0} \frac{-U(\tau)}{\tau^2}&=& 0 \label{small1}\\
  \lim_{\tau \to \tau_0} \, \mbox{Im} \, z(\tau) &=& \cases{0 \quad or \cr
  \infty } \label{small2}
\end{eqnarray}
Opposite to the case of small black-hole fixed points reached at $\tau=-\infty$ the five special K\"ahler geometry invariants
$i_1, \dots ,i_5$ calculated at the breaking point are all equal to $\infty$. It follows that:
\begin{equation}\label{finocchius}
    i_1 \, = \, i_2 \, = \, i_3 \, = \, i_4 \, = \, i_5 \, = \, \infty
\end{equation}
can be considered the hall-mark of \textit{broken small black hole solutions}.
\subsection{The very large non BPS nilpotent orbit $\mathrm{NO}_5$}
\label{frintola}
As standard representative of the fifth orbit at vanishing Taub-NUT charge we take the following Lax operator:
\begin{equation}\label{lno5}
    \mathcal{L}^{NO_5} \, \equiv \, 2 \sqrt{\frac{2}{3}} \, \exp \left[ - \,\frac{\pi}{6}\, \left(L_+^{(I)} - L_-^{(I)} \right)\right ] \,\left ( E_1 \, + \, E_2\right ) \, \exp \left[   \,\frac{\pi}{6}\,\left(L_+^{(I)} - L_-^{(I)} \right)\right ]
\end{equation}
where $E_i$ are the new step operators defined in eq.(\ref{nuovabasa}) and the standard generators of $\mathbb{H}^\star$ are those defined in eq.(\ref{phenix})
\begin{equation}\label{fortusquintus}
    \mathcal{L}^{NO_5} \, = \,
  \left(
\begin{array}{lllllll}
 -1 & -\frac{1}{\sqrt{3}} &
   \frac{1}{2} &
   \frac{1}{\sqrt{6}} &
   -\frac{1}{2} & 0 & 0 \\
 -\frac{1}{\sqrt{3}} & -1 &
   -\frac{\sqrt{3}}{2} &
   -\frac{1}{\sqrt{2}} &
   -\frac{1}{2 \sqrt{3}} & 0 & 0
   \\
 -\frac{1}{2} &
   \frac{\sqrt{3}}{2} & 0 &
   -\sqrt{\frac{2}{3}} & 0 &
   \frac{1}{2 \sqrt{3}} &
   -\frac{1}{2} \\
 -\frac{1}{\sqrt{6}} &
   \frac{1}{\sqrt{2}} &
   -\sqrt{\frac{2}{3}} & 0 &
   -\sqrt{\frac{2}{3}} &
   -\frac{1}{\sqrt{2}} &
   -\frac{1}{\sqrt{6}} \\
 \frac{1}{2} & \frac{1}{2
   \sqrt{3}} & 0 &
   -\sqrt{\frac{2}{3}} & 0 &
   \frac{\sqrt{3}}{2} &
   \frac{1}{2} \\
 0 & 0 & -\frac{1}{2 \sqrt{3}} &
   \frac{1}{\sqrt{2}} &
   -\frac{\sqrt{3}}{2} & 1 &
   -\frac{1}{\sqrt{3}} \\
 0 & 0 & \frac{1}{2} &
   \frac{1}{\sqrt{6}} &
   -\frac{1}{2} &
   -\frac{1}{\sqrt{3}} & 1
\end{array}
\right)
   \end{equation}
If we consider the case of initial conditions corresponding to the
identity element of the $\mathrm{G_{(2,2)}}$ group, \textit{i.e.}
$\mathbb{L}_0 \, = \, \mathbf{1}$, we can easily calculate all the
corresponding physical charges and we obtain:
\begin{equation}
\begin{array}{rclcl}
  \mathbf{n} &\equiv& - \, \mbox{Tr} \left (\mathcal{L}^{NO_5} \, L_1^E\right) &=& 0 \\
  \mbox{Mass}  &\equiv & \mbox{Tr} \left (\mathcal{L}^{NO_5} \, L_0^E\right)&=& - 2\\
  \left(\begin{array}{c}
          s_+ \\
          s_0
        \end{array}
  \right) &\equiv& \mbox{Tr} \left (\mathcal{L}^{NO_5} \, \begin{array}{c}L_1\\L_0\\ \end{array}\right)&=& \left(
\begin{array}{l}
 - 2 \sqrt{3} \\
 0
\end{array}
\right)
\end{array} \label{fantastico1bis}
\end{equation}
and
\begin{equation}\label{carichetopotris}
  \begin{array}{rclcl}
    \left( \begin{array}{c}
             p_1 \\
             p_2\\
             q_1 \\
             q_2
           \end{array}
    \right)  & \equiv &\mbox{Tr} \left(\mathcal{L}^{NO_5} \, \begin{array}{c}
                                                      W^{1,1}\\
                                                     W^{1,2} \\
                                                      W^{1,3}\\
                                                      W^{1,4}
                                                    \end{array}
    \right) & = &\left(
\begin{array}{l}
 - \sqrt{3}\\
 \sqrt{3} \\
 1 \\
 -1
\end{array}
\right)
  \end{array}
\end{equation}
As we demonstrated in section \ref{larghezza} with the set of
charges (\ref{carichetopotris}), which correspond to a negative value
of the quartic invariant $\mathfrak{I}_4(p,q) \, = \, - \ft 43$,
there are no BPS attractor points and there is on the other hand a
non-BPS attractor point of type II.
\par
If the fixed point were reached by a solution generated by the above
Lax operator of nilpotency degree seven we would have a clash.
Indeed the asymptotic behavior of the warp factor should be
$\exp[-U(\tau)] \sim \mbox{const}\times \tau^2 $ for very large
negative values of $\tau$ which is incompatible with the higher
nilpotency degree. The resolution of the puzzle resides in that the
fixed point is not attained. On the contrary at a finite value of
$\tau$ the warp factor runs into a root of the various higher degree
polynomials generated by the integration algorithm and goes to zero.
The curvature goes to infinity and the solution breaks down. The
explicit solutions generated by the integration algorithm in the
case of the fifth orbit are too complicated to be displayed and
require also a considerable amount of computer time and memory to be
constructed. Yet we have numerically verified the above statements
and they appear to be generically true.
\par
The lessons taught by this example are three. First we learn how
the regular solutions attaining their fixed point can arise only
from Lax operators of nilpotency degree three. Secondly, recalling
the discussion of section \ref{brekka} we realize that even the
condition ${\mathrm{Lax}}^3 \, = \, 0$ is not sufficient, since
also with such operators broken solutions can arise. Thirdly the
distinction between fixed points and Lax orbits is emphasized. No
Lax operator of nilpotency degree seven can generate solutions
that attain the fixed point determined by their electromagnetic
charges. Yet there are other operators of nilpotency degree three
that have the same electromagnetic charge and generate solution
flowing to the corresponding fixed point.
\section{Other members of the Schwarzschild non-extremal orbit and their extremal limits}
In order to give a more in depth analysis of the relation between extremal and non extremal black holes, in this final section we consider other solutions belonging to the Schwarzschild regular orbits. In particular we show how two different versions of the classical Reissner Nordstr\"om non extremal solution of General Relativity can be embedded into supergravity by means of our algorithm, the first version corresponding to in the extremal limit to a BPS finite area black-hole, the second a non BPS one, also of finite area.
\par
Given the Schwarzschild Lax operator $\mathfrak{S}_\alpha$ (see eq.(\ref{schwarzwaldo})), we consider two different one-parameter orbits departing from it, namely:
\begin{eqnarray}
  \mathfrak{N}^+_{(\alpha,\lambda)} &=& \exp\left[ \log[\lambda]\,\left(L^I_+ + L^I_-\right)\right] \, \mathfrak{S}_\alpha
  \exp\left[ -\log[\lambda]\,\left(L^I_+ + L^I_-\right)\right]\nonumber\\
  \mathfrak{N}^-_{(\alpha,\lambda)} &=& \exp\left[ \log[\sqrt{\lambda}]\,\left(L^I_+ + L^I_- + L^{II}_+ + L^{II}_-\right)\right] \, \mathfrak{S}_\alpha \,\exp\left[ -\log[\sqrt{\lambda}]\,\left(L^I_+ + L^I_- + L^{II}_+ + L^{II}_-\right)\right] \,
  \nonumber\\
\end{eqnarray}
Explicitly we obtain:
\begin{equation}\label{laxpiumeno}
 \mathfrak{N}^\pm_{(\alpha,\lambda)}    \, = \, \left(
\begin{array}{lllllll}
 \frac{\alpha  \left(\lambda
   ^2+1\right)}{4 \lambda } &
   0 & 0 & \frac{\alpha
   \left(\lambda
   ^2-1\right)}{4 \sqrt{2}
   \lambda } & 0 & 0 & 0 \\
 0 & \frac{\alpha
   \left(\lambda
   ^2+1\right)}{4 \lambda } &
   \pm\frac{\alpha  \left(\lambda
   ^2-1\right)   }{8 \lambda
   } & 0 & \frac{\alpha
   -\alpha  \lambda ^2}{8
   \lambda } & 0 & 0 \\
 0 &\pm \frac{\left(\alpha
   -\alpha  \lambda ^2\right)
     }{8 \lambda } & 0 & 0 &
   0 & \frac{\alpha
   \left(\lambda
   ^2-1\right)}{8 \lambda } &
   0 \\
 -\frac{\alpha  \left(\lambda
   ^2-1\right)}{4 \sqrt{2}
   \lambda } & 0 & 0 & 0 & 0 &
   0 & -\frac{\alpha
   \left(\lambda
   ^2-1\right)}{4 \sqrt{2}
   \lambda } \\
 0 & \frac{\alpha
   \left(\lambda
   ^2-1\right)}{8 \lambda } &
   0 & 0 & 0 &
  \pm \frac{\left(\alpha -\alpha
   \lambda ^2\right)   }{8
   \lambda } & 0 \\
 0 & 0 & \frac{\alpha -\alpha
   \lambda ^2}{8 \lambda } & 0
   & \pm\frac{\alpha
   \left(\lambda ^2-1\right)
     }{8 \lambda } &
   -\frac{\alpha
   \left(\lambda
   ^2+1\right)}{4 \lambda } &
   0 \\
 0 & 0 & 0 & \frac{\alpha
   \left(\lambda
   ^2-1\right)}{4 \sqrt{2}
   \lambda } & 0 & 0 &
   -\frac{\alpha
   \left(\lambda
   ^2+1\right)}{4 \lambda }
\end{array}
\right)
\end{equation}

If we use $L_0\,=\,\mathfrak{N}^\pm_{(\alpha,\lambda)}$ as initial Lax operator and the identity matrix $\mathbb{L}_0 \, = \, 1_{7\times 7}$ as initial  coset representative at $\tau =0$, we obtain that the physical charges are the following ones:
\begin{eqnarray}
  \mathbf{n} &=& 0 \\
  \mbox{mass} &=& \frac{\alpha  \left(\lambda ^2+1\right)}{2 \lambda } \\
  \left(\begin{array}{c}
     s_+ \\
     s_0
   \end{array}\right )
   &=& \left(\begin{array}{c}
     0 \\
     0
   \end{array}\right ) \\
  \left( \begin{array}{c}
             p_1 \\
             p_2\\
             q_1 \\
             q_2
           \end{array}
    \right) &=& \left( \begin{array}{c}
             0 \\
             \frac{\alpha -\alpha  \lambda ^2}{4 \lambda }\\
             \mp \,\sqrt{3} \, \frac{\alpha -\alpha  \lambda ^2}{4 \lambda } \\
             0
           \end{array}
    \right) \label{lecarichediRN}
\end{eqnarray}
Running the integration algorithm, we obtain the following result:
\begin{equation}\label{rnUfunzia}
    U(\tau) \, = \, \log \left(\frac{16
   \sqrt{e^{-3 \alpha  \tau }}
   \lambda
   ^2}{\left(e^{-\alpha  \tau
   }\right)^{5/2}
   \left((\lambda
   +1)^2-e^{\alpha  \tau }
   (\lambda
   -1)^2\right)^2}\right)
\end{equation}
for the warp factor and
\begin{equation}\label{rnscalar}
    z (\tau)\, = \,- {\rm i}
\end{equation}
for the scalar field. When $z=-{\rm i}$ the matrix $\mathcal{M}_4 \, = \, -1_{4\times 4}$ is just minus the identity matrix. Hence we immediately get the form of the $\dot{Z}^A$ determining the field strengths:
\begin{equation}\label{theZfieldi}
    \dot{Z}^A \, = \, - \frac{1}{\sqrt{2}}\exp[U(\tau)] \, Q^A \, = \, \, - \frac{1}{\sqrt{2}}\exp[U(\tau)] \, \left( \begin{array}{c}
             0 \\
             \frac{\alpha -\alpha  \lambda ^2}{4 \lambda }\\
             \mp \,\sqrt{3} \, \frac{\alpha -\alpha  \lambda ^2}{4 \lambda } \\
             0
           \end{array}
    \right)
\end{equation}
Naming:
\begin{equation}\label{fortedeimarmi}
    q \, = \, \frac{\alpha -\alpha  \lambda ^2}{4 \lambda }
\end{equation}
we immediately obtain the final expressions for the two electro-magnetic field strengths:
\begin{eqnarray}
  F^1 &=& \pm \, \sqrt{\frac{3}{2}} \, q \, \exp[U] \, d\tau \, \wedge \, dt \nonumber \\
  F^2 &=& 2 \, q \, \sin\theta \, d\theta \, \wedge \, d\phi \label{frequentazioni}
\end{eqnarray}
Let us then consider the following renaming of the parameters
\begin{eqnarray}\label{reparatro}
   \alpha & = & 2
   \sqrt{m^2 -q^2}\nonumber\\
   \lambda & = &
   \sqrt{\frac{m-q}{m+q}} \nonumber\\
\end{eqnarray}
which is consistent with eq.(\ref{fortedeimarmi}) and the following coordinate transformation from $\tau$ to the standard radial coordinate $r$:
\begin{eqnarray}
\tau & = & \frac{\log
   \left(\frac{-\sqrt{\frac{m-
   q}{m+q}} q-m
   \left(\sqrt{\frac{m-q}{m+q}
   }+1\right)+r}{\sqrt{\frac{m
   -q}{m+q}} q+m
   \left(\sqrt{\frac{m-q}{m+q}
   }-1\right)+r}\right)}{2
   \sqrt{m^2-q^2}}\nonumber\\
   \null & \Downarrow& \null \nonumber\\
d\tau & = &
   \frac{dr}{q^2+r^2-2
   m r}
   \label{tauradtransfo}
   \end{eqnarray}
Upon these transformations, recalling that $v^2 = \alpha^2/4$, the metric (\ref{metricona},\ref{funzia}) becomes the standard non-extremal Reissner Nordstr\"om metric:
\begin{equation}\label{standard}
    ds^2_{RN} \, = \, -\left( 1 - \frac{2m}{r} + \frac{q^2}{r^2}\right) \, dt^2 \, + \, \left( 1 - \frac{2m}{r} + \frac{q^2}{r^2}\right)^{-1} \, dr^2 \, +\, r^2 \left ( d\theta^2 + \sin^2 \, d\phi^2 \right)
\end{equation}
while the two field strengths (\ref{frequentazioni}) assume the following form:
\begin{eqnarray}
  F^1 &=& \pm \, \sqrt{\frac{3}{2}} \, \frac{q}{r^2} dr \, \wedge \, dt \nonumber \\
  F^2 &=& 2 \, q \, \sin\theta \, d\theta \, \wedge \, d\phi \label{berlusvat}
\end{eqnarray}
In this way we have demonstrated how the classical non extremal Reissner Nordstr\"om solution of General Relativity can be embedded in supergravity using two vector fields, one of which carries a magnetic charge, while the other carries a static electric one. We have actually two solutions, distinguished only by the relative signs of the electric and magnetic charges.
We have also shown that the these two Reissner-Nordstr\"om solutions correspond to two different  $\mathrm{H}^\star$ rotations of the Schwarzschild Lax operator. Our main interest in this respect is to understand the extremality limit ($m\to q$) from the Lax point of view. The whole catch of such a limit is encoded in eq.s(\ref{reparatro}): when $m\to q$, both $\alpha$ and $\lambda$ go to zero. Setting $m=q+\epsilon^2$ where $\epsilon$ is an infinitesimal parameter, we realize that $\alpha$ and $\lambda$ go to zero with the same power of $\epsilon$:
\begin{equation}\label{scaling}
   \lambda \, \sim \, \frac{1}{\sqrt{2 \,q}} \, \epsilon \quad ; \quad \alpha \, \sim \, 2\, \sqrt{2 \,q} \, \epsilon
\end{equation}
So we can calculate the finite limit:
\begin{equation}\label{scalalimita}
    \widehat{\mathfrak{N}}^\pm(q) \, = \, \lim_{\epsilon \to 0} \,\mathfrak{N}^\pm_{2\, \sqrt{2 \,q} \, \epsilon\, ,\, \frac{1}{\sqrt{2 \,q}} \, \epsilon}
\end{equation}
The explicit result is:
\begin{equation}\label{gringo}
    \widehat{\mathfrak{N}}^\pm(q) \, = \, \left(
\begin{array}{lllllll}
 q & 0 & 0 &
   -\frac{q}{\sqrt{2}} & 0 & 0
   & 0 \\
 0 & q & \mp\frac{q   }{2} & 0
   & \frac{q}{2} & 0 & 0 \\
 0 & \pm\frac{q   }{2} & 0 & 0
   & 0 & -\frac{q}{2} & 0 \\
 \frac{q}{\sqrt{2}} & 0 & 0 &
   0 & 0 & 0 &
   \frac{q}{\sqrt{2}} \\
 0 & -\frac{q}{2} & 0 & 0 & 0
   & \pm\frac{q   }{2} & 0 \\
 0 & 0 & \frac{q}{2} & 0 &
   \mp\frac{q   }{2} & -q & 0
   \\
 0 & 0 & 0 &
   -\frac{q}{\sqrt{2}} & 0 & 0
   & -q
\end{array}
\right)
\end{equation}
Comparing now the above equation with eq.s (\ref{lassandobps}) and (\ref{bpslaxrep}) we realize that:
\begin{eqnarray}
  \mathfrak{N}^+(q)&=& \widehat{\mathfrak{L}}_0^{q|-q)}\nonumber\\
  \mathfrak{N}^-(q)&=& {\mathfrak{L}}_0^{(q|q)}
\end{eqnarray}
namely the two limiting Lax operators are nilpotent and respectively belong to the $\mathrm{NO_4}$ and $\mathrm{NO_3}$ orbits.
In particular they coincide with the standard representatives we have previously used for the construction of BPS and non BPS regular Black-Holes when the electric and magnetic charges are taken equal in absolute value ($p=q$).
\par
The above discussion illustrates at the level of Lax operator the mechanism by means of which extremal regular Black-Hole solutions can be obtained as appropriate limits of regular non extremal ones.
\section{Generating solutions for regular and small black holes}\label{generat}
In \cite{Bergshoeff:2008be} and \cite{marioetal}  representatives of the regular extremal black hole orbits
with the least number of parameters (generating solutions) were explicitly constructed in symmetric supergravities. In particular it  was shown that  these were dilatonic solutions described by null geodesics in
a characteristic submanifold of the form:
\begin{eqnarray}\label{genman}
\left(\frac{{\rm SL}(2,\mathbb{R})}{{\rm SO}(1,1)}\right)^p\subset \frac{\mathrm{U_{D=3}}}{\mathrm{H}^*}\,,
\end{eqnarray}
where $p$ is the non-compact rank of the coset $H^\star/H_c$, $H_c$ being the maximal compact subgroup of $H^\star$.
In our case $\mathrm{H}^*/\mathrm{H}_c={\rm SO}(2,2)/[{\rm SO}(2)\times {\rm SO}(2)]$ and $p=2$. One can show \cite{Bergshoeff:2008be} that $p$ is related to the electric and magnetic charges:
In fact the normal form of the electric-magnetic charge vector with respect to the action of $\mathrm{H_{D=4}}\times {\rm U}(1)$ is a $p$-charge vector.
In the $S^3$ model $\mathrm{H_{D=4}}={\rm SO}(2)$ and the normal form of the electric-magnetic charge vector with respect to ${\rm SO}(2)^2$ has indeed two parameters, which can be chosen as either the D0-D4 charges $Q_0,\,P^1$ or the
D2-D6 charges $Q_1,\,P^0$ (see section \ref{tqi} for the relation between $Q_\Lambda,\,P^\Lambda$ and the $p,q$- charges used throughout the paper). The generators of the ${\rm SL}(2,\mathbb{R})^2$ group on the left hand side of eq. (\ref{genman}) are constructed out of the nilpotent matrices $W^{1M}$ corresponding to these charges. Since solutions with $D6-D2$ charges were  studied earlier, we can now consider black holes originating from a  $D0-D4$ system and thus choose the normal form to correspond to the charges $Q_0,\,P^1$ and denote ${\rm SL}(2,\mathbb{R})^2={\rm SL}(2,\mathbb{R})_{Q_0}\times {\rm SL}(2,\mathbb{R})_{P^1}$, so that the two factor groups are generated by the following algebras:
\begin{eqnarray}
\mathfrak{sl}(2,\mathbb{R})_{Q_0}&\equiv&\{J_{Q_0},\,K_{Q_0},\mathcal{H}_{Q_0}\}\,\,:\,\,\cases{
J_{Q_0}=e_2+f_2\cr K_{Q_0}=e_2-f_2\cr \mathcal{H}_{Q_0}=\mathcal{H}_2} \nonumber\\
\mathfrak{sl}(2,\mathbb{R})_{P^1}&\equiv&\{J_{P^1},\,K_{P^1},\mathcal{H}_{P^1}\}\,\,:\,\,\cases{
J_{P^1}=e_4+f_4\cr K_{P^1}=e_4-f_4\cr
\mathcal{H}_{P^1}=\frac{2}{3}\,\mathcal{H}_1+\mathcal{H}_2}
\end{eqnarray}
where the normalizations are chosen so that, defining for each algebra the nilpotent generators $N^\pm_{Q_0},\,N^\pm_{P^1}$ as:
\begin{eqnarray}
N^\pm_{Q_0}&\equiv & \frac{1}{2}\,(\mathcal{H}_{Q_0}\mp K_{Q_0})\,\,;\,\,\,N^\pm_{P^1}\equiv  \frac{1}{2}\,(3\mathcal{H}_{P^1}\mp K_{P^1})\,,
\end{eqnarray}
the following commutation relations hold:
\begin{eqnarray}
[J,N^\pm]&=&\pm 2\,N^\pm\,\,;\,\,\,[N^+,\,N^-]=J\,,
\end{eqnarray}
where we have suppressed the charge subscripts and it is easily verified that generators with different subscripts, thus pertaining to different $\mathfrak{sl}(2)$ algebras, commute.
We can choose $\{N^\pm_{Q_0},\,N^\pm_{P^1}\}$ as a basis of generators for the  coset manifold on the left hand side of eq. (\ref{genman}). We can find from different combinations of these nilpotent generators representatives of the orbits $NO_1,\,NO_2,\,NO_3,\,NO_4$
\begin{itemize}
\item{{\bf Orbit $NO_4$.} The representative is given by the following Lax generator at infinity:
\begin{eqnarray}
& \mathcal{L}_{BPS}\,=\, \sqrt{2}\,|Q_0|\,N^-_{Q_0}+\sqrt{2}\,|P^1|\,N^+_{P^1}=\label{repno4}\\&
\left(
\begin{array}{lllllll}
 \sqrt{2}  {|P^1|} & 0 & 0 &  {|P^1|} & 0 & 0 & 0 \\
 0 & \frac{ {|P^1|}+ {|Q_0|}}{\sqrt{2}} & \frac{ {|Q_0|}}{\sqrt{2}} & 0 &
   -\frac{ {|P^1|}}{\sqrt{2}} & 0 & 0 \\
 0 & -\frac{ {|Q_0|}}{\sqrt{2}} & \frac{ {|P^1|}- {|Q_0|}}{\sqrt{2}} & 0
   & 0 & \frac{ {|P^1|}}{\sqrt{2}} & 0 \\
 - {|P^1|} & 0 & 0 & 0 & 0 & 0 & - {|P^1|} \\
 0 & \frac{ {|P^1|}}{\sqrt{2}} & 0 & 0 &
   \frac{ {|Q_0|}- {|P^1|}}{\sqrt{2}} & -\frac{ {|Q_0|}}{\sqrt{2}} & 0 \\
 0 & 0 & -\frac{ {|P^1|}}{\sqrt{2}} & 0 & \frac{ {|Q_0|}}{\sqrt{2}} &
   -\frac{ {|P^1|}+ {|Q_0|}}{\sqrt{2}} & 0 \\
 0 & 0 & 0 &  {|P^1|} & 0 & 0 & -\sqrt{2}  {|P^1|}
\end{array}
\right)\,.&
\end{eqnarray}
If we choose the fields at infinity to correspond to the origin of the manifold, the solution generated by the above Lax matrix is the dilatonic BPS black hole with charges $Q_0>0,\,P^1>0$ and, computing the tensors classifiers on $\mathcal{L}_{BPS}$ we find:
\begin{eqnarray}
\mathcal{T}^{xy}&\equiv&{\bf 0}\,\Rightarrow\,\,\{0,0,0\}\,,\nonumber\\
\mathbb{T}^{ab}&=&72\,|Q_0|\,|P^1|^3\,\left(\matrix{1 & 1 & 1\cr 1 & 1 & 1\cr1 & 1 & 1}\right)\Rightarrow\,\,\{0,0,216\,|Q_0|\,|P^1|^3\}=\{0,0,54\,\mathfrak{I}_4(P,Q)\}\,,\nonumber\\
\mathfrak{T}^{xy}&=&{\bf 0}\,\Rightarrow\,\,\{0,0,0\}\,,\nonumber\\
W^{x|q}&=&\left(
\begin{array}{lll}
 -3  {|P^1|} ( {|P^1|}+ {|Q_0|}) & 3  {|P^1|}
   ( {|Q_0|}- {|P^1|}) & -3  {|P^1|}
   ( {|P^1|}+ {|Q_0|}) \\
 -3  {|P^1|} ( {|P^1|}+ {|Q_0|}) & 3  {|P^1|}
   ( {|Q_0|}- {|P^1|}) & -3  {|P^1|}
   ( {|P^1|}+ {|Q_0|}) \\
 -3  {|P^1|} ( {|P^1|}+ {|Q_0|}) & 3  {|P^1|}
   ( {|Q_0|}- {|P^1|}) & -3  {|P^1|}
   ( {|P^1|}+ {|Q_0|})
\end{array}
\right)\neq 0\,.\nonumber
\end{eqnarray}
We thus verify that $\mathcal{L}_{BPS}$ indeed belongs to the $NO_4$ orbit. The explicit dilatonic solution corresponding to $z_{\infty}=S_{\infty}=-i$ can be written in a closed form in terms of the  harmonic functions $H_0=1-\sqrt{2}\,Q_0\tau\,,H^1=1-\sqrt{2}\,P^1\tau$:
\begin{eqnarray}
S&=&-\frac{1}{z}=-i\,\sqrt{\frac{H_0}{H^1}}\,\,;\,\,\,\,e^{-U}=\sqrt{H_0\,(H^1)^3}\,.\label{solno4}
\end{eqnarray}
}
\item{{\bf Orbit $NO'_4$.} In section \ref{classificazia}, it was shown that the orbit $\mathrm{NO}'_4$ differs from   $\mathrm{NO}_4$ only in the sign of the non-vanishing eigenvalue of $\mathbb{T}^{ab}$. A Lax-representative of it can thus be obtained from (\ref{repno4}) by changing the relative sign of the two nilpotent terms, which amounts in the previous discussion to replacing $|P^1|\rightarrow -|P^1|$. As a consequence $P^1=-|P^1|<0$ and the quartic invariant will now be negative. The corresponding solution will still have the form (\ref{solno4}) but  with $H^1=1+\sqrt{2}\,|P^1|\tau$. The zero of this harmonic function for finite  $\tau$ implies a corresponding zero for $e^{-U}$ and thus the four-dimensional solution will be singular.
    This is consistent with the fact that there is no regular BPS four dimensional solution with negative quartic invariant.
}
\item{{\bf Orbit $NO_3$.} We choose the corresponding representative Lax as follows:
\begin{eqnarray}
&\mathcal{L}_{non-BPS}\, = \, \sqrt{2}\,|Q_0|\,N^-_{Q_0}+\sqrt{2}\,|P^1|\,N^-_{P^1}=\label{repno3}\\
&\left(
\begin{array}{lllllll}
 \sqrt{2}  {|P^1|} & 0 & 0 & - {|P^1|} & 0 & 0 & 0 \\
 0 & \frac{ {|P^1|}+ {|Q_0|}}{\sqrt{2}} & \frac{ {|Q_0|}}{\sqrt{2}} & 0 &
   \frac{ {|P^1|}}{\sqrt{2}} & 0 & 0 \\
 0 & -\frac{ {|Q_0|}}{\sqrt{2}} & \frac{ {|P^1|}- {|Q_0|}}{\sqrt{2}} & 0
   & 0 & -\frac{ {|P^1|}}{\sqrt{2}} & 0 \\
  {|P^1|} & 0 & 0 & 0 & 0 & 0 &  {|P^1|} \\
 0 & -\frac{ {|P^1|}}{\sqrt{2}} & 0 & 0 &
   \frac{ {|Q_0|}- {|P^1|}}{\sqrt{2}} & -\frac{ {|Q_0|}}{\sqrt{2}} & 0 \\
 0 & 0 & \frac{ {|P^1|}}{\sqrt{2}} & 0 & \frac{ {|Q_0|}}{\sqrt{2}} &
   -\frac{ {|P^1|}+ {|Q_0|}}{\sqrt{2}} & 0 \\
 0 & 0 & 0 & - {|P^1|} & 0 & 0 & -\sqrt{2}  {|P^1|}
\end{array}
\right)\,. &\nonumber
\end{eqnarray}
The solution generated by the above Lax matrix is the dilatonic non-BPS black hole with charges $Q_0>0,\,P^1<0$ and, computing the tensors classifiers we find:
 \begin{eqnarray}
\mathcal{T}^{xy}&\equiv&\left(
\begin{array}{lll}
 \frac{2  {|P^1|}  {|Q_0|}}{3} & -\frac{2  {|P^1|}  {|Q_0|}}{3} &
   \frac{2  {|P^1|}  {|Q_0|}}{3} \\
 -\frac{2  {|P^1|}  {|Q_0|}}{3} & \frac{2  {|P^1|}  {|Q_0|}}{3} &
   -\frac{2  {|P^1|}  {|Q_0|}}{3} \\
 \frac{2  {|P^1|}  {|Q_0|}}{3} & -\frac{2  {|P^1|}  {|Q_0|}}{3} &
   \frac{2  {|P^1|}  {|Q_0|}}{3}
\end{array}
\right)\,\Rightarrow\,\,\{0,0,-2\,P^1\,Q_0\}\,,\nonumber\\
\mathbb{T}^{ab}&=&{\bf 0}\Rightarrow\,\,\{0,0,0\}\,,\nonumber\\
\mathfrak{T}^{xy}&=&{\bf 0}\,\Rightarrow\,\,\{0,0,0\}\,,\nonumber\\
W^{x|q}&=&\left(
\begin{array}{lll}
 -3  {|P^1|}^2 & 3  {|P^1|}^2 & -3  {|P^1|}^2 \\
 3  {|P^1|}^2 & -3  {|P^1|}^2 & 3  {|P^1|}^2 \\
 -3  {|P^1|}^2 & 3  {|P^1|}^2 & -3  {|P^1|}^2
\end{array}
\right)\neq {\bf 0}\nonumber
\end{eqnarray}
which implies that $\mathcal{L}_{non-BPS}$ indeed belongs to the $NO_3$ orbit;
The explicit dilatonic solution corresponding to $z_{\infty}=S_{\infty}=-i$ can be written in a closed form in terms of the  harmonic functions $H_0=1-\sqrt{2}\,Q_0\tau\,,H^1=1+\sqrt{2}\,P^1\tau$:
\begin{eqnarray}
S&=&-\frac{1}{z}=-i\,\sqrt{\frac{H_0}{H^1}}\,\,;\,\,\,\,e^{-U}=\sqrt{H_0\,(H^1)^3}\,.\label{solno3}
\end{eqnarray}
}
\item{{\bf Orbit $NO'_3$.} According to our discussion of section \ref{classificazia}, the orbit $\mathrm{NO}'_3$ differs from   $\mathrm{NO}_3$ in the sign of the non-vanishing eigenvalue of $\mathcal{T}^{xy}$. A Lax-representative of it can thus be obtained from (\ref{repno3}) by changing the relative sign of the two nilpotent terms, which amounts in the previous discussion to replacing $|P^1|\rightarrow -|P^1|$. As a consequence $P^1=|P^1|>0$ and the quartic invariant will now be positive. The corresponding solution will still have the form (\ref{solno3}) but with $H^1=1+\sqrt{2}\,|P^1|\tau$. Since this harmonic function now has a zero in $\tau$, so will $e^{-U}$ and thus the corresponding four-dimensional solution will be singular.
}
\item{{\bf Orbit $NO_2$.} The corresponding representative is obtained from any of the two above by setting $Q_0=0$ and corresponds to a \emph{lightlike} small black hole;}
\item{{\bf Orbit $NO_1$.} The corresponding representative is obtained from any of the first two representatives  by setting $P^1=0$ and corresponds to a \emph{doubly-critical} small black hole;}
\end{itemize}
Let us comment on the relation between the regular BPS and non-BPS representatives. It is important to note that both $\mathcal{L}_{BPS}$ and $\mathcal{L}_{non-BPS}$ are expressed as combinations with positive coefficients
of commuting nilpotent matrices. Note that the relative sign of the two terms can be changed by an $H^\mathbb{C}$ transformation of the form $\exp(i \pi\,J/2)$, for instance:
\begin{eqnarray}
e^{-\frac{i\pi}{2}\,J_{P^1} }\, N^\pm_{P^1}\, e^{\frac{i\pi}{2}\,J_{P^1} }&=& -N^\pm_{P^1}\,.
\end{eqnarray}
Such transformation will therefore not alter the $H^\mathbb{C}$-
orbit (and thus the $\mathfrak{g}_{2(2)}$-orbit) of the matrix.
Instead the difference in the two Lax representatives is in
\emph{the grading} of one of the two nilpotent components (in our
case $N^+_{P^1}\rightarrow N^-_{P^1}$). Changing this grading
amounts to changing the sign on the antisymmetric component
$K_{P^1}$ of $N^+_{P^1}$, associated with the magnetic charge
$P^1$, \emph{without changing the sign of the symmetric component
$\mathcal{H}_{P^1}$, associated with the scalar charges}. The
operation mapping the BPS into the non-BPS representative thus
consists in changing the sign of one of the two charges leaving
the scalar charges unaltered. The transformation which does the
job is the following:
\begin{eqnarray}
S&\equiv &e^{\frac{\pi}{2}\,K_{P^1} }\,e^{\frac{i\pi}{2}\,J_{P^1} }\,.
\end{eqnarray}
We indeed find:
\begin{eqnarray}
S^{-1} N^\pm_{P^1}\,S&=& N^\mp_{P^1}\,\,;\,\,\,\,S^{-1} N^\pm_{Q_0}\,S= N^\pm_{Q_0}\,.
\end{eqnarray}
The transformation $S$ is clearly  in $G_{2}^\mathbb{C}/H^\mathbb{C}$ and thus its effect is to alter the $\mathfrak{g}_{2(2)}$-orbit of the matrix it acts on.\par
As a byproduct of the above analysis we also find the following general property. Suppose $E$ ad $E'$ are two shift generators in $\mathbb{K}$,
in the Chevalley basis. Suppose they  correspond to two orthogonal roots (as $N^+_{Q_0},\,N^+_{P^1}$ were),  so that $[E,E']=[E,F']=[F,E']=[F,F']={\bf 0}$, then $E+E'$ and $E-E'$ belong to the \emph{same} $H^\mathbb{C}$-orbit.
To show this one considers the matrix $H'\equiv[E',F']\in \mathbb{H}^*$ and the $H^{\mathbb{C}}$ transformation
$\mathcal{O}=\exp(i\frac{\pi}{2} H')$, whose adjoint action flips the sign in front of $E'$ leaving $E$ unaltered, so that:
\begin{eqnarray}
\mathcal{O}^{-1}(E+E')\mathcal{O}&=& E-E'\,.
\end{eqnarray}
This $\mathcal{O}$-transformation will possibly alter the signs of the eigenvalues of the tensor classifiers $\mathcal{T},\,\mathfrak{T},\,\mathbb{T}$, but not their being zero or not. This implies that either both $E\pm E'$  correspond to BPS solutions or none of them does.
\section{Conclusions}
As pointed out in the introduction, the main goal of the present paper has been the classification of both non-extremal and extremal spherical symmetric black-holes of supergravity, according to orbits of the $\mathrm{H}^\star$ isotropy group. This latter appears in the time-like dimensional reduction of $D=4$ supergravity to $D=3$ when the scalar fields span a symmetric coset manifold $\mathrm{U_{D=4}}/\mathrm{H_{D=4}}$. In this case the the black-hole solutions are identified with the geodesics of a Lorentzian coset manifold $\mathrm{U_{D=3}}/\mathrm{H}^\star$ and the corresponding geodesic equations are best approached when they are cast into the Lax form (\ref{primolasso}). In this way the central object of investigation becomes the Lax operator $L(\tau)$ which, by definition, is an element of the complementary subspace $\mathbb{K}$ in the orthogonal decomposition: $\mathbb{U}_{D=3}\, = \, \mathbb{H}^\star \oplus \mathbb{K}$. In all $\mathcal{N}=2$ supergravities based on Special Geometries that are symmetric coset manifolds, the stability subgroup $\mathrm{H}^\star$ is of the following form:
\begin{equation}\label{isotropystar}
    \mathrm{H}^\star \, \sim \, \mathrm{SL(2,\mathbb{R})} \times \mathrm{U_{D=4}}
\end{equation}
and the complementary  subspace $\mathbb{K}$ falls into a universal irreducible representation:
\begin{equation}\label{univrep}
    \mathbb{K} \, \simeq \, \left(2,\mathbf{W}_{(2n+2)}\right)
\end{equation}
where $2$ denotes the fundamental defining representation of $\mathrm{SL(2,\mathbb{R})}$ while $\mathbf{W}_{(2n+2)}$ denotes the $2n+2$-dimensional symplectic representation of $\mathrm{U_{D=4}}$ which enters the definition of the special K\"ahler structure of
$\mathrm{U_{D=4}}$. By  $n$ we denote the complex dimension of $\mathrm{U_{D=4}}/\mathrm{H_{D=4}}$ which is the number of vector multiplets coupled to supergravity. Because of this fundamental algebraic property the  Lax operator $L$ is a two index tensor:
\begin{equation}\label{twoindessi}
    L \, \Leftrightarrow \, \Delta^{\alpha|A}
\end{equation}
where $\alpha$ takes two values and spans the fundamental representation of $\mathrm{SL(2,\mathbb{R})}$ while $A$ takes $\mathbf{W}_{(2n+2)}$-values and spans the symplectic representation $\mathbf{W}_{(2n+2)}$. Classifying black-hole orbits amounts to the classification of such tensors. The corresponding Lax operator may be a diagonalizable matrix, in which case we deal with non-extremal solutions, or a nilpotent one, this case corresponding to  extremal black-holes.
\par
The classification of nilpotent orbits within Lie algebras can be
addressed with the mathematical techniques related with the
Kostant-Sekiguchi theorem \cite{nilorbits} and this was done, for
the case of the $\mathfrak{g}_{(2,2)}$ Lie algebra in
\cite{bruxelles}. The central point of our paper resides in the
observation that the analysis of orbits (regular or nilpotent) can
be done in an allied way using some universal tensor structures
that can be constructed starting from the two index tensor
$\Delta^{\alpha|A}$.  In the body of the paper we presented these
structures for the specific case of the
$\mathfrak{g}_{(2,2)}$-model. Here we show their immediate
generalization to all the other series of homogeneous symmetric
special geometries.
\par
The first and most fundamental tensor classifier that separates supersymmetric from non supersymmetric orbits is the  antisymmetric tensor
$\mathcal{T}^{[AB]} $ which emerges from the following decomposition:
\begin{equation}\label{tensorstorto}
 \Delta^{\alpha|A} \, \Delta^{\beta|B} \, \epsilon_{\alpha\beta} \, = \,  \mathcal{T}^{[AB]} \, + \, \mathbb{C}^{AB} \, \mathfrak{H}_{quad}
\end{equation}
The above equation is understood, recalling that the quadratic invariant of $\mathrm{H}^\star$ which defines the hamiltonian $\mathfrak{H}_{quad}$ of the dynamical system and, in the black-hole interpretation, the extremality parameter $v^2$, is universally given by:
\begin{equation}\label{quadram}
    \mathrm{Tr} L^2 \, \varpropto \, \mathfrak{H}_{quad} \, \varpropto \, \Delta^{\alpha|A} \, \Delta^{\beta|B} \, \epsilon_{\alpha\beta} \,\mathbb{C}^{AB}
\end{equation}
where $\mathbb{C}^{AB}$ is the invariant symplectic tensor of the $\mathbf{W}_{(2n+2)}$ representation.
\par
For all extremal black-holes and, hence, for all nilpotent orbits, the quadratic invariant $\mathfrak{H}_{quad}$ vanishes. As we showed in the text the supersymmetric orbits are further characterized by the vanishing of the antisymmetric tensor $\mathcal{T}^{[AB]}$. In the $\mathfrak{g}_{(2,2)}$-model $\mathcal{T}^{[AB]}$ is an irreducible representation, the spin $\mathrm{j}=2$. Reducible or irreducible, the vanishing of this representation is the condition characterizing all supersymmetric orbits.
\par
Next, considering the symmetric product of Lax operators, we construct the following object:
\begin{equation}\label{woget}
    \mathcal{W}^{a|(AB)} \, \equiv \, \Pi^{a}_{\alpha\beta} \, \Delta^{\alpha|A} \, \Delta^{\beta|B}
\end{equation}
where $\Pi^{a}_{\alpha\beta}$, defined in eq.(\ref{paffuto}), are the projectors onto the symmetric $\mathrm{j=1}$ representation of $\mathrm{SL(2,\mathbb{R})}$. Utilizing the mixed quadratic tensor $\mathcal{W}^{a|(AB)}$ we can always construct the following quartic tensors:
\begin{equation}\label{fantasia}
    \mathfrak{T}^{AB,CD} \, \equiv \, \mathcal{W}^{a|(AB)} \, \mathcal{W}^{b|(CD)}\, \eta_{ab}
\end{equation}
and
\begin{equation}\label{felicione}
    \mathbb{T}^{(ab)} \, = \, \mathcal{W}^{a|(AB)} \, \mathcal{W}^{b|(CD)} \, d_{ABCD}
\end{equation}
where $d_{ABCD}$ are the coefficients defining the always existing quartic  symplectic invariant of the representation $\mathbf{W}_{(2n+2)}$, namely:
\begin{equation}\label{quartinva}
    \mathfrak{I}_4(q,p) \, = \, d_{ABCD} \, \mathcal{Q}^A \, \mathcal{Q}^B \, \mathcal{Q}^C \, \mathcal{Q}^D
\end{equation}
having used $\mathcal{Q}^A=\left \{q,p\right\}$ to denote the vector of electro-magnetic charges.
\par
Those in eq.s (\ref{fantasia}) and (\ref{felicione})  are the generalizations of the tensors defined respectively in eq.s (\ref{gluppo}) and (\ref{farnetico}) for the $\mathfrak{g}_{(2,2)}$-model. In that  model the regular BPS orbit of extremal black-holes is characterized by the vanishing not only of $\mathcal{T}^{[AB]}$ but also of $\mathfrak{T}^{AB CD}$, while the universal symmetric tensor
$\mathbb{T}^{ab}$ has signature $(\bullet,0,0)$. On the contrary the regular non-BPS solutions have vanishing $\mathbb{T}^{(ab)}$ and a non vanishing  $\mathcal{T}^{[AB]}$. The tensor $\mathfrak{T}^{AB CD}$ is zero for all regular nilpotent orbits. An urgent question to be answered is to what extent  these properties apply to more general symmetric Special Geometries. This investigation is postponed to a forthcoming publication by the present authors.\par
We wish now to point out an interesting mathematical analogy between the problem of classifying the $H^\star$-orbits  of the Lax operator  $\{\Delta^{\alpha A}\}$, addressed in the present paper,  and that addressed in \cite{Ferrara:2010ug} and \cite{Andrianopoli:2011gy} of studying the duality orbits, in the four-dimensional theory, of  two-centered black-hole solutions. In the latter case one deals with two symplectic charge-vectors ${\bf Q}^\alpha\equiv (\mathcal{Q}^{\alpha A})$ which transform in the representation $\left(2,\mathbf{W}_{(2n+2)}\right)$ of
an ${\rm SL}(2,\mathbb{R})\times {\rm U}_{D=4}$ group, just as the Lax components do.
In the two-charge problem, however,  the role of the group ${\rm SL}(2,\mathbb{R})\times {\rm U}_{D=4}$ is different:
It is not the symmetry group of the theory (as $H^\star$ is for the $D=3$ model), but rather contains the true on-shell global symmetry ${\rm U}_{D=4}$
of the four-dimensional theory, times an extra \emph{horizontal symmetry} group ${\rm SL}(2,\mathbb{R})$. The analysis in \cite{Ferrara:2010ug} and \cite{Andrianopoli:2011gy}, in other words, aimed at the definition of the two-charge orbit with respect to the group ${\rm U}_{D=4}$ alone. Nevertheless  the approach defined in the present paper may be relevant in order to characterize properties of the two-centered solutions which are both
invariant under the $D=4$ duality and the horizontal symmetry.
 \par
The allied weapon of the tensor classifiers was used in our paper in comparison with the mathematical classification of nilpotent and regular orbits and it allowed us to refine existing classifications, improving some statements appearing in the current literature.  In particular, concerning non-extremal black-holes, we were able to show that the condition of vanishing of the higher order Chevalley hamiltonians, corresponding to the enforcement of the cubic equation (\ref{cubicequa}), is a necessary but not sufficient condition to single out the orbit of regular black-holes. The locus (\ref{cubicequa}) tipically splits into distinct orbits characterized by different properties of the tensors classifiers $\mathcal{T}^{[AB]}, \mathfrak{T}^{AB,CD}$ and $\mathbb{T}^{(ab)}$. In the $\mathfrak{g}_{(2,2)}$ model the true regular black-hole orbit is characterized by the non-negativeness of all eigenvalues of all the classifiers. Once again it is urgent to verify whether a similar characterization holds true also in the other symmetric Special Geometries.
\par
By means of the explicit integration algorithm derived by us in previous publications we were able to explore the properties of the supergravity solutions occurring in all orbits and we shew that those generated by Lax operators with degree of nilpotency higher than three, always correspond to \textit{broken solutions}, where a true space-time singularity is developed at finite euclidian time $\tau \simeq \frac{1}{r}$ before the scalars can flow to the fixed points of the corresponding geodesic potential. From the point of view of the tensor classifiers the regular black-hole solutions (extremal and non extremal) appear to fulfill, in the $\mathfrak{g}_{(2,2)}$ model, the following condition:
\begin{equation}\label{conietta}
    \mbox{rank} \, \left [\mathbb{T}^{(ab)} \right ]\, \le \, 1 \quad ; \quad \mbox{eigenvalues} \, \left [ \mathbb{T}^{(ab)}\right ] \, \ge \, 0
\end{equation}
It is tempting to conjecture that eq.(\ref{conietta}) is of general validity for all symmetric Special Geometries. We plan to investigate this point together with the other ones mentioned above in our next coming paper.

{}~

{}~

{\bf Acknowledgments} The work of A.S. was partially supported by
the RFBR Grants No. 09-02-12417-$\mathrm{ofi\_m}$,
09-02-91349-NNIO\_a; DFG grant No 436 RUS/113/669, the
Heisenberg-Landau and CERN-BLTP JINR Programs.

\newpage
\appendix
\section{The Kostant hamiltonians in involution}
\label{papponaenorme}
The first hamiltonian, which is quadratic and which determines the flow equations has the following form:
\begin{equation}\label{hamil1}
  \mathfrak{K}_{1} \, = \, 2\, \mathfrak{H}_{quad} \, = \,  6 \Phi _1^2+6 \Phi _2^2+3 \Phi _3^2-3 \Phi _4^2-3 \Phi _5^2-3
   \Phi _6^2-3 \Phi _7^2+3 \Phi _8^2
\end{equation}
The second independent polynomial hamiltonian is homogeneous of order six and contains 246 terms. It  can be displayed as follows. We write it as the sum of three addends:
\begin{equation}\label{addenda}
    \mathfrak{K}_{2} \, = \, \mathfrak{K}_{(2,1)} \, + \, \mathfrak{K}_{(2,2)}\, + \, \mathfrak{K}_{(2,3)}
\end{equation}
whose explicit form is given in the following three formulae.
\begin{equation}\label{hamil21}
 \begin{array}{l}
 \mathfrak{K}_{(2,1)} \, = \,\\
 4 \Phi _1^6-24 \Phi _2^2 \Phi _1^4+6 \Phi _3^2 \Phi _1^4+36
   \Phi _2^4 \Phi _1^2-24 \Phi _2^2 \Phi _3^2 \Phi _1^2 \\
 +\frac{\Phi _3^6}{2}+3 \Phi _1^2 \Phi _3^4-6 \Phi _2^2 \Phi
   _3^4+18 \Phi _2^4 \Phi _3^2-6 \Phi _1^4 \Phi _4^2 \\
 -6 \sqrt{3} \Phi _2 \Phi _4^2 \Phi _1^3+18 \Phi _2^2 \Phi
   _4^2 \Phi _1^2-6 \Phi _3^2 \Phi _4^2 \Phi _1^2+18 \sqrt{3}
   \Phi _2^3 \Phi _4^2 \Phi _1-9 \sqrt{3} \Phi _2 \Phi _3^2
   \Phi _4^2 \Phi _1 \\
 -\frac{3}{2} \Phi _4^2 \Phi _3^4-9 \Phi _2^2 \Phi _4^2 \Phi
   _3^2+\frac{9}{4} \Phi _1^2 \Phi _4^4+\frac{27}{4} \Phi
   _2^2 \Phi _4^4+\frac{9}{2} \sqrt{3} \Phi _1 \Phi _2 \Phi
   _4^4 \\
 -18 \sqrt{2} \Phi _3 \Phi _4 \Phi _5 \Phi _2^3-24 \sqrt{6}
   \Phi _1 \Phi _3 \Phi _4 \Phi _5 \Phi _2^2+9 \sqrt{2} \Phi
   _3^3 \Phi _4 \Phi _5 \Phi _2-18 \sqrt{2} \Phi _1^2 \Phi _3
   \Phi _4 \Phi _5 \Phi _2-3 \sqrt{\frac{3}{2}} \Phi _1 \Phi
   _3 \Phi _4^3 \Phi _5 \\
 -6 \Phi _5^2 \Phi _1^4-18 \sqrt{3} \Phi _2 \Phi _5^2 \Phi
   _1^3-30 \Phi _2^2 \Phi _5^2 \Phi _1^2+6 \sqrt{3} \Phi _2^3
   \Phi _5^2 \Phi _1-\frac{9 \Phi _2 \Phi _3 \Phi _4^3 \Phi
   _5}{\sqrt{2}} \\
 -\frac{3}{2} \Phi _5^2 \Phi _3^4-6 \Phi _1^2 \Phi _5^2 \Phi
   _3^2+3 \Phi _2^2 \Phi _5^2 \Phi _3^2+9 \sqrt{3} \Phi _1
   \Phi _2 \Phi _5^2 \Phi _3^2+\frac{3}{2} \Phi _1^2 \Phi
   _4^2 \Phi _5^2 \\
 -9 \sqrt{\frac{3}{2}} \Phi _1 \Phi _3 \Phi _4 \Phi
   _5^3-\frac{9 \Phi _2 \Phi _3 \Phi _4 \Phi
   _5^3}{\sqrt{2}}+\frac{27}{2} \Phi _2^2 \Phi _4^2 \Phi
   _5^2-\frac{3}{2} \Phi _3^2 \Phi _4^2 \Phi _5^2+6 \sqrt{3}
   \Phi _1 \Phi _2 \Phi _4^2 \Phi _5^2 \\
 -\frac{15}{4} \Phi _1^2 \Phi _5^4+\frac{3}{4} \Phi _2^2 \Phi
   _5^4+\frac{3}{2} \Phi _3^2 \Phi _5^4-\frac{3}{2} \Phi _4^2
   \Phi _5^4-\frac{9}{2} \sqrt{3} \Phi _1 \Phi _2 \Phi _5^4
   \\
 -\frac{\Phi _5^6}{2}+36 \sqrt{6} \Phi _1^2 \Phi _2 \Phi _3
   \Phi _6 \Phi _5+3 \sqrt{3} \Phi _3^2 \Phi _4^3 \Phi _6+12
   \sqrt{3} \Phi _2^2 \Phi _3^2 \Phi _4 \Phi _6+36 \Phi _1
   \Phi _2 \Phi _3^2 \Phi _4 \Phi _6 \\
 -12 \sqrt{6} \Phi _3 \Phi _5 \Phi _6 \Phi _2^3+12 \sqrt{3}
   \Phi _4 \Phi _5^2 \Phi _6 \Phi _2^2+36 \Phi _1 \Phi _4
   \Phi _5^2 \Phi _6 \Phi _2-3 \sqrt{6} \Phi _3 \Phi _4^2
   \Phi _5 \Phi _6 \Phi _2+9 \sqrt{2} \Phi _1 \Phi _3 \Phi
   _4^2 \Phi _5 \Phi _6 \\
 -6 \Phi _6^2 \Phi _1^4+18 \sqrt{3} \Phi _2 \Phi _6^2 \Phi
   _1^3-30 \Phi _2^2 \Phi _6^2 \Phi _1^2-6 \sqrt{3} \Phi _2^3
   \Phi _6^2 \Phi _1+3 \sqrt{3} \Phi _4^3 \Phi _5^2 \Phi _6
   \\
 -\frac{3}{2} \Phi _6^2 \Phi _3^4-6 \Phi _1^2 \Phi _6^2 \Phi
   _3^2+3 \Phi _2^2 \Phi _6^2 \Phi _3^2-9 \sqrt{3} \Phi _1
   \Phi _2 \Phi _6^2 \Phi _3^2+\frac{21}{2} \Phi _1^2 \Phi
   _4^2 \Phi _6^2 \\
 -\frac{9}{2} \Phi _6^2 \Phi _4^4-\frac{9}{2} \Phi _2^2 \Phi
   _6^2 \Phi _4^2-\frac{3}{2} \Phi _3^2 \Phi _6^2 \Phi
   _4^2-15 \sqrt{3} \Phi _1 \Phi _2 \Phi _6^2 \Phi _4^2-9
   \sqrt{\frac{3}{2}} \Phi _1 \Phi _3 \Phi _5 \Phi _6^2 \Phi
   _4 \\
 -\frac{15}{2} \Phi _1^2 \Phi _5^2 \Phi _6^2+\frac{21}{2}
   \Phi _2^2 \Phi _5^2 \Phi _6^2+3 \Phi _3^2 \Phi _5^2 \Phi
   _6^2+\frac{3}{2} \Phi _4^2 \Phi _5^2 \Phi _6^2-\frac{27
   \Phi _2 \Phi _3 \Phi _4 \Phi _5 \Phi _6^2}{\sqrt{2}} \\
 -\frac{3}{2} \Phi _6^2 \Phi _5^4-\frac{15}{4} \Phi _1^2 \Phi
   _6^4+\frac{3}{4} \Phi _2^2 \Phi _6^4+\frac{3}{2} \Phi _3^2
   \Phi _6^4+\frac{9}{2} \sqrt{3} \Phi _1 \Phi _2 \Phi _6^4
   \\
 -\frac{\Phi _6^6}{2}+3 \Phi _4^2 \Phi _6^4-\frac{3}{2} \Phi
   _5^2 \Phi _6^4+6 \sqrt{6} \Phi _2 \Phi _3^3 \Phi _4 \Phi
   _7+36 \Phi _1 \Phi _2 \Phi _3^2 \Phi _5 \Phi _7 \\
 -3 \sqrt{3} \Phi _4^2 \Phi _7 \Phi _5^3-9 \sqrt{2} \Phi _1
   \Phi _3 \Phi _4 \Phi _7 \Phi _5^2+9 \sqrt{6} \Phi _2 \Phi
   _3 \Phi _4 \Phi _7 \Phi _5^2-12 \sqrt{3} \Phi _2^2 \Phi
   _3^2 \Phi _7 \Phi _5-3 \sqrt{3} \Phi _3^2 \Phi _4^2 \Phi
   _7 \Phi _5 \\
 +18 \sqrt{2} \Phi _3 \Phi _6 \Phi _7 \Phi _2^3-24 \sqrt{6}
   \Phi _1 \Phi _3 \Phi _6 \Phi _7 \Phi _2^2-9 \sqrt{2} \Phi
   _3^3 \Phi _6 \Phi _7 \Phi _2+18 \sqrt{2} \Phi _1^2 \Phi _3
   \Phi _6 \Phi _7 \Phi _2-3 \sqrt{\frac{3}{2}} \Phi _1 \Phi
   _3 \Phi _4^2 \Phi _6 \Phi _7 \\
 +9 \Phi _5 \Phi _6 \Phi _7 \Phi _4^3-\frac{45 \Phi _2 \Phi _3
   \Phi _6 \Phi _7 \Phi _4^2}{\sqrt{2}}-18 \Phi _1^2 \Phi _5
   \Phi _6 \Phi _7 \Phi _4-18 \Phi _2^2 \Phi _5 \Phi _6 \Phi
   _7 \Phi _4-9 \sqrt{\frac{3}{2}} \Phi _1 \Phi _3 \Phi _5^2
   \Phi _6 \Phi _7
\end{array}
\end{equation}
\begin{equation}\label{hamil22}
   \begin{array}{l}
   \mathfrak{K}_{(2,2)} \, = \,\\
 -9 \Phi _4 \Phi _6 \Phi _7 \Phi _5^3+\frac{27 \Phi _2 \Phi
   _3 \Phi _6 \Phi _7 \Phi _5^2}{\sqrt{2}}+36 \Phi _1 \Phi _2
   \Phi _6^2 \Phi _7 \Phi _5+9 \sqrt{2} \Phi _1 \Phi _3 \Phi
   _4 \Phi _6^2 \Phi _7+9 \sqrt{6} \Phi _2 \Phi _3 \Phi _4
   \Phi _6^2 \Phi _7 \\
 -9 \sqrt{\frac{3}{2}} \Phi _1 \Phi _3 \Phi _7 \Phi
   _6^3+\frac{9 \Phi _2 \Phi _3 \Phi _7 \Phi
   _6^3}{\sqrt{2}}-9 \Phi _4 \Phi _5 \Phi _7 \Phi _6^3-12
   \sqrt{3} \Phi _2^2 \Phi _5 \Phi _7 \Phi _6^2+6 \sqrt{3}
   \Phi _4^2 \Phi _5 \Phi _7 \Phi _6^2 \\
 -6 \Phi _7^2 \Phi _1^4+6 \sqrt{3} \Phi _2 \Phi _7^2 \Phi
   _1^3+18 \Phi _2^2 \Phi _7^2 \Phi _1^2-6 \Phi _3^2 \Phi
   _7^2 \Phi _1^2-18 \sqrt{3} \Phi _2^3 \Phi _7^2 \Phi _1 \\
 -\frac{3}{2} \Phi _7^2 \Phi _3^4-9 \Phi _2^2 \Phi _7^2 \Phi
   _3^2+9 \sqrt{3} \Phi _1 \Phi _2 \Phi _7^2 \Phi
   _3^2+\frac{9}{2} \Phi _1^2 \Phi _4^2 \Phi
   _7^2-\frac{27}{2} \Phi _2^2 \Phi _4^2 \Phi _7^2 \\
 +\frac{21}{2} \Phi _1^2 \Phi _5^2 \Phi _7^2-\frac{9}{2} \Phi
   _2^2 \Phi _5^2 \Phi _7^2+15 \sqrt{3} \Phi _1 \Phi _2 \Phi
   _5^2 \Phi _7^2-3 \sqrt{\frac{3}{2}} \Phi _1 \Phi _3 \Phi
   _4 \Phi _5 \Phi _7^2+\frac{45 \Phi _2 \Phi _3 \Phi _4 \Phi
   _5 \Phi _7^2}{\sqrt{2}} \\
 +3 \Phi _7^2 \Phi _5^4-\frac{3}{2} \Phi _3^2 \Phi _7^2 \Phi
   _5^2-\frac{9}{2} \Phi _4^2 \Phi _7^2 \Phi _5^2-9 \sqrt{2}
   \Phi _1 \Phi _3 \Phi _6 \Phi _7^2 \Phi _5+3 \sqrt{3} \Phi
   _3^2 \Phi _4 \Phi _6 \Phi _7^2 \\
 +\frac{3}{2} \Phi _1^2 \Phi _6^2 \Phi _7^2+\frac{27}{2} \Phi
   _2^2 \Phi _6^2 \Phi _7^2-6 \sqrt{3} \Phi _1 \Phi _2 \Phi
   _6^2 \Phi _7^2-6 \sqrt{3} \Phi _4 \Phi _5^2 \Phi _6 \Phi
   _7^2-3 \sqrt{6} \Phi _2 \Phi _3 \Phi _5 \Phi _6 \Phi _7^2
   \\
 -\frac{3}{2} \Phi _7^2 \Phi _6^4+3 \sqrt{3} \Phi _4 \Phi
   _7^2 \Phi _6^3-\frac{3}{2} \Phi _3^2 \Phi _7^2 \Phi
   _6^2-\frac{9}{2} \Phi _4^2 \Phi _7^2 \Phi _6^2+\frac{3}{2}
   \Phi _5^2 \Phi _7^2 \Phi _6^2 \\
 -3 \sqrt{3} \Phi _5 \Phi _6^2 \Phi _7^3-3 \sqrt{3} \Phi _3^2
   \Phi _5 \Phi _7^3-3 \sqrt{\frac{3}{2}} \Phi _1 \Phi _3
   \Phi _6 \Phi _7^3+\frac{9 \Phi _2 \Phi _3 \Phi _6 \Phi
   _7^3}{\sqrt{2}}+9 \Phi _4 \Phi _5 \Phi _6 \Phi _7^3 \\
 +\frac{9}{4} \Phi _1^2 \Phi _7^4+\frac{27}{4} \Phi _2^2 \Phi
   _7^4-\frac{9}{2} \Phi _5^2 \Phi _7^4-\frac{9}{2} \sqrt{3}
   \Phi _1 \Phi _2 \Phi _7^4+3 \sqrt{3} \Phi _3^3 \Phi _4^2
   \Phi _8 \\
 +9 \sqrt{2} \Phi _1 \Phi _4 \Phi _8 \Phi _5^3+3 \sqrt{3} \Phi
   _3 \Phi _4^2 \Phi _8 \Phi _5^2+18 \sqrt{3} \Phi _1^2 \Phi
   _3 \Phi _8 \Phi _5^2-6 \sqrt{3} \Phi _2^2 \Phi _3 \Phi _8
   \Phi _5^2+18 \sqrt{2} \Phi _1 \Phi _3^2 \Phi _4 \Phi _8
   \Phi _5 \\
 -9 \Phi _4 \Phi _6 \Phi _8 \Phi _3^3-9 \Phi _4^3 \Phi _6
   \Phi _8 \Phi _3+18 \Phi _1^2 \Phi _4 \Phi _6 \Phi _8 \Phi
   _3+18 \Phi _2^2 \Phi _4 \Phi _6 \Phi _8 \Phi _3+3 \sqrt{6}
   \Phi _2 \Phi _4 \Phi _5^3 \Phi _8 \\

 +18 \sqrt{6} \Phi _5 \Phi _6 \Phi _8 \Phi _1^3-6 \sqrt{6}
   \Phi _2^2 \Phi _5 \Phi _6 \Phi _8 \Phi _1-9 \sqrt{6} \Phi
   _3^2 \Phi _5 \Phi _6 \Phi _8 \Phi _1-21 \sqrt{\frac{3}{2}}
   \Phi _4^2 \Phi _5 \Phi _6 \Phi _8 \Phi _1-\frac{27 \Phi _2
   \Phi _4^2 \Phi _5 \Phi _6 \Phi _8}{\sqrt{2}} \\
 +9 \sqrt{\frac{3}{2}} \Phi _1 \Phi _6 \Phi _8 \Phi
   _5^3+\frac{9 \Phi _2 \Phi _6 \Phi _8 \Phi
   _5^3}{\sqrt{2}}+6 \sqrt{3} \Phi _3 \Phi _4^2 \Phi _6^2
   \Phi _8-18 \sqrt{3} \Phi _1^2 \Phi _3 \Phi _6^2 \Phi _8+6
   \sqrt{3} \Phi _2^2 \Phi _3 \Phi _6^2 \Phi _8 \\
 +9 \Phi _3 \Phi _4 \Phi _8 \Phi _6^3+9 \sqrt{\frac{3}{2}}
   \Phi _1 \Phi _5 \Phi _8 \Phi _6^3-\frac{9 \Phi _2 \Phi _5
   \Phi _8 \Phi _6^3}{\sqrt{2}}-9 \sqrt{2} \Phi _1 \Phi _4
   \Phi _5 \Phi _8 \Phi _6^2-9 \sqrt{6} \Phi _2 \Phi _4 \Phi
   _5 \Phi _8 \Phi _6^2 \\
 +6 \sqrt{6} \Phi _4 \Phi _7 \Phi _8 \Phi _1^3-9
   \sqrt{\frac{3}{2}} \Phi _4^3 \Phi _7 \Phi _8 \Phi _1-18
   \sqrt{6} \Phi _2^2 \Phi _4 \Phi _7 \Phi _8 \Phi _1+9
   \sqrt{6} \Phi _3^2 \Phi _4 \Phi _7 \Phi _8 \Phi
   _1-\frac{27 \Phi _2 \Phi _4^3 \Phi _7 \Phi _8}{\sqrt{2}}
   \\
 -9 \Phi _5 \Phi _7 \Phi _8 \Phi _3^3+18 \Phi _1^2 \Phi _5
   \Phi _7 \Phi _8 \Phi _3+18 \Phi _2^2 \Phi _5 \Phi _7 \Phi
   _8 \Phi _3+18 \Phi _4^2 \Phi _5 \Phi _7 \Phi _8 \Phi _3+9
   \sqrt{\frac{3}{2}} \Phi _1 \Phi _4 \Phi _5^2 \Phi _7 \Phi
   _8 \\
 +9 \Phi _3 \Phi _7 \Phi _8 \Phi _5^3-\frac{27 \Phi _2 \Phi _4
   \Phi _7 \Phi _8 \Phi _5^2}{\sqrt{2}}+9 \sqrt{2} \Phi _1
   \Phi _6 \Phi _7 \Phi _8 \Phi _5^2-9 \sqrt{6} \Phi _2 \Phi
   _6 \Phi _7 \Phi _8 \Phi _5^2-18 \sqrt{2} \Phi _1 \Phi _3^2
   \Phi _6 \Phi _7 \Phi _8 \\
 -3 \sqrt{3} \Phi _7^2 \Phi _8 \Phi _3^3-9 \sqrt{2} \Phi _1
   \Phi _6^3 \Phi _7 \Phi _8+3 \sqrt{6} \Phi _2 \Phi _6^3
   \Phi _7 \Phi _8+9 \sqrt{\frac{3}{2}} \Phi _1 \Phi _4 \Phi
   _6^2 \Phi _7 \Phi _8+\frac{27 \Phi _2 \Phi _4 \Phi _6^2
   \Phi _7 \Phi _8}{\sqrt{2}} \\
 -6 \sqrt{3} \Phi _3 \Phi _5^2 \Phi _8 \Phi _7^2-3 \sqrt{3}
   \Phi _3 \Phi _6^2 \Phi _8 \Phi _7^2+18 \Phi _3 \Phi _4
   \Phi _6 \Phi _8 \Phi _7^2-21 \sqrt{\frac{3}{2}} \Phi _1
   \Phi _5 \Phi _6 \Phi _8 \Phi _7^2+\frac{27 \Phi _2 \Phi _5
   \Phi _6 \Phi _8 \Phi _7^2}{\sqrt{2}}
\end{array}
\end{equation}
\begin{equation}\label{hamil23}
    \begin{array}{l}
    \mathfrak{K}_{(2,3)} \, = \,\\
 -12 \Phi _8^2 \Phi _1^4+36 \Phi _2^2 \Phi _8^2 \Phi _1^2-9
   \sqrt{\frac{3}{2}} \Phi _4 \Phi _7^3 \Phi _8 \Phi
   _1+\frac{27 \Phi _2 \Phi _4 \Phi _7^3 \Phi _8}{\sqrt{2}}-9
   \Phi _3 \Phi _5 \Phi _7^3 \Phi _8 \\
 -3 \Phi _8^2 \Phi _3^4-12 \Phi _1^2 \Phi _8^2 \Phi _3^2+18
   \Phi _2^2 \Phi _8^2 \Phi _3^2+9 \Phi _1^2 \Phi _4^2 \Phi
   _8^2+9 \sqrt{3} \Phi _1 \Phi _2 \Phi _4^2 \Phi _8^2 \\
 -\frac{9}{2} \Phi _3^2 \Phi _4^2 \Phi _8^2-15 \Phi _1^2 \Phi
   _5^2 \Phi _8^2+3 \sqrt{3} \Phi _1 \Phi _2 \Phi _5^2 \Phi
   _8^2-12 \sqrt{6} \Phi _1 \Phi _3 \Phi _4 \Phi _5 \Phi
   _8^2-9 \sqrt{2} \Phi _2 \Phi _3 \Phi _4 \Phi _5 \Phi _8^2
   \\
 +\frac{3}{2} \Phi _8^2 \Phi _5^4+\frac{3}{2} \Phi _3^2 \Phi
   _8^2 \Phi _5^2-3 \sqrt{3} \Phi _4 \Phi _6 \Phi _8^2 \Phi
   _5^2-6 \sqrt{6} \Phi _2 \Phi _3 \Phi _6 \Phi _8^2 \Phi
   _5+6 \sqrt{3} \Phi _3^2 \Phi _4 \Phi _6 \Phi _8^2 \\
 -15 \Phi _1^2 \Phi _6^2 \Phi _8^2+\frac{3}{2} \Phi _3^2 \Phi
   _6^2 \Phi _8^2+\frac{9}{2} \Phi _4^2 \Phi _6^2 \Phi
   _8^2-\frac{3}{2} \Phi _5^2 \Phi _6^2 \Phi _8^2-3 \sqrt{3}
   \Phi _1 \Phi _2 \Phi _6^2 \Phi _8^2 \\
 +\frac{3}{2} \Phi _8^2 \Phi _6^4+3 \sqrt{3} \Phi _4 \Phi _8^2
   \Phi _6^3-12 \sqrt{6} \Phi _1 \Phi _3 \Phi _7 \Phi _8^2
   \Phi _6-3 \sqrt{3} \Phi _5^3 \Phi _7 \Phi _8^2-6 \sqrt{3}
   \Phi _3^2 \Phi _5 \Phi _7 \Phi _8^2 \\
 +9 \Phi _1^2 \Phi _7^2 \Phi _8^2-9 \sqrt{3} \Phi _1 \Phi _2
   \Phi _7^2 \Phi _8^2+3 \sqrt{3} \Phi _5 \Phi _6^2 \Phi _7
   \Phi _8^2+9 \sqrt{2} \Phi _2 \Phi _3 \Phi _6 \Phi _7 \Phi
   _8^2+18 \Phi _4 \Phi _5 \Phi _6 \Phi _7 \Phi _8^2 \\
 -3 \sqrt{3} \Phi _3 \Phi _5^2 \Phi _8^3+9 \Phi _3 \Phi _4
   \Phi _6 \Phi _8^3-\frac{9}{2} \Phi _3^2 \Phi _7^2 \Phi
   _8^2+\frac{27}{2} \Phi _4^2 \Phi _7^2 \Phi
   _8^2+\frac{9}{2} \Phi _5^2 \Phi _7^2 \Phi _8^2 \\
 +9 \Phi _1^2 \Phi _8^4+3 \sqrt{3} \Phi _3 \Phi _6^2 \Phi
   _8^3-3 \sqrt{6} \Phi _1 \Phi _5 \Phi _6 \Phi _8^3-9
   \sqrt{6} \Phi _1 \Phi _4 \Phi _7 \Phi _8^3+9 \Phi _3 \Phi
   _5 \Phi _7 \Phi _8^3 \, + \, \frac{9}{2} \Phi _3^2 \Phi _8^4
\end{array}
\end{equation}
The third and the fourth hamiltonians in involution are rational functions.
In particular  we have:
\begin{eqnarray}
  \mathfrak{K}_{3} &=& - \, \frac{1}{4 \, \Phi_8} \, \mathfrak{P}_5 \label{hamil3}\\
  \mathfrak{K}_{4} &=& - \, \frac{1}{8 \, \Phi_8^2}\, \, \mathfrak{P}_4 \label{hamil4}
\end{eqnarray}
where $\mathfrak{P}_{5,4}$ are homogeneous polynomials of order
five and four respectively:
\begin{equation}\label{hamil3bis}
    \begin{array}{l}
    \mathfrak{P}_5 \, = \, \\
 6 \sqrt{3} \Phi _4^2 \Phi _3^3+36 \sqrt{2} \Phi _1 \Phi _4
   \Phi _5 \Phi _3^2+36 \sqrt{3} \Phi _1^2 \Phi _5^2 \Phi
   _3-12 \sqrt{3} \Phi _2^2 \Phi _5^2 \Phi _3+6 \sqrt{3} \Phi
   _4^2 \Phi _5^2 \Phi _3 \\
 -18 \Phi _4 \Phi _6 \Phi _3^3+36 \Phi _1^2 \Phi _4 \Phi _6
   \Phi _3+36 \Phi _2^2 \Phi _4 \Phi _6 \Phi _3+18 \sqrt{2}
   \Phi _1 \Phi _4 \Phi _5^3+6 \sqrt{6} \Phi _2 \Phi _4 \Phi
   _5^3 \\
 +36 \sqrt{6} \Phi _5 \Phi _6 \Phi _1^3-12 \sqrt{6} \Phi _2^2
   \Phi _5 \Phi _6 \Phi _1-18 \sqrt{6} \Phi _3^2 \Phi _5 \Phi
   _6 \Phi _1-21 \sqrt{6} \Phi _4^2 \Phi _5 \Phi _6 \Phi
   _1-18 \Phi _3 \Phi _4^3 \Phi _6 \\
 +9 \sqrt{6} \Phi _1 \Phi _6 \Phi _5^3+9 \sqrt{2} \Phi _2 \Phi
   _6 \Phi _5^3-27 \sqrt{2} \Phi _2 \Phi _4^2 \Phi _6 \Phi
   _5-36 \sqrt{3} \Phi _1^2 \Phi _3 \Phi _6^2+12 \sqrt{3}
   \Phi _2^2 \Phi _3 \Phi _6^2 \\
 +18 \Phi _3 \Phi _4 \Phi _6^3+9 \sqrt{6} \Phi _1 \Phi _5 \Phi
   _6^3+12 \sqrt{3} \Phi _3 \Phi _4^2 \Phi _6^2-18 \sqrt{2}
   \Phi _1 \Phi _4 \Phi _5 \Phi _6^2-18 \sqrt{6} \Phi _2 \Phi
   _4 \Phi _5 \Phi _6^2 \\
 +12 \sqrt{6} \Phi _4 \Phi _7 \Phi _1^3-9 \sqrt{6} \Phi _4^3
   \Phi _7 \Phi _1-36 \sqrt{6} \Phi _2^2 \Phi _4 \Phi _7 \Phi
   _1+18 \sqrt{6} \Phi _3^2 \Phi _4 \Phi _7 \Phi _1-9
   \sqrt{2} \Phi _2 \Phi _5 \Phi _6^3 \\
 -18 \Phi _5 \Phi _7 \Phi _3^3+36 \Phi _1^2 \Phi _5 \Phi _7
   \Phi _3+36 \Phi _2^2 \Phi _5 \Phi _7 \Phi _3+36 \Phi _4^2
   \Phi _5 \Phi _7 \Phi _3-27 \sqrt{2} \Phi _2 \Phi _4^3 \Phi
   _7 \\
 +18 \Phi _3 \Phi _7 \Phi _5^3+9 \sqrt{6} \Phi _1 \Phi _4 \Phi
   _7 \Phi _5^2-27 \sqrt{2} \Phi _2 \Phi _4 \Phi _7 \Phi
   _5^2+18 \sqrt{2} \Phi _1 \Phi _6 \Phi _7 \Phi _5^2-36
   \sqrt{2} \Phi _1 \Phi _3^2 \Phi _6 \Phi _7 \\
 -18 \sqrt{2} \Phi _1 \Phi _7 \Phi _6^3+6 \sqrt{6} \Phi _2
   \Phi _7 \Phi _6^3+9 \sqrt{6} \Phi _1 \Phi _4 \Phi _7 \Phi
   _6^2+27 \sqrt{2} \Phi _2 \Phi _4 \Phi _7 \Phi _6^2-18
   \sqrt{6} \Phi _2 \Phi _5^2 \Phi _7 \Phi _6 \\
 -6 \sqrt{3} \Phi _7^2 \Phi _3^3-12 \sqrt{3} \Phi _5^2 \Phi
   _7^2 \Phi _3+36 \Phi _4 \Phi _6 \Phi _7^2 \Phi _3-21
   \sqrt{6} \Phi _1 \Phi _5 \Phi _6 \Phi _7^2+27 \sqrt{2}
   \Phi _2 \Phi _5 \Phi _6 \Phi _7^2 \\
 -48 \Phi _8 \Phi _1^4-9 \sqrt{6} \Phi _4 \Phi _7^3 \Phi
   _1+27 \sqrt{2} \Phi _2 \Phi _4 \Phi _7^3-18 \Phi _3 \Phi
   _5 \Phi _7^3-6 \sqrt{3} \Phi _3 \Phi _6^2 \Phi _7^2 \\
 -12 \Phi _8 \Phi _3^4-48 \Phi _1^2 \Phi _8 \Phi _3^2+72 \Phi
   _2^2 \Phi _8 \Phi _3^2+144 \Phi _1^2 \Phi _2^2 \Phi _8+36
   \Phi _1^2 \Phi _4^2 \Phi _8 \\
 -18 \Phi _3^2 \Phi _8 \Phi _4^2+36 \sqrt{3} \Phi _1 \Phi _2
   \Phi _8 \Phi _4^2-48 \sqrt{6} \Phi _1 \Phi _3 \Phi _5 \Phi
   _8 \Phi _4-36 \sqrt{2} \Phi _2 \Phi _3 \Phi _5 \Phi _8
   \Phi _4-60 \Phi _1^2 \Phi _5^2 \Phi _8 \\
 +6 \Phi _8 \Phi _5^4+6 \Phi _3^2 \Phi _8 \Phi _5^2+12
   \sqrt{3} \Phi _1 \Phi _2 \Phi _8 \Phi _5^2-24 \sqrt{6}
   \Phi _2 \Phi _3 \Phi _6 \Phi _8 \Phi _5+24 \sqrt{3} \Phi
   _3^2 \Phi _4 \Phi _6 \Phi _8 \\
 -12 \sqrt{3} \Phi _4 \Phi _6 \Phi _8 \Phi _5^2-60 \Phi _1^2
   \Phi _6^2 \Phi _8+6 \Phi _3^2 \Phi _6^2 \Phi _8+18 \Phi
   _4^2 \Phi _6^2 \Phi _8-12 \sqrt{3} \Phi _1 \Phi _2 \Phi
   _6^2 \Phi _8 \\
 +6 \Phi _8 \Phi _6^4+12 \sqrt{3} \Phi _4 \Phi _8 \Phi _6^3-6
   \Phi _5^2 \Phi _8 \Phi _6^2-12 \sqrt{3} \Phi _5^3 \Phi _7
   \Phi _8-24 \sqrt{3} \Phi _3^2 \Phi _5 \Phi _7 \Phi _8 \\
 +12 \sqrt{3} \Phi _5 \Phi _7 \Phi _8 \Phi _6^2-48 \sqrt{6}
   \Phi _1 \Phi _3 \Phi _7 \Phi _8 \Phi _6+36 \sqrt{2} \Phi
   _2 \Phi _3 \Phi _7 \Phi _8 \Phi _6+72 \Phi _4 \Phi _5 \Phi
   _7 \Phi _8 \Phi _6+36 \Phi _1^2 \Phi _7^2 \Phi _8 \\
 -18 \Phi _3^2 \Phi _8 \Phi _7^2+54 \Phi _4^2 \Phi _8 \Phi
   _7^2+18 \Phi _5^2 \Phi _8 \Phi _7^2-36 \sqrt{3} \Phi _1
   \Phi _2 \Phi _8 \Phi _7^2-18 \sqrt{3} \Phi _3 \Phi _5^2
   \Phi _8^2 \\
 +18 \sqrt{3} \Phi _3 \Phi _6^2 \Phi _8^2+54 \Phi _3 \Phi _4
   \Phi _6 \Phi _8^2-18 \sqrt{6} \Phi _1 \Phi _5 \Phi _6 \Phi
   _8^2-54 \sqrt{6} \Phi _1 \Phi _4 \Phi _7 \Phi _8^2+54 \Phi
   _3 \Phi _5 \Phi _7 \Phi _8^2\\
   +72 \Phi _1^2 \Phi _8^3+36 \Phi _3^2 \Phi _8^3\\
\end{array}
\end{equation}
and
\begin{eqnarray}\label{p3}
    \mathfrak{P}_4 & = & 12 \sqrt{3} \Phi _7 \Phi _5^3+9 \Phi _6^2 \Phi _5^2+12
   \sqrt{3} \Phi _3 \Phi _8 \Phi _5^2\nonumber\\
   &&-54 \Phi _4 \Phi _6 \Phi
   _7 \Phi _5+12 \sqrt{6} \Phi _1 \Phi _6 \Phi _8 \Phi _5-36
   \Phi _3 \Phi _7 \Phi _8 \Phi _5\nonumber\\
   &&-12 \sqrt{3} \Phi _4 \Phi
   _6^3-27 \Phi _4^2 \Phi _7^2\nonumber\\
   &&-72 \Phi _1^2 \Phi _8^2-36 \Phi
   _3^2 \Phi _8^2-12 \sqrt{3} \Phi _3 \Phi _6^2 \Phi _8-36
   \Phi _3 \Phi _4 \Phi _6 \Phi _8+36 \sqrt{6} \Phi _1 \Phi
   _4 \Phi _7 \Phi _8
\end{eqnarray}
\begin{landscape}
\section{Tables}
\subsection{The Lax operator in the first nilpotent orbit}
{\scriptsize{\begin{eqnarray}
&\mathfrak{L}^{NO_1}(\sigma,y)=&\nonumber\\
   & \, \sigma\, \left(
\begin{array}{lllllll}
 -y^6+y^4+y^2-1 & -2 y
   \left(y^4-1\right) & 4 y^2
   \left(y^2-1\right) & -2
   \sqrt{2} y \left(y^2-1\right)^2
   & \left(y^2-1\right)^3 & 0 & 0
   \\
 -2 y \left(y^4-1\right) & -4 y^2
   \left(y^2+1\right) & 8 y^3 & -4
   \sqrt{2} y^2 \left(y^2-1\right)
   & 2 y \left(y^2-1\right)^2 & 0
   & 0 \\
 -4 y^2 \left(y^2-1\right) & -8
   y^3 & -\left(y^2+1\right)
   \left(y^4-6 y^2+1\right) & -2
   \sqrt{2} y \left(y^4-1\right) &
   0 & -2 y \left(y^2-1\right)^2 &
   \left(y^2-1\right)^3 \\
 2 \sqrt{2} y \left(y^2-1\right)^2
   & 4 \sqrt{2} y^2
   \left(y^2-1\right) & -2
   \sqrt{2} y \left(y^4-1\right) &
   0 & -2 \sqrt{2} y
   \left(y^4-1\right) & -4
   \sqrt{2} y^2 \left(y^2-1\right)
   & 2 \sqrt{2} y
   \left(y^2-1\right)^2 \\
 -\left(y^2-1\right)^3 & -2 y
   \left(y^2-1\right)^2 & 0 & -2
   \sqrt{2} y \left(y^4-1\right) &
   \left(y^2+1\right) \left(y^4-6
   y^2+1\right) & -8 y^3 & 4 y^2
   \left(y^2-1\right) \\
 0 & 0 & 2 y \left(y^2-1\right)^2
   & 4 \sqrt{2} y^2
   \left(y^2-1\right) & 8 y^3 & 4
   y^2 \left(y^2+1\right) & -2 y
   \left(y^4-1\right) \\
 0 & 0 & -\left(y^2-1\right)^3 &
   -2 \sqrt{2} y
   \left(y^2-1\right)^2 & -4 y^2
   \left(y^2-1\right) & -2 y
   \left(y^4-1\right) &
   \left(y^2-1\right)^2
   \left(y^2+1\right)
\end{array}
\right)&\nonumber\\
\label{orbacortissima}
\end{eqnarray}}}
The expression of the parameters $\sigma,y$ in terms of the charges
{\scriptsize{\begin{eqnarray}
\sigma & = & \left(\sqrt{3} \kappa  \xi
   \left(4 \kappa ^2+3 \xi
   ^2\right) p_1^4+\left(\sqrt{3}
   \xi  \sqrt{\left(\kappa ^2+\xi
   ^2\right) p_1^2+2 \sqrt{3} \xi
   p_2 p_1+3 p_2^2} \left(4 \kappa
   ^2+\xi ^2\right)+3 \left(4
   \kappa ^3+9 \xi ^2 \kappa
   \right) p_2\right) p_1^3\right. \nonumber\\
   &&\left. +3 p_2
   \left(\sqrt{\left(\kappa ^2+\xi
   ^2\right) p_1^2+2 \sqrt{3} \xi
   p_2 p_1+3 p_2^2} \left(4 \kappa
   ^2+3 \xi ^2\right)+9 \sqrt{3}
   \kappa  \xi  p_2\right) p_1^2 \right.\nonumber\\
   &&\left. +9
   p_2^2 \left(\sqrt{3}
   \sqrt{\left(\kappa ^2+\xi
   ^2\right) p_1^2+2 \sqrt{3} \xi
   p_2 p_1+3 p_2^2} \xi +3 \kappa
   p_2\right) p_1+9 p_2^3
   \sqrt{\left(\kappa ^2+\xi
   ^2\right) p_1^2+2 \sqrt{3} \xi
   p_2 p_1+3 p_2^2}\right)\left(144 \kappa
   ^{3/2} p_2^2 \left(\xi
   p_1+\sqrt{3} p_2\right)\right)^{-1}\nonumber \\
  y &=& \frac{\kappa
   p_1-\sqrt{\left(\kappa ^2+\xi
   ^2\right) p_1^2+2 \sqrt{3} \xi
   p_2 p_1+3 p_2^2}}{\xi
   p_1+\sqrt{3} p_2}\label{razzolin}
\end{eqnarray}
}}
\end{landscape}


\newpage
\begin{thebibliography}{99}
\bibitem{ferrarakallosh} S.~Ferrara, R.~Kallosh,
  \emph{Supersymmetry and attractors},
  Phys.\ Rev.\  {\bf D54}, 1514-1524 (1996).
  [hep-th/9602136].\\
  S.~Ferrara, R.~Kallosh, A.~Strominger,
  \emph{N=2 extremal black holes}
  Phys.\ Rev.\  {\bf D52 } (1995)  5412-5416.
[hep-th/9508072].
%
\bibitem{Gibbons:1996af}
  G.~W.~Gibbons, R.~Kallosh, B.~Kol,
 \emph{Moduli, scalar charges, and the first law of black hole thermodynamics},
  Phys.\ Rev.\ Lett.\  {\bf 77 } (1996)  4992-4995.
  [hep-th/9607108].
\bibitem{stromingerBH}  A.~Strominger and C.~Vafa,
  \emph{Microscopic Origin of the Bekenstein-Hawking Entropy},
  Phys.\ Lett.\  B {\bf 379} (1996) 99
  [arXiv:hep-th/9601029].
%
\bibitem{dualitiessg}
C.~M.~Hull and P.~K.~Townsend,
 \emph{Unity of superstring dualities},
  Nucl.\ Phys.\  B {\bf 438} (1995) 109
  [arXiv:hep-th/9410167];
    L.~Borsten, D.~Dahanayake, M.~J.~Duff and W.~Rubens,
 \emph{Black holes admitting a Freudenthal dual},
  Phys.\ Rev.\  D {\bf 80} (2009) 026003
  [arXiv:0903.5517 [hep-th]].
\bibitem{olderBHliterature}
Older references on susy black holes:
\par
G. Gibbons, in ``{\it Unified theories of Elementary Particles. Critical
Assessment
and Prospects}'', Proceedings of the Heisemberg Symposium, M\"unchen, West
 Germany, 1981,
 ed. by P. Breitenlohner and H. P. D\"urr, Lecture Notes in Physics Vol.
 160 (Springer-Verlag, Berlin, 1982);\\
G. W. Gibbons and C. M. Hull, Phys. lett. {\bf 109B} (1982) 190;\\
G. W. Gibbons, in ``{\it Supersymmetry, Supergravity and Related Topics}'',
 Proceedings
of the XVth GIFT International Physics, Girona, Spain, 1984, ed. by F. del
 Aguila, J.
de Azc\'arraga and L. Ib\'a\~nez, (World Scientific, 1995), pag. 147;\\
R. Kallosh, A. Linde, T. Ortin, A. Peet and A. Van Proeyen, Phys. Rev.
 {\bf D46} (1992) 5278;\\
R. Kallosh, T. Ortin and A. Peet, Phys. Rev. {\bf D47} (1993) 5400;
R. Kallosh, Phys. Lett. {\bf B282} (1992) 80;\\
R. Kallosh and A. Peet, Phys. Rev. {\bf D46} (1992) 5223;\\
A. Sen, Nucl. Phys. {\bf B440} (1995) 421; Phys. Lett. {\bf B303} (1993) 221;
 Mod. Phys. Lett. {\bf A10}
(1995) 2081;\\
J. Schwarz and A. Sen, Phys. Lett. {\bf B312} (1993) 105;
%
\bibitem{firstorderSUSY}
 G.~W.~Moore,
 \emph{Arithmetic and attractors},
  arXiv:hep-th/9807087.
L.~Andrianopoli, R.~D'Auria, S.~Ferrara {\it et al.},
  \emph{E(7)(7) duality, BPS black hole evolution and fixed scalars}
  Nucl.\ Phys.\  {\bf B509 } (1998)  463-518.
  [hep-th/9707087].\\
 L.~Andrianopoli, R.~D'Auria, S.~Ferrara {\it et al.},
  \emph{Solvable Lie algebras in type IIA, type IIB and M theories}
  Nucl.\ Phys.\  {\bf B493 } (1997)  249-280.
  [hep-th/9612202].
\\
L.~Andrianopoli, R.~D'Auria, S.~Ferrara {\it et al.},
  \emph{RR scalars, U duality and solvable Lie algebras},
  Nucl.\ Phys.\  {\bf B496 } (1997)  617-629.
  [hep-th/9611014].
\\
G.~Arcioni, A.~Ceresole, F.~Cordaro {\it et al.},
  \emph{N=8 BPS black holes with 1/2 or 1/4 supersymmetry and solvable Lie algebra decompositions}
  Nucl.\ Phys.\  {\bf B542 } (1999)  273-307.
  [hep-th/9807136].
\\
M. J. Duff, J. T. Liu, J. Rahmfeld, \emph{ Four
Dimensional String/String/String Triality }, Nucl.Phys. B459 (1996) 125,
hep-th/9508094;\\
 R. Kallosh, M. Shmakova, W. K. Wong, \emph{ Freezing of Moduli
by N=2 Dyons}, Phys.Rev. D54 (1996) 6284, hep-th/9607077;\\
K. Behrndt, R.
Kallosh, J. Rahmfeld, M. Smachova and W. K. Wond, \emph{ STU Black-holes and
string triality}, hep-th/9608059;\\
 G. Lopes Cardoso, D. Lust and T. Mohaupt,
\emph{ Modular Symmetries of N=2 Black Holes }, Phys.Lett. B388 (1996) 266,
hep-th/9608099;\\
\bibitem{Behrndt:1997ny}
  K.~Behrndt, D.~Lust, W.~A.~Sabra,
  ``Stationary solutions of N=2 supergravity,''
  Nucl.\ Phys.\  {\bf B510 } (1998)  264-288.
  [hep-th/9705169];
 \bibitem{iomatteo}   M.~Bertolini, M.~Trigiante,
 \emph{Regular BPS black holes: Macroscopic and microscopic description of the generating solution},
  Nucl.\ Phys.\  {\bf B582 } (2000)  393-406.
  [hep-th/0002191].
    M.~Bertolini, M.~Trigiante,
  \emph{Regular RR and NS NS BPS black holes},
  Int.\ J.\ Mod.\ Phys.\  {\bf A15 } (2000)  5017.
  [hep-th/9910237].

\bibitem{ortino} R. Kallosh, T. Ortin, "{\it Charge quantization
of Axion Dilaton black--holes}" Phys. rev. {\bf D48} (1993) 742,
hep-th 9302109, \\ E. Bergshoeff, R. Kallosh, T. Ortin "{\it
Stationary Axion Dilaton solutions and supersymmetry} Nuc. Phys.
{\bf B478} (1996) 156, hep-th 9605059.
%
\bibitem{harmfunpbrane}
  K.~S.~Stelle,
\emph{Lectures on supergravity p-branes},
  [hep-th/9701088].
%
\bibitem{microstateliterature}
A.~Strominger,
\emph{Black hole entropy from near horizon microstates}
JHEP {\bf 9802 } (1998)  009.[hep-th/9712251].\\
G.~T.~Horowitz, J.~M.~Maldacena, A.~Strominger,
\emph{Nonextremal black hole microstates and U duality}
Phys.\ Lett.\  {\bf B383 } (1996)  151-159.
[hep-th/9603109].\\
A.~Dabholkar,
\emph{Microstates of nonsupersymmetric black holes}
Phys.\ Lett.\  {\bf B402 } (1997)  53-58.
[hep-th/9702050].\\
F.~Larsen,
\emph{A String model of black hole microstates}
Phys.\ Rev.\  {\bf D56 } (1997)  1005-1008.[hep-th/9702153].\\
A.~Ghosh, P.~Mitra,
  \emph{Counting of black hole microstates}
  Indian J.\ Phys.\  {\bf 80 } (2006)  867.
  [gr-qc/0603029];
%
 J.~M.~Maldacena, A.~Strominger, E.~Witten,
 \emph{Black hole entropy in M theory},
  JHEP {\bf 9712 } (1997)  002.
  [hep-th/9711053].
   M.~Bertolini, M.~Trigiante,
 \emph{Microscopic entropy of the most general four-dimensional BPS black hole},
  JHEP {\bf 0010 } (2000)  002.
  [hep-th/0008201].
\bibitem{SKG}
{\it Pioneering papers on vector multiplet coupling to Supergravity:}\\
B. de Wit, P.G. Lauwers, R. Philippe, Su S.Q.
and A. Van Proeyen, Phys. Lett. 134B (1984) 37;\\
S. J. Gates, Nucl. Phys. B238 (1984) 349;\\
B. de Wit and A. Van Proeyen, Nucl. Phys. B245 (1984) 89.\\
E. Cremmer, C. Kounnas, A. Van Proeyen, J. P. Derendinger,
S. Ferrara, B. de Wit and L. Girardello, Nucl. Phys. {B250} (1985)
385;\\
{\it Definition of Special K\"ahler Geometry:}\\
%
A.~Strominger,
Comm. Math. Phys. 133 (1990) 163.\\
%
L. Castellani, R. D'Auria and S. Ferrara,
Phys. Lett. 241B (1990) 57; Class. Quantum Grav. 7 (1990) 1767.\\
%
R. D'Auria, S. Ferrara and P. Fr\'e, Nucl. Phys. B359 (1991) 705.\\
%
\bibitem{Andrianopoli:1996cm}
  L.~Andrianopoli, M.~Bertolini, A.~Ceresole, R.~D'Auria, S.~Ferrara, P.~Fre and T.~Magri,
  \emph{N = 2 supergravity and N = 2 super Yang-Mills theory on general scalar
  manifolds: Symplectic covariance, gaugings and the momentum map},
  J.\ Geom.\ Phys.\  {\bf 23} (1997) 111
  [arXiv:hep-th/9605032].
%
\bibitem{specHomgeo}
{\it Homogeneous Special K\"ahler Geometry}:\\
B. de Wit, A. Van Proeyen, \emph{Special geometry, cubic polynomials
and homogeneous quaternionic spaces}, Commun.Math.Phys.149:307-334
(1992), hep-th/9112027;\\
B. de Wit, A. Van Proeyen, \emph{Broken
sigma model isometries in very special geometry},
Phys.Lett.B293:94-99 (1992), hep-th/9207091;
B. de Wit, F.
Vanderseypen, A. Van Proeyen, \emph{Symmetry structure of special
geometries}, Nucl.Phys. B400 (1993) 463-524, hep-th/9210068;\\
B. de Wit, A. Van Proeyen, \emph{Hidden symmetries, special geometry and
quaternionic manifolds}, Int.J.Mod.Phys. D3 (1994) 31-48,
hep-th/9310067; .
%
\bibitem{CYandBH}
M.~Bertolini, P.~Fre, R.~Iengo and C.~A.~Scrucca,
\emph{Black holes as D3-branes on Calabi-Yau threefolds}
Phys.\ Lett.\  B {\bf 431} (1998) 22
[arXiv:hep-th/9803096].
%
\bibitem{orbifoldlimits}.
M.~Cvetic, J.~Louis, B.~A.~Ovrut,
\emph{A String Calculation of the Kahler Potentials for Moduli of Z(N) Orbifolds},
  Phys.\ Lett.\  {\bf B206 } (1988)  227.\\
S.~Ferrara, P.~Fre and P.~Soriani,
  \emph{On the moduli space of the T**6 / Z(3) orbifold and its modular group,}
  Class.\ Quant.\ Grav.\  {\bf 9} (1992) 1649
  [arXiv:hep-th/9204040].\\
  G.~Lopes Cardoso, D.~Lust and T.~Mohaupt,
  \emph{Moduli spaces and target space duality symmetries in (0,2) Z(N) orbifold
  theories with continuous Wilson lines}
  Nucl.\ Phys.\  B {\bf 432} (1994) 68
  [arXiv:hep-th/9405002].\\
%

\bibitem{criticalrefs}
K.~Goldstein, N.~Iizuka, R.~P.~Jena and S.~P.~Trivedi,
  \emph{Non-supersymmetric attractors}
  Phys.\ Rev.\  D {\bf 72}, 124021 (2005)
  [arXiv:hep-th/0507096].\\
P.~K.~Tripathy and S.~P.~Trivedi,
  \emph{Non-Supersymmetric Attractors in String Theory}
  JHEP {\bf 0603}, 022 (2006)
  [arXiv:hep-th/0511117].\\
R.~Kallosh,
  \emph{New Attractors}
  JHEP {\bf 0512}, 022 (2005)
  [arXiv:hep-th/0510024].\\
A.~Giryavets,
 \emph{New attractors and area codes}
  JHEP {\bf 0603}, 020 (2006)
  [arXiv:hep-th/0511215].\\
R.~Kallosh, N.~Sivanandam and M.~Soroush,
\emph{The non-BPS black hole attractor equation}
JHEP {\bf 0603}, 060 (2006)
  [arXiv:hep-th/0602005].\\
S.~Bellucci, S.~Ferrara and A.~Marrani,
 \emph{On some properties of the attractor equations}
Phys.\ Lett.\  B {\bf 635}, 172 (2006)
[arXiv:hep-th/0602161].\\
  S.~Bellucci, S.~Ferrara, M.~Gunaydin, A.~Marrani,
\emph{Charge orbits of symmetric special geometries and attractors},'
  Int.\ J.\ Mod.\ Phys.\  {\bf A21 } (2006)  5043-5098.
  [hep-th/0606209].
 L.~Andrianopoli, R.~D'Auria, S.~Ferrara, M.~Trigiante,
  \emph{Extremal black holes in supergravity},
  Lect.\ Notes Phys.\  {\bf 737 } (2008)  661-727.
  [hep-th/0611345].
%
\bibitem{fakeprepo}
  A.~Ceresole and G.~Dall'Agata,
  \emph{Flow Equations for Non-BPS Extremal Black Holes}
  JHEP {\bf 0703}, 110 (2007)
  [arXiv:hep-th/0702088].

\bibitem{fakeprepoHJ}
L.~Andrianopoli, R.~D'Auria, E.~Orazi and M.~Trigiante,
\emph{First Order Description of Black Holes in Moduli Space}
  JHEP {\bf 0711}, 032 (2007)
  [arXiv:0706.0712 [hep-th]];\\
  L.~Andrianopoli, R.~D'Auria, E.~Orazi and M.~Trigiante,
\emph{First Order Description of D=4 static Black Holes and the Hamilton-Jacobi
equation}
Nucl.\ Phys.\  B {\bf 833}, 1 (2010) [arXiv:0905.3938 [hep-th]];
  L.~Andrianopoli, R.~D'Auria, S.~Ferrara, M.~Trigiante,
\emph{Fake Superpotential for Large and Small Extremal Black Holes},
  JHEP {\bf 1008 } (2010)  126.
  [arXiv:1002.4340 [hep-th]].
\bibitem{Lopes Cardoso:2007ky}
  G.~Lopes Cardoso, A.~Ceresole, G.~Dall'Agata, J.~M.~Oberreuter and J.~Perz,
\emph{First-order flow equations for extremal black holes in very special
  geometry},
  JHEP {\bf 0710} (2007) 063
  [arXiv:0706.3373 [hep-th]].
\bibitem{fakeprepo2}
A.~Ceresole, G.~Dall'Agata, S.~Ferrara and A.~Yeranyan,
 \emph{First order flows for N=2 extremal black holes and duality invariants}
  Nucl.\ Phys.\  B {\bf 824}, 239 (2010)
  [arXiv:0908.1110 [hep-th]].\\
 G.~Bossard, Y.~Michel and B.~Pioline,
\emph{Extremal black holes, nilpotent orbits and the true fake superpotential}
JHEP {\bf 1001}, 038 (2010) [arXiv:0908.1742 [hep-th]].\\
A.~Ceresole, G.~Dall'Agata, S.~Ferrara and A.~Yeranyan,
 \emph{Universality of the superpotential for d = 4 extremal black holes,}
  Nucl.\ Phys.\  B {\bf 832}, 358 (2010)
  [arXiv:0910.2697 [hep-th]].\\
  %

\bibitem{HJLi} V.~Arnold,  {\em {Mathematical Methods of Classical Mechanics}}, Springer
  (1997); \\
  J.~McCauley,  {\em {Classical Mechanics: Transformations, Flows, Integrable and
  Chaotic Dynamics}}, Cambridge University Press (1997);\\
  K.~R. Meyer, G.~R. Hall and D.~Offin,  {\em {Introduction to Hamiltonian
  Dynamical Systems and the N-Body Problem}}, Springer (2009).
%


\bibitem{Breitenlohner:1987dg}
P.~Breitenlohner, D.~Maison  and G.~W. Gibbons,
\emph{Four-dimensional black
  holes from Kaluza-Klein theories}, Commun. Math. Phys. {\bf 120} (1988)
295;
\bibitem{pioline}
M.~Gunaydin, A.~Neitzke, B.~Pioline  and A.~Waldron, \emph{BPS
black holes,
  quantum attractor flows and automorphic forms}, Phys. Rev. {\bf D73} (2006)
  084019;
  B.~Pioline,
  \emph{Lectures on black holes, topological strings and quantum attractors},
  Class.\ Quant.\ Grav.\  {\bf 23 } (2006)  S981.
  [hep-th/0607227];
  M.~Gunaydin, A.~Neitzke, B.~Pioline, A.~Waldron,
 \emph{Quantum Attractor Flows},
  JHEP {\bf 0709 } (2007)  056;
\bibitem{Gaiotto:2007ag} D.~Gaiotto, W.~W. Li  and M.~Padi, \emph{{Non-Supersymmetric
Attractor Flow in
  Symmetric Spaces}}, JHEP {\bf 12} (2007) 093,
\bibitem{Bergshoeff:2008be}
  E. Bergshoeff, W. Chemissany, A.~Ploegh, M.~Trigiante and T.~Van Riet,
  \emph{Generating Geodesic Flows and Supergravity Solutions,}
  Nucl.\ Phys.\ {\bf B 812} (2009) 343
  [arXiv:0806.2310].
\bibitem{Bossard:2009at}
  G.~Bossard, H.~Nicolai, K.~S.~Stelle,
  \emph{Universal BPS structure of stationary supergravity solutions,}
  JHEP {\bf 0907 } (2009)  003.
\bibitem{nilorbits} The main reference on this issue is D. H. Collingwood and
W.M. McGovern, \emph{Nilpotent Orbits in Semisimple Lie Algebras},  Van Nostrand Reinhold 1993.
%
%
\bibitem{cmappa}
S. Ferrara,  S. Sabharwal, \emph{Quaternionic
manifolds for type II superstring vacua of Calabi-Yau spaces},
Nucl.Phys.B332:317 (1990)\\
S.~Cecotti,
  \emph{Homogeneous Kahler Manifolds And T Algebras In N=2 Supergravity And Superstrings,''
  Commun.\ Math.\ Phys.\  {\bf 124 } (1989)  23-55.}\\
R.~D'Auria, S.~Ferrara, M.~Trigiante,
  \emph{C - map, very special quaternionic geometry and dual Kahler spaces}
  Phys.\ Lett.\  {\bf B587 } (2004)  138-142.
  [hep-th/0401161].\\
P.~Fre, F.~Gargiulo, J.~Rosseel {\it et al.},
  \emph{Tits-Satake projections of homogeneous special geometries},''
  Class.\ Quant.\ Grav.\  {\bf 24 } (2007)  27-78.
  [hep-th/0606173].
%
\bibitem{myparis} P. Fr\`e \emph{Gaugings and other supergravity tools for $p$-brane
physics}, Lectures given at the RTN School Recent Advances in
M-theory, Paris February 1-8 IHP, hep-th/0102114.
%
\bibitem{Chemissany:2010zp}
  W.~Chemissany, P.~Fre, J.~Rosseel, A.~S.~Sorin, M.~Trigiante and T.~Van Riet,
  \emph{Black holes in supergravity and integrability},  JHEP {\bf 1009} (2010) 080.
%

%
\bibitem{Fre:2003ep}
  P.~Fre, V.~Gili, F.~Gargiulo, A.~S.~Sorin, K.~Rulik and M.~Trigiante,
  \emph{Cosmological backgrounds of superstring theory and solvable algebras:
  Oxidation and branes}
  Nucl.\ Phys.\  B {\bf 685} (2004) 3
  [arXiv:hep-th/0309237].
%
\bibitem{Fre:2003tg}
  P.~Fre, K.~Rulik and M.~Trigiante,
  \emph{Exact solutions for Bianchi type cosmological metrics, Weyl orbits of
  E(8(8)) subalgebras and p-branes,}
  Nucl.\ Phys.\  B {\bf 694} (2004) 239
  [arXiv:hep-th/0312189].
%
\bibitem{Fre':2005si}
  P.~Fre', F.~Gargiulo, K.~Rulik and M.~Trigiante,
  \emph{The general pattern of Kac Moody extensions in supergravity and the  issue
  of cosmic billiards},
  Nucl.\ Phys.\  B {\bf 741} (2006) 42
  [arXiv:hep-th/0507249].
%
\bibitem{Fre':2005sr}
  P.~Fre', F.~Gargiulo and K.~Rulik,
  \emph{Cosmic billiards with painted walls in non-maximal supergravities: A
  worked out example,}
  Nucl.\ Phys.\  B {\bf 737} (2006) 1
  [arXiv:hep-th/0507256].
%
\bibitem{Bellucci:2006ib}
  S.~Bellucci, S.~Ferrara, A.~Marrani and A.~Yeranyan,
  \emph{Mirror Fermat Calabi-Yau Threefolds and Landau-Ginzburg Black Hole
  Attractors,}
  Riv.\ Nuovo Cim.\  {\bf 29N5} (2006) 1
  [arXiv:hep-th/0608091].
%
\bibitem{Bellucci:2007eh}
  S.~Bellucci, S.~Ferrara, A.~Marrani and A.~Shcherbakov,
  \emph{Splitting of Attractors in 1-modulus Quantum Corrected Special Geometry},
  JHEP {\bf 0802} (2008) 088
  [arXiv:0710.3559 [hep-th]].
%
\bibitem{Ferrara:2010ug}
  S.~Ferrara, A.~Marrani, E.~Orazi, R.~Stora and A.~Yeranyan,
  \emph{Two-Center Black Holes Duality-Invariants for stu Model and its lower-rank Descendants},
  arXiv:1011.5864 [hep-th].
\bibitem{Andrianopoli:2011gy}
  L.~Andrianopoli, R.~D'Auria, S.~Ferrara, A.~Marrani and M.~Trigiante,
 \emph{Two-Centered Magical Charge Orbits},
  arXiv:1101.3496 [hep-th].
\bibitem{specspec1}B. de Wit and A. Van Proeyen, Nucl. Phys. { B245}
(1984) 89;
%
\bibitem{gaizum} M.K: Gaillard, B. Zumino, Nucl. Phys. { B193} (1981)
221
%
\bibitem{helgason}
S. Helgason, \emph{Differential geometry, Lie groups, and symmetric spaces}, American Mathematical
Society, 2001.
%
\bibitem{sahaedio}
P. Fr\'e and A.S. Sorin, \emph{Integrability of Supergravity Billiards
and the generalized Toda lattice equations}, Nucl. Phys. {\bf B 733} (2006) 334
[arXiv:hep-th/0510156].
%
\bibitem{Fre':2007hd}
  P. Fr\'e and A.S. Sorin,
  \emph{The arrow of time and the Weyl group: all supergravity billiards are
  integrable,} Nucl. Phys. {\bf B 815} (2009) 430
  [arXiv:0710.1059].
%
\bibitem{noiultimo} P. Fr\'e, A.S. Sorin,
    \emph{ Supergravity Black Holes and Billiards and Liouville integrable structure of dual Borel algebras},
       JHEP 03 (2010) 066
       [arXiv:0903.2559].
%
\bibitem{Fre:2009dg}
  P.~Fr\'e and A.S.~Sorin,
  \emph{The Integration Algorithm for Nilpotent Orbits of $G/H^\star$ Lax systems: for
  Extremal Black Holes}
  [arXiv:0903.3771].
  %
\bibitem{CFS}
W. Chemissany, P. Fr\'e, A.S. Sorin, \emph{The Integration Algorithm
of Lax equation for both Generic Lax matrices and Generic Initial
Conditions}, Nucl. Phys. {\bf B 833} (2010) 220
[arXiv:0904.0801].
%
\bibitem{marioetal} W. Chemissany, J. Rosseel, M. Trigiante, T. Van Riet,
     \emph{The full integration of black hole solutions to symmetric supergravity theories},
 Nucl. Phys. {\bf B 830} (2010) 391 [arXiv:0903.2777].
%

\bibitem{bruxelles}S. Kim, J. Lindman H\"ornlund, J. Palmkvist and A.
Virmani,
\emph{Extremal solutions of the $S^3$ model and nilpotent $G_{2(2)}$ orbits}
[arXiv::1004.5242v2].

\bibitem{richardandme} Riccardo D'Auria and Pietro Fr\'e \emph{BPS Black-Holes in supergravity:Duality Groups, p-Branes, Central Charges and Entropy} in Classical and Quantum Black Holes, IOP Publishing Ltd 1999, P.Fr\'e, V. Gorini, G. Magli and U. Moschella Editors, pages 137-272.
\bibitem{itanteA} A. Ceresole, S. Ferrara, A. Marrani, \emph{Small N=2 Extremal Black Holes in Special Geometry}, [arXiv::1006.2007v1]
\bibitem{itanteB} A. Ceresole, G. Dall'Agata, S. Ferrara, A. Yeranyan, \emph{First order flows for N=2 extremal black holes and duality invariants}, [arXiv::0908.1110v2]
\bibitem{GekShap} M. Gekhtman, M. Shapiro, \emph{Non-commutative and commutative integrability of generic Toda
flows in simple Lie algebras}, [arXiv:solv-int/9704011].
\end{thebibliography}
\end{document}